\documentclass[rmp,aps,twocolumn,nofootinbib,floatfix]{revtex4}

\usepackage{graphicx} 
\usepackage{epsfig}
\usepackage{dcolumn}  
\usepackage{bm}       

\newcommand{\Dzkpi}{\Dz\to\Km\pip}
\newcommand{\Dzkpipiz}{\Dz\to\Km\pip\piz}
\newcommand{\Dzkpipipi}{\Dz\to\Km\pip\pip\pim}
\newcommand{\Dpkpipi}{\Dp\to\Km\pip\pip}
\newcommand{\Dpkpipipiz}{\Dp\to\Km\pip\pip\piz}
\newcommand{\Dpkspi}{\Dp\to\KS\,\pip}
\newcommand{\Dpkspipiz}{\Dp\to\KS\,\pip\piz}
\newcommand{\Dpkspipipi}{\Dp\to\KS\,\pip\pip\pim}
\newcommand{\Dpkkpi}{\Dp\to\Kp\Km\pip}

\newcommand{\Dzbarkpi}{\Dzbar\to\Kp\pim}
\newcommand{\Dzbarkpipiz}{\Dzbar\to\Kp\pim\piz}
\newcommand{\Dzbarkpipipi}{\Dzbar\to\Kp\pim\pim\pip}
\newcommand{\Dmkpipi}{\Dm\to\Kp\pim\pim}
\newcommand{\Dmkpipipiz}{\Dm\to\Kp\pim\pim\piz}
\newcommand{\Dmkspi}{\Dm\to\KS\,\pim}
\newcommand{\Dmkspipiz}{\Dm\to\KS\,\pim\piz}
\newcommand{\Dmkspipipi}{\Dm\to\KS\,\pim\pim\pip}
\newcommand{\Dmkkpi}{\Dm\to\Km\Kp\pim}

\newcommand{\epem}{\elp\elm}

\newcommand{\elp}{e^+}
\newcommand{\elm}{e^-}

\newcommand{\pip}{\pi^+}
\newcommand{\pim}{\pi^-}
\newcommand{\piz}{\pi^0}
\newcommand{\pipm}{\pi^\pm}

\newcommand{\Kp}{K^+}
\newcommand{\Km}{K^-}

\newcommand{\KS}{K^0_S}
\newcommand{\KL}{K^0_L}
\newcommand{\Kpm}{K^\pm}

\newcommand{\Dp}{D^+}
\newcommand{\Dm}{D^-}
\newcommand{\Dz}{D^0}
\newcommand{\Dbar}{\overline{D}}
\newcommand{\Dzbar}{\overline{D}{}^0}
\newcommand{\barD}{\overline{D}{}^0}

\newcommand{\Jpsi}{J/\psi}
\newcommand{\psiprime}{\psi(2S)}
\newcommand{\psidprime}{\psi(3770)}

\newcommand{\BDzkpivalue}{3.891 \pm 0.035 \pm 0.059 \pm 0.035}
\newcommand{\BDpkpipivalue}{9.15 \pm 0.10 \pm 0.16 \pm 0.07}

\newcommand{\Ecm}{E_\mathrm{cm}}

\newcommand{\Gev}{\mathrm{GeV}}
\newcommand{\Gevc}{\mathrm{GeV}/c}
\newcommand{\Gevcsq}{\mathrm{GeV}/c^2}
\newcommand{\Mevc}{\mathrm{MeV}/c}
\newcommand{\Mevcsq}{\mathrm{MeV}/c^2}

\newcommand{\DeltaE}{\Delta E}

\newcommand{\MD}{M_D}

\newcommand{\Mbc}{M_\mathrm{BC}}
\newcommand{\Mbcbar}{\overline{M}_\mathrm{BC}}

\newcommand{\pbar}{\bar{p}}
\newcommand{\sigmabar}{\bar{\sigma}}

\newcommand{\gE}{g_E}
\newcommand{\gp}{g_p}
\newcommand{\fee}{f_{\elp\elm}}
\newcommand{\fBW}{f_\mathrm{BW}}
\newcommand{\fpsiE}{f_{\psi}}
\newcommand{\uDq}{u_D}
\newcommand{\vDp}{v_D}
\newcommand{\wDM}{w_D}
\newcommand{\vDDbar}{v_{D\Dbar}}
\newcommand{\wDDbar}{w_{D\Dbar}}
\newcommand{\Mpsi}{M_\psi}
\newcommand{\Gammapsi}{\Gamma_\psi}
\newcommand{\Gammaz}{\Gamma_0}
\newcommand{\Gammap}{\Gamma_+}
\newcommand{\calBz}{\calB_0}
\newcommand{\calBp}{\calB_+}

\newcommand{\vecp}{\mathbf{p}}
\newcommand{\vecq}{\mathbf{q}}

\newcommand{\calB}{\mathcal{B}}

\newcommand{\pbinv}{pb$^{-1}$}
\newcommand{\fbinv}{fb$^{-1}$}

\newcommand{\eff}{\epsilon}

\newcommand{\NDzDzbar}{N_{\Dz\Dzbar}}
\newcommand{\NDpDm}{N_{\Dp\Dm}}

\newcommand{\ie}{\textit{i.e.}}
\newcommand{\vs}{\textit{vs.}}

\newcommand{\Mmisssq}{M^2_\mathrm{miss}}

\newcommand{\epsilonmc}{\epsilon_\mathrm{MC}}
\newcommand{\epsilondata}{\epsilon_\mathrm{data}}

\newcommand{\Begitem}{\begin{itemize}}
\newcommand{\Enditem}{\end{itemize}}

\newcommand{\Eqn}[1]{Eq.~(\ref{#1})}

\newcommand{\Tab}[1]{Table~\ref{#1}}

\newcommand{\Begeqn}{\begin{equation}}
\newcommand{\Endeqn}{  \end{equation}}
\newcommand{\beq}{\begin{equation}}
\newcommand{\eeq}{  \end{equation}}
\newcommand{\bea}{\begin{eqnarray}}
\newcommand{\eea}{  \end{eqnarray}}

\def\Dzero{D$\oslash$}

\def\D0bar{\overline D{}^0}
\def\K0bar{\overline K{}^0}

\def\3bar{\overline{3}}

\def\15bar{\overline{15}}
\def\24bar{\overline{24}}
\def\42bar{\overline{42}}
\def\60bar{\overline{60}}
\def\cO{{\cal O}}

\newlength{\Plotwidth}
\begin{document}

\title{Hadronic $D$ and $D_s$ Meson Decays}

\author{Anders Ryd}
\email{Anders.Ryd@cornell.edu}
\affiliation{Laboratory of Elementary-Particle Physics,
             Cornell University,
             Ithaca, NY 14853 USA}
\author{Alexey A Petrov}
\email{apetrov@wayne.edu}
\affiliation{Department of Physics and Astronomy, 
 Wayne State University,
 Detroit, MI 48201}
\affiliation{Michigan Center for Theoretical Physics, 
 University of Michigan,
 Ann Arbor, MI 48109}

\begin{abstract}
We provide a comprehensive review of hadronic decays of $D$ and $D_s$ mesons.
We discuss current theoretical and experimental challenges and successes in 
understanding of hadronic transitions of those mesons. A brief overview of the theoretical 
and experimental tools are given before discussing the absolute branching fractions for 
$D$ and $D_s$ mesons. Cabibbo suppressed and rare hadronic decays are discussed 
and compared with theory before discussing our understanding of hadronic multibody
decays.
\end{abstract}

\maketitle
\tableofcontents

\section{INTRODUCTION}
\label{sec:intro}

The discovery of charmed meson states in 1974 signaled a new era in particle physics. The arrival of
the first heavy quark has solidified the evidence that the Standard Model (SM) provides a correct 
low-energy description of
particle physics. Three decades later, the charm quark still plays an important role in studies of strong and weak 
interactions. It also serves as an important tool for exploring physics beyond the Standard Model, indirectly 
probing energy scales well above several TeV, which will be
directly probed by the Large Hadron Collider (LHC). In some cases, 
charm transitions provide possibilities for almost background-free studies of low-energy signals of new physics (NP). 
For example, signals of $CP$ violation in the charm system predicted within the Standard Model are very small, so any 
observation of $CP$ violation in the current round of experiments would rather unambiguously signal presence of 
new physics. Charm is also rather unique in that it is the only up-type quark that can have flavor oscillations.

A distinctive feature of all charmed hadrons is that their masses, ${\cal O}(2\mbox{~GeV})$, place them in the 
middle of the region where non-perturbative hadronic physics is operative. While this fact does not markedly 
affect theoretical description of leptonic and semileptonic decays of charmed hadrons, it poses significant 
challenges in the analyses of their hadronic transitions. There is a great deal of optimism, however, that abundant 
experimental data would provide some hints on the structure of charm hadronic decays, so those problems 
will eventually be overcome.

The data on charm transitions originate from several different types of experiments. Experiments at 
$e^+e^-$ machines operating at the $\psi(3770)$ and $\psi(4140)$ resonances, such as CLEO-c and BES III, 
have several important advantages. First, the final state is extremely simple, being essentially just a 
$D\bar{D}$ pair. Second, the cross-section for charm production is relatively high, 
$\sigma (D^0\bar{D}^0)=3.66 \pm 0.03\pm 0.06$ nb and 
$\sigma (D^+D^-)=2.91 \pm 0.03\pm 0.05$ nb. In conjunction with low multiplicity of the final state, 
this allows for measurements of absolute branching fractions for several reference modes. Finally, 
in those experiments, the $D\bar{D}$ pairs are produced in a quantum-coherent state, which allows for 
unique probes of the structure of decay amplitudes and phases, as well as novel measurements of mixing 
and $CP$ violation. 

The $B$ factory $e^+e^-$ experiments BABAR and Belle, operating at the $\Upsilon\mbox{(4S)}$ center-of-mass energy,
produce significant amount of charm data. In fact, at the resonance center-of-mass energy, $\sigma(b\bar{b})\sim
1.1$ nb, while $\sigma(c\bar{c})\sim 1.3$ nb. The very large integrated 
luminosities of these experiments have produced 
large samples of reconstructed charm. The higher operating energy makes 
possible the production
of charmed baryons.

Experiments at hadron machines, such as CDF and \Dzero,
and fixed targed facilities are plagued by even higher backgrounds. However,
much higher production cross-section, combined with a relatively long lifetime of charmed hadrons, provides a
possibility to trigger on charm decay events with displaced vertices. This technique allowed for hadron machines 
to be major players in charm physics. New results from the Large Hadron Collider (LHC) experiments
LHCb, ATLAS, and CMS will continue to supply us with new data. 

This paper provides a comprehensive review of hadronic decays of $D$ and $D_s$ mesons.
In this review we adopt the averages performed  by the Particle Data Group~\cite{Amsler:2008zzb}. 
Only if there are newer measurements that are not included in the review by the Particle Data Group we will 
do our own averaging.

This review is organized as follows.
Section~\ref{sect:opencharm} contains a brief discussion of the discovery of open charm followed in
Section~\ref{sect:exp} by a discussion of the experimental techniques used for studying 
charm decays. This includes a brief discussion of the main experiments that have contributed to
our understanding of $D$ decays and the production mechanisms employed 
in these studies. Final state radiation is discussed in this section as it is an important 
effect in many of the precision measurements discussed in this review.
In Section~\ref{sect:theory} the theoretical description of hadronic $D$ decays is provided. This
includes discussion of SU(3)$_{\rm F}$ flavor symmetry, the flavor-flow-diagram approach, and factorization.
These are common tools used to analyze and interpret hadronic $D$ decay data.  
Sections~\ref{sect:dref} and~\ref{sect:dsref} discuss the determination of the absolute
branching fractions for for $D$ and $D_s$ decays. Rare and suppressed
modes are discussed in Section~\ref{sect:cabibbo}. Multibody decays and Dalitz plot studies are
discussed in Section~\ref{sect:multi}. This review concludes in Section~\ref{sect:conclusions} with a 
summary and outlook.

\section{DISCOVERY OF OPEN CHARM}
\label{sect:opencharm}

The arrival of the quark model in 1964~\cite{GellMann:1964nj,Zweig:1964jf} 
greatly simplified the description 
of elementary particles. The idea that all observed particles 
are made of the three quarks, $u$, $d$, and $s$, was gaining acceptance. 
By the early 1970's, the proton structure was probed and 
the quarks were found to
be real particles. Further development of perturbative Quantum 
Chromodynamics and the concept of asymptotic freedom allowed 
consistent explanation of those experiments in terms of those
three quark flavors. The possible existence of a fourth 
quark had been theoretically discussed in the 60's~\cite{Bjorken:1964gz}, 
however it was not required. 

Hints of the incompleteness of the current picture 
came after experimental observation of rare, 
electroweak, decays of kaons. The observed rate for 
$K^0_L\to \mu^+\mu^-$ turned out to be smaller than predicted. 
Similarly, the $K^0_S$--$K^0_L$ mass difference did not agree with 
predictions based on only having the $u$, $d$, and $s$ quarks. 
To solve those problems, Glashow, Iliopoulos, and Maiani (GIM) proposed 
an elegant mechanism~\cite{Glashow:1970gm}, which involved adding 
the forth quark, $c$. The resulting mechanism not only established the
absence of the tree-level flavor-changing neutral currents in the Standard Model,
but also provided for reduced rates for $K^0_L\to \mu^+\mu^-$ decays by requiring 
cancelations with additional diagrams involving intermediate charm quarks. Using the 
observed rate for $K^0_L\to \mu^+\mu^-$ and $K^0_S$--$K^0_L$ mass difference, 
it was estimated that the charm 
quark would have a mass in the range 1 to 3 GeV~\cite{Gaillard:1974hs,Gaillard:1974mw}. 
The existence of the new quark implied that it would form bound states 
with its own anti-quark, as well as with the lighter quarks,
which could be observable experimentally.

\begin{figure}[tb]
\begin{center}
\includegraphics[width=0.95\linewidth]{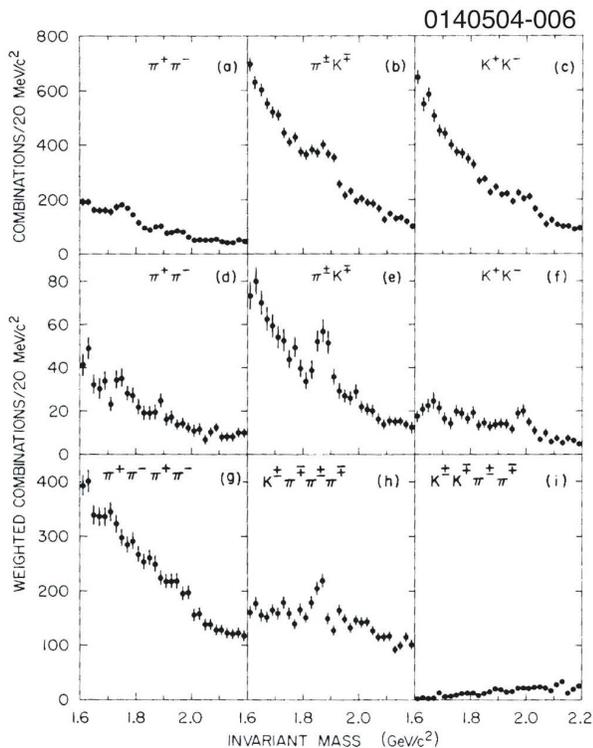}
\caption{The invariant mass distributions observed by the 
SLAC-LBL experiment for two and four hadrons in the final state.
(a) $\pi^+\pi^-$ assigning pion mass to all tracks, 
(b) $K^{\mp}\pi^{\pm}$ assigning kaon and pion masses to all tracks, 
(c) $K^+K^-$ assigning kaon mass to all tracks, 
(d) $\pi^+\pi^-$ weighted by $\pi\pi$ time of flight probability, 
(e) $K^{\mp}\pi^{\pm}$ weighted by $K\pi$ time of flight, 
(f) $K^+K^-$ weighted by $KK$ time of flight, 
(g) $\pi^+\pi^-\pi^+\pi^-$ weighted by $4\pi$ time of flight probability
(h) $K^{\pm}\pi^{\mp}\pi^+\pi^-$ weighted by $K3\pi$ time of flight probability
(i) $K^{\pm}K^{\mp}\pi^+\pi^-$ weighted by $KK\pi\pi$ time of flight 
probability.
From~\textcite{Goldhaber:1976xn}.
}
\label{fig:SLAC_LBL_Dz_observation}
\end{center}
\end{figure}

These bound states were experimentally discovered in November 1974 
by two independent research 
groups at SLAC~\cite{Aubert:1974js} and BNL~\cite{Augustin:1974xw}. The 
mass of the observed $J/\psi$ 
resonance of about 3.1 GeV was in the range where a $c\bar c$ bound 
state was expected. 
In addition, the very small width, of about 93 keV, was very different 
from other high mass resonances observed. 
The interpretation of the $J/\psi$ as a $c\bar c$ bound state was 
confirmed when "open charm" states were discovered 
a little later, first  the $\Dz$~\cite{Goldhaber:1976xn} and then 
the $\Dp$~\cite{Peruzzi:1976sv}. The first observation of 
the $D^0$ was made in the final states $K^-\pi^+$ and $K^-\pi^+\pi^-\pi^+$. 
The observed invariant mass 
distributions are shown in Fig.~\ref{fig:SLAC_LBL_Dz_observation}.

After the observation of the $D^0$ and $D^+$ mesons it took a 
little longer to establish the $D^+_s$. There
were several candidates observed before the $D^+_s$, originally 
called the $F$ meson, was observed
by CLEO~\cite{Chen:1983kr}.

It is also interesting to note that there were hints of 
the existence 
of open charm states in photoemulsion experiments 
even before the $J/\psi$ had been discovered~\cite{Niu:1971xu,Hoshino:1975fc}.

\section{GENERAL REMARKS ON EXPERIMENTAL FACILITIES AND TECHNIQUES}
\label{sect:exp}

Charm has been studied in a large number of different experiments.
In $\epem$ collisions charm decays have been studied from threshold
to the $Z$ pole. There has also been a number of fixed target 
experiment, either using hadroproduction or photoproduction.
The $e^+e^-$ and fixed target experiments dominate the literature
on charm meson decays. In addition, there are 
also studies using proton--anti-proton collisions.

In this section we review some of the basic properties
of the different types of production mechanisms and the
experiments used to collect the data. First, $e^+e^-$
experiments are discussed and then fixed target. For
$e^+e^-$ experiments, where typically triggering is very 
open and most of the produced events are recorded, we compare
the luminosity and the produced number of $c\bar c$ events.
A summary of $e^+e^-$ experiments is given in 
Table~\ref{tab:charmepem}.
For fixed target experiments a similar comparison 
is made in Table~\ref{tab:charmreco} for
the number of exclusively reconstructed $D$ mesons.
At threshold the final state
charm mesons are produced without any additional hadrons.
The CLEO-c experiment is
described in some more detail as it is the experiment
operating near threshold with the largest data samples
to date. 
At higher $\epem$ center-of-mass energy the charm hadrons 
are produced either in fragmentation or in decays of heavier
particles such as hadrons containing $b$-quarks.
Last, fixed target experiments are reviewed. 
Fixed target
experiments can be categorized as photoproduction
or hadroproduction experiments based on the particle type
incident on the target.

\begin{table*}[btp]
\caption{
Summary of charm samples produced in $e^+e^-$ colliding 
beam experiments.
}
\label{tab:charmepem}
\begin{center}
\begin{tabular}{lcccc}
\hline\hline
Experiment & Year & $\sqrt{s}$ & $\int{\cal L}$ & Produced Charm  \\
\hline
Mark III   & 1982-1988 &  3.77 GeV &   9 \pbinv\  &   28,000 $D^0\bar D^0$   \\
           &      &           &      &   20,000 $D^+D^-$   \\
           &      &  4.14 GeV &   6.3 \pbinv\   &      \\
BES        &      &  4.04 GeV &  9 \pbinv   & 6,000 $D_s^+D_s^-$    \\
BES II     &      &           &      & \\
CLEO-c     & 2003-2008     &  3.77 GeV &   818 \pbinv  &   $3.0\times10^6$ $D^0\bar D^0$ \\
           &      &           &               &   $2.4\times10^6$ $D^+D^-$   \\
           &      &  4.17 GeV &   589 \pbinv  &     $0.58\times10^6$ $D^{\pm}_sD^{*\mp}_s$ \\
\hline
CLEO       & 1979-1988 &  $\approx 10.5$ GeV   & 314 \pbinv & $0.41 \times 10^6\ c\bar c$   \\
CLEO II    & 1989-1994 &  $\approx 10.5$ GeV   & 4.7 \fbinv & $6.1 \times 10^6\ c\bar c$     \\
CLEO II.V  & 1995-1999 &  $\approx 10.5$ GeV   & 9.1 \fbinv & $12 \times 10^6\ c\bar c$ \\
CLEO III   & 2000-2003 &  $\approx 10.5$ GeV   & 15 \fbinv  & $19 \times 10^6\ c\bar c$ \\
ARGUS      & 1982-1992 &  $\approx 10.5$ GeV  & 514 \pbinv   & $0.67\times 10^6\ c\bar c$ \\
BABAR      & 1999-2008     &  $\approx 10.5$ GeV    &  $531$ \fbinv  & $0.69 \times 10^9\ c\bar c$  \\
Belle\footnote{As of Sept. 1, 2009}      & 1999-     &  $\approx 10.5$ GeV         &  $>700$ \fbinv  & $0.91 \times 10^9\ c\bar c$  \\
\hline
HRS        & 1982-1986   &  29 GeV   &  300 \pbinv   & 52,000 $c\bar c$    \\       
LEP        &  1989-1996    &  $\approx 91$ GeV   &      $4.2\times 10^6\ Z's$ & 220,000 $c\bar c$\\
           &      &           &   per experiment   &  per experiment \\ 
\hline\hline
\end{tabular}
\end{center}
\end{table*}

\begin{table}[bt]
\small{
\caption{
The number of reconstructed charm mesons for different fixed target
experiments.
}
\label{tab:charmreco}
\begin{center}
\begin{tabular}{lccc}
\hline\hline
Experiment & Year    & Events            &     Reconstructed   \\
           &         & Recorded/$10^6$   & Charm Decays  \\
\hline
Photoproduction:     & & &\\
E691       & 1985    & 100      &  10,000\\
E687       & 1992    & 500      &  100,000\\
FOCUS (E831)&1996    & 7,000    &  $1.2\times 10^6$\\
Hadroproduction:     & & &\\
WA75       & 1984    & 2        &  350\\
NA32       & 1986    & 17       &  1,300\\
WA82       & 1989    & 10       &  3,000\\
E653       & 1988    & 10       &  1,000\\
E769       & 1988    & 500      &  4,000\\
E791       & 1992    & 20,000   &  200,000\\
\hline\hline
\end{tabular}
\end{center}
}
\end{table}

\subsection{Experiments using $\epem$ annihilation near threshold}
\label{sect:exp_threshold}

At threshold $D$ meson pairs are produced without any
additional hadrons. 
This provides the experiments operating at threshold 
with a very clean environment for studying charm decays. 
As will be discussed in 
Section~\ref{sect:lineshape} the initial electron or positron
may radiate low energy photons, initial state radiation (ISR), 
such that the total
energy of the produced charm hadrons is less than the
center-of-mass energy in the $e^+e^-$ initial state.

Experiments that studied charm decays at threshold include
the Mark I, II, and III experiments~\cite{Augustin:1974xq, 
Abrams:1979xk, Bernstein:1983wk}
at SPEAR; BES I, BES II, 
BES III~\cite{Bai:1994zm, Bai:2001dw, BES:2009zza} at
BEPC, and CLEO-c~\cite{Kubota:1991ww,Peterson:2002sk,
Artuso:2002ya} 
at CESR-c.
For studies of $D^0$ and $D^+$
decays experiments have run at the $\psidprime$. 
The total hadronic cross-section at the $\psidprime$ 
resonance has been measured by CLEO-c~\cite{Besson:2005hm}
$$
\sigma(e^+e^- \to {\rm hadrons})=(6.38\pm 0.08^{+0.41}_{-0.30})\ {\rm nb}.
$$
The cross-sections for $D^0\bar D^0$ and
$D^+D^-$ production has been measured by CLEO-c~\cite{Dobbs:2007zt}
\begin{eqnarray*}
\sigma(e^+e^- \to   D^0\bar D^0 ) & = &  (3.66\pm 0.03\pm 0.06)\ {\rm nb},\\
\sigma(e^+e^- \to   D^+ D^- ) & = & (2.91\pm 0.03\pm 0.05)\ {\rm nb}.
\end{eqnarray*}
The total cross-section for $D\bar D$ production at the $\psidprime$
is $\sigma(e^+e^- \to   D\bar D )=(6.57\pm0.04\pm0.10)\ {\rm nb}$.
This is larger than, but consistent with, the inclusive hadronic 
cross-section discussed above. These results indicates that 
the majority of the $\psidprime$ decays to $D\bar D$.
 CLEO-c~\cite{Adam:2005mr} and BES~\cite{Bai:2003hv} have
observed some non-$D\bar D$ decays of the $\psidprime$. 
The largest of these decays is the radiative transition
$\psidprime\to \gamma\chi_{c0}$ with a branching fraction 
of $(0.73\pm0.09)\%$. Summing the observed branching 
fractions for non-$D\bar D$ decays we 
obtain $1.4\pm 0.1\%$, consistent with the cross-section measurements
above.
BES~\cite{Ablikim:2006aj,Ablikim:2006zq,Ablikim:2007zz,Ablikim:2008zzb} 
has performed
direct measurements of the cross-section
for $\psi(3770)\to {\rm non-}D\barD$ final states as well
as measurements of the $D\bar D$ cross-sections. The 
PDG~\cite{Amsler:2008zzb} average these measurements and finds that
$(14.7\pm3.2)\%$ of $\psi(3770)$ resonances decays to
non-$D\bar D$ final states. This result is inconsistent with the 
CLEO-c results at the $2\sigma$ level.

\begin{figure}[bt]
\begin{center}
\includegraphics[width=0.7\linewidth]{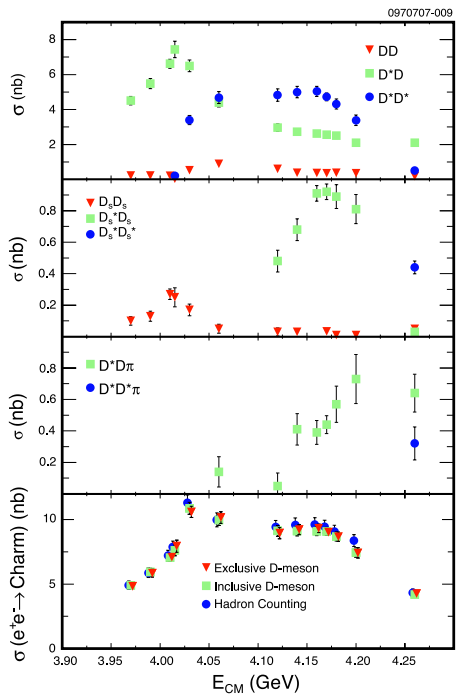}
\caption{The measured cross-sections for different $D\bar D$ final states.
From~\textcite{CroninHennessy:2008yi}.
}
\label{fig:cleoc_D_crosssections}
\end{center}
\end{figure}

For studies of $D_s$ mesons different $e^+e^-$ 
center-of-mass energies have been used.
The cross-sections for producing $D_{(s)}$, or $D^*_{(s)}$ mesons,
as measured by CLEO-c~\cite{CroninHennessy:2008yi}, are
shown in Fig.~\ref{fig:cleoc_D_crosssections}.
BES collected data at 4.03 $\Gev$. At this energy 
$D_s^+D_s^-$ mesons pairs are produced. CLEO-c on the other 
hand ran at a higher energy, about 4.17 $\Gev$. At this 
energy pairs of $D_s^{\pm}D_s^{*\mp}$ mesons are produced. The $D_s^*$
meson decays to either $D_s\gamma$ or $D_s\pi^0$, with
branching fractions of $(94.2\pm 0.7)\%$ and $(5.8\pm 0.7)\%$,
respectively~\cite{Aubert:2005ik,Amsler:2008zzb}. The
advantage of the higher energy is the larger cross-section.
CLEO-c reports~\cite{CroninHennessy:2008yi} a cross-section of 
$(0.27\pm0.03)$ nb at 4.03 $\Gev$ for $D^+_sD^-_s$ production
and $(0.92\pm0.05)$ nb at 4.17 $\Gev$ for $D^{*\pm}_sD^{\mp}_s$
production.
For most analyses the larger cross-section outweighs the 
complication of the additional particles in the final state.

\subsubsection{Quantum coherence}
\label{sect:coherence}

Threshold production of $D\overline{D}$ pairs can be explored to 
understand the phase structure of hadronic decay amplitudes of $D^0$ mesons.
Here one can use the fact that neutral charm mesons $D^0$ and $\Dzbar$ mix. 
$D^0-\Dzbar$ mixing arises from electroweak or New Physics $|\Delta C|=2$ interactions 
that generate off-diagonal terms in the neutral $D$ mass matrix (see, 
e.g.~\cite{Artuso:2008vf,Bergmann:2000id} for more information) 
\beq\label{MixingMatrix}
\left[{\bf M} - i \frac{{\bf \Gamma}}{2} \right] = 
\left(
\begin{array}{cc}
A & p^2 \\
q^2 & A 
\end{array} 
\right),
\eeq
where $A$ parameterizes masses and lifetimes of $D^0$ and $\Dzbar$ states and
the complex parameters $p^2$ and $q^2$ parameterize contributions from 
$|\Delta C|=2$ interactions. The non-diagonal structure of the mixing matrix of 
Eq.~(\ref{MixingMatrix}) leads to the (physical) mass eigenstates of a Hamiltonian of 
Eq.~(\ref{MixingMatrix}) $D_1$ and $D_2$ becoming superpositions of the flavor 
eigenstates $D^0$ and $\Dzbar$,
\begin{equation} \label{definition1}
| D_{{\rm 1}\atop{\rm 2}}  \rangle  = p\,  | D^0  \rangle  ~ \pm ~ q\, | \Dzbar  \rangle  \ ,
\end{equation}
where $|p|^2 + |q|^2=1$.  A simplified assumption can be made that
in the studies of strong phases described below $CP$ violation may be 
neglected. This could be justified in the Standard Model by noting that 
$CP$-violating contributions are always suppressed by small values of 
the third-generation Cabibbo-Kobayashi-Maskawa (CKM) matrix elements. 
In general, smallness of $CP$-violating contributions in charm transitions can be 
deduced from tight experimental constraints on $CP$-violating 
asymmetries~\cite{Artuso:2008vf}. In such case $p=q$, so mass eigenstates 
also become eigenstates of $CP$, 
\begin{eqnarray} \label{defCP}
| D_{\pm} \rangle =
\frac{1}{\sqrt{2}} \left[
| D^0 \rangle \pm | \bar D^0 \rangle \right].
\end{eqnarray}
It follows then that these $CP$ eigenstates $| D_{\pm} \rangle$ do
not evolve with time. Their mass and lifetime differences can be observed,
\beq
x=\frac{\Delta M_D}{\Gamma},\quad y= \frac{\Delta \Gamma_D}{2\Gamma},
\eeq
where $\Gamma = \left(\Gamma_++\Gamma_-\right)/2$ is the average lifetime of 
mass and $CP$ eigenstates.

At threshold $e^+e^-$ experiments, such as BES and CLEO-c, 
$D^0\Dzbar$ pairs are produced through resonances of specific charge conjugation.  
The $D^0\Dzbar$ will therefore be in an entangled state with the same quantum numbers as the
parent resonance. In particular, since both mesons are pseudoscalars, charge
conjugation reads $C=(-1)^L$, if the produced resonance has angular 
momentum $L$. This implies that the quantum mechanical state at the time of
$D^0\Dzbar$ production is
\begin{eqnarray}
\Psi 
 =  \frac{1}{\sqrt{2}}
\left \{
| D^0 ({\bf k_1})\barD ({\bf k_2}) \rangle +
C | D^0 ({\bf k_2})\barD ({\bf k_1}) \rangle
\right \}.
\label{entangle1}
\end{eqnarray}
where ${\bf k_1}$ and ${\bf k_2}$ are the momenta of the mesons. Rewriting this in terms of the 
$CP$ basis we arrive at
\begin{eqnarray} 
\Psi_{C=+1} &=& 
\frac{1}{\sqrt{2}} \left \{ | D_+ ({\bf k_1}) D_+ ({\bf k_2}) \rangle - | D_- ({\bf k_1})
D_- ({\bf k_2}) \rangle \right \}
\nonumber\\
\Psi_{C=-1} &=& 
\frac{1}{\sqrt{2}} \left \{ | D_- ({\bf k_1}) D_+ ({\bf k_2}) \rangle + | D_+ ({\bf k_1})
D_- ({\bf k_2}) \rangle \right \}
\nonumber\\
\label{entangle2}
\end{eqnarray}
Thus in the $L=$ odd; $C=-1$ case, which would apply to the
experimentally important $\psi(3770)$ resonance, the
$CP$ eigenstates of the $D$ mesons are anti-correlated while if $L=$ even;
$C=+1$ the eigenstates are correlated. This can happen when $D^0\Dzbar$
pair is produced in the decays $\psi(4140) \to D {\overline D} \gamma$
of the more massive charmonium state $\psi(4140)$. In either case the $CP$ 
conservation implies that correlation between the eigenstates is independent 
of when they decay. In this way, if $D(k_1)$ decays to the final state which is 
also a $CP$-eigenstate, then the $CP$ eigenvalue of the meson $D(k_2)$ is therefore  
{\it determined}: it is either the same as $D(k_1)$ for $C=+1$ or opposite, as  
in the case of $C=-1$.
The use of this eigenstate correlation as a tool to investigate $CP$ violation
has been suggested in $B$-physics~\cite{Falk:2000ga}. In charm physics this
method of {\it $CP$-tagging} can be used to study relative strong phases of
$D^0$-meson amplitudes. Such measurements are needed for studies of
$D^0\Dzbar$-mixing.

To illustrate the method, the amplitude for the $CP$-tagged eigenstate decaying to, say,
$K\pi$ final state can be written as
\begin{equation}
\sqrt{2} {\cal A} (D_\pm \to K^-\pi^+) = {\cal A} (D^0 \to K^-\pi^+) \pm {\cal A} (\Dzbar \to K^-\pi^+)
\end{equation}
which follows from Eq.~(\ref{defCP}). This relation implies that 
\begin{equation}\label{CosDelta}
1 \pm 2 \cos\delta \sqrt{R} = 2 \frac{{\cal B} (D_\pm \to K^-\pi^+)}{{\cal B}(D^0 \to K^-\pi^+)},
\end{equation}
where $R$ is a small ratio of doubly-Cabibbo suppressed (DCS) decay rate to 
Cabibbo favored (CF) one (see Section~\ref{sect:theory}), and $\delta$ is the 
strong phase difference between those amplitudes,  
${\cal A} (\Dzbar \to K^-\pi^+)/{\cal A} (D^0 \to K^-\pi^+) =-\sqrt{R} e^{-i \delta}$.
Eq.~(\ref{CosDelta}) can be used to extract $\delta$ if the
$CP$-tagged branching ratio is measured~\cite{Gronau:2001nr,Atwood:2002ak}.

The method of quantum correlations can be used to study the multitude of parameters of 
$D^0$ decay and mixing~\cite{Atwood:2002ak,Asner:2005wf}. 
\begin{table}[bt]
\caption{
Correlated branching ratios for various processes. Correlated results are presented for
$C=1$ and normalized to the multiple of the uncorrelated branching fractions. CP-violation is 
neglected.}
\label{tab:correlated_br}
\begin{center}
\begin{tabular}{lc}
\hline\hline
Decay modes   &    Correlated  branching fractions  \\
\hline
$K^- \pi^+$ vs. $K^- \pi^+$      &  $R_m$    \\
$K^- \pi^+$ vs.  $K^+ \pi^-$      &  $(1+R_{\mbox{ws}})^2-4r_\delta(r_\delta + y)$  \\ 
$K^- \pi^+$ vs.  $S_\pm$          &  $1+R_{\mbox{ws}} \pm 2r_\delta \pm y$  \\
$K^- \pi^+$ vs.  $L^\pm$                &  $1-\sqrt{R}(y\cos\delta+x\sin\delta)$   \\
$S_\pm$ vs.  $S_\pm$              &  $0$  \\
$S_\pm$ vs.  $S_\mp$              &  $4$  \\
$S_\pm$ vs.  $L^\pm$                    &  $1 \pm y$  \\
\hline\hline
\end{tabular}
\end{center}
\end{table}
In particular, correlated decays of $D$-mesons into $CP$-mixed final states (such as  $K^- \pi^+$), 
CP-specific final states $S_\pm$ (such as $S_+=K^+K^-$ or $S_-=K_S\pi^0$), or a flavor specific 
semi-leptonic decay $L^\pm$ into a state containing $\ell^\pm$ can probe various combinations of 
mixing and decay parameters (see Table~\ref{tab:correlated_br}). Note that $R_m=(x^2+y^2)/2$ and 
$R_{\mbox{ws}}=R+ \sqrt{R} \left(y \cos\delta - x \sin\delta\right) + R_m$, and $r_\delta = \sqrt{R} \cos\delta$.
The quantum-correlated rates are clearly different from the singly-tagged (ST) rates, i.e. when only one of the 
$D^0$s is reconstructed. For example, the ST rate for the wrong-sign (e.g. $D^0 \to K^+ \pi^-$) decay is 
given by $R_{\mbox{ws}}$.

Besides the discussed studies of the phases of hadronic decay amplitudes, the results summarized in 
Table~\ref{tab:correlated_br} can be used to extract $D^0-{\overline D}^0$-mixing parameters. The current
status of charm mixing will be discussed elsewhere (see also \cite{Artuso:2008vf}).

\subsubsection{Experiments at threshold}
\label{sect:thresholdexp}

\begin{figure}[bt]
\begin{center}
\includegraphics[width=0.99\linewidth]{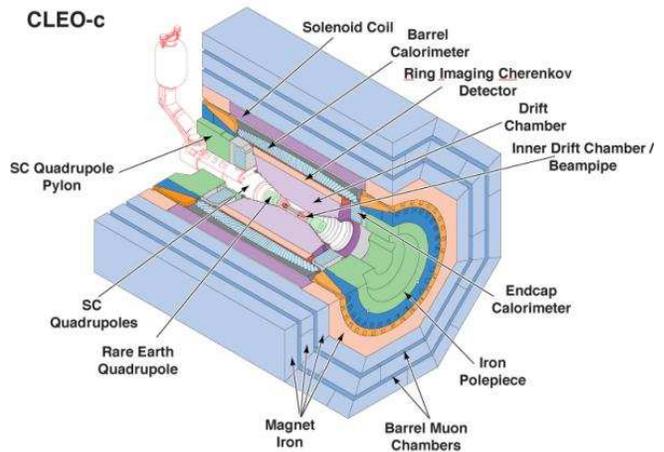}
\caption{The CLEO-c detector. The charged particle tracking
system consists of an inner drift chamber near the interaction
point and the main drift chamber for the momentum 
measurement. Radially outside the main drift chamber is the
CLEO-c RICH detector for charged hadron identification
followed by the CsI electromagnetic calorimeter. The
instrumented flux return for muon detectors is outside
the super conducting solenoid coil.}
\label{fig:cleocdetector}
\end{center}
\end{figure}

The CLEO-c experiment plays a unique role here as it has a 
very large data sample collected at threshold. The CLEO-c
detector is an evolution of the CLEO III detector where
the silicon-strip vertex detector has been replaced with
a low-mass inner six-layer drift chamber~\cite{Kubota:1991ww, 
Peterson:2002sk, Artuso:2002ya}. 
The CLEO-c experiment is shown schematically in 
Fig.~\ref{fig:cleocdetector}. The wires in the 
inner drift chamber are at a small stereo angle with respect
to the drift chamber axis. This allows determining the 
$z$ position of charged particles. The charged particle
tracking system in CLEO-c also includes the
47-layer main drift chamber, operating
in a 1.0 T magnetic field along the drift chamber axis.
The CLEO-c tracking system provides a momentum resolution
of about 0.6\% for tracks with a momentum of $1$ GeV that traverses all
layers of the drift chamber. CLEO-c has excellent electromagnetic
calorimetry from the approximately 7800 CsI(Tl) crystal
calorimeter. For energies of 1 GeV the calorimeter
has an energy resolution of about 2\%. For energies of 
100 MeV the resolution is about 5\%. 
The excellent energy resolution and coverage allow CLEO-c
to efficiently reconstruct $\pi^0$ and $\eta$ mesons
in the $\gamma\gamma$ final state. The $\pi^0$
mass resolution obtained is about 6 MeV. Charged hadrons are identified 
by a combination of specific ionization, $dE/dx$, in
the drift chamber for particles with momenta below about 700 MeV. 
For higher momenta, where $dE/dx$ is less powerful, CLEO-c
uses the RICH detector to separate kaons from pions.

The BESIII~\cite{BES:2009zza} detector constitutes a
substantial upgrade of the earlier BES II detector. Among
the new features are a 1 T magnetic field generate by a
superconducting coil, a new drift chamber, and a CsI(Tl)
dopped electromagnetic calorimeter. The time-of-flight 
system provides $\pi$-$K$ separation at 0.9 GeV with a
$2\sigma$ separation. The operation of the BES experiment
has just started with a first run at the $\psi(2S)$.

\subsubsection{Experimental features at threshold}
\label{sect:lineshape}

\begin{figure}[bt]
\begin{center}
\includegraphics[width=0.79\linewidth]{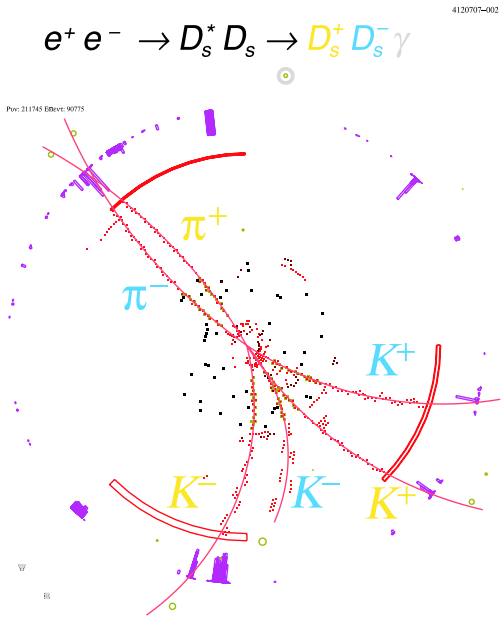}
\caption{Event display from CLEO-c showing a candidate $D^{*+}_sD^-_s$ event
with $D^{*+}_s\to D^{+}_s\gamma$ and both  $D^+_s$ candidates
decaying to $\phi\pi^+$. }
\label{fig:cleoceventdisplay}
\end{center}
\end{figure}

At threshold $D$ mesons are produced in pairs. A very powerful
analysis technique involves reconstructing one $D$ meson 
exclusively. This allows us to infer the existence of another
$\bar D$ mesons in the event. This 'tagging' technique,
or 'double tag' technique, was first used
by MARK III~\cite{Baltrusaitis:1985iw, Adler:1987as}, 
but due to their relatively
small sample of tags the technique was of limited use. With
much larger samples, and a more modern detector, the CLEO-c 
experiment has made great use of this tagging technique. 
The event environment at threshold is very clean. The $D\bar D$
signal is produced with no additional hadrons.
An example from CLEO-c of a fully reconstructed $D^{*\pm}_sD^{\mp}_s$
is shown in
Fig.~\ref{fig:cleoceventdisplay}.

Many analyses make use of fully reconstructed $D$ candidates. The
$D$ candidates are built from charged kaons and pions, neutral
pions, $\eta$ and $K^0_S$ mesons. 
CLEO-c typically
require that kaon and pion candidates are consistent with
charged hadron particle identification based on energy loss
in the drift chamber and Cherenkov radiation in the RICH detector.
The $K^0_S$ candidates are
reconstructed in the $\pi^+\pi^-$ final state. 
For the $\pi^+\pi^-$ pairs used to form
$K^0_S$ candidates the usual usual track quality criteria are 
relaxed and no particle identification criteria are
applied.

To extract the signal in fully reconstructed hadronic 
$D$ decays it is typically required that the reconstructed
$D$ candidate energy is consistent with the beam energy, as
each $D$ in the final state will carry half of the 
center-of-mass energy. Specifically, 
$$
\Delta E\equiv E_{\rm cand}-
E_{\rm beam},
$$
where 
$$
E_{\rm cand}=\sum_i{\sqrt{{\bf p}_i^2+m_i^2}}
$$
is the energy of the $D$ candidate. 
For correctly reconstructed $D$ candidates the $\Delta E$ distribution
peak at zero.
The resolution on $\Delta E$ is
mode dependent and the actual criteria applied varies between
different analyses depending on the backgrounds and cleanliness
of the signal that is desired. 

After applying a mode dependent 
$\Delta E$ selection criteria the beam constrained mass is formed
$$
M_{\rm BC}\equiv \sqrt{E^2_{\rm beam}-(\sum_i {\bf p}_i)^2}.
$$
Here the candidate energy has been replaced by the beam energy 
which typically is much better known. 

\begin{figure}[bt]
\begin{center}
\includegraphics[width=0.99\linewidth]{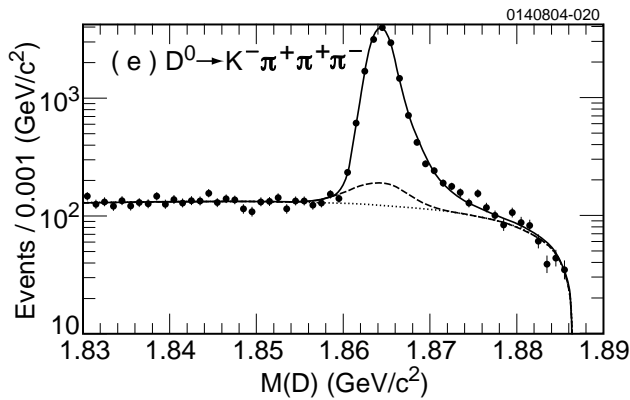}
\caption{The $M_{\rm BC}$ distribution. The dotted line shows
the contribution from the ARGUS function that describes the combinatorial
background. The solid and the dashed lines show the contributions
to the signal shape for two different detector resolution functions.
The tail on the high side for the signal shape is due to initial 
state radiation that lowers the energy of the produced $D$ mesons.}
\label{fig:cleoc_mbc}
\end{center}
\end{figure}

A typical plot of the $\Mbc$ distribution is shown in 
Fig.~\ref{fig:cleoc_mbc}.
The signal yield is 
determined by fitting the $M_{\rm BC}$ distribution to a
background shape plus a signal shape. 
The background shape
is due to combinatorial backgrounds either from other $D$
decays or from continuum. The background is typically fit
using an 'ARGUS' function~\cite{Albrecht:1990am}
\begin{equation}
a(M_{\rm BC};m_0,\xi,\rho)=A\, M_{\rm BC} \left(1-{M^2_{\rm BC} 
\over m_0^2}\right)^\rho e^{\xi\left(1-{M^2_{\rm BC} \over m_0^2}\right)}. \label{eq:argusf}
\end{equation}
This function describes the phase space distribution expected near
threshold for $\rho=1/2$ and $\xi=0$. By allowing
$\rho$ and $\xi$ to take on different values a more general function
which can describe the data better is obtained.

For the signal shape CLEO-c has used several different 
parameterizations. The most detailed description is that used
for example in~\textcite{Dobbs:2007zt}. This form incorporates the effects of
detector resolution, beam energy distribution, initial state radiation,
and the line shape of the $\psidprime$. 
The beam energy distribution, initial state radiation, and 
the $\psi(3770)$ lineshape control the energy of the
produced $D$-mesons. The effect of ISR is to produce the 
$\psi(3770)$ with an energy below
the nominal $e^+e^-$ center-of-mass energy. This produces a tail
on the high side of the $M_{\rm BC}$ distribution as seen in 
Fig.~\ref{fig:cleoc_mbc}.
The detector resolution
effects lead to a smearing of the measured momentum. 
Following~\textcite{Dobbs:2007zt},
a brief description of the lineshape is given below.

The distribution\footnote{For simplicity, these distribution functions 
are not normalized to 1. The RooFit~\cite{RooFit} fitting package 
used used by CLEO-c takes care of the overall normalization of the distribution functions 
used in fits.} of the energy of ISR photons is taken to be~\cite{kuraev-fadin}
\Begeqn
h(E_{\gamma}) = E_{\gamma}^{\beta-1},
\Endeqn
where
\Begeqn
\beta\equiv{2\alpha\over \pi}\left[2\ln\left({\Ecm\over m_e}\right)-1\right].
\Endeqn
At the $\psidprime$ resonance, $\beta\approx 0.078$.
The energy distribution $\fee(E)$ of the $\elp$ and $\elm$ when they 
collide is obtained from an integration of the beam energy spread and 
the ISR photon energy distribution,
\Begeqn
\fee(E)= \int_0^{\infty} h(E_{\gamma})\gE(E+E_{\gamma})\, dE_{\gamma}.
\Endeqn
CLEO-c has taken the $\psidprime$ natural line shape to be 
\Begeqn
\fBW(E) = {\Gamma(E) \over (E^2-\Mpsi^2)^2+(\Mpsi\,\Gamma_T(E))^2}, \label{eq:bwfunction}
\Endeqn
where $\Mpsi$ is the mass of the $\psidprime$.  The total width 
$\Gamma_T(E)$ is the sum of the partial widths for neutral and 
charged $D\Dbar$ pairs, $\Gamma_T(E) \equiv \Gammaz(E)+\Gammap(E)$.  
The numerator $\Gamma(E)$ is either $\Gammaz(E)$ or $\Gammap(E)$ 
depending on whether $\Dz\Dzbar$ or $\Dp\Dm$ events are being fit.  
The partial widths are 
\begin{eqnarray}
\Gammaz(E) &=& \Gammapsi\, \calBz \frac{q_0^3}{q_{0M}^3} \frac{1+(rq_{0M})^2}{1+(rq_{0})^2} ~\textrm{and} \nonumber\\[1ex]
\Gammap(E) &=& \Gammapsi\, \calBp \frac{q_+^3}{q_{+M}^3} \frac{1+(rq_{+M})^2}{1+(rq_{+})^2}, \label{eq:Edependwidths}
\end{eqnarray}
respectively. In these expressions, $\Gammapsi$ is the measured width 
of the $\psidprime$, $\calBz$($\calBp)$ is the branching fraction for 
the decay of the $\psidprime$ to $\Dz\Dzbar$($\Dp\Dm$) pairs, $q_0$ ($q_+$) 
is the momentum of a $\Dz$($\Dp$) of energy $E/2$, and $q_{0M}$($q_{+ M}$) 
is the momentum of a $\Dz$($\Dp$) of energy $\Mpsi/2$. CLEO-c used 
$\calBz = 0.57$ and $\calBp = 0.43$.  The 
parameter $r$ is the Blatt-Weisskopf interaction radius taken to be
$r = 12.3~\mathrm{GeV}^{-1} = 2.4~\mathrm{fm}$.

The energy distribution of the $\psidprime$ mesons that are produced is 
obtained by multiplying the $\elp\elm$ energy distribution $f_{e^+e^-}(E)$ 
with the cross-section for $\psidprime$ production,
\Begeqn
\fpsiE(E)=\fBW(E)\int_0^{\infty} h(E_{\gamma})\gE(E+E_{\gamma})\, dE_{\gamma}.
\label{eq:energydist}
\Endeqn
The $\psidprime$ energy $E$ is related to $q$, the magnitude ($|\bf q|$) 
of the 
momentum of the produced $D$ and $\Dbar$, by $E = 2\sqrt{q^2 + m_D^2}$.  
Hence, $\fpsiE(E)$ can be transformed into a distribution function $\uDq(q)$ 
for the $D$ momentum,
\Begeqn\label{eq:fDq}
\uDq(q) = \fpsiE(E) \left| {dE \over dq} \right|.
\Endeqn

The measured $D$ momentum $\vecp$ differs from $\vecq$ due to detector 
resolution.
The resolution distribution is 
described by the sum of three-dimensional Gaussian resolution 
functions.  Each term in this sum is given by
\begin{equation}
\gp(\vecp;\vecq,\sigma_p)=
{1\over (2\pi)^{3/2}\sigma_p^3}
e^{-(\vecp-\vecq)^2/(2\sigma_p^2)},
\label{eq:momresolution}
\end{equation}
where $\vecq$ is the momentum of the $D$ meson, $\vecp$ is the 
reconstructed momentum, and $\sigma_p$ is the momentum resolution, 
assumed to be the same for both longitudinal and transverse components 
of $\vecp$ relative to the direction of $\vecq$.  The $D$ is reconstructed 
from multiple final-state particles, and the vector sum of their momenta 
tends to average out any directional dependence.
In the discussion below, we 
consider smearing with a single Gaussian, for simplicity.  

Since the line shape distribution $\uDq(q)$ depends only on the 
magnitude $q = |\vecq|$ of the $D$ meson momentum, we reduce the 
three-dimensional momentum resolution function $\gp(\vecp;\vecq,\sigma_p)$ 
to a one-dimensional resolution function $r(p;q,\sigma_p)$ for the 
probability distribution of the measured value of $p \equiv |\vecp|$ 
given the produced value of $q$.  This requires 
integrating $p^2\, \gp(\vecp;\vecq,\sigma_p)\,dp\, d\Omega$ over angles
transverse to $\vecq$.  In this expression, $p^2\, dp\, d\Omega$ is the 
usual spherical coordinate volume element and the polar and azimuthal 
angles of $d\Omega$ are relative to the vector $\vecq$.  Therefore,
\begin{eqnarray}
& & r(p;q,\sigma_p) = p^2 \int \gp(\vecp;\vecq,\sigma_p)\, d\Omega = \nonumber\\
& & {p \over q}{1\over \sqrt{2\pi}\sigma_p}
\left[e^{-(p-q)^2/(2\sigma_p^2)}-e^{-(p+q)^2/(2\sigma_p^2)}\right].~~~~~
\end{eqnarray}
The distribution of the reconstructed $D$ momentum, $\vDp(p)$, is then 
determined by smearing the distribution of the true $D$ momentum, 
$\uDq(q)$ of \Eqn{eq:fDq}, with $r(p;q,\sigma_p)$,
\begin{eqnarray}
& & \vDp(p) = \int_0^\infty r(p;q,\sigma_p)\uDq(q)\, dq =\nonumber\\
& & \int_{2m_D}^\infty r(p;q(E),\sigma_p)\fpsiE(E)\, dE.  \label{eq:fp_ST}
\end{eqnarray}
Since the measured value of $\Mbc$ is a function of the reconstructed
momentum $p$, the distribution function $\wDM(\Mbc)$ of $\Mbc$ is related 
to $\vDp(p)$ by
\begin{eqnarray}
& &\wDM(\Mbc) = \left| {dp\over d\Mbc}\right| \vDp(p) = \nonumber\\
& &{\Mbc\over p}\int_{2m_D}^\infty r(p;q(E),\sigma_p)\fpsiE(E)\, dE.
\label{eq:mbcsingletag}
\end{eqnarray}

An example of a fit using this form is shown in Fig.~\ref{fig:cleoc_mbc}.
The tail on the high side is due to ISR and the two components 
used for the momentum resolution are shown.

The distribution for double tags, \ie~for $\Mbc \equiv \Mbc(D)$ and 
$\Mbcbar \equiv \Mbc(\Dbar)$, is similar to the form developed above for 
a single $\Mbc$ distribution. Since both $D$ mesons are produced with 
the same momentum $q$, \Eqn{eq:fp_ST} generalizes to the following 
probability distribution for reconstructing the $D\Dbar$ pair with 
measured momenta $p$ and $\pbar$ given resolutions $\sigma_p$ 
and $\sigmabar_p$,
\Begeqn
\vDDbar(p,\pbar)=\int r(p;q,\sigma_p)r(\pbar;q,\sigmabar_p) \uDq(q)\, dq.
\Endeqn
Written in terms of $\Mbc$ and $\Mbcbar$, we have
\begin{eqnarray}
& & \wDDbar(\Mbc,\Mbcbar)={\Mbc\over p}{\Mbcbar\over \pbar}\times\nonumber\\
& & \int_{2m_D}^\infty r(p;q(E),\sigma_p)r(\pbar;q(E),\sigmabar_p)\fpsiE(E)\, 
dE. ~~~~
\label{eq:mbcdoubletag}
\end{eqnarray}

In the single tag fits it is hard to separate the effects
of beam energy smearing and detector resolution. In
the double tag fits these contributions can be separated
as the effects of detector resolution is uncorrelated amongst
the two $D$ candidates, while the beam energy smearing 
is strongly correlated among the two $D$ candidates.

\subsubsection{Systematic uncertainties}
\label{sect:cleoc_syst}

Many of the analyses discussed in this review are limited by
systematic uncertainties. 
This applies in particular to the determination of the 
Cabibbo favored $D^0$ and $D^+$ reference branching fractions
that are discussed in Sect.~\ref{sect:dref}.
A substantial effort has been put into
understanding the systematic uncertainties associated with 
track finding, $K^0_S$ reconstruction, particle identification,
and $\pi^0$ reconstruction. At the $\psi(3770)$ resonance
many of these uncertainties can
be evaluated using hadronic decays in an event environment
very similar to the channels studied. This gives confidence in
the sometimes small systematic uncertainties obtained in these
studies. The most detailed systematic studies carried out 
by CLEO-c are described in Ref.~\cite{Dobbs:2007zt}. 
As the results of these studies are important for many
results discussed in this review some of these studies
are discussed below. 

Track finding has been studied in CLEO-c using a missing mass
technique where all particles in an event are reconstructed 
except for one particle which we are interested in studying. 
As an example consider the use of the kaon in $D^0\to K^-\pi^+$
to measure the kaon tracking efficiency. In this case 
the opposite $\bar D^0$ in the event would be
fully reconstructed 
in some channel and the
$\pi^+$ from $D^0$ decay looked for. 
Given the $\bar D^0$ and $\pi^+$ candidates
the missing mass in the event can be calculated
\begin{equation}
\Mmisssq = (p_\mathrm{tot} - p_{\Dbar} - p_\mathrm{other})^2,
\end{equation}
where $p_{\Dbar}$ is the four-momentum of the reconstructed 
$\Dbar$, $p_\mathrm{other}$ is the four-momentum of the other 
particles that were combined with the tag $\Dbar$, in this 
example the $\pi^+$, and $p_\mathrm{tot}$ 
is the four-momentum of the initial $\elp\elm$ pair.  In the missing 
mass squared calculation, 
the $\Dbar$ momentum is rescaled to the
momentum magnitude expected from the 
beam energy, but its direction is left unchanged.  This constraint 
improves the $\Mmisssq$ resolution.

Candidates where the missing momentum vector fails the polar
angle requirement $|\cos\theta|<0.9$ are rejected. This eliminates
candidates in which the missing particle is expected to be
outside the tracking fiducial volume. This requirement is
tighter than the angular acceptance of the CLEO-c detector,
$|\cos\theta|<0.93$. A tighter requirement is used to 
compensate for the resolution in the predicted direction 
of the missing momentum. Later an uncertainty is added
to compensate for the extrapolation to the full tracking 
fiducial.

For each missing mass candidate all remaining 
tracks in the event are considered, applying the standard track quality 
criteria. If a track candidate is found such that it forms a
good $D$ candidate when combined with the other particles
in the missing mass candidate the missing particle is said
to have been found. The requirements for a good $D$ candidate
are $|\Mbc-M_D|<0.01$ $\Gevcsq$ and $|\Delta E|<0.05$ $\Gevc$.
If a good $D$ candidate is not found the 
missing particle is said not to be found.

\begin{figure}[htb]
\begin{center}
\includegraphics[width=0.49\linewidth]{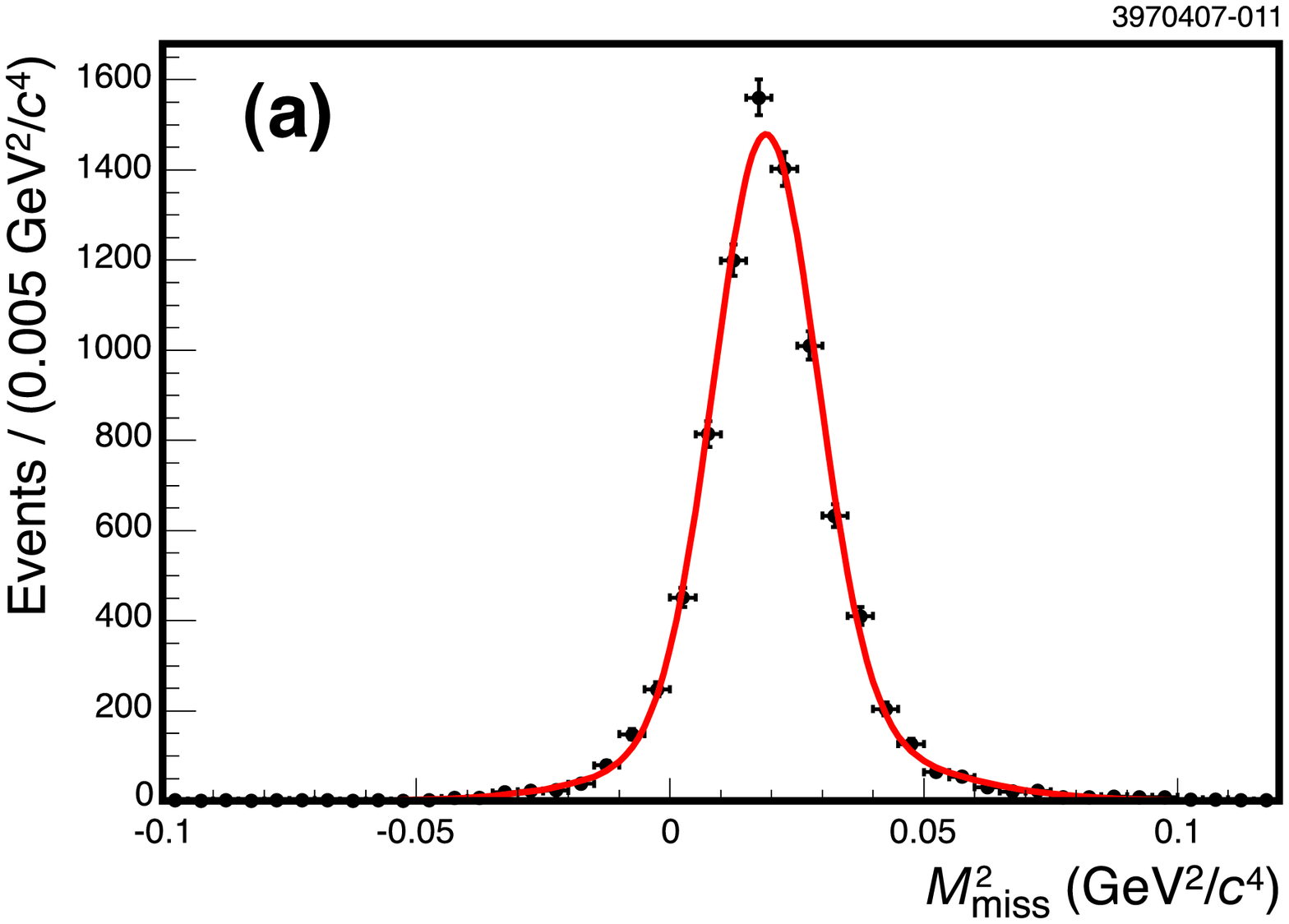}
\includegraphics[width=0.49\linewidth]{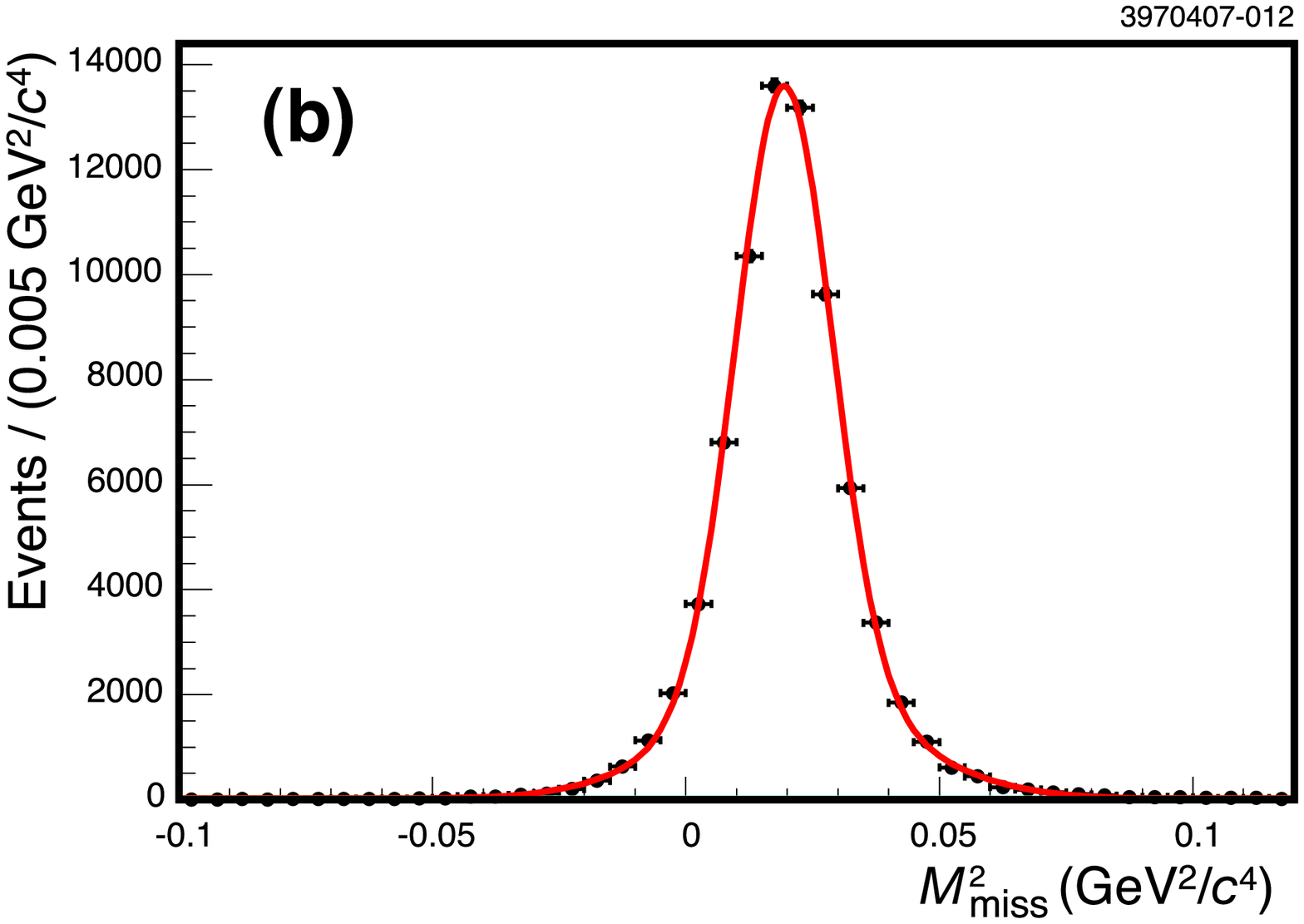}\\
\includegraphics[width=0.49\linewidth]{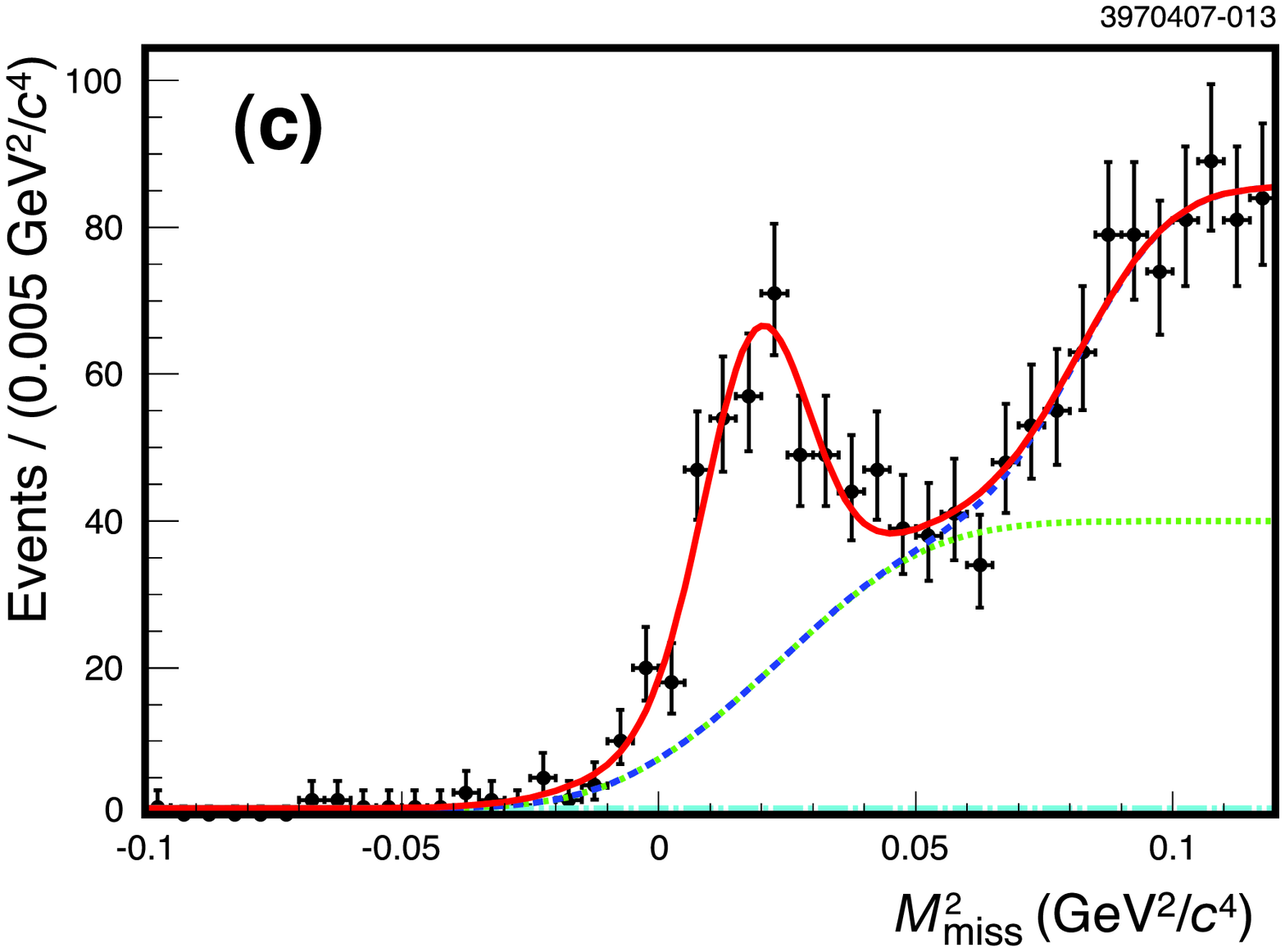}
\includegraphics[width=0.49\linewidth]{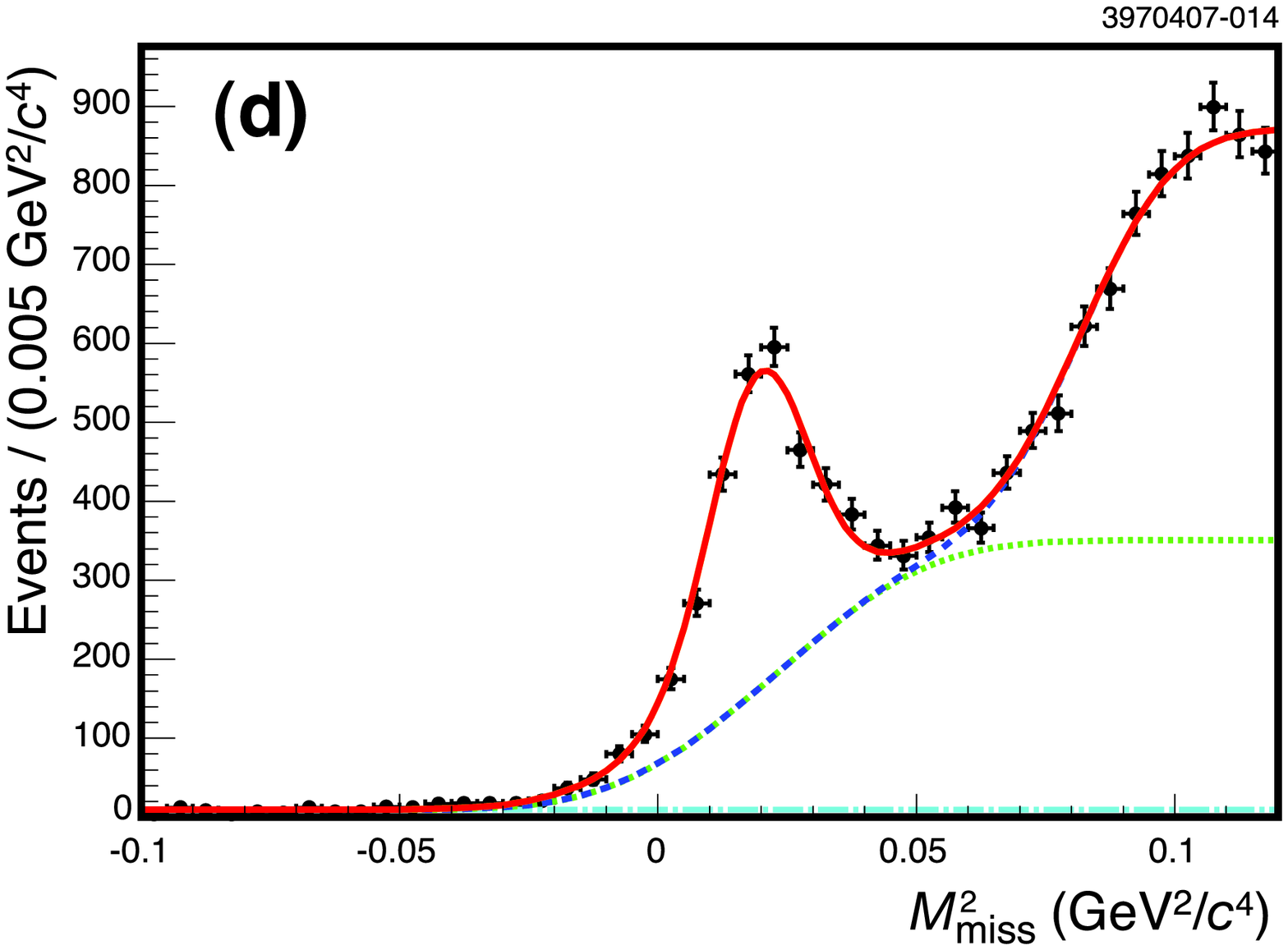}
\end{center}
\caption{Histograms of and fits to $\Mmisssq$ distributions 
from $\Dp\to\Km\pip\pip$ decays to determine the charged pion 
efficiency for $p_{\pip} > 0.2~\Gevc$.  Figures (a) and (c) are 
from events in data, and (b) and (d) are from events in Monte Carlo 
simulation.  Figures (a) and (b) are from decays in which the pion 
was found, while (c) and (d) are from decays in which the pion was 
not found.  The solid curves are fits to the data or Monte Carlo 
sample; the dashed curves in (c) and (d) are background 
contributions.
From~\textcite{Dobbs:2007zt}.
\label{fig:trackingeff}}
\end{figure}

The missing mass candidates are separated
into two samples; the sample where the missing particle was
found and the remaining events where the missing 
particle was not found. An example is shown in
Fig.~\ref{fig:trackingeff}. The case where the missing
particle is found corresponds to a fully reconstructed
$\psidprime$ event and is very clean. The events in this
sample are fit to a signal shape using a sum of two 
Gaussians. A small background component is also included
in the fit. For the sample where the missing particle is
not found a clear peak can be seen corresponding to the
events where we had an inefficiency.
In addition to this peak there are also substantial backgrounds.
These backgrounds include semileptonic decays as well
as higher multiplicity hadronic $D$ decays. These backgrounds
are parameterized using Monte Carlo simulated events.

As described in detail in~\textcite{Dobbs:2007zt} CLEO-c
measures the tracking efficiency for both kaons and pions in
three momentum ranges ($0.2<p<0.5$ $\Gevc$, $0.5<p<0.7$ $\Gevc$,
and $p>0.7$ $\Gevc$). CLEO-c evaluates the tracking
efficiency and find agreement between data and the Monte
Carlo simulation and assigns a per track systematic uncertainty
of $\pm 0.3\%$ per track for pions. For kaons an additional
uncertainty of $\pm 0.6\%$ is added due to evidence for a 
tracking efficiency difference between $K^+$ and $K^-$.

\begin{figure}
\begin{center}
\includegraphics[width=0.49\linewidth]{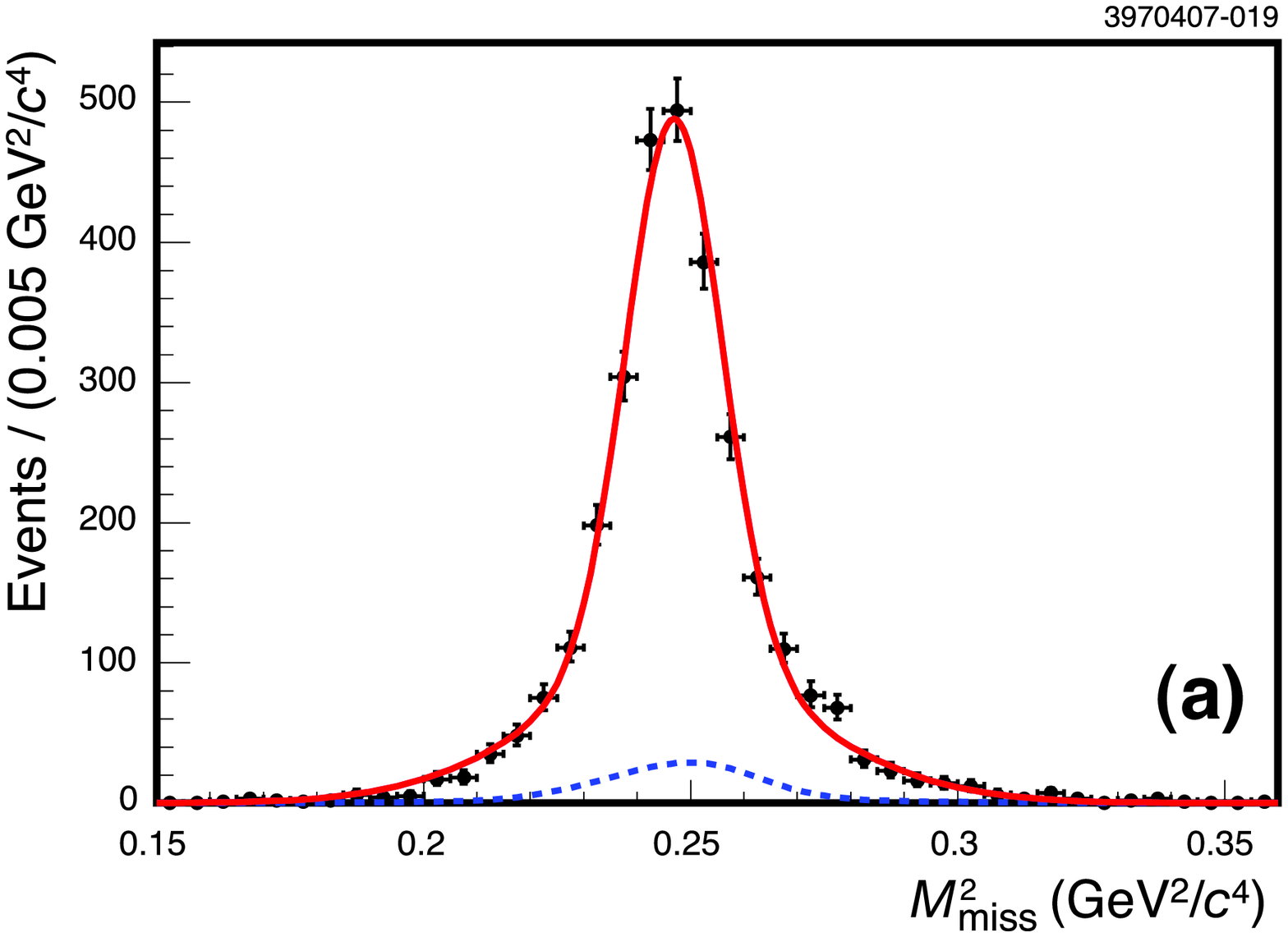}
\includegraphics[width=0.49\linewidth]{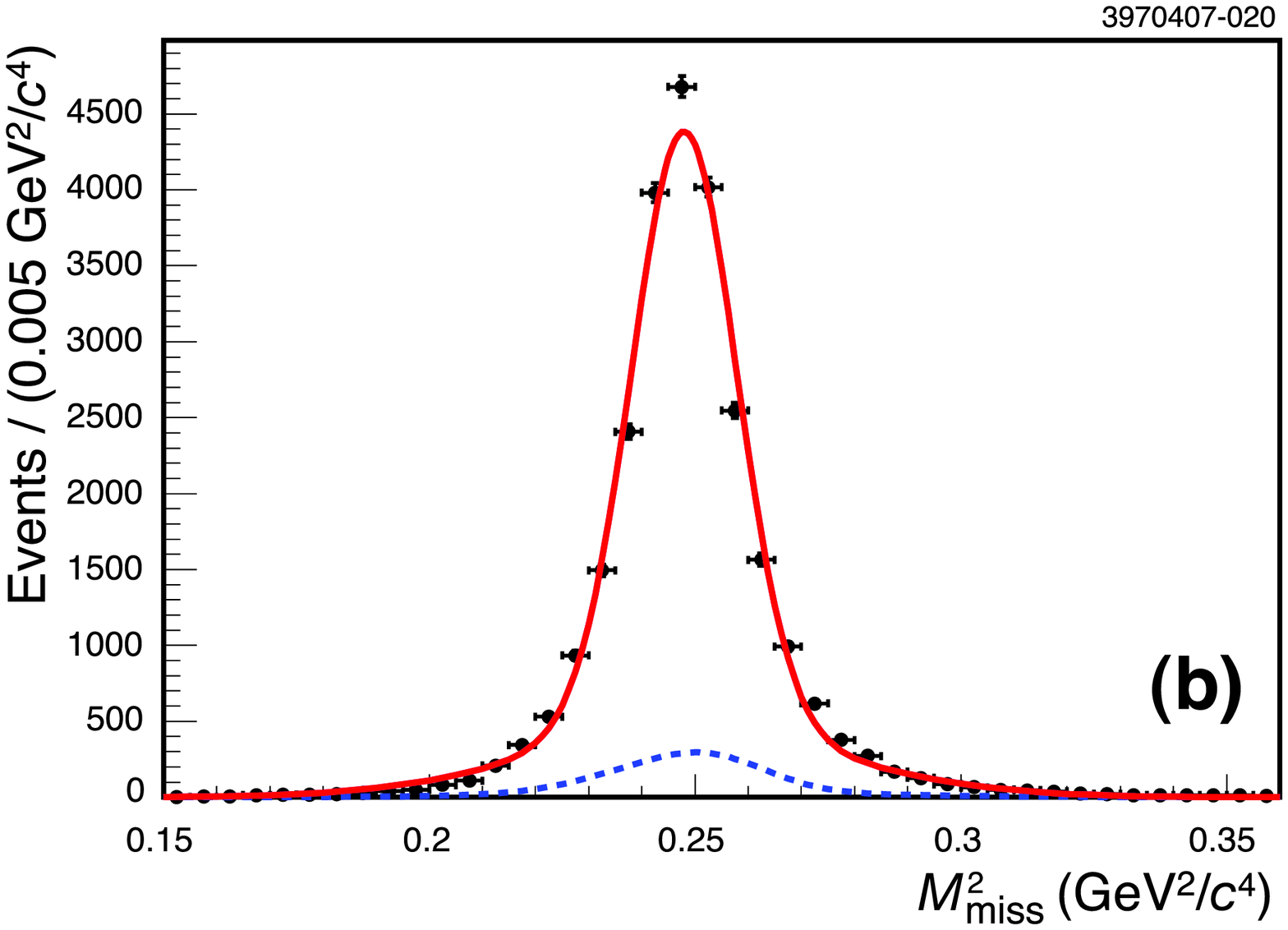}\\
\includegraphics[width=0.49\linewidth]{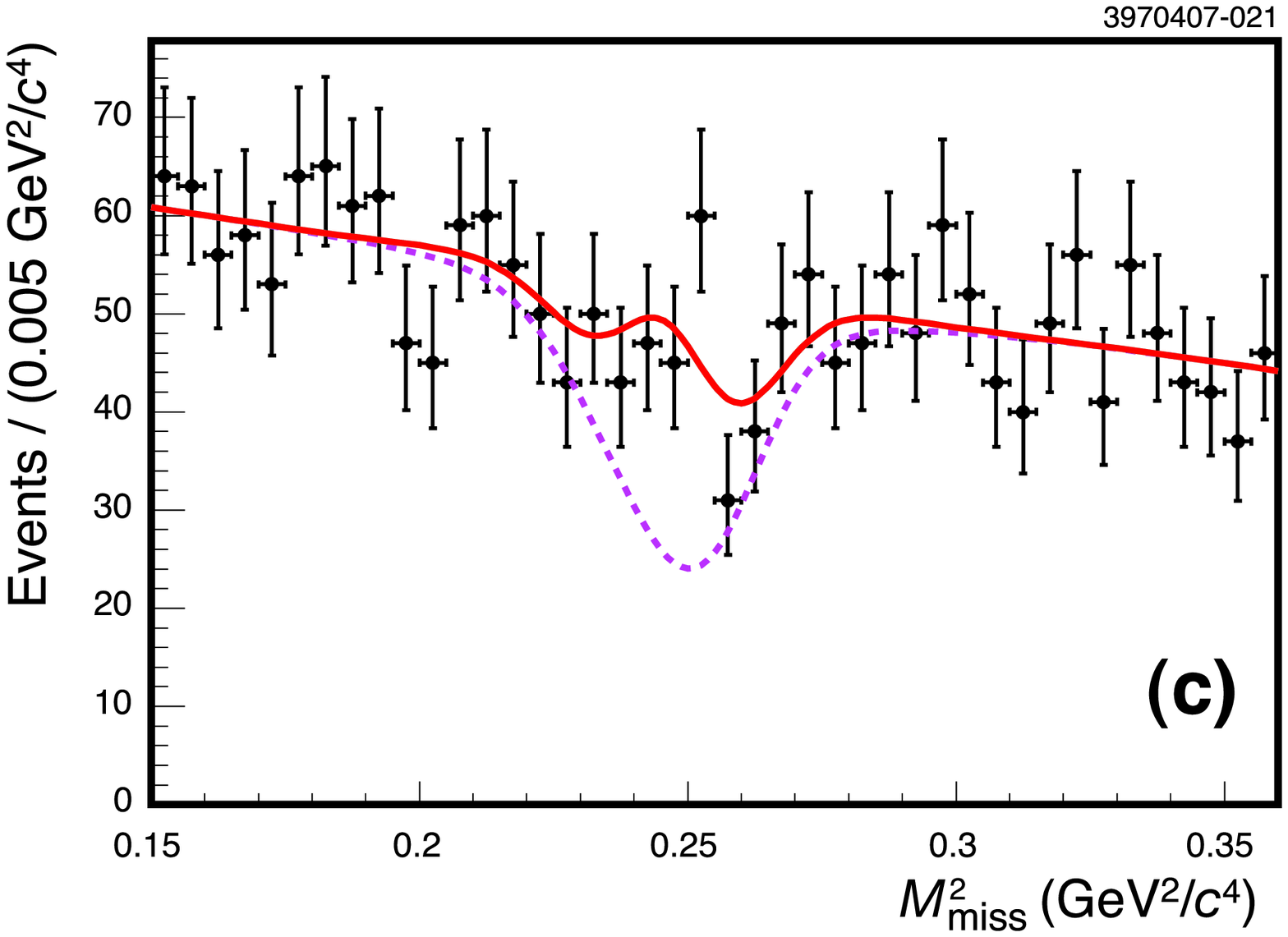}
\includegraphics[width=0.49\linewidth]{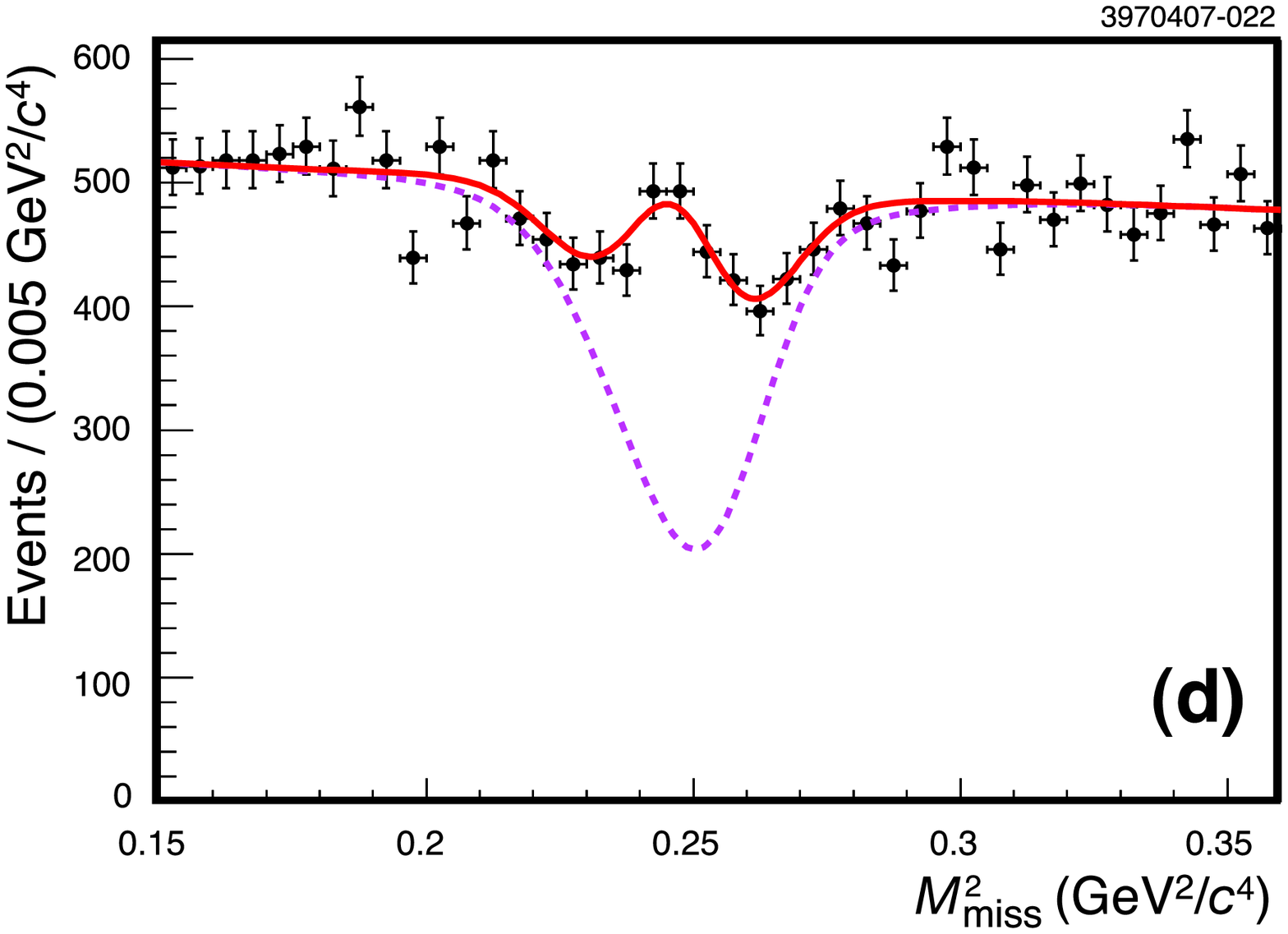}
\end{center}
\caption{Histograms of and fits to $\Mmisssq$ distributions to determine 
the $\KS$ efficiency.  Figures (a) and (c) are from events in data, 
and (b) and (d) are from events in Monte Carlo simulation.  Figures (a) 
and (b) are from decays in which the $\KS$ was found, while (c) and (d) 
are from decays in which the $\KS$ was not found.  The background peak 
and deficit are determined by searching for $\KS$ candidates in high and 
low sidebands of the $\KS$ mass.  In Figs.~(a) and (b), the dashed curves 
are the contributions from fake $\KS$ candidates.  In Figs.~(c) and (d), 
the dashed  curve is the background --- a linear function with a deficit 
due to events in which a fake $\KS$ candidate was found --- and the solid 
curve is the total fit function including the signal peak.  The area between 
the curves is proportional to the number of $\KS$ mesons not found.
From~\textcite{Dobbs:2007zt}.
\label{fig:CLEOcKSeff}}
\end{figure}

The $K^0_S\to\pip\pim$ reconstruction efficiency is studied in
$\Dz$ or $\Dzbar$ decays to $\KS\pip\pim$ decays
using a technique similar to what was used for the
tracking efficiencies.
One tag $D$ is
fully reconstructed and two charged pions are required to
be found. To factor
out the track finding efficiency and also to reject 
$\KL\pip\pim$ and $K^0_S\to\piz\piz$ decays it is required that
two additional tracks are found in the event. These tracks are
required to satisfy loose consistency requirements with coming 
from a $K^0_S$ decay. The invariant mass of the two tracks
are required to be in the range from 0.2 to 0.7 $\Gevcsq$. In addition
the difference between the missing momentum vector and the
momentum vector of the sum of the two charged tracks is required
to be less than 60 $\Mevc$. Candidates that satisfy these
requirements are searched for a $K^0_S$ candidate found using the
standard $K^0_S$ vertex finder. Similar to the tracking studies 
the candidates are separated into two categories; where the $K^0_S$
was found and where it was not found. Compared to the tracking
systematics study described above the $K^0_S$ study is more
complicated because there are fake $K^0_S$ candidates from
wrong $\pip\pim$ tracks in either $\KS\pip\pim$ or
$\pip\pim\pip\pim$ events. This gives rise to a 'hole' in the
events where the $K^0_S$ candidate was not found because combinatorial
background got promoted to signal. This is illustrated in 
Fig.~\ref{fig:CLEOcKSeff}. Using this technique CLEO-c 
assigns a systematic uncertainty of $\pm 1.8\%$ for the
$\KS$ finding efficiency.

The efficiency for $\piz\to\gamma\gamma$ reconstruction has 
been studied using a missing mass technique in $\psiprime\to\psi\piz\piz$
events recorded at $E_\mathrm{CM}=m_{\psiprime}$. 
There are not any really useful hadronic $D$ decays for 
$\piz$ efficiency studies. One candidate is the $\Dz\to K^-\pip\piz$
decay. If the same missing mass technique as described above for the
charged particle tracking efficiency is applied where
all particles except for the $\piz$ are reconstructed there is
a background from the semileptonic decay $\Dz\to K^-\ell^+\nu$
which peaks at $M^2_\mathrm{miss}=m^2_{\nu}=0$ which can not
be separated from the signal peaking at $M^2_\mathrm{miss}=m^2_{\piz}$.
The only mode of some use is $\Dz\to K^0_S\piz$. But the 
relatively small branching fraction and reconstruction efficiency
makes it marginally useful with the data samples currently 
available. 

\begin{figure}
\begin{center}
\includegraphics[width=0.99\linewidth]{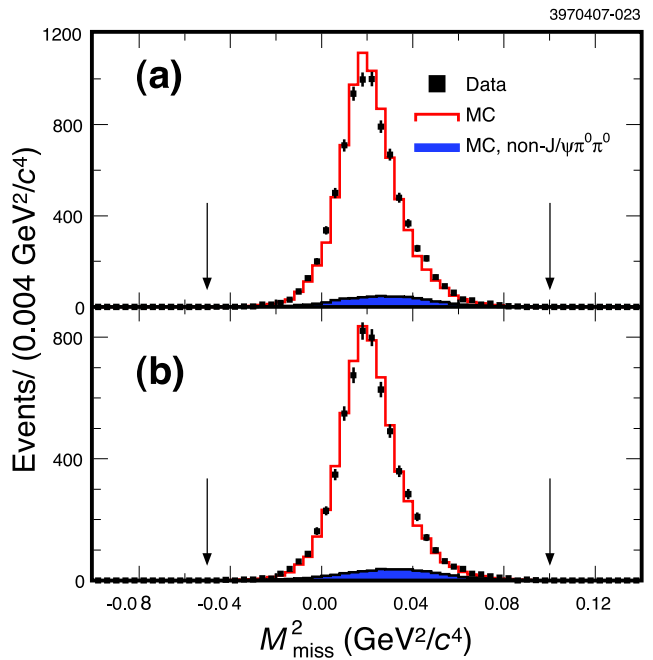}
\end{center}
\caption{Distributions of $\piz$ missing mass squared in candidate
$\psiprime\to \Jpsi\piz\piz$ events for data (points) and Monte Carlo events (histogram). 
The predicted background level is also shown.  The vertical arrows demarcate the
signal region. Events in which the second $\piz$ was found are
shown in (a) whereas the events where the second $\piz$ was
not found are shown in (b).
From~\textcite{Dobbs:2007zt}.
\label{fig:CLEOcpi0eff}}
\end{figure}

To study the $\piz$ efficiency using the $\psiprime\to J/\psi\piz\piz$
channel CLEO-c reconstruct the $J/\psi$ and one $\piz$. The $J/\psi$ is
reconstructed in the $e^+e^-$ and $\mu^+\mu^-$ final states. The
di-lepton invariant mass is required to be within 50 MeV of
the known $J/\psi$ mass. The $J/\psi$ candidate is combined with a
$\piz$ candidate and the $M^2_\mathrm{miss}$ is calculated. For
signal events, $\psiprime\to J/\psi\piz\piz$, $M^2_\mathrm{miss}$
peaks at $m^2_{\piz}$. To remove backgrounds, primarily from
$\psiprime\to\psi\piz$, it is required that $p_\psi<500$ $\Mevc$ and
$p_{\piz}<500$ $\Mevc$. In addition, to select the kinematic region
populated by $\psiprime\to J/\psi\piz\piz$ it is required that
$$
(p_{\piz}^2+p_\mathrm{miss}^2) - (p_{\piz}^2-p_\mathrm{miss}^2)^2 / (0.5~\mathrm{GeV}^2) > 0.10~\mathrm{GeV}^2$$ 
and
$$(p_{\piz}^2+p_\mathrm{miss}^2) - (p_{\piz}^2-p_\mathrm{miss}^2)^2 / (2~\mathrm{GeV}^2) < 0.17~\mathrm{GeV}^2.$$
Next a second $\piz$ is looked for in the event. If a second $\piz$
exists that satisfies $M(\Jpsi\piz\piz)-M(\Jpsi)$ within $50~\Mevcsq$
of the nominal $\psiprime$--$\Jpsi$ mass difference, the $\piz$ 
is considered to be found. The distributions of the $M^2_\mathrm{miss}$
is shown in Fig.~\ref{fig:CLEOcpi0eff} for the case (a) where
the second $\piz$ was found and for the case (b) where the
second $\piz$ was not found. Using this study CLEO-c finds a 
difference between the data and the Monte Carlo simulation in
this sample of 
$\eta\equiv(\epsilondata/\epsilonmc)-1=(-4.37\pm 0.72\pm 0.41)\%$.
However the average momentum of the $\piz$s in the $\psiprime\to\psi\piz\piz$
sample is about 250 $\Mevc$ whereas the typical momentum in
$D$ decays such as $\Dz\to\Km\pip\piz$, $\Dp\to\Km\pip\pip\piz$,
and $\Dp\to\KS\pip\piz$ is higher, typically around 450 $\Mevc$.
CLEO-c study the $\piz$ efficiency as a function of the $\piz$
momentum. The relative data to Monte Carlo efficiency, $\eta$,
as a function of momentum was fit to a straight line in order
to extrapolate the efficiency correction to a $\piz$ momentum
of 450 $\Mevc$. CLEO-c finds a 3.9\% correction and assigns
a $\pm 2.0\%$ uncertainty to the corrected efficiency.

\subsection{$c\bar c$ production in $e^+e^-$ above threshold}

At energies above charm threshold, charm hadrons are produced in 
fragmentation and are part of a jet, or are produced as 
secondary particles in decays of $b$-hadrons. The largest
charm samples are those produced at the $B$ factories at
$\epem$ center-of-mass energies near 10.58 $\Gev$ corresponding
to the $\Upsilon(4S)$ resonance. The large cross-section, about 
1.3 nb, combined with the large integrated luminosities recorded by 
CLEO, BABAR, and Belle have produced these very large samples.

At even higher energy, the LEP operated near the $Z$ resonance and
produced over $4$ million $Z$ bosons per experiment. The jet nature
of the events here is more clear than at the $\Upsilon(4S)$.

The CLEO, BABAR, and Belle experiments were designed to study
$B$ meson decays
but they are also well suited for studying charm.
These experiments all have excellent charged particle
tracking capabilities and vertex detectors capable of 
detecting the separated vertices from the relatively 
long lived charm and beauty hadrons. 
All three experiments have CsI(Tl) electromagnetic calorimeters
with excellent photon detection capabilities and electron
identification using $E/p$. Detection of muons in all three 
experiments are done using an instrumented flux return.
Also key for these experiments is the identification of 
charged hadrons, particularly $K$--$\pi$ separation. The
three experiments chose different technologies here.
BABAR used a DIRC (Detector of Internally Reflected Cherenkov light),
CLEO-III used a RICH (Ring Imagine Cherenkov Detector),
and Belle uses aerogel Cherenkov counters. All three 
different types of charged hadron particle identification
detectors have worked well.

The BABAR and Belle experiments
were built around an energy asymmetric collision designed
to allow resolving the time evolution of the produced $B$ 
mesons, as discussed in the BABAR Physics Book~\cite{Harrison:1998yr}.
The energy asymmetric collisions are reflected in the design
of the detector; the interaction point is offset to optimize
the acceptance due to the boost of the collision center-of-mass.

\subsection{Fixed target experiments}

\begin{figure}[bt]
\begin{center}
\includegraphics[width=0.99\linewidth]{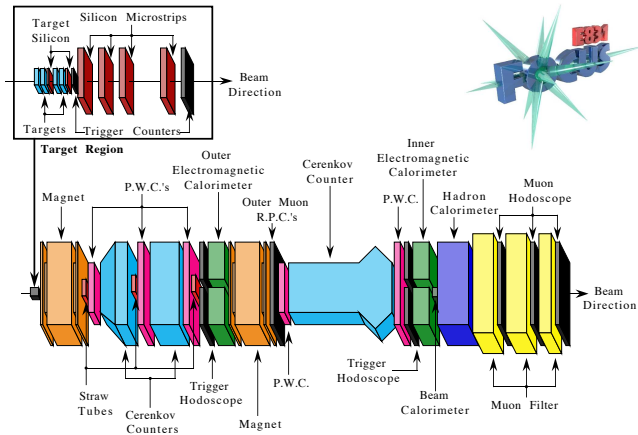}
\caption{The FOCUS (E831) spectrometer.}
\label{fig:focus_detector}
\end{center}
\end{figure}

Charm mesons are sufficiently light that they can be produced
efficiently in fixed target experiments. The main experimental 
challenge is to separate charm production from the large non-charm
rate. The development of silicon based tracking detectors
enabled experiments to effectively identify the long lived
charmed hadrons. 
The pioneering Fermilab photoproduction experiment E691 
was the first experiment
to produced large samples of reconstructed charm hadrons. In this
experiment a beam of photons with an average energy around
180 GeV was incident on a Beryllium target.
The cross-section for charm production was measured to be
about 0.5 $\mu$b. This is about 0.5\% of the 100 $\mu$b total
hadronic cross-section. 
The most powerful tool
for identifying the charm signal is to make use of the relatively
long charm-hadron lifetimes, from $(410.1\pm1.5)$ fs for the $D^0$ 
to $(1040\pm7)$ fs for the $D^+$.
Using the silicon vertex detectors it is possible to separate
the long lived charm-hadrons from the prompt backgrounds.
A series of fixed target experiments for charm physics are
summarized in Table~\ref{tab:charmreco}.
The latest
of these experiments at Fermilab, FOCUS or E831,
reconstructed over 1.2 million exclusive charm decays.
The FOCUS spectrometer is shown in Fig.~\ref{fig:focus_detector}.

\subsection{Final state radiation}
\label{sect:FSR}

The treatment of final state radiation (FSR) is common to
many analyses and will be discussed here. In many earlier 
measurements the effects of final state radiation was often
omitted, but as the measurements have become increasingly more
precise this has become an important effect that can not
be ignored.  In the latest measurements of the branching
fraction for $D^0\to K^-\pi^+$ the size of the radiative
correction is larger than the combined statistical and
systematic uncertainties.

Any reaction involving charged particles will also
radiate photons~\cite{Bloch:1937pw}. In fact, an arbitrary
number of photons will be produced, though most of these
are very soft. In general, when we discuss a branching fraction
for a process, like for example ${\cal B}(D^0\to \Km\pip)$,
this includes final states with additional (soft) photons.
Experimentally, if photons are emitted with an energy that
is smaller than the experimental resolution these events
are automatically included in the measurement. However,
sometimes the photon energies are larger, and the energy carried
away by the photon will make the event fail the selection
criteria. In order to account for this, and provide a
measurement of a physically meaningful quantity, experiments
simulate the effect of final state radiation in their
Monte Carlo simulations. This has been a common practice
for semileptonic decays, in particular with electrons in
the final state for quite some time. For hadronic final states
this is not yet universally done. In $D$ decays the first 
experiment that considered FSR corrections was 
CLEO~\cite{Akerib:1993pm}. Today most measurements of hadronic
$D$ decays include FSR corrections.

For simulation of final state radiation in hadronic decays the
most commonly used tool is the PHOTOS package~\cite{Barberio:1993qi}.
In the measurement of the $D^0\to \Km\pip$ branching fraction
CLEO-c uses version 2.15 with interference enabled. The effect of
interference, here referring to interference between photons radiated
from different charged particles in the final state, is
important. For the final state $D^0\to \Km\pip$ the effect of 
including interference changes the fraction of events that
radiate more than 30 MeV from 2.0\% to 2.8\%.
Earlier versions of PHOTOS were only able to simulate
the interference for decays to final state with a particle---anti-particle
pair.
PHOTOS has been compared with 
calculations to higher order in $\alpha$ and found to produce
the amount of energy radiated very well in semileptonic
decays of $B$ mesons and decays of $\tau$ leptons~\cite{RichterWas:1992qb}.
However, for hadronic final states there is an additional
uncertainty introduced by the fact that the final state 
particles, kaons and pions, are not point like. This uncertainty
affects in particular higher energy photons that probe the
structure of the final state particles. Higher energy photons
could also be radiated directly from the quarks; this effect is
not included in the simulation. CLEO-c includes a 30\% 
systematic uncertainty
on the correction to the branching fraction due to including
final state radiation. Given the excellent agreement between
exact calculations and next order calculations in $\alpha$ this systematic
uncertainty is probably conservative. 

For many earlier measurements it is not always clear what was
done to correct for the FSR effects. If the effect is not 
included it is hard to correct for it after-the-fact as the
effect of FSR depends on the selection criteria used and how
strongly they would reject events with radiation.

\section{THEORETICAL DESCRIPTION OF $D$ DECAYS}
\label{sect:theory}


Hadronic decays of D-mesons involve transitions of the initial-state D-meson into
several final state mesons or baryons. Thus, they are described by an effective
Hamiltonian containing four-quark operators. The theoretical description of hadronic 
decays of charmed mesons is significantly more complicated than leptonic or 
semileptonic ones, although relevant effective Hamiltonians look similar. 

Charmed hadronic decays are usually classified by the degree of 
Cabibbo-Kobayashi-Maskawa (CKM)  matrix element suppression.  Least 
suppressed, where the quark level transitions are $c\to su\bar d$ are labeled 
``Cabibbo favored" (CF) decays and governed by
\bea\label{Hcf}
{\cal H}_{CF} &=& \frac{G_F}{\sqrt{2}} V_{ud} V^*_{cs}
\left[
C_1(\mu) {\cal O}_1 + C_2(\mu) {\cal O}_2 \right] + {\rm h.c},
\nonumber \\
{\cal O}_1 &=& \left(\overline{s}_i \Gamma_\mu c_i\right)
\left(\overline{u}_k \Gamma^\mu d_k\right), \\
{\cal O}_2 & = & \left(\overline{s}_i \Gamma_\mu c_k\right)
\left(\overline{u}_k \Gamma^\mu d_i\right),
\eea
where $C_n(\mu)$ are the Wilson coefficients obtained by
perturbative QCD running from $M_W$ scale to the scale $\mu$
relevant for hadronic decay, and the Latin indices denote quark
color.  $G_F$ is a Fermi constant, and
$\Gamma_\mu=\gamma_\mu \left(1-\gamma_5\right)$.

The ``Cabibbo suppressed" (CS) transitions are driven by $c\to du\bar d$ or $c\to
su\bar s$ quark processes. Due to the presence of the quark-antiquark pair of the
same flavor in the final state, the effective Hamiltonian takes much more elaborate
form,
\bea\label{Hcs}
{\cal H}_{CS} &=& \frac{G_F}{\sqrt{2}} \sum_{q=s,d} V_{uq} V^*_{cq}
\left[
C_1(\mu) {\cal O}_1^q + C_2(\mu) {\cal O}_2^q \right]
\nonumber \\
&-& \frac{G_F}{\sqrt{2}} V_{ub} V^*_{cb} \sum_{n=3}^6 C_n (\mu) {\cal O} +  {\rm h.c},
\nonumber \\
{\cal O}_1 &=& \left(\overline{q}_i \Gamma_\mu c_i\right)
\left(\overline{u}_k \Gamma^\mu q_k\right), \\
{\cal O}_2 &= &\left(\overline{q}_i \Gamma_\mu c_k\right)
\left(\overline{u}_k \Gamma^\mu q_i\right),
\nonumber
\eea
where $q=d,~s$, and ${\cal O}_{3-6}$ are the so-called ``penguin'' operators of the
type $(\overline{u} c)_{V-A}\sum_q (\overline{q} q)_{V\pm A}$
(see, e.g. Ref.~\cite{Buccella:1994nf}). It is often easy to denote the degree of 
suppression by powers of the Wolfenstein parameter 
$\lambda = \sin\theta_C = V_{us} \simeq 0.22$, there $\theta_C$ is a 
Cabibbo angle.

The ``Doubly Cabibbo suppressed" (DCS) decay is the one in which $c\to du\bar s$
quark transition drives the decay. The effective Hamiltonian for
DCS decay can be obtained from Eq.~(\ref{Hcf}) by interchanging
$s\leftrightarrow d$.

Calculations of hadronic decay rates governed by these transitions are quite
complicated and model-dependent. Most often, simplified assumptions, such as
factorization~\cite{Bauer:1986bm,Buras:1985xv} are used to estimate the needed
branching ratios. Some dynamical approaches, such as QCD sum rules, have been
used to justify those assumptions~\cite{Blok:1992hw}. The main problem with reliable 
calculations of charmed meson decays is that they populate the energy range where 
non-perturbative quark dynamics is active. This leads to resonance effects that affect the 
phases of hadronic decay amplitudes~\cite{Falk:1999ts}, which makes predictions 
based on factorization quite unreliable. 

Instead of predicting an absolute decay rate, it is often useful to obtain relations among several 
decay rates. These relations are helpful when some decay rates in a relation are measured, and 
some are unknown. This allows for a relation to be used to predict the unknown transition rate(s). 
The relations can be built based on some symmetries, such as standard flavor $SU(3)$~\cite{Savage:1991wu}, 
or on overcomplete set of universal quark-level amplitudes~\cite{Gronau:1994rj,Rosner:1999xd}. 
We shall discuss those methods below.

The partial width for a specific two-body decay of a charmed meson depends on both the
invariant amplitude ${\cal A}$ and a phase space factor. For a specific two-body 
decay into a $PP$ final state,
\beq
\Gamma(D \to PP) = \frac{\left| \ {\bf p} \ \right|}{8 \pi M_D^2} 
\left| {\cal A} (D \to PP)\right|^2,
\eeq
where $\left|\vecp\right|$ is a center-of-mass 3-momentum of each final state particle.
For a decay into a $PV$ final state,
\beq
\Gamma(D \to PV) = \frac{\left| \ {\bf p} \ \right|^3}{8 \pi M_D^2} 
\left| {\cal A} (D \to PV)\right|^2.
\eeq
Note that in the case of $PP$ final state the final state mesons are in the S-wave, while in
the case of $PV$ final state they are in a P-wave. This is why $\left| {\cal A} (D \to PP)\right|$
has dimension of energy, while $\left| {\cal A} (D \to PV)\right|$ is dimensionless.

\subsection{$SU(3)_F$ flavor symmetries}

One popular approach that was adopted for studies of hadronic charm
decays involves application of approximate flavor symmetries, such as
flavor $SU(3)_F$. This approach is based on the fact that the QCD Lagrangian
acquires that symmetry in the limit where masses of all light quarks are the same.
The $SU(3)_F$ analysis of decay amplitudes cannot predict their absolute values. However,
at least in the symmetry limit, this approach can relate transition amplitudes for 
different decays, which could prove quite useful for an experimental analysis. 
One potential difficulty with this approach is related to the fact that available
experimental data show that flavor $SU(3)_F$ symmetry is broken in charm transitions, 
so symmetry-breaking corrections should be taken into 
account~\cite{Savage:1991wu, Hinchliffe:1995hz}.

In the flavor-symmetry approach all particles are denoted by their $SU(3)_F$ representations.
Charm quark transforms as singlet under flavor $SU(3)$. The fundamental representation 
of $SU(3)_F$ is a triplet, {\bf 3}, so the light quarks $u$, $d$, and $s$ belong to this 
representation with $(1,2,3)=(u,d,s)$. The operator $D_i$ that creates a $\overline D$-meson 
is of the form $\bar cu$, so it also transforms in the fundamental representation of $SU(3)$.
In the hadronic decay of a charm meson the final state mesons are made of $u$, $d$, and $s$ 
quarks, so they either form an octet {\bf 8} representation of $SU(3)_F$ (pseudoscalars 
$\pi^\pm$, $\pi^0$, $K^\pm$, $K^0$, $\overline{K}^0$, $\eta_8$ and vectors
$\rho^\pm$, $\rho^0$, $K^{*\pm}$, $K^{*0}$, $\overline{K}^{*0}$, $\omega_8$), e.g.
\beq\label{octet}
P_i^k = 
\left(
\begin{array}{ccc}
\frac{1}{\sqrt{6}} \eta_8 + \frac{1}{\sqrt{2}} \pi^0   &    \pi^+    & K^+ \\
\pi^-    &    \frac{1}{\sqrt{6}} \eta_8 - \frac{1}{\sqrt{2}} \pi^0     & K^0 \\
K^-  &  \overline{K}^0 & - \sqrt{\frac{2}{3}} \eta_8
\end{array}
\right),
\eeq
or an $SU(3)_F$ singlet ($\eta_1$ and $\omega_1$). The physical states $\eta$, $\eta'$, $\phi$, and $\omega$ 
are linear combinations of $\eta_{1,8}$ and $\omega_{1,8}$ states respectively. 

The $\Delta C=-1$ part of the weak Hamiltonian has the flavor structure
 $(\bar q_ic)(\bar q_jq_k)$ (see Eq.~(\ref{Hcf})), so its matrix representation is written with a 
fundamental index and two antifundamentals, $H^{ij}_k$.  This operator is a sum of irreducible
representations contained in the product $3 \times \3bar \times \3bar =\15bar+ 6 + \3bar + \3bar$.  
In the limit in which the third generation is neglected,
$H^{ij}_k$ is traceless, so only the $\15bar$ (symmetric on $i$ and $j$)
and 6 (antisymmetric on $i$ and $j$) representations appear.  
That is, the $\Delta C=-1$ part of ${\cal H}_w$ may be decomposed as 
${1\over2} (\cO_{\15bar} + \cO_6)$, where
\bea
\cO_{\15bar} &=& (\bar sc)(\bar ud) + (\bar uc)(\bar sd)
    + s_1(\bar dc)(\bar ud) \nonumber\\
&&    + s_1(\bar uc)(\bar dd) - s_1(\bar sc)(\bar us) 
    - s_1(\bar uc)(\bar ss) \nonumber\\
&&  - s_1^2(\bar dc)(\bar us) - s_1^2(\bar uc)(\bar ds) \,, \nonumber\\
\cO_6 &=& (\bar sc)(\bar ud) - (\bar uc)(\bar sd)
    + s_1(\bar dc)(\bar ud) \nonumber\\
&& - s_1(\bar uc)(\bar dd) - s_1(\bar sc)(\bar us) 
    + s_1(\bar uc)(\bar ss) \nonumber\\
&&  - s_1^2(\bar dc)(\bar us) + s_1^2(\bar uc)(\bar ds) \,,
\eea
and $s_1=\sin\theta_C\approx0.22$.  The matrix representations
$H(\15bar)^{ij}_k$ and $H(6)^{ij}_k$ have nonzero elements
\begin{equation}
\begin{tabular}{rll}
$H(\15bar)^{ij}_k:\quad$
    &  $H^{13}_2 = H^{31}_2=1$\,,  &  $H^{12}_2 = H^{21}_2 = s_1$\,,\\
&  $H^{13}_3 = H^{31}_3 = -s_1$\,,  &  $H^{12}_3 =
H^{21}_3=-s_1^2$\,,\\[4pt]
$H(6)^{ij}_k:\quad$
    &  $H^{13}_2 = -H^{31}_2=1$\,,  &  $H^{12}_2 = -H^{21}_2 = s_1$\,,\\
&  $H^{13}_3 = -H^{31}_3 = -s_1$\,,$\qquad$
    &  $H^{12}_3 = -H^{21}_3 = -s_1^2$\,.
\end{tabular}
\end{equation}
In the $SU(3)_F$ limit the effective Hamiltonian for the hadronic decays to 
two pseudoscalars $D \to PP$  can be written as
\bea\label{EffHam}
{\cal H}_{\mbox{eff}~SU(3)} & = & a_{\15bar} D_i H(15)^{ij}_k P^l_j P^k_l +
b_{\15bar} D_i P^i_l H(15)^{lj}_k P^k_j \nonumber \\
 & + & c_6 D_i H(6)^{ij}_k P^l_j P^k_l 
\eea
There are a number of amplitude relations that can be obtained from Eq.~(\ref{EffHam}).
In particular, it can be seen that it implies that 
$\left| A_{D^0 \to K^+ K^-} \right| = \left| A_{D^0 \to \pi^+ \pi^-} \right|$. In practice,
the corresponding branching fractions differ by a factor of three (see Table~\ref{tab:scskpi} 
below). Clearly, $SU(3)_F$ symmetry is broken in $D$-decays.  

A consistent approach should then include $SU(3)_F$-breaking corrections, which could 
consistently be included in the analysis. For example, one could assume that 
$SU(3)_F$  breaking is proportional to light quark masses. In this case, it can be
included in the analysis as a perturbation that transforms as {\bf 8} + {\bf 1}, as the quark 
mass operator belongs to the matrix representation $M^i_j={\rm diag}(m_u,m_d,m_s)$,
which is an {\bf 8}. Note that the $SU(3)_F$ breaking term that transforms as a triplet 
{\bf 3} also breaks isospin, so it is usually neglected in all analyses. A complete 
analysis with broken $SU(3)_F$ is possible~\cite{Savage:1991wu, Hinchliffe:1995hz}, although 
is not quite useful due to a large number of unknown amplitudes.

In some cases one does not need to employ the full formalism of $SU(3)_F$, but only rely on 
its subgroups. An example of such subgroup is isospin. Isospin relations among decay amplitudes 
are much more robust, as isospin breaking is believed to be quite small in charm decays. 
For example, the di-pion modes, $D^+\to\pip\piz$, $\Dz\to\pip\pim$ and $\Dz\to\piz\piz$ are 
related by two isospin amplitudes $A_0$ and $A_2$ corresponding, respectively, to the $S$-wave 
di-pion isospin $I=0$ and $I=2$ states produced
\bea\label{isospinDdecays}
  A^{+0} &=& \sqrt{3\over 2}A_2, \qquad
  A^{+-} = \sqrt{2\over 3}A_0 + \sqrt{1\over 3}A_2 \nonumber\\
  A^{00} &=& \sqrt{1\over 3}A_0 - \sqrt{2\over 3}A_2.
\eea
Some conclusions about strong interaction dynamics in D-meson decays can be reached by
extracting these amplitudes from experimental information. The phases of amplitudes 
in Eq.~(\ref{isospinDdecays}) give an indication of the size of strong interactions among 
decay products in those decays. Following the procedure outlined in \cite{Selen:1993pt}, 
CLEO obtains~\cite{Rubin:2005py} from their results $|A_2/A_0|=0.420\pm 0.014\pm 0.01$ and 
$\arg(A_2/A_0)=(86.4\pm 2.8\pm3.3)^{\circ}$. As one can see, the phase is rather large. It is thus 
clear that final state interactions play an important role in $D$-decays.

Other subgroups of the $SU(3)_F$ also offer useful predictions. For example, the $U$-spin, a 
symmetry of the Lagrangian with respect to $s \to d$ quark interchange, can be employed to 
obtain several useful relations. For example, for the decays of $D^0$ meson into final 
states containing $M^0 =  \pi^0$, $\eta$, and $\eta^\prime$, one can obtain
\beq\label{UspinD0}
\frac{{\cal A}(D^0 \to K^0 M^0)}{{{\cal A}(D^0 \to \overline K^0 M^0)}} = - \tan^2 \theta_C.
\eeq
Equation~(\ref{UspinD0}) derives from the following argument. The initial state, $D^0$ contains 
$c$ and $\bar u$ quarks, and so is a $U$-spin singlet. The CF transition $c \to s u \bar d$
and DCS transition $c \to d u \bar s$ produce $U=1$ finals states with opposite $U_3=1$
in the decays of $D^0$ meson. The final state meson $M^0$ form a linear combination of 
$U$-spin singlet and triplet states, while neutral kaons are $U=1$ , $U_3=\pm 1$ states. Thus, 
$U$-spin triplet part of $M^0$ cannot be produced, as it leads to the $U=2$ final state. Thus, only the
singlet part of $M^0$ can contribute to the transition, which leads to Eq.~(\ref{UspinD0}).

\subsection{Flavor-flow (topological) diagram approach}

Another useful approach to tackle hadronic decays of charmed mesons, equivalent
to the $SU(3)_F$ amplitude method described above, is the flavor-flow (or topological $SU(3)$ 
approach), which involves an overcomplete set of quark diagrams~\cite{Gronau:1994rj,Rosner:1999xd}. 
The application of this method to $D$-decays can even prove advantageous compared to 
flavor $SU(3)$ approach, as the number of unknown amplitudes grows rapidly if $SU(3)_F$-breaking 
is taken into account.

\begin{figure}[tb]
\begin{center}
\includegraphics[width=0.48\linewidth]{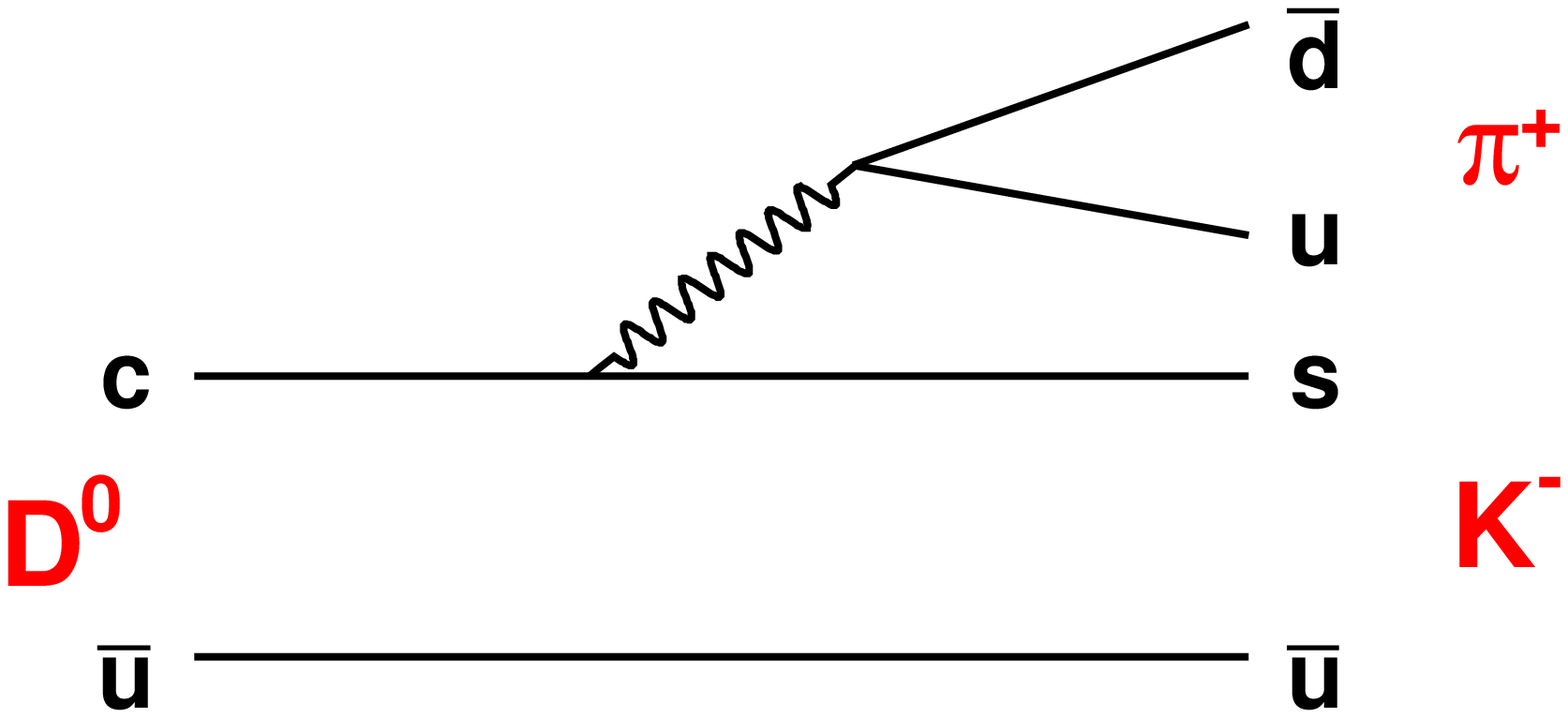}
\includegraphics[width=0.48\linewidth]{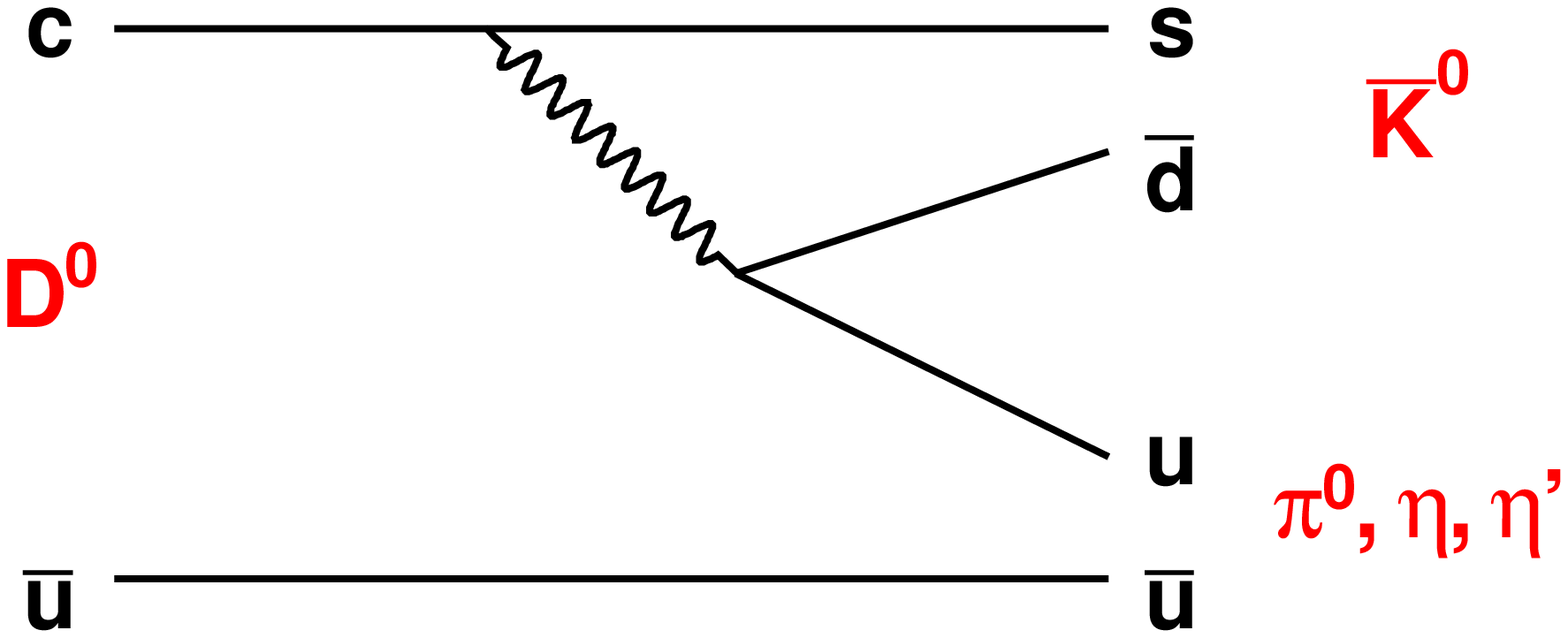}
\includegraphics[width=0.48\linewidth]{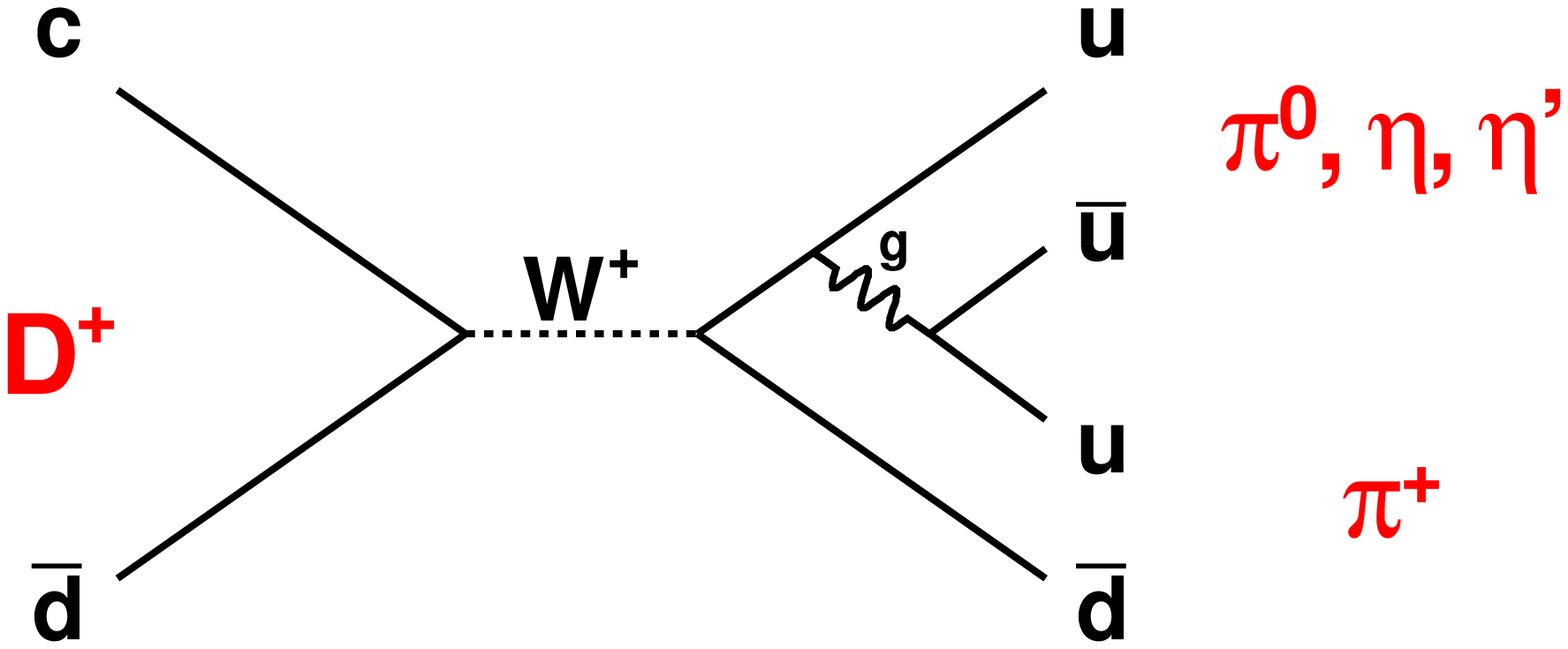}
\includegraphics[width=0.48\linewidth]{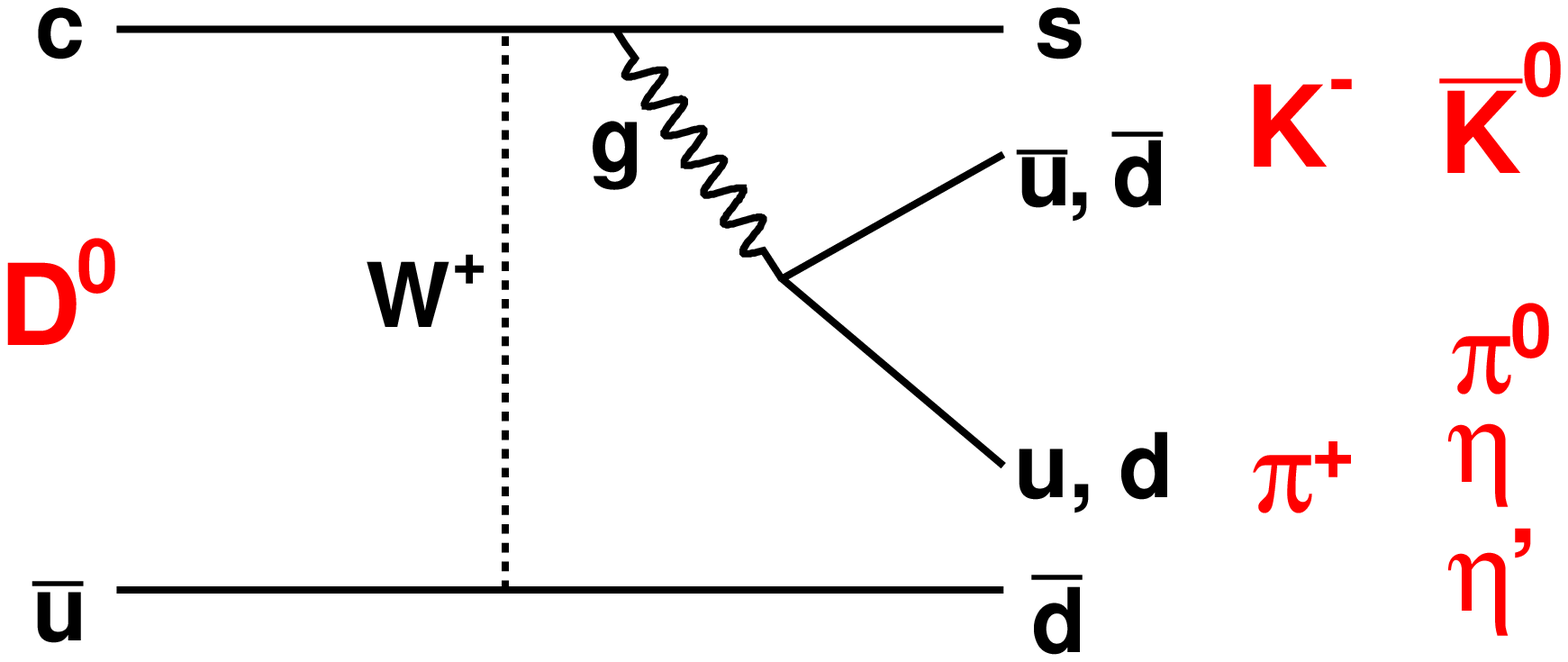}
\includegraphics[width=0.48\linewidth]{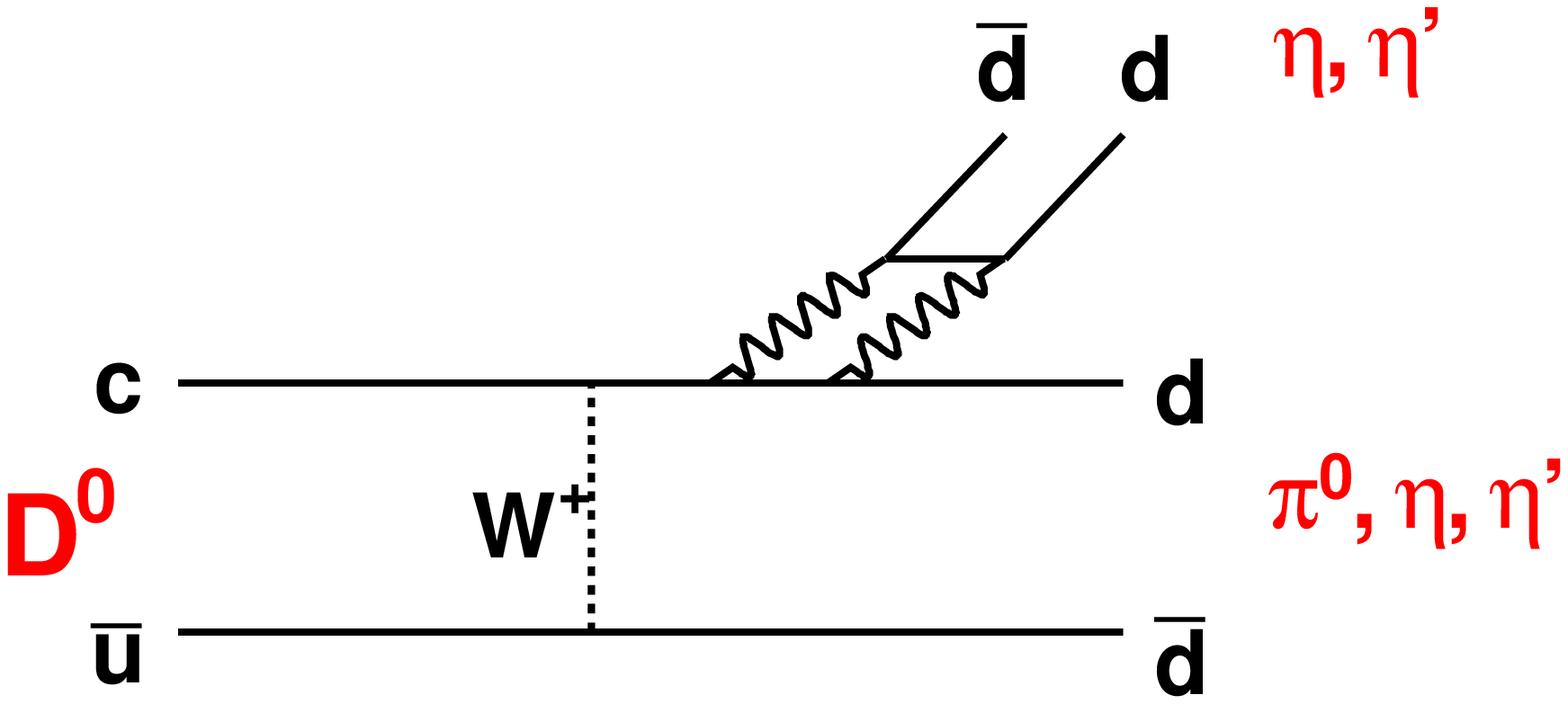}
\caption{
Basic topological amplitudes for D-meson decays. Top row:
tree $T^{(\prime)}$ and color-suppressed $C^{(\prime)}$, middle row:
weak annihilation $A^{(\prime)}$ and weak exchange $E^{(\prime)}$,
bottom row: singlet weak exchange $SE^{\prime}$. CSC amplitudes are
usually denoted by primes.
}
\label{fig:flavor_flow}
\end{center}
\end{figure}

In the topological flavor-flow approach each decay amplitude is parametrized according to the
topology of Feynman diagrams (see Fig.~\ref{fig:flavor_flow}): a color-favored tree amplitude 
(usually denoted by $T$), a color-suppressed tree amplitude ($C$), an exchange amplitude ($E$), 
and an annihilation amplitude ($A$). This set of amplitudes is sufficient for description of CF and DCS 
decays. For SCS decays other amplitudes must be added~\cite{Chiang:2002mr}.

In order to describe charm meson decays in terms of these amplitudes, it is convenient to 
decompose initial and final states according to their isospin structure. For instance, in the
notation of ~\cite{Rosner:1999xd}, the following phase conventions are used:
\begin{enumerate}\addtolength{\itemsep}{-0.5\baselineskip}
\item
Charmed mesons: 
$D^0 = -c \overline u$, $D^+ = c \overline d$,  and $D_s = c \overline s$.
\item
Pseudoscalar mesons:
$\pi^+ = u \overline d$, $\pi^0 = \left(u \overline u - d \overline d\right)/\sqrt{2}$,
$\pi^- = -d \overline u$, $K^+ = u \overline s$, $K^0 = d \overline s$, 
$\overline K^0 = s \overline d$, $K^- = - s \overline u$,
$\eta = \left(s \overline s - u \overline u - d \overline d\right)/\sqrt{3}$, and
$\eta^\prime = \left(u \overline u + d \overline d - 2 s \overline s\right)/\sqrt{6}$.
\item
Vector mesons:
$\rho^+ = u \overline d$, $\rho^0 = \left(u \overline u - d \overline d\right)/\sqrt{2}$,
$\rho^- = -d \overline u$,  $\omega^0 = \left(u \overline u + d \overline d\right)/\sqrt{2}$,
$K^{*+} = u \overline s$, $K^{*0} = d \overline s$, 
$\overline K^{*0} = s \overline d$, $K^{*-} = - s \overline u$, and
$\phi = s \overline s$.
\end{enumerate}
As with the $SU(3)_F$ approach, this method does not provide absolute predictions for the 
branching fractions in D-meson decays. However, it provides relations among several 
decay amplitudes by matching the quark-level "flavor topology" graphs with the final 
states defined above. For example, a DCS transition $D^0 \to K^+\pi^-$ can proceed via 
a tree-level amplitude $T (c \to u \overline s d)$ and an exchange amplitude 
$E (c \overline u \to \overline s d)$. Matching those with the initial state meson
$D^0 = -c \overline u$ and final state mesons $K^+ = u \overline s$ and $\pi^- = -d \overline u$, 
one obtains the following amplitude relation,
\beq\label{FlavorFlow1}
A(D^0 \to K^+\pi^-) = T+E \equiv \frac{G_F}{\sqrt{2}} V_{ud} V_{cs}^* 
\left( {\cal T} + {\cal E}\right),
\eeq
where we use calligraphic notation for the amplitudes with $G_F/\sqrt{2}$ and CKM-factors removed.
Similarly, for other transitions one obtains
\bea\label{FlavorFlow2}
A(D^0 \to K^0 \pi^0) &=& \frac{1}{\sqrt{2}} (C-E) \nonumber \\
   & = &
 \frac{1}{\sqrt{2}} \frac{G_F}{\sqrt{2}} V_{us} V_{cd}^*  \ ({\cal C}^{\prime\prime}- {\cal E}^{\prime\prime}), 
\nonumber \\
A(D^0 \to \overline K^0 \pi^0) &=& \frac{1}{\sqrt{2}} (C-E) \nonumber \\
 & = & 
 \frac{1}{\sqrt{2}} \frac{G_F}{\sqrt{2}} V_{ud} V_{cs}^*  \ ({\cal C}- {\cal E}), 
\nonumber \\
A(D^+ \to K^0 \pi^+) &=& C+A = 
 \frac{G_F}{\sqrt{2}} V_{us} V_{cd}^*  \ ({\cal C}^{\prime\prime} + {\cal A}^{\prime\prime}), 
\nonumber \\
A(D^+ \to \overline K^0 \pi^+) &=& T+C = 
\frac{G_F}{\sqrt{2}} V_{ud} V_{cs}^* \  ({\cal T}+ {\cal C}), 
\nonumber \\
A(D^0 \to K^0 \eta) &=& \frac{1}{\sqrt{3}} C  = 
\frac{1}{\sqrt{3}} \frac{G_F}{\sqrt{2}} V_{us} V_{cd}^* \  {\cal C}^{\prime\prime},
\eea
and so on. Note that in Eq.~(\ref{FlavorFlow2}) we denoted DCS amplitudes with double primes. 
Singly-Cabibbo-suppressed amplitudes are conventionally denoted by a single prime. CF amplitudes
can be related to SCS and DCS amplitudes by proper scaling with $\tan\theta_C$. We shall give
particular examples below. 

One reason for the employed phase convention is a requirement that $SU(3)_F$ sum rules 
are satisfied. For example, for transitions $D^+ \to K^+\pi^0$, $D^+ \to K^+\eta$, and 
$D^+ \to K^+\eta^\prime$, a sum rule
\beq
3 \sqrt{2} A(K^+\pi^0) + 4 \sqrt{3} A(K^+\eta)+\sqrt{6} A(K^+\eta^\prime) = 0
\eeq
can be written. With the flavor-flow parameterization,
\bea
A(D^+ \to \overline K^+ \pi^0) &=& \frac{1}{\sqrt{2}} \left(T-A\right),
\nonumber \\
A(D^+ \to \overline K^+ \eta) &=& - \frac{1}{\sqrt{3}} \ T
\\
A(D^+ \to \overline K^+ \eta^\prime) &=& \frac{1}{\sqrt{6}} \left(T + 3A\right)
\nonumber
\eea
the above sum rule gives $3(T-A)-4T+(T+3A)=0$.

Thus, provided that a sufficient number of decay modes is measured, one 
can predict both branching fractions and amplitude phases for a number of transitions.
Still, no prediction for absolute branching ratios are possible in this approach.

\subsection{Factorization ansatz}

The simplest way to estimate an absolute decay rate of a charmed meson is to employ a 
factorization ansatz. This ansatz implies that the amplitude for the hadronic transition 
can be written as a product of known form-factors. Schematically, 
\bea\label{NaiveFact}
{\cal A}(D_q \to M_1 M_2) & = & \langle M_1, M_2 | {\cal H} | D_q \rangle
\\ 
& \sim & \langle M_1 | \left(\overline{u}_k \Gamma^\mu q_k\right) | 0 \rangle \times
\langle M_2 | \left(\overline{q}_i \Gamma_\mu c_i\right) | D_q \rangle
\nonumber
\eea
This is a clear simplification, as the first non-perturbative parameter
$\langle M_1 | \left(\overline{u}_k \Gamma^\mu q_k\right) | 0 \rangle$  can be written  
in terms of a meson decay constant $f_{M_1}$,
\beq\label{DecayConst}
\langle M_1 | \bar u\gamma^\mu \gamma_5 q | 0 \rangle = i f_{M_1} p_{M_1}^\mu,
\eeq
which parameterizes the amplitude of probability for quarks to ``find each other'' in 
a light mesons and can be measured in leptonic decays of $M_1$, 
\begin{equation}
\Gamma(M_1\to \ell\nu) = {G_F^2\over
8\pi}f_{M_1}^2m_{\ell}^2 m_{M_1}  \left(1-{m_{\ell}^2\over
m_{M_1}^2}\right)^2 \left|V_{uq}\right|^2~~~,
\end{equation}
where $m_{M_1}$ is the $M_1$ mass, $m_{\ell}$ is the mass of the final state lepton,
and $|V_{uq}|$ is the CKM matrix element associated with the $q \to u$ transition. 
The decay constants can also be computed in lattice gauge theories or using other
non-perturbative approaches (see \cite{Artuso:2008vf} for review).

The second non-perturbative parameter, 
$\langle M_2 | \left(\overline{q}_i \Gamma_\mu c_i\right) | D_q \rangle$,
is related to form-factors that can be extracted from semileptonic $D_q$ decays,
\begin{equation}
\frac{d\Gamma(D\to M_2 e \nu_e)}{dq^2} =
\frac{G_F^2|V_{cq}|^2}{24\pi^3} \left |{\bf p}_{M_2} \right |^3 |f_+(q^2)|^2
\end{equation}
where ${\bf p}_{M_2}$ is the hadron 3-momentum in the $D$ rest frame.

Theoretical parameterizations of semileptonic decays involve two
non-perturbative quantities parameterizing the matrix element of a
single hadronic current. Traditionally, the hadronic matrix elements
for transitions to pseudoscalar hadrons are described in terms of
two form factors, $f_+(q^2)$ and $f_-(q^2)$,
\beq
\langle M_2 | \bar q \Gamma^\mu c | D \rangle =
f_+^{D\to M_2}(q^2) P^\mu + f_-^{D\to M_2}(q^2) q^\mu,
\eeq
where $P=p_D+p_{M_2}$ and $q=p_D-p_{M_2}$. An alternative
parameterization is often used,
\bea
\langle M_2 | \bar q \Gamma^\mu c | D \rangle & = &
\left(P^\mu-\frac{m_D^2-m_{M_2}^2}{q^2} q^\mu \right) f_+^{D\to M_2}(q^2)
\nonumber \\
& + &
\frac{m_D^2-m_{M_2}^2}{q^2} q^\mu f_0^{D\to M_2}(q^2),
\eea
with $f_0^{D\to M_2}(q^2)=f_+^{D\to M_2}(q^2)+f_-^{D\to M_2}(q^2) q^2/(m_D^2-m_{M_2}^2)$.
Form factors have been evaluated at specific $q^2$ points in a variety of
phenomenological models, where the shape is
typically assumed from some model arguments~\cite{Artuso:2008vf}. 

Clearly, naive factorization of Eq.~(\ref{NaiveFact}), while convenient,  cannot be correct, as
it assumes that scale and scheme dependence of a product of quark bilinears is the same as 
that of a four-fermion operator, which it is not. The situation can in principle be corrected, at 
least in the heavy-quark limit. In $B$-decays, a QCD factorization formula has been written that 
takes into account perturbative QCD corrections~\cite{Beneke:1999br}. It is however not clear that 
this approach is applicable to charm decays, as charm quark might be too light for this approach 
to be applicable. Nevertheless, even naive factorization provides a convenient way to estimate 
$D$-meson decay rates.

Besides decay amplitudes for $D$-mesons, which can be computed using the factorization arguments 
above, both flavor-flow and $SU(3)_F$ amplitudes can also be estimated. For example, 
contrary to the relation Eq.~(\ref{UspinD0}), the corresponding relation for charged $D$-meson decays,
\beq\label{DplusRatio}
\frac{{\cal A}(D^+ \to K^0 \pi^+)}{{{\cal A}(D^+ \to \overline K^0 \pi^+)}} = - \tan^2 \theta_C
\frac{{\cal C}^{\prime\prime}  + {\cal A}^{\prime\prime}}{{\cal C} + {\cal T}} =
\frac{{\cal C} + \left(C_2/C_1\right) {\cal E}}{{\cal C} + {\cal T}},
\eeq
cannot be fixed by symmetry arguments alone. However, the factorization approach can be used
to estimate this ratio. In particular,
\bea\label{FactorAmps}
{\cal T} &=& f_\pi \left(m_D^2-m_K^2\right) f_+^{D\to K}(m_\pi^2) \ a_1,
\nonumber \\
{\cal C} & = & f_K \left(m_D^2-m_\pi^2\right) f_+^{D\to \pi}(m_\pi^2) \ a_2,
\nonumber \\
{\cal T}^{\prime\prime}  &=& f_K \left(m_D^2-m_\pi^2\right) f_+^{D\to \pi}(m_K^2) \ a_1,
\nonumber \\
{\cal C}^{\prime\prime}  & = & f_\pi \left(m_D^2-m_K^2\right) f_+^{D\to K}(m_\pi^2) \ a_2
\eea
where $a_{1,2}=C_{1,2}+C_{2,1}/N_c$. 
Note that some analyses employ $a_{1,2} \to a_{1,2}^{eff}$, which are fitted from the data and treated as 
universal fit parameters. This way of calculating charm hadronic decay matrix elements is sometimes 
called "modified factorization" approach. The argument for doing this is an attempt to include unknown 
non-perturbative corrections to Eq.~(\ref{FactorAmps}). While this approach defines a convenient model to 
deal with hadronic decays, there is no reason to believe that soft contributions are universal in all transitions. 

Calculations of ${\cal E}$ and ${\cal A}$ amplitudes in factorization are much more complicated. It has been 
argued~\cite{Gao:2006nb} that they can be estimated using methods similar to those employed in 
B-decays~\cite{Beneke:1999br}. Numerically, the calculation of the ratio of Eq.~(\ref{DplusRatio}) amounts to
\beq\label{DplusRatioNum}
\frac{{\cal A}(D^+ \to K^0 \pi^+)}{{{\cal A}(D^+ \to \overline K^0 \pi^+)}} = - \tan^2 \theta_C r_s e^{i \phi_s},
\eeq
with $r_s \approx 1.521$ and $\phi_s \approx 103^o$ for $C_2/C_1 \approx -0.5$.  This ratio will be used to
estimate decay asymmetries with kaons later in this paper.

\section{CABIBBO FAVORED $D^0$ AND $D^+$ DECAYS AND REFERENCE BRANCHING FRACTIONS}
\label{sect:dref}

The absolute branching fractions for decays of the ground
state charmed mesons are important as they are used to normalize
many $B$ and $D$ meson decays. For example, the determination
of $|V_{cb}|$ from $B\to D^*\ell\nu$~\cite{Richman:1995wm} depends
directly on the determination of the $D$ branching fractions 
used to reconstruct the final state. 

To measure the absolute branching fractions we need to have 
a mechanism to determine the number of $D$ mesons produced. As
the cross-sections for producing $D$ mesons are not directly calculable
we have to count the $D$ mesons in the data sample. Broadly
speaking there are two methods employed for this $D$ counting.
At threshold MARK III and CLEO-c have used a tagging technique described in
Sect.~\ref{sect:exp_threshold}, where one $D$ meson is fully reconstructed
and tag the existence of another $\bar D$ in the event. 
At higher energies the presence of a
$D^{*+}$ meson can be tagged using the slow pion in the $D^{*+}\to D^0\pi^+$
decay. The slow pion in this decay is often denoted $\pi_s$. 
This 'slow pion' tagging technique has been used by several
experiments including CLEO and ALEPH to count the number of 
$D^{*+}\to D^0\pi^+$ decays
in charm jets produced in $e^+e^-$ collisions. 
A variation of this idea has been used by ARGUS, CLEO,
and BABAR where $D^*$ mesons produced in semileptonic $B$ decays,
$\bar B^0\to D^{*+}\ell^-\bar\nu$, are tagged by the presence of a slow
pion and a lepton. 
These different techniques are discussed in this section.

Before the CLEO-c measurement of the $D^+\to K^-\pi^+\pi^+$
branching fraction
using tagging as described in Sect.~\ref{sect:cleocdhad} 
there was a statistics limited study by MARK III~\cite{Adler:1987as}
and model dependent analyses. CLEO~\cite{Balest:1994gs} 
reconstructed the two decay chains $D^{*+}\to D^0\pi^+,\ D^0\to K^-\pi^+$
and $D^{*+}\to D^+\pi^0,\ D^+\to K^-\pi^+\pi^+$. This allowed
CLEO to measure the ratio of produced $K^-\pi^+\pi^+$ to $K^-\pi^+$ final
states, which can be expressed as
$$
{N_{K\pi\pi}\over N_{K\pi}} = { 
{\cal B}(D^{*+}\to D^+\pi^0){\cal B}(D^+\to K^-\pi^+\pi^+)\epsilon(K\pi\pi) \over
{\cal B}(D^{*+}\to D^0\pi^+){\cal B}(D^0\to K^-\pi^+)\epsilon(K\pi)
},
$$
where $\epsilon(K\pi\pi)$ and $\epsilon(K\pi)$
are the efficiencies for reconstructing the $D^+\to K^-\pi^+\pi^+$
and $D^0\to K^-\pi^+$ final states, respectively, including the $D^{*+}$.
To extract the $D^+\to K^-\pi^+\pi^+$ branching fraction CLEO used
the measured $D^0\to K^-\pi^+$ branching fraction and the ratio
$$
{{\cal B}(D^{*+}\to D^+\pi^0) \over {\cal B}(D^{*+}\to D^0\pi^+)}.
$$
The determination of this ratio is discussed by 
\textcite{Butler:1992rp,Bartelt:1997yu} and ultimately 
relies on isospin conservation. Though the errors are expected to 
be small they are hard to quantify.

\subsection{Absolute $D^0$ branching fractions using slow pion tagging}

The method of tagging $D^{*+}\to D^0\pi^+$ decays in
jets produced in $e^+e^-\to c\bar c$ interactions by the presence of 
a slow pion from the $D^*$ decay is sometimes referred to as the
HRS technique after the first experiment that used this
method. 
As the 
$Q$ value of the $D^{*+}\to D^0\pi^+$ decay is only about 
5 MeV and the produced pion has a momentum of only 39 MeV
in the $D^*$ restframe it can at most contribute this amount
to the transverse momentum with respect to the thrust axis.
Experimentally, the slow pion from the $D^{*+}$
decay closely follows the original $D^*$ direction. 
Due to the soft track associated with this decay, the pion tends
to bend out from the jet in the 
magnetic field of the tracking system.

\begin{figure}[tb]
\begin{center}
\includegraphics[width=0.9\linewidth]{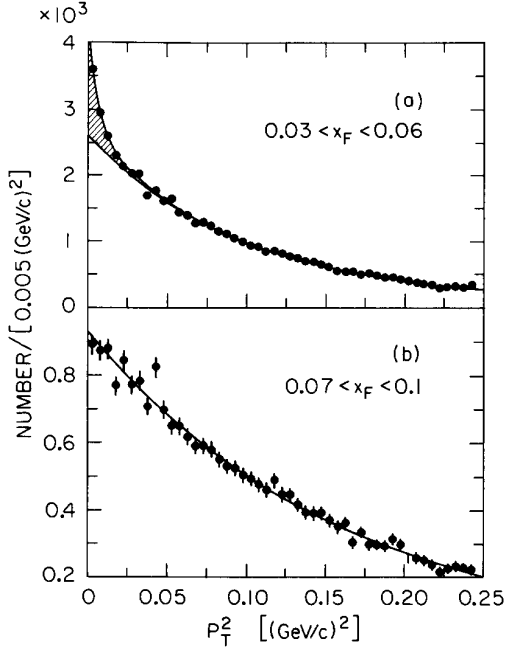}
\caption{The $p_t^2$ distribution for data from the HRS
collaboration. In (a) the fractional slow pion momentum in
the range $0.03<x_F<0.06$ is shown while in (b) the 
fractional momentum range $0.07<x_F<0.1$ is shown. In
the low momentum range where we expect slow pions from
$D^*$ decays a clear excess at very low $p_t^2$ is
seen. 
From~\textcite{Abachi:1988uc}. 
}
\label{fig:HRS_pt}
\end{center}
\end{figure}

The HRS experiment~\cite{Abachi:1988uc} used 300 \pbinv\ of data collected at 
$E_{\rm cm}=29$ GeV. For candidate slow pions the transverse momentum,
$p_T$, is calculated with respect to the thrust axis
determined from the particles
in the opposite hemisphere with respect to the slow pion candidate under
consideration. The choice of using only tracks in the opposite
hemisphere for the calculation of the thrust axis is to avoid
any possible bias due to the decay of the $D$ meson.
In Fig.~\ref{fig:HRS_pt} the $p_T^2$
distribution is shown in two ranges of the fractional momentum
$x_F=2p_{\parallel}/E_\mathrm{cm}$ of the slow pion, where $p_{\parallel}$
is the component of the slow pion momentum that is parallel to the
thrust axis.
In the low fractional
momentum range ($0.03<x_F<0.06$) a clear excess is seen at very low
values of the transverse momentum due to slow pions
from $D^{*+}\to D^0\pi^+$ decays. This excess is not present in the 
higher $x_F$ range as slow pions from $D^{*+}$ decays do not
populate this range. The HRS collaboration use the excess
at low $p^2_t$ to determine that they had $1584\pm 110$
$D^{*+}\to D^0\pi^+$ decays in their sample. Next a $D^0$ is
reconstructed in the $D^0\to \Km\pip$ channel. The $\Dz$
candidate is combined with the slow pion and the mass difference
$M_{K\pi\pi_s}-M_{K\pi}$ is required to be in the range 0.143
to 0.148 $\Gevcsq$. The yield is determined by fitting the 
$M_{K\pi}$ mass distribution. A total of $56\pm 9$ events
were observed. The efficiency for finding the $K\pi$ pair,
given that the $\pi_s$ is found, is determined to be 79\%
giving a branching fraction of 
${\cal B}(\Dz\to\Km\pip)=(4.5\pm0.8\pm0.5)\%$. The largest
systematic uncertainty quoted is bias due to event selection
criteria. This uncertainty is evaluated by changing the
event selection criteria to remove the thrust and collinearity 
criteria
used. The analysis was limited by statistics.

The same technique as pioneered above by the HRS collaboration
has been used by ALEPH~\cite{Decamp:1991jw,Barate:1997mm}, 
CLEO~\cite{Akerib:1993pm},
and ARGUS~\cite{Albrecht:1994nb}.
ALEPH used a sample of $e^+e^-$ data collected from
1991 to 1994 at LEP near the $Z$ pole. CLEO and ARGUS used 
samples of $1.79$ \fbinv\ and $355$ \pbinv\ respectively 
of $e^+e^-$ data collected near the $\Upsilon(4S)$ resonance. 

\begin{figure}[tb]
\begin{center}
\includegraphics[width=0.99\linewidth]{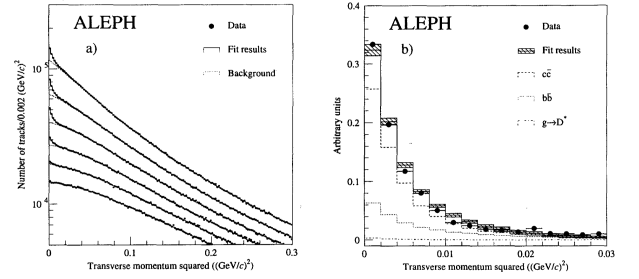}
\caption{The $p_t^2$ distribution for data from the ALEPH
experiment. In (a) the transverse slow pion momentum squared in
six equal momentum bins from 1.0 to 4.0 $\Gevc$. The 1.0 to 1.5 $\Gevc$
momentum bin is the uppermost and the 3.5 to 4.0 $\Gevc$ bin
is the lowest. The slow pion from $D^{*+}\to D^0\pip$ is
clearly visible in the lower momentum range.
In (b) the transverse momentum distributions from different sources
of $D^*$ mesons are shown. 
From~\textcite{Barate:1997mm}.
}
\label{fig:ALEPH_pt}
\end{center}
\end{figure}

ALEPH followed the HRS approach closely. They analyzed the data in
six ranges of the slow pion momentum, from 1.0 to 4.0 $\Gevc$.
The transverse momentum squared distributions in the six
momentum bins are shown in Fig.~\ref{fig:ALEPH_pt}. A 
$\Dz\to\Km\pip$ candidate is searched for in events with
a slow pion, and candidates where $0.1435<M_{K\pi\pi_s}-M_{K\pi}<0.1475$
GeV are accepted. In Table~\ref{tab:ALEPH_DtoKpi} the
yields and branching fractions from the ALEPH analysis
are summarized. The results from the different momentum bins
are combined, including correlations, to obtain the 
final result
$$
{\cal B}(\Dz\to\Km\pip)=(3.90\pm 0.09\pm 0.12)\%.
$$
This result includes corrections (1.9\%) due to final state radiation.
The largest systematic uncertainties come from the background
shape in extracting the inclusive $D^*$ yield and the modeling
of the angle between the $D^*$ and the jet thrust axis.

\begin{table*}[bt]
\caption{Event yields and branching fractions for the ALEPH
study of the $D^0\to K^-\pi^+$ decay in bins of the slow
pion momentum. The first column is the momentum range, the
second and third columns show the $D^{*+}\to \Dz\pip$ yield
determined from the slow pion transverse momentum and the
$\Dz\to \Km\pip$  yields, respectively. The last column shows
the $D^0\to K^-\pi^+$ branching fraction.
}
\label{tab:ALEPH_DtoKpi}
\begin{center}
\begin{tabular}{lccc}
\hline\hline
Momentum Range & $N_{D^{*+}\to \Dz\pip}$ & $N_{\Dz\to \Km\pip}$ &
   ${\cal B}(D^0\to K^-\pi^+)$ (\%) \\
($\Gevc$) &  & & \\
\hline
1.0---1.5     &   $79,038.2\pm 2,021.9 \pm 12,018.0$ & $2,472.9\pm55.5\pm 11.0$
 &  $4.400\pm 0.150\pm 1.041$ \\
1.5---2.0     &   $56,393.2\pm 1,140.4 \pm    921.6$ & $1,558.3\pm41.4\pm  5.4$
 &  $3.990\pm 0.133\pm 0.139$ \\
2.0---2.5     &   $35,303.4\pm   855.8 \pm    842.2$ & $  913.8\pm30.9\pm  2.8$
 &  $3.768\pm 0.157\pm 0.150$ \\
2.5---3.0     &   $12,287.8\pm   674.7 \pm    535.1$ & $  321.5\pm18.2\pm  1.3$
 &  $3.758\pm 0.296\pm 0.206$ \\
3.0---3.5     &   $ 3,497.4\pm   499.2 \pm    630.4$ & $  115.7\pm10.9\pm  0.7$
 &  $5.010\pm 0.857\pm 1.228$ \\
3.5---4.0     &   $   192.4\pm   366.8 \pm    401.5$ & $    9.8\pm 3.3\pm  0.4$
 &  $7.44\pm 14.2\pm 19.4$ \\
\hline\hline
\end{tabular}
\end{center}
\end{table*}

\begin{figure}[tb]
\begin{center}
\includegraphics[width=0.9\linewidth]{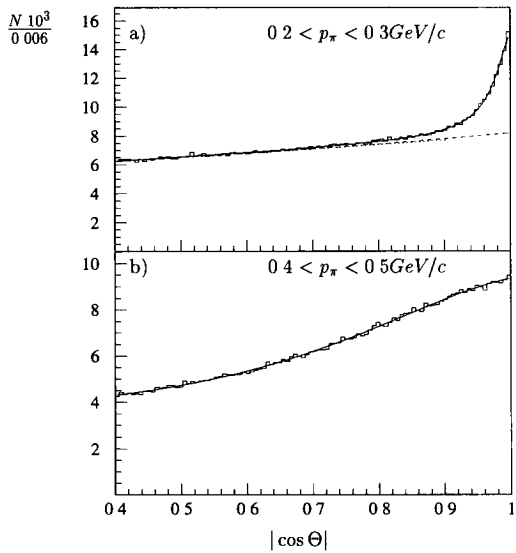}
\caption{The $|\!\cos\theta|$ distribution for data from the ARGUS
experiment. In (a) the distribution is shown for the slow
pion momentum in the range 0.2 to 0.3 $\Gevc$ and in 
(b) for the range 0.4 to 0.5 $\Gevc$.
From~\textcite{Albrecht:1994nb}.
}
\label{fig:ARGUS_costheta}
\end{center}
\end{figure}

ARGUS used the same technique to count $D^{*+}\to \Dz\pip$
decays. To extract the $D^{*+}\to \Dz\pip$ yield ARGUS
plot the distributions of $|\!\cos\theta|$ where $\theta$
is the angle between the slow pion candidate and the thrust
axis of the jet in the opposite hemisphere. 
Figure~\ref{fig:ARGUS_costheta} shows the $|\!\cos\theta|$
distribution in two ranges of the slow pion momentum.
In the momentum range 0.2 to 0.3 $\Gevc$ a clear excess of events
near $|\!\cos\theta|=1$ is seen from $D^{*+}\to \Dz\pip$ decays.
In the range 0.4 to 0.5 $\Gevc$ no excess is seen as this
is above the momentum where we have slow pions from $D^{*+}$
decays. From a fit to the $|\!\cos\theta|$ distribution ARGUS
determines a yield of $51,327\pm757$ $D^{*+}\to \Dz\pip$ decays
in the sample. The systematic uncertainty on this yield
is estimated to be 5.9\% by varying the signal shape parameterization.
ARGUS reconstructs the $D^0$ in three channels and
determines the following branching fractions
\begin{eqnarray*}
{\cal B}(\Dz\to \Km\pip)&=&(3.41\pm0.12\pm0.28)\% \\
{\cal B}(\Dz\to \Km\pip\pim\pip)&=&(6.80\pm0.27\pm0.57)\% \\
{\cal B}(\Dz\to \bar{K^0}\pim\pip)&=&(5.03\pm0.39\pm0.49)\% \\
\end{eqnarray*}

The CLEO~\cite{Akerib:1993pm} study is very similar to the
ARGUS analysis. CLEO only studied the final state 
$D^0\to K^-\pi^+$. They tagged $165,658\pm1,149$ $D^{*+}\to D^0\pi^+$
decays and measured the branching fraction
$$
{\cal B}(D^0\to K^-\pi^+)=3.95\pm0.08\pm0.17\%.
$$
This includes a correction of about 1\% for the effects of
final state radiation. The largest contribution to the
systematic uncertainty $(\pm 4.0\%)$ comes from the track
reconstruction efficiency for the final $K\pi$ system.

These measurements are limited by systematic uncertainties
on the determination of the number of $D^{*+}\to D^0\pi^+$ 
decays in the data sample. The yield is extracted by extrapolating
the background into the signal region based on shapes
determined from Monte Carlo simulations.

\subsection{Tagging with $\bar B^0\to D^{*+}\ell^-\bar\nu$}

Tagging semileptonic $B$ decays with the presence of
a lepton plus a slow pion was first used by ARGUS~\cite{Albrecht:1993gr}
and has since been used by CLEO~\cite{Artuso:1997mc} and
most recently BABAR~\cite{Aubert:2007wn}. The BABAR analysis uses the largest
data sample, 210 fb$^{-1}$ of $e^+e^-$ data
collected at the $\Upsilon(4S)$. 

In the first study that used this technique ARGUS used
a sample of 246 \pbinv\ of $e^+e^-$ data
collected at the $\Upsilon(4S)$ containing $209,000\pm 9,500$
$B\bar B$ pairs. They obtained the branching fractions
\begin{eqnarray*}
{\cal B}(D^0\to K^-\pi^+) & = & (4.5\pm 0.6\pm 0.4)\%,\\
{\cal B}(D^0\to K^-\pi^+\pi^+\pi^-) & = & (7.9\pm 1.5\pm 0.9)\%.
\end{eqnarray*}
This measurement is clearly statistics limited, ARGUS
reconstructed a sample of $2,693\pm 183\pm 105$ $D^{*+}\to D^0\pi^+$
candidates. 

CLEO used a sample of 3.1 \fbinv\ of $e^+e^-$ data
collected at the $\Upsilon(4S)$ containing $3.3\times 10^6$
$B\bar B$ events. A sample of 1.6 \fbinv\ of data collected
below the $\Upsilon(4S)$ resonance was used for continuum
subtraction. CLEO reconstructs $44,504\pm360$ inclusive
events and $1,165\pm45$ exclusive $D^0\to K^-\pi^+$
decays and determines a branching fraction
$$
{\cal B}(D^0\to K^-\pi^+)=(3.81\pm0.15\pm0.16)\%.
$$
This branching fraction does not include radiative corrections.

\begin{figure}[tb]
\begin{center}
\includegraphics[width=0.9\linewidth]{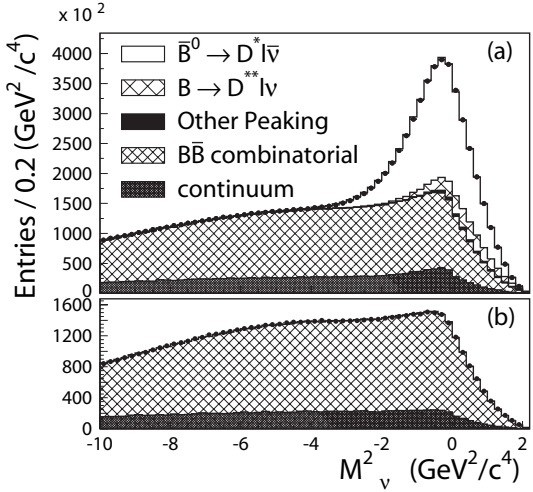}
\caption{The distribution of the missing mass squared, $M^2_{\nu}$
for (a) right sign events and (b) wrong sign events. The wrong sign
events show that the simulation of the background shape is good. 
From~\textcite{Aubert:2007wn}.
}
\label{fig:babar_m2nu}
\end{center}
\end{figure}

BABAR used 210 fb$^{-1}$ of $e^+e^-$ data
collected at the $\Upsilon(4S)$ resonance, corresponding to $230\times 10^6$
$B\bar B$ pairs, and 22 \fbinv\ collected 40 MeV below the resonance.
The offresonance sample is used to subtract non-$B\bar B$ backgrounds. 
In this analysis the semileptonic 
$B$ decay, $\bar B^0\to D^{*+}\ell^-\bar\nu$ followed
by $D^{*+}\to D^0\pi^+$ is used. BABAR use the lepton
in the $B$ decay and the slow pion from the $D^*$ to 
count $\bar B^0\to D^{*+}\ell^-\bar\nu$ decays followed
by $D^{*+}\to D^0\pi^+$. BABAR used both electrons and muons
in the momentum range $1.4<|{\bf p}|_{\ell}<2.3$ GeV/$c$. For the soft
pion candidate the momentum is in the range $60<|{\bf p}|_{\pi_s}<190$ MeV/$c$.
As the energy release in the $D^{*+}\to D^0\pi^+$ decay
is very small the reconstructed slow pion direction
is used to approximate the direction of the $D^{*+}$.
The momentum magnitude of the $D^{*+}$ is parameterized as
a linear function of the slow pion momentum. Using this
estimate of the $D^{*+}$ momentum, the missing mass squared of the
neutrino is approximated as
$$
M^2_{\nu}=(E_{\rm beam}-E_{D^*}-E_{\ell})^2-({\bf p}_{D^*}+{\bf p}_{\ell})^2,
$$
where $E_{\rm beam}$ is half the center-of-mass energy and the
momentum of the $B$ is taken to be zero. 
The energies and momenta in this expression are evaluated 
in the $e^+e^-$ center-of-mass frame.
For signal candidates
it is required that the charge of the slow pion and the lepton
are opposite. For background studies BABAR considers same-charge
candidates.
BABAR extracts the number of $\bar B^0\to D^{*+}\ell^-\bar\nu$
decays using the missing mass squared, $M^2_{\nu}$, against the
$D^*$ and the lepton. 
Besides the $\bar B^0\to D^{*+}\ell^-\bar\nu$ signal events there
are a few additional sources of events that peaks near zero in the 
missing mass squared. BABAR includes these events as signal
candidates, they include 
1) $\bar B\to D^{*+}(n\pi)\ell^-\bar\nu$ (``$D^{**}$'') where $n\ge 1$; 
2) $\bar B\to D^{*+}\bar D$, $\bar D\to \ell^-X$;
3) $\bar B^0\to D^{*+}\tau^-\bar\nu$, 
$\tau^-\to \ell^-\bar\nu_{\ell}\nu_{\tau}$ (``cascade''); 
4) $\bar B^0\to D^{*+}h^-$ (``fake-lepton'') , 
where $h^-$ is a kaon or pion that
has been misidentified as a lepton.
The $M^2_{\nu}$ distributions are shown in Fig.~\ref{fig:babar_m2nu}.
A clear signal is observed for $M^2_{\nu}>-2.0$ GeV$^2$. 
However, there are substantial backgrounds from combinatorics
in $B\bar B$ events and in continuum production that need
to be subtracted. The continuum
background is modeled using offresonance data and the 
$B\bar B$ combinatorial background, as well as the signal components,
are modeled using Monte Carlo simulations. The signal yields 
are extracted from fits to the $M^2_{\nu}$ distributions in 
the range from $-10.0$ to 2.5 GeV$^2$. The data are divided into ten
different lepton momentum ranges to reduce sensitivity to the 
Monte Carlo simulation. In each lepton momentum bin the 
continuum yields are fixed by scaling the off-resonance sample
to the luminosity of the on-resonance sample; while the
number of events from primary signal, $D^{**}$, and
combinatorial $B\bar B$ are independently varied. The contributions
from cascades and fake-leptons are fixed from the simulation.
These two contributions account for about 3\% of the total
inclusive signal.

Table~\ref{tab:babar_bkgd}
summarizes the event yields for the inclusive 
$\bar B^0\to D^{*+}\ell^-\bar\nu$ reconstruction in the column 'Inclusive'.
BABAR finds $N^{\rm incl}=2,170,640\pm 3,040\pm 18,100$ $\bar B^0\to D^{*+}\ell^-\bar\nu$
decays followed by $D^{*+}\to D^0\pi^+$ in their data sample.

\begin{table}[bt]
\caption{Event yields for the inclusive $\bar B^0\to D^{*+}\ell^-\bar\nu$
reconstruction and the exclusive analysis where the $D^0\to K^-\pi^+$
final state is reconstructed in the BABAR analysis to determine the 
branching fraction for $D^0\to K^-\pi^+$ decay.
Errors are only statistical.
}
\label{tab:babar_bkgd}
\begin{center}
\begin{tabular}{lcc}
\hline\hline
Source & Inclusive ($\times 10^6$) & Exclusive ($\times 10^4$)  \\
\hline
Data                       & $4.4124\pm 0.0021$         & $4.727\pm 0.022$ \\
Continuum                  & $0.46\pm 0.0021)$          & $0.309\pm 0.017$ \\
Combinatorial $B\bar B$    & $1.7817\pm 0.0007$         & $0.819\pm 0.005$ \\
Peaking                    &                            & $0.163\pm 0.008$ \\
Cabibbo suppressed         &                            & $0.055\pm 0.001$ \\
\hline
Signal                     & $2.1706\pm 0.0030$         & $3.381\pm 0.029$ \\
\hline\hline
\end{tabular}
\end{center}
\end{table}

The next step in this analysis is to
reconstruct the $D^0\to K^-\pi^+$ decay. 
All reconstructed charged tracks in the event are considered
except for the tracks associated with the lepton and slow pion
candidates. Pairs of tracks with opposite charge are combined,
and the track with the opposite charge with respect to the
slow pion candidate is assigned the kaon mass. The kaon candidate
is required to satisfy loose kaon identification criteria that
retain more than 80\% of real kaons while rejecting 95\% of pions.
The kaon plus pion invariant mass is required to satisfy
$1.82<M_{K\pi}<1.91$ GeV. Each $D^0$ candidate is combined
with the slow pion and the mass difference
$\Delta M=M(K^-\pi^+\pi_s^+)-M(K^-\pi^+)$ is computed. The signal 
is looked for in the range $142.4<\Delta M<149.9$ MeV.

Besides the signal events, the exclusive sample contains: continuum, 
combinatorial $B\bar B$, uncorrelated peaking $D^{*+}$, and Cabibbo
suppressed decays. As for the inclusive sample, the continuum background
is subtracted using the off-resonance sample. The combinatorial $B\bar B$
background is determined from simulated $B\bar B$ events, normalized
in the $\Delta M$ sideband $153.5<\Delta M<162.5$ MeV. The background
from uncorrelated peaking $D^{*+}$ arises from events where the $D^{*+}$
and lepton comes from different $B$ mesons. This background peaks 
in $\Delta M$ but not in $M^2_{\nu}$. This background is estimated
using the sideband in $M^2_{\nu}$. The backgrounds from Cabibbo
suppressed $D^0\to K^-K^+$ and $D^0\to \pi^-\pi^+$ decays are 
subtracted using simulated events.

\begin{figure}[tb]
\begin{center}
\includegraphics[width=0.7\linewidth]{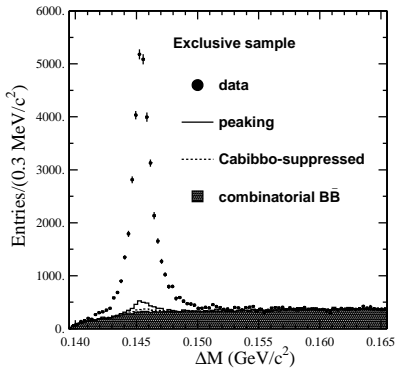}
\caption{The $\Delta M$ distribution for the reconstructed $D^0\to K^-\pi^+$
candidates in events with a $\bar B^0\to D^{*+}\ell^-\bar\nu$ tag.
From~\textcite{Aubert:2007wn}.
}
\label{fig:babar_dm}
\end{center}
\end{figure}

The mass difference, $\Delta M$, is 
shown in Fig.~\ref{fig:babar_dm}. The yields
for this 'exclusive' sample are given in Table~\ref{tab:babar_bkgd}.
After background subtraction BABAR 
finds $N^{\rm excl}=(3.381\pm 0.029)\times 10^4$
events, where the uncertainty is only statistical.
The branching fraction for $D^0\to K^-\pi^+$ is 
calculated using
$$
{\cal B}(D^0\to K^-\pi^+)={ N^{\rm excl} \over 
                 N^{\rm incl} \xi\epsilon_{K\pi}},
$$
where $\epsilon_{K\pi}=(36.96\pm0.09)\%$ from simulation and
$\xi=1.033\pm0.002$ is the selection bias for the partial reconstruction.
The selection bias stems from the fact that the reconstruction efficiency
for the slow pion is larger in events where the $D^0\to K^-\pi^+$ 
than
in generic $D$ decays with more tracks.

BABAR has considered many sources of systematic uncertainties that
affects the measured $D^0\to K^-\pi^+$ branching fraction. The most
important uncertainties include: selection bias ($\pm 0.35\%$), 
nonpeaking combinatorial background ($\pm 0.89\%$), peaking
combinatorial background ($\pm 0.34\%$), tracking efficiency for
kaon and pion ($\pm 1.00\%$), $K^-$ identification ($\pm 0.70\%$),
$D^0$ invariant mass selection ($\pm 0.56\%$), and final state
radiation in the $D^0\to K^-\pi^+$ decay ($\pm 0.50\%$). The total systematic 
uncertainty is estimated to be $\pm 1.80\%$. BABAR obtains the final
result 
$$
{\cal B}(D^0\to K^-\pi^+)=(4.007\pm0.037\pm0.072)\%.
$$

\subsection{Absolute $D$ hadronic branching fractions using double tags}
\label{sect:cleocdhad}

CLEO-c~\cite{He:2005bs, Dobbs:2007zt}  
has used a double tag technique, where by 
reconstructing one $D$ in the event the presence
of an additional $\bar D$ in the event is tagged. By determining
how often the other $D$ meson can be reconstructed
in the event the branching fraction for the $D$ 
decays can be calculated. This type of analysis
was first pioneered by the Mark III 
collaboration~\cite{Baltrusaitis:1985iw, Adler:1987as}.
The CLEO-c analysis described here uses the
same basic idea.

The CLEO-c analysis 
determines the number of single tags,
separately for $D$ and $\bar D$ decays,
$$N_i=\epsilon_i{\cal B}_i N_{D\bar D}$$ 
and 
$${\bar N}_j=\bar \epsilon_j{\cal B}_j N_{D\bar D}$$
where $\epsilon_i$ and ${\cal B}_i$ are the efficiencies
and branching fractions for mode $i$ and $N_{D\bar D}$
is the number of produced $D\bar D$ pairs. 
Though the yields are determined separately for 
$D$ and $\bar D$ decays it is assumed that the
branching fractions are the same.
Similarly,
CLEO-c reconstructs double tags where both $D$ mesons are
reconstructed. The number of double tags found is given
by 
$$N_{ij}=\epsilon_{ij}{\cal B}_i{\cal B}_j N_{D\bar D}$$
where $i$ and $j$ label the $D$ and $\bar D$ mode used
to reconstruct the event and $\epsilon_{ij}$ is the
efficiency for reconstructing the final state.
Combining the two equations above allow us to solve for $N_{D\bar D}$
as 
$$
N_{D\bar D}={{N_i}{\bar N_j}\over N_{ij}}{\epsilon_{ij}\over \epsilon_i\bar\epsilon_j}.
$$
This gives us the number of produced $D\bar D$ events. 
Note that many systematic uncertainties 
cancel in the ratio of efficiencies. This includes for 
example track finding efficiencies
and particle identification that are common to efficiencies
in the denominator and numerator. However, systematic 
uncertainties from, for example, the determination of the yields
do not cancel as they are not correlated.
In this analysis CLEO-c 
determines all the
single tag and double tag yields in data and the efficiencies
from Monte Carlo simulations. 
The branching
fractions and $D\bar D$ yields are extracted from a combined fit to all 
measured data yields and efficiencies.

\begin{figure}[tb]
\begin{center}
\includegraphics[width=\linewidth]{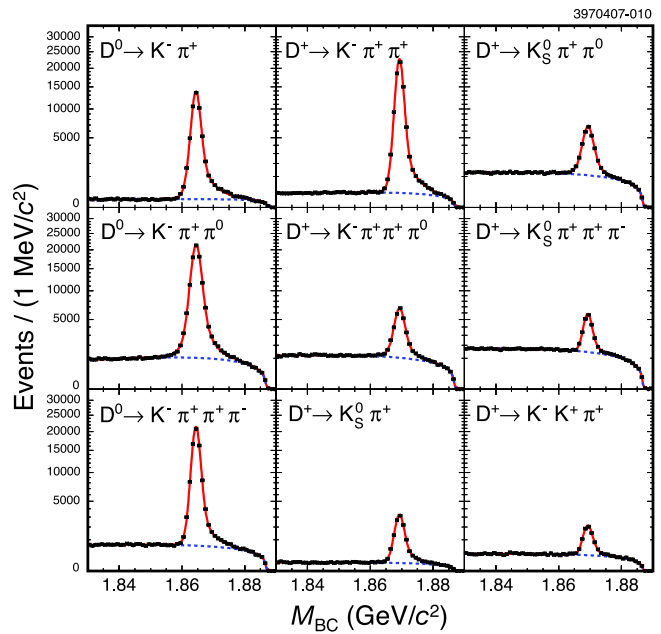}
\caption{Distributions of measured $\Mbc(D)$ and $\Mbc(\bar D)$ 
values for single tag $D^0$ and $D^+$ candidates with 
$D$ and $\bar D$ candidates combined in each plot. The points
are data and the curves are fits to the data. In each plot, the
dashed curve shows the background contributions and the solid
curve shows the sum of the background and signal function.
The number of events is shown on a square-root scale. 
From~\textcite{Dobbs:2007zt}.  
}
\label{fig:cleoc_dhad_st}
\end{center}
\end{figure}

This analysis uses three $D^0$ decay modes
($K^-\pi^+$, $K^-\pi^+\pi^0$, and $K^-\pi^+\pi^-\pi^+$) 
and six $D^+$ decay modes
($K^-\pi^+\pi^+$, $K^-\pi^+\pi^+\pi^0$, $K^0_S\pi^+$, $K^0_S\pi^+\pi^0$,
 $K^0_S\pi^+\pi^-\pi^+$, and $K^-K^+\pi^+$).
The $\pi^0$ candidates are reconstructed in the $\gamma\gamma$ final 
state, and the $K^0_S$ candidates are reconstructed in the
$\pi^+\pi^-$ final state. Particle identification criteria
are applied on kaons and pions (excluding pions in $K^0_S$
candidates). A mode dependent selection criteria on $\Delta E$, 
the candidate
energy minus the beam energy, is applied. The precise 
criteria are listed in Table~\ref{tab:cleoc_deltaecuts}.
To extract the signal yields fits are performed to the
$\Mbc$ distributions for the candidates that pass the 
selection criteria. The fit is described in Sect.~\ref{sect:lineshape}.
The fit is performed separately for
$D$ and $\bar D$ candidates in each mode. These fits
are shown in Fig.~\ref{fig:cleoc_dhad_st} where the
$D$ and $\bar D$ decays have been combined. In
Table~\ref{tab:cleoc_st-eff-yield} the signal efficiencies
and data yields are are shown for all single tag modes.
Many backgrounds have been considered in this analysis
and are discussed in detail in~\textcite{Dobbs:2007zt}.
These backgrounds are summarized in Table~\ref{tab:cleoc_st-eff-yield}.

\begin{table}[tbh]
\caption{Requirements on $\DeltaE$ for $D$ candidates
in the CLEO-c analysis for the absolute $D^0$ and $D^+$
branching fractions.  The limits are set at
approximately 3 standard deviations of the resolution.
From~\textcite{Dobbs:2007zt}.
\label{tab:cleoc_deltaecuts}}
\begin{tabular}{lc} \hline\hline
Mode              &  Requirement (GeV)\\ \hline
$\Dzkpi$       &  $|\Delta E|<0.0294$      \\
$\Dzkpipiz$    &  $-0.0583<\Delta E<0.0350$      \\
$\Dzkpipipi$   &  $|\Delta E|<0.0200$    \\
$\Dpkpipi$     &  $|\Delta E|<0.0218$    \\
$\Dpkpipipiz$  &  $-0.0518<\Delta E<0.0401$    \\
$\Dpkspi$      &  $|\Delta E|<0.0265$    \\
$\Dpkspipiz$   &  $-0.0455<\Delta E<0.0423$    \\
$\Dpkspipipi$  &  $|\Delta E|<0.0265$    \\
$\Dpkkpi$      &  $|\Delta E|<0.0218$    \\
\hline\hline
\end{tabular}
\end{table}

\begin{table*}[htb]
\begin{center}
\caption{Single tag efficiencies, yields from data, and peaking 
background expectations.  The efficiencies include the branching 
fractions for 
$\piz\to\gamma\gamma$ and $\KS\to\pip\pim$ decays.  
The entries in the column 
labeled ``Background'' are the number of events in the signal 
peak produced by non-signal events and the associated systematic 
uncertainty.  The quoted yields include 
these background events.
From~\textcite{Dobbs:2007zt}.
\label{tab:cleoc_st-eff-yield}}
\begin{tabular}{lc r@{~$\pm$~}l r@{~$\pm$~}l} \hline \hline
Single Tag Mode  & ~~Efficiency~(\%)~~~ &   \multicolumn{2}{c}{~~Data Yield~~~~~~}     & \multicolumn{2}{c}{\hspace*{-0.5em}Background}   \\ \hline
$\Dzkpi$  & $64.18 \pm 0.19$  & $25,760 $&$ 165$  & $96  $&$27$\\
$\Dzbarkpi$  & $64.90 \pm 0.19$  & $26,258 $&$ 166$  & $96  $&$27$\\
$\Dzkpipiz$  & $33.46 \pm 0.12$  & $50,276 $&$ 258$  & $114 $&$10$\\
$\Dzbarkpipiz$  & $33.78 \pm 0.12$  & $50,537 $&$ 259$  & $114 $&$10$\\
$\Dzkpipipi$  & $45.27 \pm 0.16$  & $39,709 $&$ 216$  & $889 $&$135$\\
$\Dzbarkpipipi$  & $45.81 \pm 0.16$  & $39,606 $&$ 216$  & $889 $&$135$\\
$\Dpkpipi$  & $54.07 \pm 0.18$  & $40,248 $&$ 208$  & \multicolumn{2}{c}{$<1$}\\
$\Dmkpipi$  & $54.18 \pm 0.18$  & $40,734 $&$ 209$  & \multicolumn{2}{c}{$<1$}\\
$\Dpkpipipiz$  & $26.23 \pm 0.18$  & $12,844 $&$ 153$  & \multicolumn{2}{c}{$<1$}\\
$\Dmkpipipiz$  & $26.58 \pm 0.18$  & $12,756 $&$ 153$  & \multicolumn{2}{c}{$<1$}\\
$\Dpkspi$  & $45.98 \pm 0.18$  & $5,789  $&$ ~\,82$   & $81  $&$22$\\
$\Dmkspi$  & $46.07 \pm 0.18$  & $5,868  $&$ ~\,82$   & $81  $&$22$\\
$\Dpkspipiz$  & $23.06 \pm 0.19$  & $13,275 $&$ 157$  & $113 $&$53$\\
$\Dmkspipiz$  & $22.93 \pm 0.19$  & $13,126 $&$ 155$  & $113 $&$53$\\
$\Dpkspipipi$  & $31.70 \pm 0.24$  & $8,275  $&$ 134$  & $173 $&$83$\\
$\Dmkspipipi$  & $31.81 \pm 0.24$  & $8,285  $&$ 134$  & $173 $&$83$\\
$\Dpkkpi$  & $45.86 \pm 0.36$  & $3,519  $&$ ~\,73$   & \multicolumn{2}{c}{$<1$}\\
$\Dmkkpi$  & $45.57 \pm 0.35$  & $3,501  $&$ ~\,73$   & \multicolumn{2}{c}{$<1$}\\
\hline\hline
\end{tabular}
\end{center}
\end{table*}

\begin{figure}[htb]
\begin{center}
\includegraphics[width=0.99\linewidth]{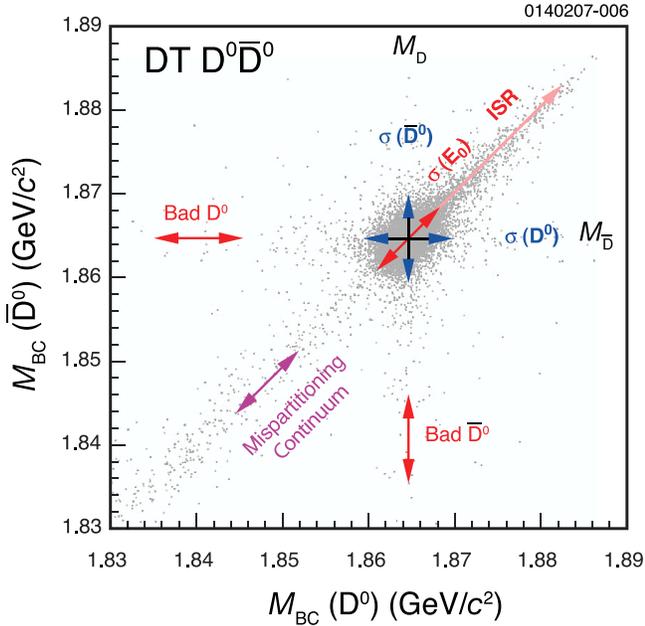}
\caption{Scatter plot of $\Mbc(\Dbar)$ \vs\ $\Mbc(D)$ 
for $\Dz\Dzbar$ double tag candidates.  Signal candidates are 
concentrated at $\Mbc(\Dbar) = \Mbc(D) = \MD$.  
The signal shape and different background contributions are 
discussed in the text.
From~\textcite{Dobbs:2007zt}.  
\label{fig:cleoc_dzdzbar-dt-scatter}}
\end{center}
\end{figure}

The double tag yields are determined separately for the $45=3^2+6^2$
double tag modes. The same criteria on $\Delta E$ that was 
applied for the single tags are applied to the double tags. This
ensures that the systematic uncertainty from the
selection in single and double tag yields cancels in the ratio
for the signal mode. To extract the
number of double tag candidates a two-dimensional unbinned
maximum likelihood fit is performed in the plane of $\Mbc(D)$
vs. $\Mbc(\bar D)$. This is illustrated in 
Fig.~\ref{fig:cleoc_dzdzbar-dt-scatter}. The signal peaks
at $\Mbc(\Dbar) = \Mbc(D) = \MD$. Beam energy smearing affects
both $\Mbc(\Dbar)$ and $\Mbc(D)$ in a correlated fashion to
spread the signal along the $\Mbc(\Dbar)$ vs. $\Mbc(D)$ diagonal.
In addition, the effects of initial state radiation will
spread the signal along the same diagonal to larger values
of $\Mbc(\Dbar)$ and $\Mbc(D)$. If all particles produced in
the $e^+e^-$ interaction are used to form the $D$ and $\bar D$
candidate, but the particles are either from continuum, or from
a $D\bar D$ event but not assigned to the right $D$ candidate
(mispartitioning) the reconstructed $\Mbc(\Dbar)$ and $\Mbc(D)$
will lie on the diagonal. There are also events in which one
of the two $D$ candidates are misreconstructed. These
events form horizontal and vertical bands in $\Mbc(\Dbar)$ vs. $\Mbc(D)$.

\begin{figure}[tb]
\begin{center}
\includegraphics[width=0.49\linewidth]{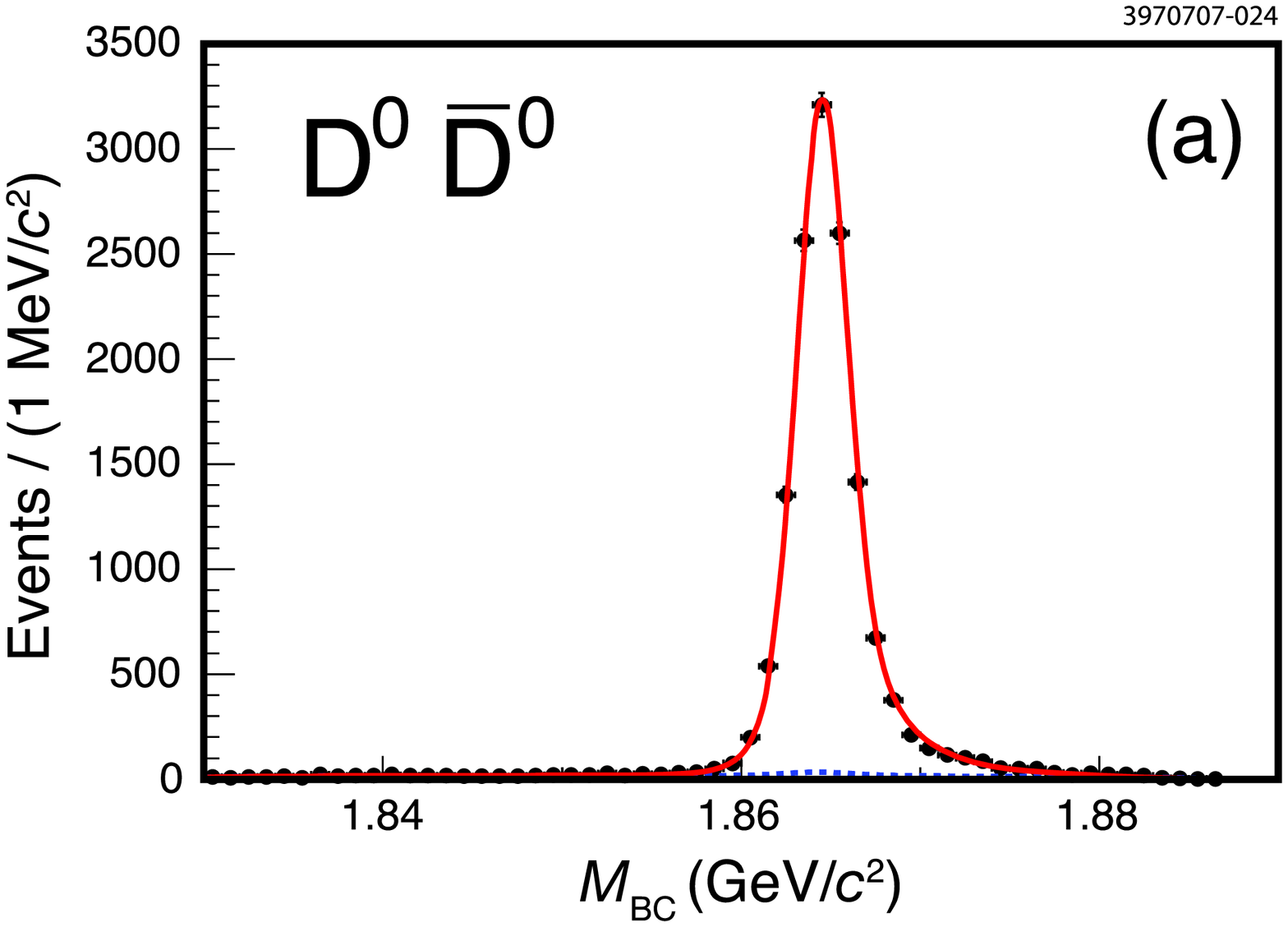}
\includegraphics[width=0.49\linewidth]{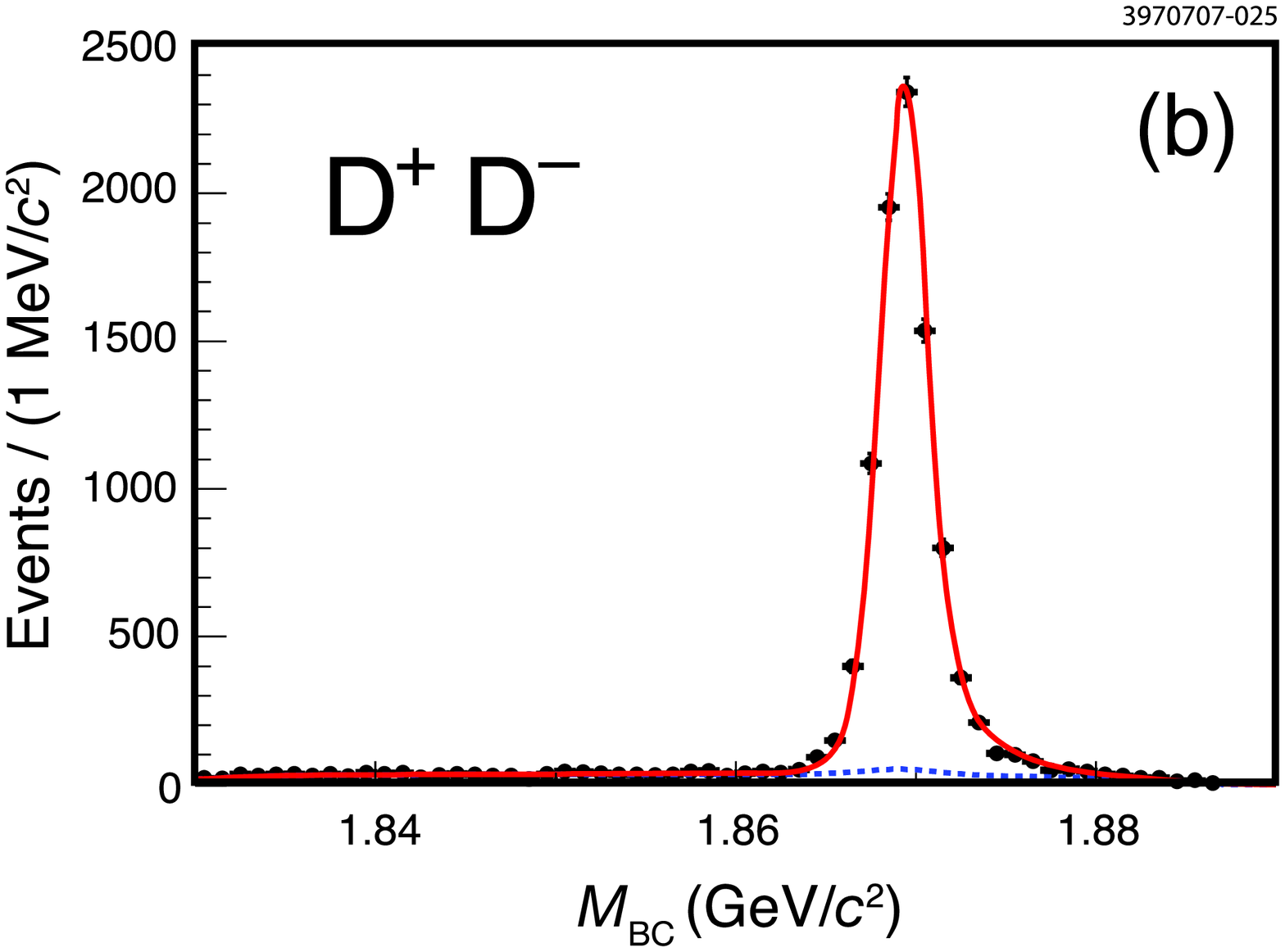}
\caption{Projections of the double tag candidate mass
on the $\Mbc(D)$ axis for (a) the nine $D^0\bar D^0$
double tag modes and (b) the 36 $D^+D^-$ double tag modes.
The points show the data and the curves the projection of
the fit results. The dashed lines shows the background
contributions and the solid line the signal shape plus the
background. 
From~\textcite{Dobbs:2007zt}.  
}
\label{fig:dhad_dt}
\end{center}
\end{figure}

The combined double tag data with the sum of the fits
 are shown in Fig.~\ref{fig:dhad_dt}
for the $D^0\bar D^0$ and $D^+D^-$ modes. 
There are a total of 
$13,591\pm 119$ $D^0\bar D^0$ double tags and $8,870\pm96$ $D^+D^-$ double
tags. For 
most of the modes studied in this analysis the statistical
uncertainty on the measured 
branching fraction is limited by the number of double tags. For
the $D^0$ modes this statistical uncertainty is $\pm 0.88\%$ and for
the $D^+$ modes this is $\pm 1.1\%$.

A detailed study of systematic uncertainties has been
performed. Some of the studies performed by CLEO-c
to determine the systematic uncertainties
are described in Sect.~\ref{sect:exp_threshold}.
The systematic uncertainties 
considered in this analysis are presented in 
Tables~\ref{tab:cleoc_systematicsI} 
and~\ref{tab:cleoc_systematicsII}, where the first table
list systematics common to all modes and the second
table list systematics that are mode specific. The signal
shape systematic uncertainty for double tags are taken
to be $\pm 0.2\%$, while for the single tags a range of systematic
uncertainties from $\pm 0.3\%$, for $D^0\to K^-\pi^+$, to $\pm 1.3\%$,
for $D^+\to K^-\pi^+\pi^+\pi^0$, are assigned. These systematic
uncertainties were assigned based on trying alternative 
signal shape parameterizations in the fit.
For the neutral $D$ decays there is an uncertainty due to 
'double Cabibbo suppressed interference'. The source of
this uncertainty comes from the interference between
signal decays and decays where both the $D^0$ and the $\bar D^0$
decays via doubly Cabibbo suppressed decays. The relative
size of this interference is $\Delta \approx 2R_{\rm ws}\cos2\delta$
where $R_{\rm ws}$ is the ratio of the doubly Cabibbo suppressed 
rate to the Cabibbo favored rate and $\delta$ is the
relative strong phase between the doubly Cabibbo suppressed
amplitude and the Cabibbo favored amplitude. CLEO-c 
assigns a systematic uncertainty of $\pm 0.8\%$ for this effect.
This covers the range of allowed values of $\Delta$ for
$R_{\rm ws}=0.004$ and incorporates the uncertainties in
$\delta$.

For the charged track reconstruction CLEO-c assigns $\pm 0.3\%$ 
uncertainty and for charged kaons an additional $\pm 0.6\%$ 
added in quadrature. In addition
CLEO-c assigns a $\pm 1.8\%$ uncertainty on the $K^0_S$ reconstruction 
in the $\pi^+\pi^-$ final state and a $\pm 2.0\%$ uncertainty
for the $\pi^0$ reconstruction in the $\gamma\gamma$ final state.
These systematic uncertainties
were discussed in Sect.~\ref{sect:cleoc_syst}. 
Kaons and pions, except for pions in the reconstruction 
of $K^0_S\to \pi^+\pi^-$
candidates, are required to satisfy particle identification 
criteria. Uncertainties of $\pm 0.25\%$ and $\pm 0.3\%$ respectively
for pions and kaons are assigned for the particle identification.

For the uncertainty of the modeling of the $\Delta E$ selection
criteria an uncertainty of $\pm 1.0\%$ is assigned for the 
$D^+\to K^0_S\pi^+\pi^0$ and $D^+\to K^+K^-\pi^+$ decays, for
all other modes an uncertainty of $\pm 0.5\%$ is assigned. These 
uncertainties are taken to be correlated across all modes.

The uncertainty in the background shape for the single tag fits
is determined by using alternative descriptions of the background
shape. CLEO-c uses sidebands in $\Delta E$ to parameterize the
combinatorial background shape described by the ARGUS function. Using the
background shapes fixed from the $\Delta E$ sidebands to fit the  
signal region
in $\Delta E$ CLEO-c determines
mode dependent systematic uncertainties from $\pm 0.4\%$ to $\pm 1.5\%$.

In this analysis only one candidate per mode per event is selected.
If there are multiple candidates the candidate with the 
smallest $\Delta E$ is selected for single tags. As this criteria
does not always pick the right candidate and the Monte Carlo
simulation might not properly simulate the rate of multiple
candidates in an event there is a systematic uncertainty 
associated with the multiple candidate modeling. CLEO-c uses
Monte Carlo simulations to determine the probability that the 
wrong candidate is selected in events with multiple candidates. This
is used, together with the difference between data and Monte
Carlo of the rate of events with multiple candidates, to 
estimate the uncertainty associated with the multiple candidate
resolution. The uncertainty for many modes are negligible due
to a very small rate of multiple candidates. The largest
uncertainty, $\pm 0.8\%$, is in the $D^0\to K^-\pi^+\pi^0$ mode due
to fake $\pi^0$ candidates. 

Multibody final states suffer from an uncertainty in
the simulation of the efficiency due to imperfect modeling
of the resonant substructure. The uncertainties associated
with the three- or four-body final states were estimated
by comparing the kinematic distributions in these decays
between data and Monte Carlo simulations. Many three-body
final stats has been studied using Dalitz plot fits and
are well described in the Monte Carlo~\cite{Lange:2001uf}. The Dalitz plot
analyses are described in Sect.~\ref{sect:multi}.

Last, final state radiation, as discussed in Sect.~\ref{sect:FSR}
was considered. CLEO-c compared the signal efficiencies
with and without FSR included in the Monte Carlo simulation.
A systematic uncertainty of $\pm 30\%$ of the change due to not
including final state radiation was assigned. This 
gives the largest uncertainty of about 0.9\% in the 
$D^0\to K^-\pi^+$ mode. 

\begin{table*}[htb]
\caption{Systematic uncertainties and the quantities 
to which they are
applied in the branching fraction fit.  Uncertainties 
not correlated between decay modes are given in the
first section, and correlated uncertainties in the 
second. The symbols $y$ and $\eff$ denote 
\textit{yields} and \textit{efficiencies}, 
respectively.  Yield uncertainties are additive and 
efficiency uncertainties are multiplicative.  See 
the text for the distinction between $\eff(\textrm{Charged})$ 
and $\eff(\Kpm)$. The detector simulation uncertainties 
are determined per charged track or per neutral pion or 
kaon.  Uncertainties for other efficiencies are determined 
per $D$.  In addition, to the systematic uncertainties 
listed here, five more mode-dependent systematic 
uncertainties are listed in \Tab{tab:cleoc_systematicsII}.
From~\textcite{Dobbs:2007zt}.
\label{tab:cleoc_systematicsI}}
\begin{ruledtabular}
\begin{tabular}{lcc}
Source & Uncertainty (\%) & Quantity or Decay Mode \\ \hline
DT Signal Shape & 0.2 & $y(\textrm{All DT Modes})$ \\
Double DCSD Interference  & 0.8  & $y(\textrm{Neutral DT})$ \\ \hline
Detector Simulation       & 0.3  & $\eff(\textrm{Charged})$ Tracking \\
& 0.6  & $\eff(\Kpm)$ Tracking  \\
& 1.8  & $\eff(\KS)$  \\
& 2.0  & $\eff(\piz)$  \\
& \hspace*{0.2em} 0.25 & $\eff(\pipm)$ PID  \\
& 0.3  & $\eff(K^\pm)$ PID  \\
Lepton Veto               & 0.1  & $\eff(\Dzkpi)$ ST \\
Trigger Simulation        & 0.2  & $\eff(\Dzkpipiz)$  \\
& 0.1  & $\eff(\Dpkspi)$  \\
$|\Delta E|$ Requirement  & 1.0  & $\eff(\Dpkspipiz)$ and $\eff(\Dpkkpi)$  \\
& 0.5  & $\eff(\textrm{All Other Modes})$ \\
\end{tabular}
\end{ruledtabular}
\end{table*}

\begin{table*}[htb]
\caption{Mode-dependent systematic uncertainties.  The 
systematic uncertainties for the signal shapes are 
correlated among all ST modes.  The systematic uncertainties 
for FSR are correlated among all ST and DT modes.  Other 
uncertainties are uncorrelated. The background and signal 
shape uncertainties are uncertainties on the yields, the 
other uncertainties in the table are uncertainties on the 
efficiency. Yield uncertainties are additive and efficiency 
uncertainties are multiplicative.
From~\textcite{Dobbs:2007zt}. 
\label{tab:cleoc_systematicsII}}
\begin{ruledtabular}
\begin{tabular}{lccccc}
Mode  & Background   & ST Signal     & FSR (\%) & Resonant          & Multiple   \\
      & Shape (\%)   & Shape (\%)    &          & Substructure (\%) & Candidates (\%)\\ \hline
$\Dzkpi$             & 0.4 &  0.3 &  0.9 &  --- &  0.0 \\
$\Dzkpipiz$          & 1.0 &  0.5 &  0.3 &  0.3 &  0.8 \\
$\Dzkpipipi$         & 0.4 &  0.7 &  0.8 &  1.2 &  0.0 \\
$\Dpkpipi$           & 0.4 &  0.3 &  0.7 &  0.6 &  0.0 \\
$\Dpkpipipiz$        & 1.5 &  1.3 &  0.3 &  0.5 &  0.5 \\
$\Dpkspi$            & 0.4 &  0.4 &  0.5 &  --- &  0.2 \\
$\Dpkspipiz$         & 1.0 &  0.5 &  0.1 &  1.2 &  0.0 \\
$\Dpkspipipi$        & 1.0 &  0.6 &  0.6 &  0.5 &  0.0 \\
$\Dpkkpi$            & 1.0 &  0.6 &  0.3 &  1.3 &  0.2 \\
\end{tabular}
\end{ruledtabular}
\end{table*}

The signal yields for single and double tags and the 
efficiencies determined from Monte Carlo simulations
are combined in a $\chi^2$ fit~\cite{Sun:2005ip}. This
fit includes both statistical and systematic
uncertainties. The fit extracts the branching fractions
for the nine $D$ decay modes studied in this analysis 
and the produced number of $D^0\bar D^0$ and $D^+D^-$
pairs. The result of this fit is shown in
Table~\ref{tab:cleoc_dhadresults}. The correlation
coefficients for the measurement is shown in 
Table~\ref{tab:cleoc_dataCorrelationMatrix}. The $\chi^2$
of the fit is 39.2 for 52 degrees of freedom, 
corresponding to a confidence level of 98\%.
The $\chi^2$ includes systematic uncertainties.

\begin{table*}[htb]
\caption{Fitted branching fractions and $D\Dbar$ pair
yields.  For $N_{\Dz\Dzbar}$ and $N_{\Dp\Dm}$, uncertainties are 
statistical and systematic,
respectively.  For branching fractions and ratios, the 
systematic uncertainties are divided into the contribution from 
FSR (third uncertainty) and all others combined (second uncertainty).  
The column of fractional systematic errors combines all systematic 
errors, including FSR.  The last column, $\Delta_{\rm FSR}$, is the 
relative shift in the fit results when FSR is not included in the 
Monte Carlo simulations used to determine efficiencies.
From~\textcite{Dobbs:2007zt}.
\label{tab:cleoc_dhadresults}}
\begin{ruledtabular}
\begin{tabular}{lcccc}
Parameter & Fitted Value & \multicolumn{2}{c}{Fractional Error} & $\Delta_{\rm FSR}$\\[-0.6ex]
&& Stat.(\%) & Syst.(\%)  &  (\%) \\ \hline

$N_{\Dz\Dzbar}$	 & $(1.031 \pm 0.008 \pm 0.013)\times 10^6$	 & $0.8$	 & $1.3$	 & $+0.1$\\
${\cal B}(\Dzkpi)$	 & $(\BDzkpivalue)\%$	 & $0.9$	 & $1.8$	 & $-3.0$\\
${\cal B}(\Dzkpipiz)$	 & $(14.57 \pm 0.12 \pm 0.38 \pm 0.05)\%$	 & $0.8$	 & $2.7$	 & $-1.1$\\
${\cal B}(\Dzkpipipi)$	 & $(8.30 \pm 0.07 \pm 0.19 \pm 0.07)\%$	 & $0.9$	 & $2.4$	 & $-2.4$\\
$N_{\Dp\Dm}$	 & $(0.819 \pm 0.008 \pm 0.010)\times 10^6$	 & $1.0$	 & $1.2$	 & $+0.1$\\
${\cal B}(\Dpkpipi)$	 & $(\BDpkpipivalue)\%$	 & $1.1$	 & $1.9$	 & $-2.3$\\
${\cal B}(\Dpkpipipiz)$	 & $(5.98 \pm 0.08 \pm 0.16 \pm 0.02)\%$	 & $1.3$	 & $2.8$	 & $-1.0$\\
${\cal B}(\Dpkspi)$ 	 & $(1.526 \pm 0.022 \pm 0.037 \pm 0.009)\%$	 & $1.4$	 & $2.5$	 & $-1.8$\\
${\cal B}(\Dpkspipiz)$	 & $(6.99 \pm 0.09 \pm 0.25 \pm 0.01)\%$	 & $1.3$	 & $3.5$	 & $-0.4$\\
${\cal B}(\Dpkspipipi)$	 & $(3.122 \pm 0.046 \pm 0.094 \pm 0.019)\%$	 & $1.5$	 & $3.0$	 & $-1.9$\\
${\cal B}(\Dpkkpi)$	 & $(0.935 \pm 0.017 \pm 0.024 \pm 0.003)\%$ & $1.8$	 & $2.6$	 & $-1.2$\\ \hline
${{\calB}(\Dzkpipiz)}/{{\calB}(\Km\pip)}$	 & $3.744 \pm 0.022 \pm 0.093 \pm 0.021$	 & $0.6$	 & $2.6$	 & $+1.9$\\
${{\calB}(\Dzkpipipi)}/{{\calB}(\Km\pip)}$	 & $2.133 \pm 0.013 \pm 0.037 \pm 0.002$	 & $0.6$	 & $1.7$	 & $+0.5$\\
${{\calB}(\Dpkpipipiz)}/{{\calB}(\Km\pip\pip)}$	 & $0.654 \pm 0.006 \pm 0.018 \pm 0.003$	 & $0.9$	 & $2.7$	 & $+1.4$\\
${{\calB}(\Dpkspi)}/{{\calB}(\Km\pip\pip)}$	 & $0.1668 \pm 0.0018 \pm 0.0038 \pm 0.0003$	 & $1.1$	 & $2.3$	 & $+0.5$\\
${{\calB}(\Dpkspipiz)}/{{\calB}(\Km\pip\pip)}$	 & $0.764 \pm 0.007 \pm 0.027 \pm 0.005$	 & $0.9$	 & $3.5$	 & $+2.0$\\
${{\calB}(\Dpkspipipi)}/{{\calB}(\Km\pip\pip)}$	 & $0.3414 \pm 0.0039 \pm 0.0093 \pm 0.0004$	 & $1.1$	 & $2.7$	 & $+0.4$\\
${{\calB}(\Dpkkpi)}/{{\calB}(\Km\pip\pip)}$	 & $0.1022 \pm 0.0015 \pm 0.0022 \pm 0.0004$	 & $1.5$	 & $2.2$	 & $+1.1$\\
\end{tabular}
\end{ruledtabular}
\end{table*}

\newlength{\ML}
\settowidth{\ML}{$-$}
\newcommand{\HML}{\hspace*{\ML}}
\begin{table*}[htb]
\caption{The correlation matrix, including systematic uncertainties,
for the branching fractions and numbers of $D\Dbar$ events determined from the fit.
From~\textcite{Dobbs:2007zt}.
\label{tab:cleoc_dataCorrelationMatrix}}
\begin{ruledtabular}
\begin{tabular}{l|cccc|ccccccc}

& $\NDzDzbar$ & $K\pi$ & $K\pi\piz$ & $K\pi\pi\pi$ & $\NDpDm$ & $K\pi\pi$ 
& $K\pi\pi\piz$ & $\KS\,\pi$ & $\KS\,\pi\piz$ & $\KS\,\pi\pi\pi$ 
& $KK\pi$ \\ \hline
$\NDzDzbar$ & $1$ & $-0.65$ & $-0.34$ & $-0.41$ & $\HML 0.39$	 
& $-0.19$ & $\HML 0.01$ & $-0.14$ & $-0.09$ & $-0.08$ & $-0.09$\\
${\calB}(\Km\pip)$ &  & $1~$ & $\HML 0.44$ & $\HML 0.70$ & $-0.22$ 
& $\HML 0.52$ & $\HML 0.23$ & $\HML 0.28$ & $\HML 0.15$ & $\HML 0.30$ 
& $\HML 0.35$\\
${\calB}(\Km\pip\piz)$ &  &  & $1~$ & $\HML 0.38$ & $-0.11$ & $\HML 0.28$ 
& $\HML 0.66$ & $\HML 0.14$ & $\HML 0.51$	 & $\HML 0.17$ & $\HML 0.21$\\
${\calB}(\Km\pip\pim\pip)$ & & & & $1~$ & $-0.09$ & $\HML 0.51$ 
& $\HML 0.29$ & $\HML 0.28$ & $\HML 0.17$ & $\HML 0.37$ & $\HML 0.34$\\
\hline
$\NDpDm$ &  &  &  &  & $1~$ & $-0.61$ & $-0.24$ & $-0.48$ & $-0.30$ 
& $-0.33$ & $-0.38$\\
${\calB}(\Km\pip\pip)$ &  &  &  &  &  & $1~$ & $\HML 0.43$ & 
$\HML 0.52$ & $\HML 0.32$ & $\HML 0.51$ & $\HML 0.55$\\
${\calB}(\Km\pip\pip\piz)$ &  &  &  &  &  &  & $1~$ & $\HML 0.27$ 
& $\HML 0.56$ & $\HML 0.29$ & $\HML 0.32$\\
${\calB}(\KS\,\pip)$ &  &  &  &  &  &  &  & $1~$ & $\HML 0.55$ 
& $\HML 0.72$ & $\HML 0.31$\\
${\calB}(\KS\,\pip\piz)$ &  &  &  &  &  &  &  &  & $1~$ & $\HML 0.50$ 
& $\HML 0.20$\\
${\calB}(\KS\,\pip\pip\pim)$ &  &  &  &  &  &  &  &  &  & $1~$ 
& $\HML 0.30$\\
${\calB}(\Kp\Km\pip)$ &  &  &  &  &  &  &  &  &  &  & $1~$\\
\end{tabular}
\end{ruledtabular}
\end{table*}

The CLEO-c analysis obtains the main branching fraction
results
\begin{eqnarray*}
{\cal B}(D^0\to K^-\pi^+) & = & (3.891\pm0.035\pm 0.059\pm 0.035)\%,\\
{\cal B}(D^+\to K^-\pi^+\pi^+) & = & (9.15\pm0.10\pm 0.16\pm 0.07)\%,
\end{eqnarray*}
where the errors are statistical, systematic, and from final state
radiation respectively. In addition the $D\bar D$ yields determined
from this analysis are used to normalize many other CLEO-c 
measurements. The cross-sections for $D\bar D$ production are
discussed in Sect.~\ref{sect:exp_threshold}.

\subsection{Summary of $\Dz\to\Km\pip$}

The absolute branching fraction for $\Dz\to\Km\pip$ has been
measured by many different experiments, using different
techniques as discussed in this Section. The different
measurements are summarized in Table~\ref{tab:DtoKpi}.
The two most recent, and most precise, measurements are
from CLEO-c and BABAR. They use very different techniques
but finds branching fractions that are in good agreement.
We adopt the PDG average 
$$
{\cal B}(D^0\to K^-\pi^+)=3.89\pm0.05.
$$
These measurements are now limited by systematic uncertainties.
There are many sources of systematic uncertainties that
contribute. Some of these can be improved with additional data.
Both CLEO-c and BABAR can increase the data samples used in
their analyses.

\begin{table*}[bt]
\small{
\caption{
Summary of measurements of the $\Dz\to\Km\pip$ branching fraction
measurements. Only the top six measurements are used in the average
by the PDG. 
}
\label{tab:DtoKpi}
\begin{center}
\begin{tabular}{lcc}
\hline\hline
Experiment & Ref. & ${\cal B}(\Dzkpi)$ (\%) \\
\hline     
CLEO-c     & \textcite{Dobbs:2007zt}    & $3.891\pm 0.035\pm 0.059\pm0.035$  \\
BABAR      & \textcite{Aubert:2007wn}   & $4.007\pm 0.037\pm 0.072$  \\
CLEO II\footnote{This is an average of the results
                in~\textcite{Aubert:2007wn,Akerib:1993pm,Coan:1997ye}.}
           & \textcite{Artuso:1997mc}   & $3.82 \pm 0.07\pm 0.12$  \\
ALEPH      & \textcite{Barate:1997mm}   & $3.90 \pm 0.09\pm 0.12$  \\
ARGUS      & \textcite{Albrecht:1993gr} & $3.41 \pm 0.12\pm 0.28$  \\
ALEPH      & \textcite{Decamp:1991jw}   & $3.62 \pm 0.34\pm 0.44$  \\ \hline

CLEO-c     & \textcite{He:2005bs}       & $3.91 \pm 0.08\pm 0.09$  \\
CLEO II     & \textcite{Artuso:1997mc}   & $3.81 \pm 0.15\pm 0.16$  \\
CLEO II     & \textcite{Coan:1997ye}     & $3.69 \pm 0.11\pm 0.16$  \\
ARGUS      & \textcite{Albrecht:1994nb} & $4.5  \pm 0.6 \pm 0.4$  \\
CLEO II     & \textcite{Akerib:1993pm}   & $3.95 \pm 0.08\pm 0.17$  \\
HRS        & \textcite{Abachi:1988uc}   & $4.5  \pm 0.8 \pm 0.5$  \\
MARK III    & \textcite{Adler:1987as}    & $4.2  \pm 0.4 \pm 0.4$  \\
MARK II     & \textcite{Schindler:1980ws}& $4.1  \pm 0.6$   \\
LGW        & \textcite{Peruzzi:1977ms}  & $4.3  \pm 1.0$   \\ \hline
Average    &      & $3.89 \pm 0.05$ \\
\hline\hline
\end{tabular}
\end{center}
}
\end{table*}

\subsection{Modes with $K^0_L$ or $K^0_S$ in the final states}

It has commonly been assumed that $\Gamma(D\to K^0_S X)=\Gamma(D\to K^0_L X)$.
However, as pointed out by Bigi and Yamamoto~\cite{Bigi:1994aw} this is
not generally true as for many $D$ decays there are contributions from
Cabibbo favored and Cabibbo suppressed decays that interfere and
produce different rates to final states with $K^0_S$ versus $K^0_L$.
As an example consider $D^0\to K^0_{S,L}\pi^0$. Contributions to
these final states involve the Cabibbo favored decay $D^0\to \bar K^0\pi^0$
as well as the doubly Cabibbo suppressed decay $D^0\to K^0\pi^0$. However,
we don't observe the $K^0$ and the $\bar K^0$ but rather the $K^0_S$
and the $K^0_L$. As the amplitudes for $D^0\to \bar K^0\pi^0$ and
$D^0\to K^0\pi^0$ interfere constructively to
form the $K^0_S$ final state, and destructively to form
a $K^0_L$, we see a rate asymmetry between the $K^0_L$ and $K^0_S$
final states. Using SU(3),
and in particular the U-spin subgroup, one can predict the
asymmetry in $D^0\to K^0_{S,L}\pi^0$
\begin{eqnarray*}
R(D^0)& = &
{{\Gamma(D^0\to K^0_S\pi^0)-\Gamma(D^0\to K^0_L\pi^0)}\over
{\Gamma(D^0\to K^0_S\pi^0)+\Gamma(D^0\to K^0_L\pi^0)}}\\
&\approx& 2\tan^2\theta_C
=0.109\pm0.001.
\end{eqnarray*}
For the corresponding charged $D$ mode, $D^+\to K^0_{S,L}\pi^+$
a similar prediction based on SU(3) is not possible. Rather
one has to rely on calculations based on factorization. Some
of these predictions are discussed below.

Experimentally these channels are challenging as they involve final
states with a $K^0_L$. CLEO-c has studied these modes~\cite{He:2007aj}.
They infer the presence of a $K^0_L$ using a missing mass technique
after vetoing events with a $K^0_S$ decaying to either a $\pi^+\pi^-$
or $\pi^0\pi^0$ pair if there are tracks or $\pi^0$ 
candidates reconstructed 
in the event. 

In addition to the challenge with the $K^0_L$ final state, these
decays are $CP$ eigenstates and at the $\psi(3770)$ where CLEO-c
recorded the data for their analysis we need to disentangle 
the effects from quantum coherence with the rate asymmetry we are
interested in here. The effect of the coherently produced $D^0\bar D^0$
pairs at the $\psi(3770)$ was discussed in Section~\ref{sect:coherence}.

CLEO-c has studied both $D^+\to K^0_{S,L}\pi^+$ and $D^0\to K^0_{S,L}\pi^0$.
First the $D^+$ analysis is discussed as it does not involve the
complication of quantum coherence. The branching fraction for 
$D^+\to K^0_S\pi^+$ is taken from~\textcite{Dobbs:2007zt}. In this
analysis only the branching fraction for $D^+\to K^0_L\pi^+$ is
directly measured. CLEO-c uses a tag technique, in which one 
charged $D$ is fully reconstructed. Six different charged $D$ tags
are used, these modes are the same as in Ref.~\cite{Dobbs:2007zt} described
in Sect.~\ref{sect:cleocdhad}. The tag $D^-$ is combined with a $\pi^+$
and events consistent with a $K^0_S$ are vetoed. An event is vetoed
if an additional charged track or neutral pion, reconstructed
in the $\pi^0\to \gamma\gamma$ channel, was found. This veto
removes about 90\% of the $K^0_S$ background as well as many
other backgrounds while retaining 98\% efficiency for signal
events. 

\begin{figure}[tb]
\begin{center}
\includegraphics[width=0.99\linewidth]{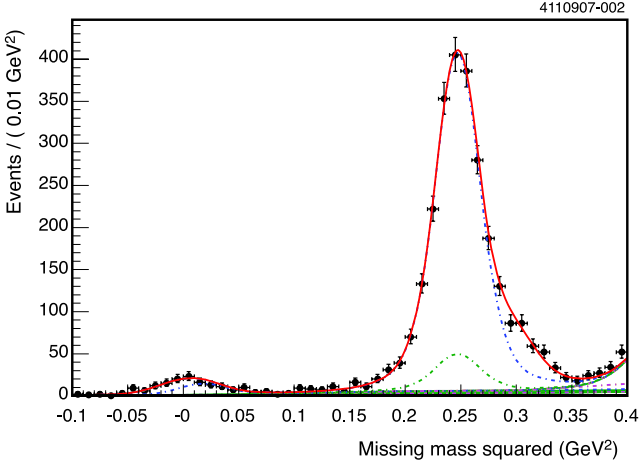}
\caption{Missing mass squared distribution for all six
tag modes for $D^+\to X\pi^+$. Events with extra tracks or
$\pi^0$ candidates have been removed.
From~\textcite{He:2007aj}.
}
\label{fig:cleoc_dptoKLpi}
\end{center}
\end{figure}

Figure~\ref{fig:cleoc_dptoKLpi} shows the invariant mass distribution
against the tag $D$ and charged pion. The signal peaks at a 
missing mass square of about 0.25 GeV$^2$ corresponding to the
$K^0_L$. From the fit to the data CLEO-c extracts a signal
of $2,023\pm 54$ events. With $165\times 10^3$ charged $D$ tags and
an efficiency of 81.6\% for finding the pion the branching fraction
is calculated to be
$$
{\cal B}(D^+\to K^0_L\pi^+)=(1.460\pm0.040\pm0.035\pm0.0005)\%,
$$
where the errors are statistical, systematic, and from the 
branching fraction for $D^+\to K^0_S\pi^+$. The largest
contributions to the systematic uncertainty come from the
extra track and $\pi^0$ veto ($\pm 1.1$\%) and the signal
peak width ($\pm 1.6$\%). The sensitivity to the peak width
comes from the $D^+\to\eta\pi^+$ events just on the high
side of the signal peak as seen in Fig.~\ref{fig:cleoc_dptoKLpi}.

Combining the $D^+\to K^0_L\pi^+$ branching fraction with the 
$D^+\to K^0_S\pi^+$ measured in~\textcite{Dobbs:2007zt} CLEO-c
obtains the asymmetry
$$
R(D^+)=0.022\pm0.016\pm0.018.
$$
There is no evidence for a significant asymmetry in the $D^+\to K^0_{S,L}\pi^+$
mode. Predictions for the asymmetry in charged $D$ decays is more 
involved than for neutral $D$ decays. 
D.-N.~Gao, based on factorization, predicts~\cite{Gao:2006nb} 
this asymmetry to be in the 
range 0.035 to 0.044, which is consistent with the observed asymmetry.

For the $D^0\to K^0_{S,L}\pi^0$ analysis the effects of the
quantum coherence has to be accounted for. In addition, experimentally
this mode is more challenging as the resolution for a $\pi^0$
is worse than for a charged pion. CLEO-c first measures the 
branching fraction for $D^0\to K^0_S\pi^0$ without using a 
$\bar D^0$ tag. Next the 'branching fraction' for $D^0\to K^0_S\pi^0$
is measured in a tagged analysis where the $\bar D^0$ is 
reconstructed in three modes. Due to the coherence the
'branching fraction' measured in the tagged analysis is
${\cal B}(D^0\to K^0_S\pi^0)(1-C_f)$, where $C_f=(r_fz_f+y)/(1+R_{ws,f})$,
as described in Sect.~\ref{sect:coherence}. For the three tag 
modes $C_f$ can now
be calculated. Finally, the 'branching fraction'
for $D^0\to K^0_L\pi^0$ is measured using the same three tag modes, each
of the tag modes give us ${\cal B}(D^0\to K^0_L\pi^0)(1+C_f)$,
and using the measured values of $C_f$ from above the
branching fraction  ${\cal B}(D^0\to K^0_L\pi^0)$ can be determined.

The
$K^0_S$ is reconstructed in the $K^0_S\to \pi^+\pi^-$ final state.
There is a background from $D^0\to \pi^+\pi^-\pi^0$. This background
is subtracted using the $K^0_S$ mass sideband. The signal yield in
this analysis is extracted using a cut-and-count technique. CLEO-c
looks in a 3 standard deviation window around the nominal values
for the beam-constrained mass and $\Delta E$. A sideband in $\Delta E$
is used to subtract the combinatorial backgrounds. The number of
$D^0\bar D^0$ pairs in the data sample is taken from~\textcite{Dobbs:2007zt}.
CLEO-c obtains the branching fraction
$$
{\cal B}(D^0\to K^0_S\pi^0)= (1.240\pm0.017\pm0.031\pm0.047)\%
$$
where the last error is due to the $\pi^0$ reconstruction efficiency.
In the asymmetry $R(D^0)$ this uncertainty will cancel.

Next the 'branching fraction' for ${\cal B}(D^0\to K^0_S\pi^0)$ is
measured with a $\bar D^0$ tag. The three tags modes used are
$\bar D^0 \to K^+\pi^-$, $\bar D^0 \to K^+\pi^-\pi^0$, and
$\bar D^0 \to K^+\pi^-\pi^+\pi^-$. The results for the tagged
analysis is summarized in Table~\ref{tab:cleoc_dtokspi0}.
Similarly the tagged 'branching fraction' for $D^0\to K^0_L\pi^0$
was studied using a missing mass technique where the event
was fully reconstructed except for the $K^0_L$. 
The results are summarized in Table~\ref{tab:cleoc_dtoklpi0}.

\begin{table}[bt]
\caption{
The efficiency is for the reconstruction of the $K^0_S\pi^0$
after the $D$ tag has been found, the tag yield is the 
number of $D$ tags reconstructed, the signal yield is the
number of $D^0\to K^0_S\pi^0$ candidates are reconstructed
against the tag $D$, and the tag bias is a correction due to
the fact that it is easier to reconstruct the tag in events
with the signal than in generic $D$ decays. From~\textcite{He:2007aj}.
}
\label{tab:cleoc_dtokspi0}
\begin{center}
\begin{tabular}{lccc}
\hline\hline
Tag mode            & $K^+\pi^-$   & $K^+\pi^-\pi^0$  & $K^+\pi^-\pi^+\pi^-$ \\
\hline
Efficiency (\%)     & $31.74$      & $31.29$          & $29.97$   \\
Tag yield           & 47,440       & 63,913           & 71,040    \\
Signal yield        & 155          & 203              & 256       \\
Tag bias correction (\%)& 1.000        & 1.014            & 1.033     \\
${\cal B}(D^0\to K^0_S\pi^0)(1-C_f)$ 
                    & $1.03\pm 0.09$ & $1.00\pm0.09$  & $1.16\pm0.08$ \\
\hline\hline
\end{tabular}
\end{center}
\end{table}

\begin{table}[bt]
\caption{
The efficiency is for the reconstruction of the $K^0_L\pi^0$,
including the $K^0_S$ veto,
after the $D$ tag has been found, the tag yield is the 
number of $D$ tags reconstructed, the signal yield is the
number of $D^0\to K^0_S\pi^0$ candidates are reconstructed
against the tag $D$, and the tag bias is a correction due to
the fact that it is easier to reconstruct the tag in events
with the signal than in generic $D$ decays. From~\textcite{He:2007aj}.
}
\label{tab:cleoc_dtoklpi0}
\begin{center}
\begin{tabular}{lccc}
\hline\hline
Tag mode            & $K^+\pi^-$   & $K^+\pi^-\pi^0$  & $K^+\pi^-\pi^+\pi^-$ \\
\hline
Efficiency (\%)     & $55.21$      & $52.72$          & $49.88$   \\
Tag yield           & 47,440       & 64,280           & 71,040    \\
Signal yield        & 334.8        & 414.5            & 466.5     \\
Tag bias correction (\%)& 1.000        & 1.037            & 1.057     \\
${\cal B}(D^0\to K^0_L\pi^0)(1+C_f)$ 
                    & $1.28\pm 0.08$ & $1.03\pm0.06$  & $1.12\pm0.06$ \\
\hline\hline
\end{tabular}
\end{center}
\end{table}

Combining these measurements CLEO-c finds an average asymmetry for the
neutral $D$ decays
$$
R(D^0)=0.108\pm0.025\pm0.024,
$$
which is in good agreement with the prediction.


\subsection{Final states with three kaons}

Final states with three kaons are not generally Cabibbo suppressed, but the
smaller branching fractions for these decays are due to the small
phase space available in these decays. The decay $D^+\to K^+K^-K^+$ 
is Cabibbo suppressed and is included in Sect.~\ref{sect:doubleCabibbo}.
The limited phase space
available has been taken advantage of to measure the $D^0$ 
mass~\cite{Cawlfield:2007dw}.

\begin{table*}[bt]
\caption{Summary of final states with three kaons. If there are more
than one measurement we quote here the PDG average.
}
\label{tab:D_to_threekaons}
\begin{center}
\begin{tabular}{lcc}
\hline\hline
Mode                                  & Ref.   & ${\calB}/ 10^{-3}$  \\
\hline
$D^0\to K^0_S K^+ K^-$                &  BABAR~\cite{Aubert:2005sm}    & ${4.72\pm0.03\pm0.15\pm0.27}$  \\
$D^0\to K^0_S K^0_S K^0_S$            &  PDG Avg.~\cite{Amsler:2008zzb} & ${0.96\pm0.12\pm0.05}$ \\
$D^0\to K^+ K^- K^- \pi^+$            &  PDG  Avg.~\cite{Amsler:2008zzb} & ${0.221\pm0.033\pm0.009}$  \\
$D^0\to K^0_S K^0_S K^{\mp}\pi^{\pm}$ &  FOCUS~\cite{Link:2004zc} & ${0.63\pm0.11\pm0.06\pm0.04}$ \\
\hline\hline
\end{tabular}
\end{center}
\end{table*}

\subsection{Summary of Cabibbo favored $D^0$ and $D^+$ decays}

In Table~\ref{tab:d_cabibbo_favored} a summary of the 
Cabibbo favored $D^0$ and $D^+$ decays are given.
Assuming that $\Gamma(D\to K^0_S X)=\Gamma(D\to K^0_L X)$
for modes where the final states with a $K^0_L$
has not been explicitly measured the Cabibbo 
favored branching fractions adds up to 
$(50.8\pm1.4)\%$ for $D^0$ meson decays and $(38.3\pm1.1)\%$
for $D^+$ decays.
The mode $D^0\to K^-\pi^+\pi^0\pi^0$  is not
included here. An early measurement by MARK III~\cite{Adler:1987as}
reported a large branching fraction of $15\pm5$\%. The
PDG is not using this result anymore in their summary
and there has not been any newer measurements. However,
CLEO-c has used this mode for tagging $D^0$ decays
in their studies of semileptonic decays~\cite{Ge:2008yi}.
They provide enough information that the
branching fraction ${\cal B}(D^0\to K^-\pi^+\pi^0\pi^0)=(7.90\pm0.14)\%$
can be calculated. The error quoted only includes the statistical error
and the uncertainty from the $D^0\to K^-\pi^+$ normalization mode. In
particular experimental systematic uncertainties are not included
and hence this is not included in the summary. But it does show
that there is a substantial branching fraction to the 
$D^0\to K^-\pi^+\pi^0\pi^0$ final state.

\begin{table}[bt]
\small{
\caption{Summary of branching fractions for
Cabibbo favored $D^0$ and $D^+$ decays.
}
\label{tab:d_cabibbo_favored}
\begin{center}
\begin{tabular}{lcc}
\hline\hline
Mode  & Ref.    & Branching Fraction \\
\hline
$D^0\to K^-\pi^+$          &           & $(3.89\pm0.05)\%$ \\
$D^0\to K^0_S\pi^0$        &           & $(1.22\pm0.06)\%$ \\
$D^0\to K^0_L\pi^0$        &           & $(1.00\pm0.07)\%$ \\
$D^0\to K^0_S\pi^+\pi^-$   &           & $(2.99\pm0.17)\%$ \\
$D^0\to K^-\pi^+\pi^0$     &           & $(13.9\pm0.5)\%$ \\
$D^0\to K^-\pi^+\pi^+\pi^-$     &      & $(8.10\pm0.20)\%$ \\
$D^0\to K^0_S\pi^+\pi^-\pi^0$   &      & $(5.4\pm0.6)\%$ \\
$D^0\to K^-\pi^+\pi^+\pi^-\pi^0$     & & $(4.2\pm0.4)\%$ \\
$D^0\to K^0_S\eta\pi^0$        &       & $(5.6\pm1.2)\times 10^{-3}$ \\
$D^0\to K^0_S\pi^+\pi^+\pi^-\pi^-$   & & $(2.84\pm0.31)\times 10^{-3}$ \\
$D^0\to K^-\pi^+\pi^+\pi^+\pi^-\pi^-$ && $(2.2\pm0.6)\times10^{-4}$ \\
\hline
$D^+\to K^0_S\pi^+$        &           & $(1.45\pm0.04)\%$ \\
$D^+\to K^0_L\pi^+$        &           & $(1.46\pm0.05)\%$ \\
$D^+\to K^-\pi^+\pi^+$     &           & $(9.22\pm0.21)\%$ \\
$D^+\to K^0_S\pi^+\pi^0$   &           & $(6.8\pm0.5)\%$ \\
$D^+\to K^-\pi^+\pi^+\pi^0$ &          & $(6.00\pm0.20)\%$ \\
$D^+\to K^0_S\pi^+\pi^+\pi^-$   &      & $(3.02\pm0.12)\%$ \\
$D^+\to K^-\pi^+\pi^+\pi^+\pi^-$ &     & $(5.6\pm0.5)\times 10^{-3}$ \\

\hline\hline
\end{tabular}
\end{center}
}
\end{table}

\section{CABIBBO FAVORED $D_s$ DECAYS AND REFERENCE BRANCHING FRACTIONS}
\label{sect:dsref}

The determination of the absolute branching fraction scale
for $D_s$ decays has been a challenge since the discovery of
the $D_s$~\cite{Chen:1983kr}. Until recently the focus has been
on the final state $D_s^+\to\phi\pi^+$, followed by $\phi\to\Km\Kp$.
This final state is easy to reconstruct with small backgrounds;
the $\phi$ is a narrow resonance and the final state consists
of all charged particles. However,
this final state is not as 'clean' as one would
wish. There are non-$\phi$ contributions, such as the $f_0(980)$, 
to the $\Kp\Km$ 
mass near the $\phi$ mass that pollutes the $D^+_s\to \phi\pi^+$ signal.
Of course, these decays are still real $D_s^+\to K^+K^-\pi^-$
decays. 
This is 
discussed further in Sect.~\ref{sect:DstoKKpiDalitz} on Dalitz plot 
analysis of $D^+_s\to K^-K^+\pi^+$.

As measurements have gotten more precise the definition
of what is measured has had to be made more precise. One of the most
recent measurements by CLEO-c~\cite{Alexander:2008cqa} does not quote a 
$D_s^+\to\phi\pip$ branching fraction, but rather a partial 
branching fraction in a $K^+K^-$ invariant mass region near the $\phi$.
The first attempts at establishing the branching fraction
scale was based on model dependent assumption about equal
partial widths for semileptonic decays of the $D^+_s$ and $D^+$.

This Section will discuss the different approaches used to measure
the $D_s$ absolute branching fractions. The early measurements 
are described very briefly and the more recent, and precise,
measurements are described in more detail.

\subsection{Model dependent approaches}

These measurements are typically no longer used in averages,
e.g. by the particle data group~\cite{Amsler:2008zzb}.
Several experiments, CLEO~\cite{Alexander:1990hp,Butler:1994gv}, 
E687~\cite{Frabetti:1993ih},
ARGUS~\cite{Albrecht:1990tu}, and
E691~\cite{Anjos:1990nm} measured
the ratio
$$
{{\cal B}(D_s^+\to \phi\ell^+\nu_{\ell})\over {\cal B}(D_s^+\to \phi\pip)}.
$$
Using theoretical predictions for the ratio
$$
\Gamma(D_s^+\to \phi\ell^+\nu_{\ell})\over 
\Gamma(D^+\to \bar K^{*0}\ell^+\nu_{\ell})
$$
and the measured $D_s^+$ and $D^+$ lifetimes 
these experiments determined the branching fraction for $D_s^+\to \phi\pip$.
Comparing these results require some care as slightly different
assumptions were made about the ratio of the semileptonic rates. 
Also, combining
these measurements require care as there are strong
systematic correlations between the measurements due to the
common, or at least similar, assumptions about partial rates for the semileptonic decays.

The NA14 experiment~\cite{Alvarez:1990wr} used 
the Lund model
to estimate the ratio of $D_s^+$ to $D^+$ production cross-sections,
which allowed them to determine the $D^+_s\to \phi\pi^+$ 
branching fractions.
The CLEO collaboration~\cite{Chen:1989tu}
used estimates of the $D_s^+$ production rate to determine
the branching fraction for $D_s^+\to \phi\pi^+$. 

All of these measurements use model dependent 
assumptions and have associated systematic uncertainties
that are hard to quantify. With larger data samples model
independent measurements became possible.

\subsection{The branching ratio for $D_s\to \phi\pi$ from $B\to D^*_sD^*$}

The first statistically significant, see Sect.~\ref{sect:ds_double_tag}, 
model independent measurement of the absolute $D^+_s$
branching fraction was performed by CLEO~\cite{Artuso:1996xr}. They
used 2.5 \fbinv\ of $e^+e^-$ data collected at the 
$\Upsilon(4S)$ resonance,
corresponding to $2.7\times 10^6$ $B\bar B$ pairs, 
to study $B\to D_s^*D^*$ decays.
The same technique has been used by BABAR~\cite{Aubert:2005xu}.
They have analyzed a sample with $(123\pm 1)\times 10^6$ $B\bar B$ pairs.

In these analyses the decay $B\to D^*_sD^*$ is reconstructed
in two different ways. First, the $D_s^*$ is fully reconstructed 
using $D_s^{*+}\to D_s^+\gamma$ followed by $D_s^+\to \phi\pi^+$
and the $D^*$ is partially reconstructed using the slow pion
from the $D^*$ decay. In the second method the $D^*$ is fully
reconstructed and the $D^{*+}_s\to D^+_s\gamma$ is only
identified through the presence of the $\gamma$. 
From this study BABAR quotes 
${\cal B}(D^+_s\to\phi\pi^+)=(4.81\pm0.52\pm0.38)\%$ and
CLEO ${\cal B}(D^+_s\to\phi\pi^+)=(3.59\pm0.77\pm0.48)\%$.

\begin{figure}[tb]
\begin{center}
\begin{minipage}{0.9\linewidth }
\includegraphics[width=0.49\linewidth]{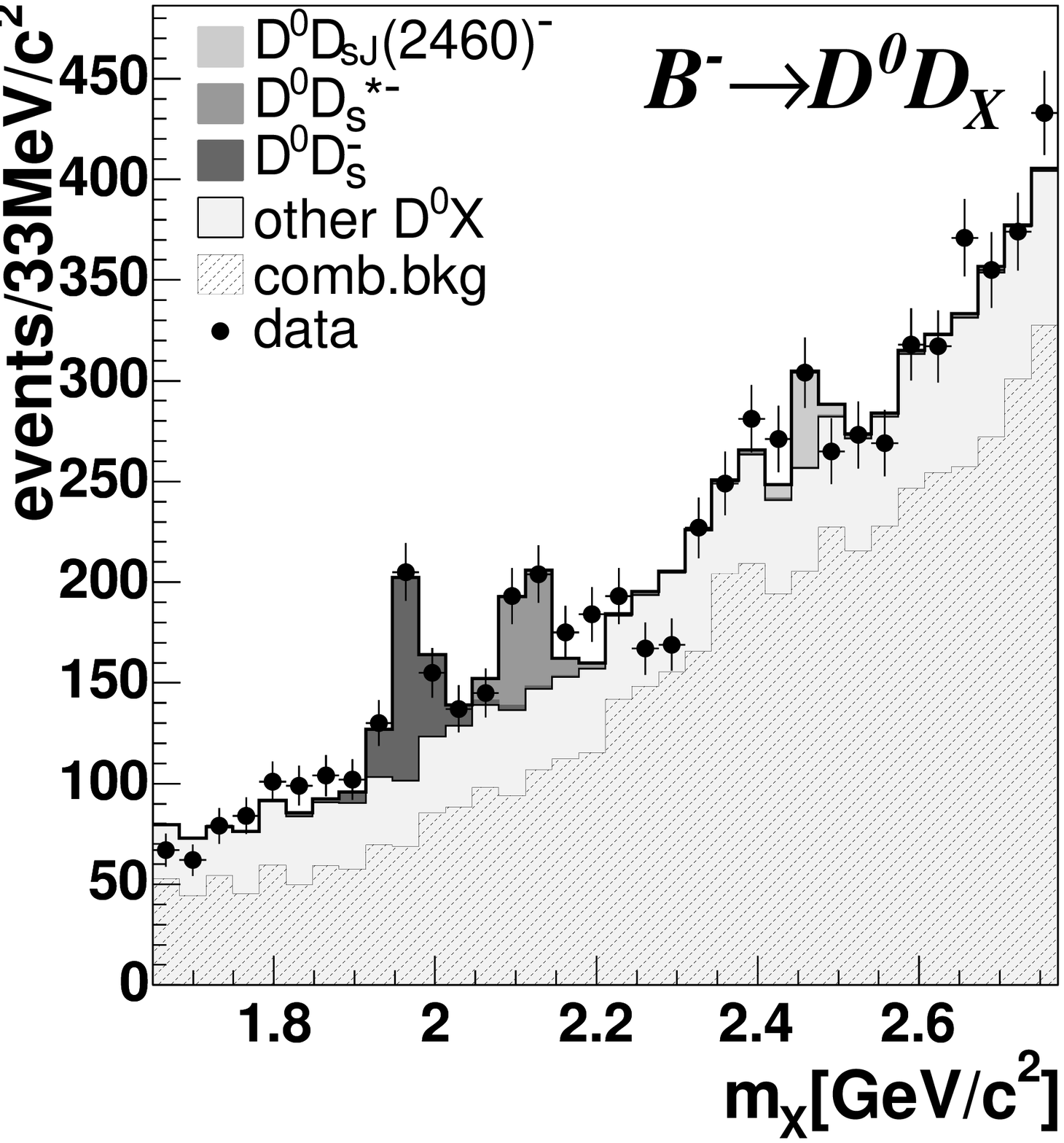}
\includegraphics[width=0.49\linewidth]{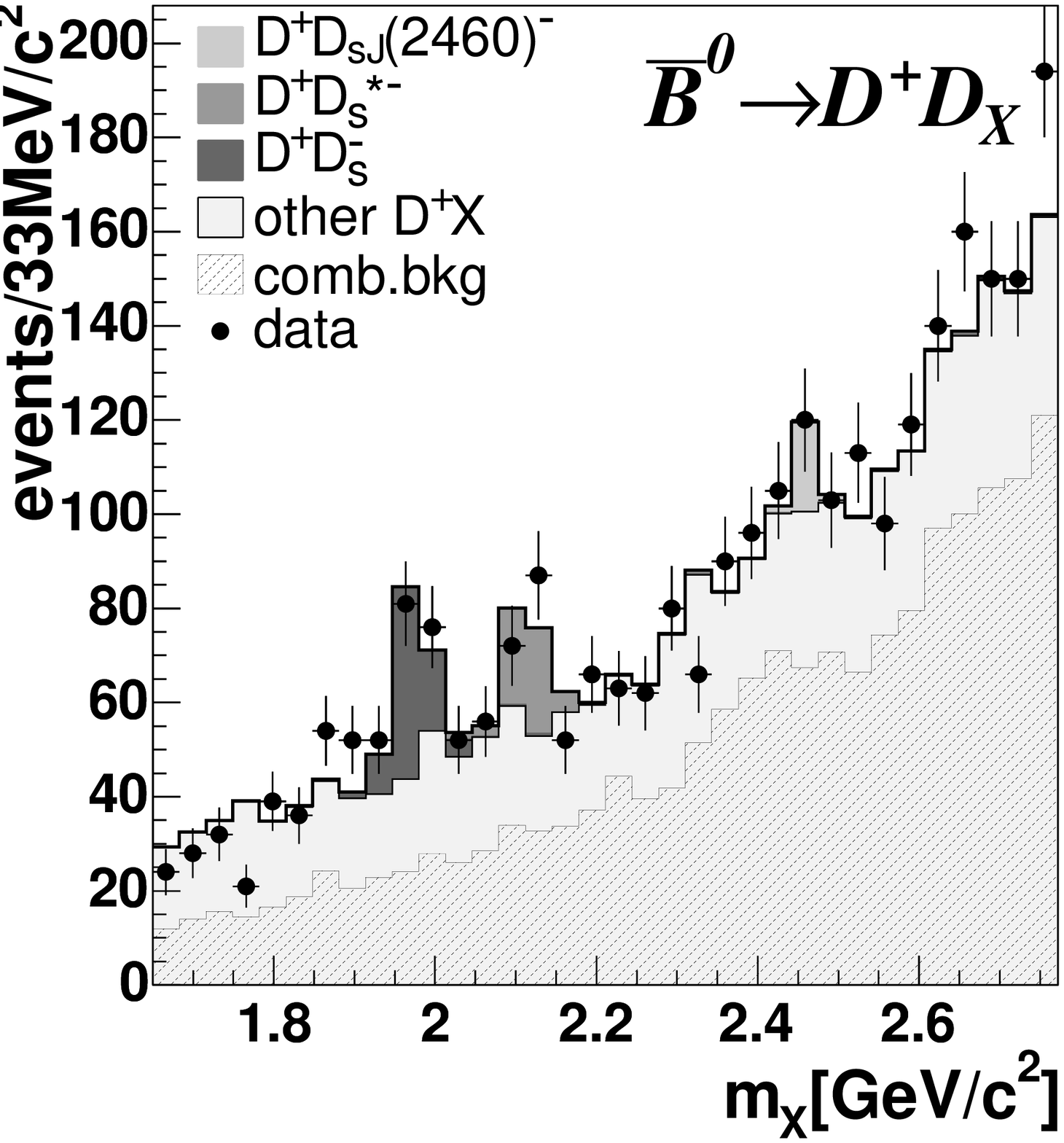}
\includegraphics[width=0.49\linewidth]{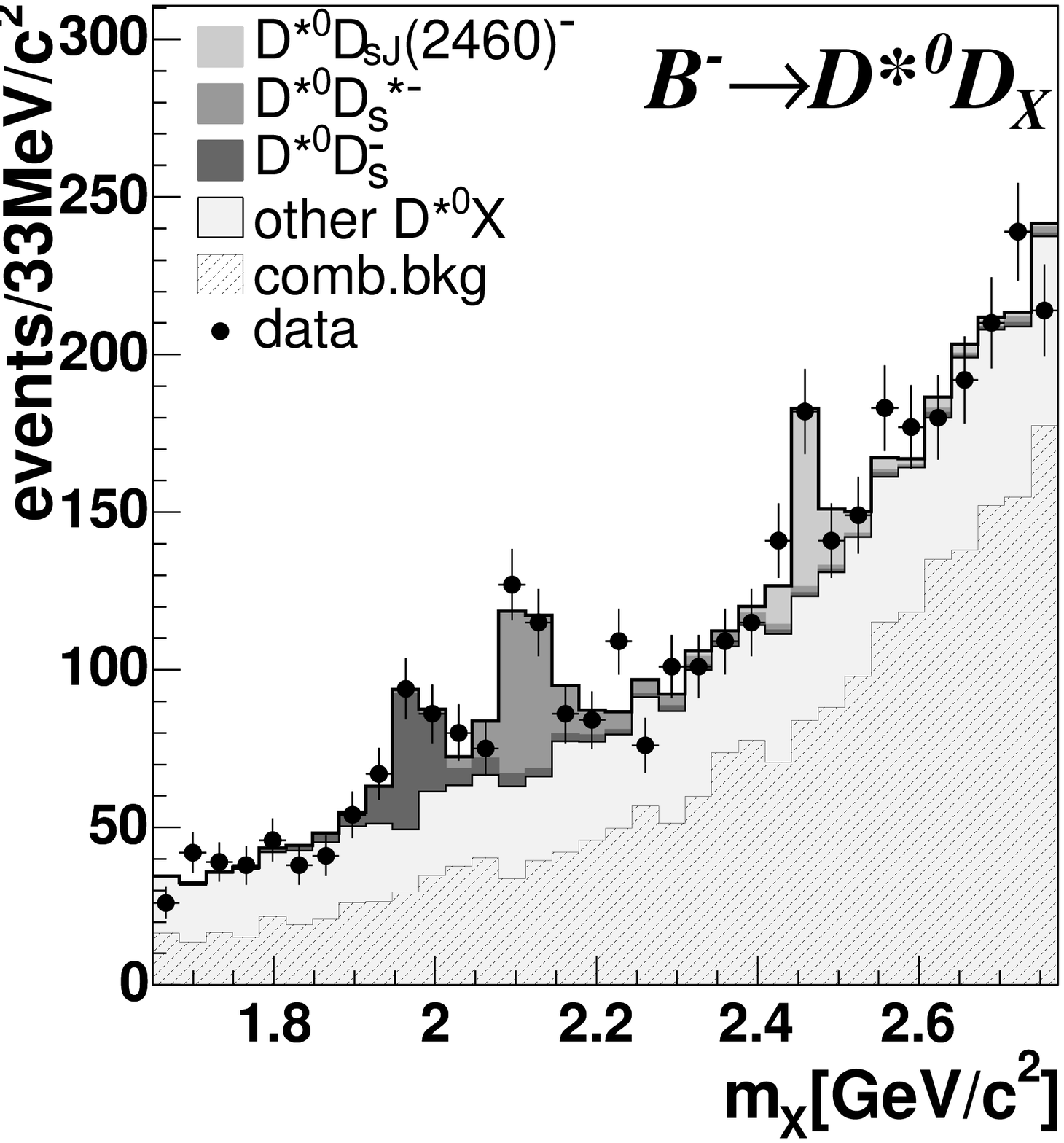}
\includegraphics[width=0.49\linewidth]{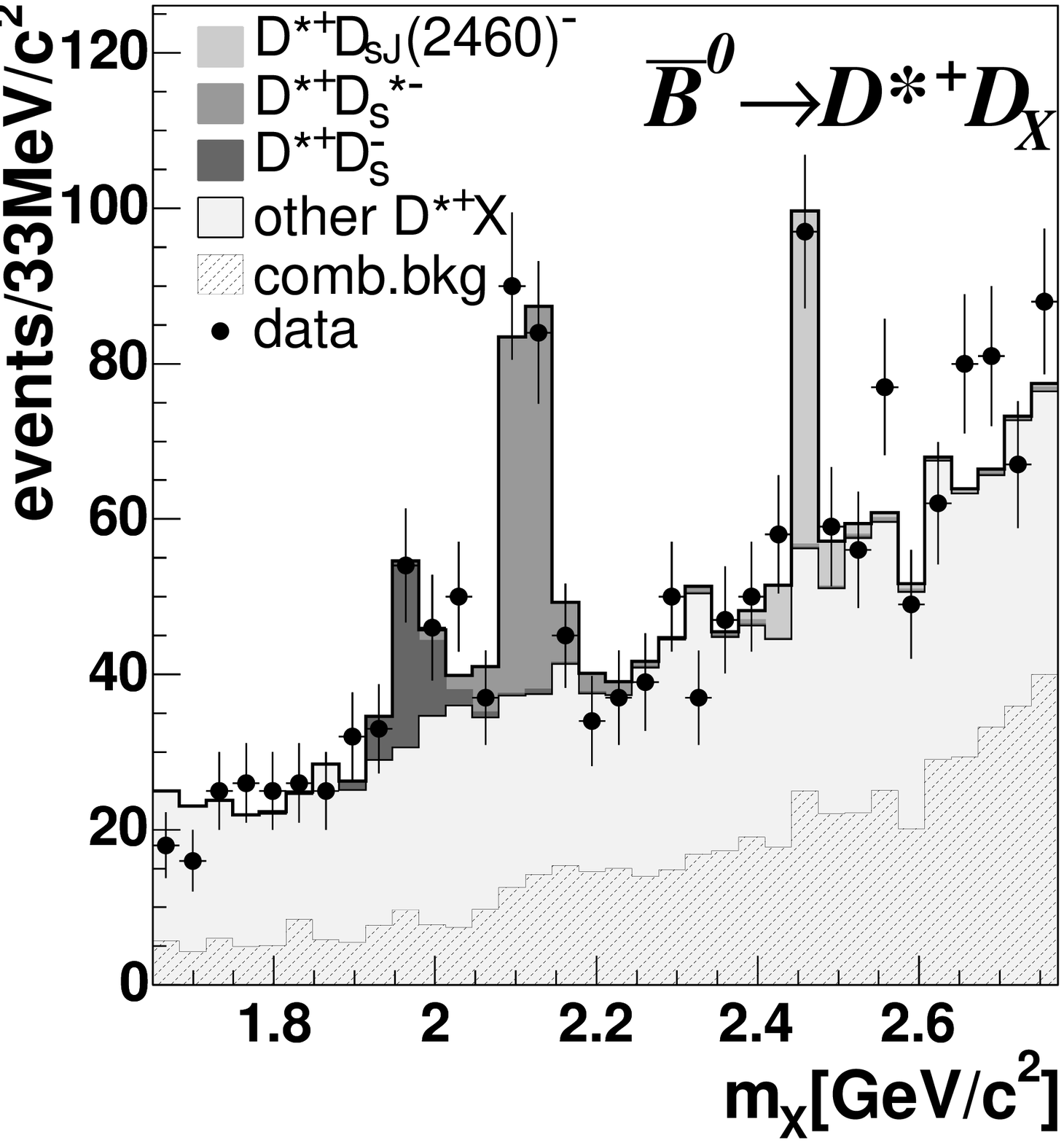}
\end{minipage}
\caption{
The recoil mass against a $D$ or  $D^*$.
From~\textcite{Aubert:2006nm}.
}
\label{fig:babar_d_recoil}
\end{center}
\end{figure}

\begin{figure}[tb]
\begin{center}
\begin{minipage}{0.9\linewidth }
\includegraphics[width=0.49\linewidth]{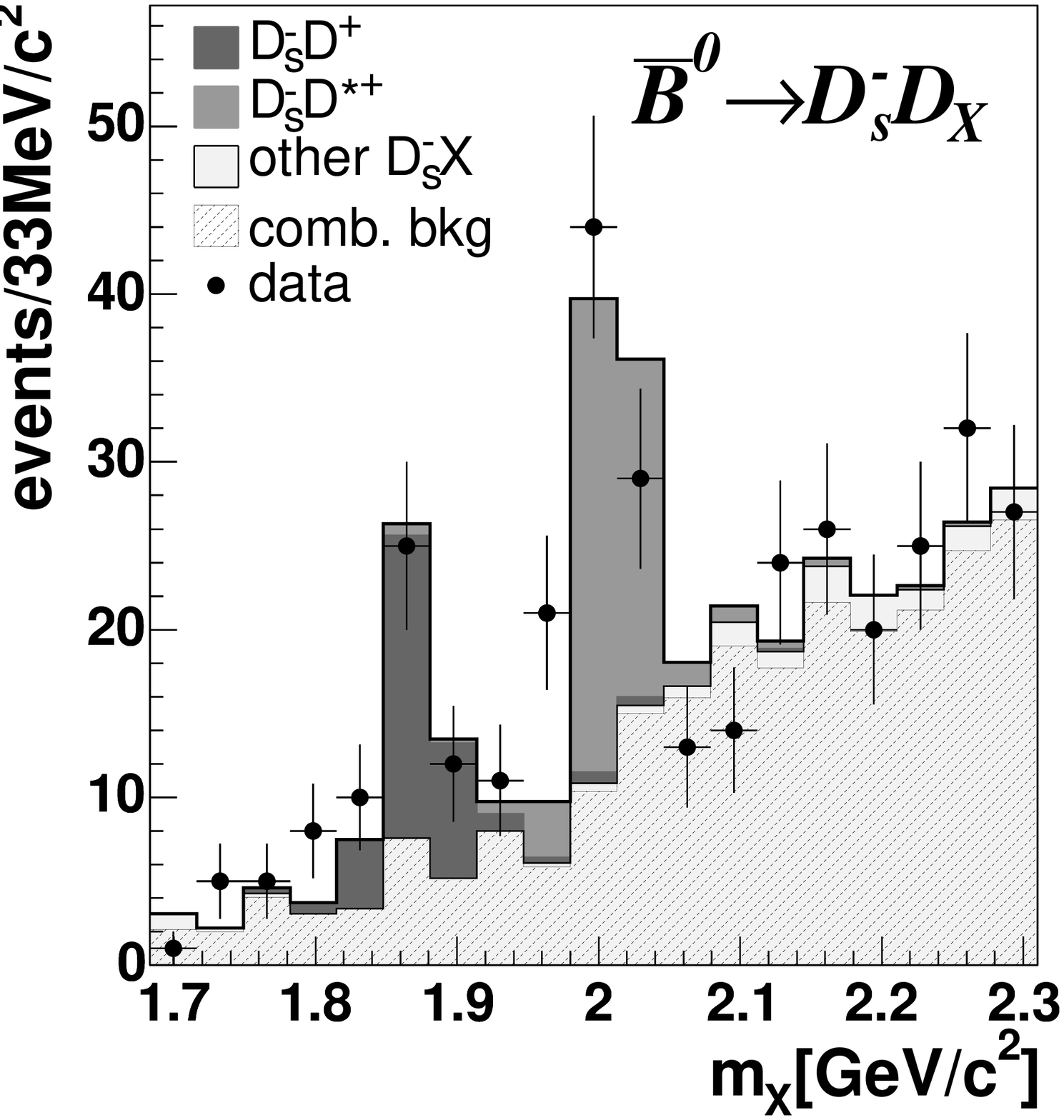}
\includegraphics[width=0.49\linewidth]{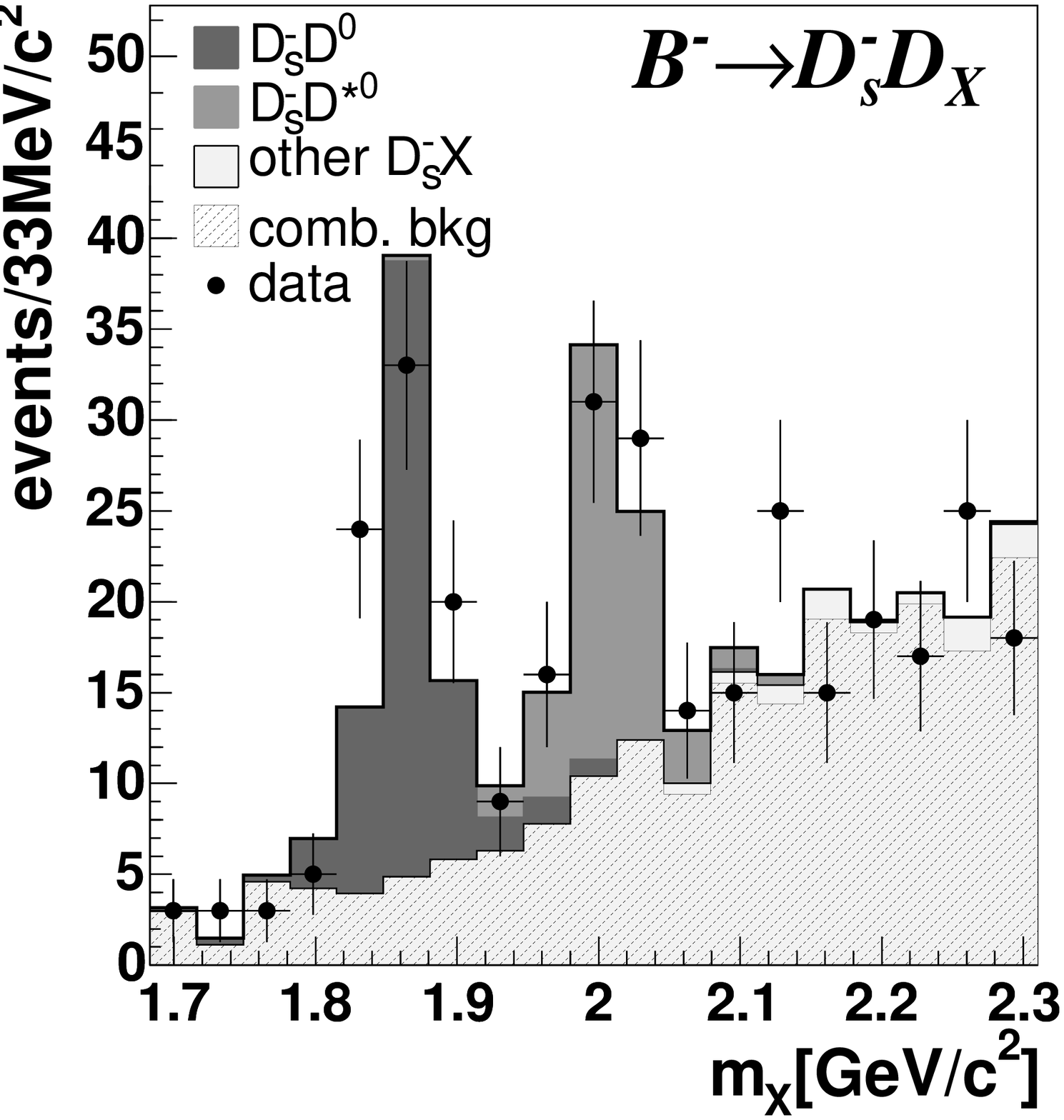}
\includegraphics[width=0.49\linewidth]{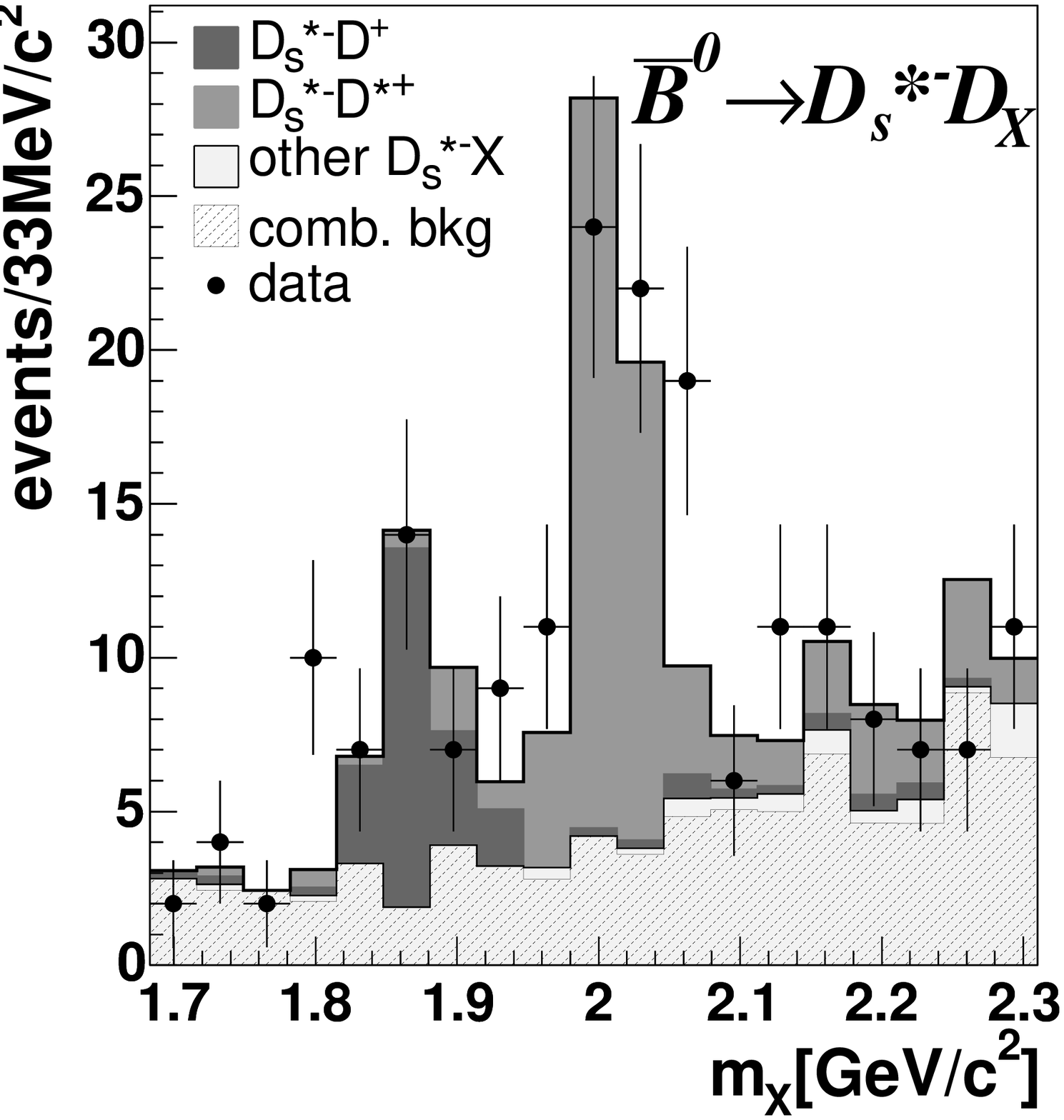}
\includegraphics[width=0.49\linewidth]{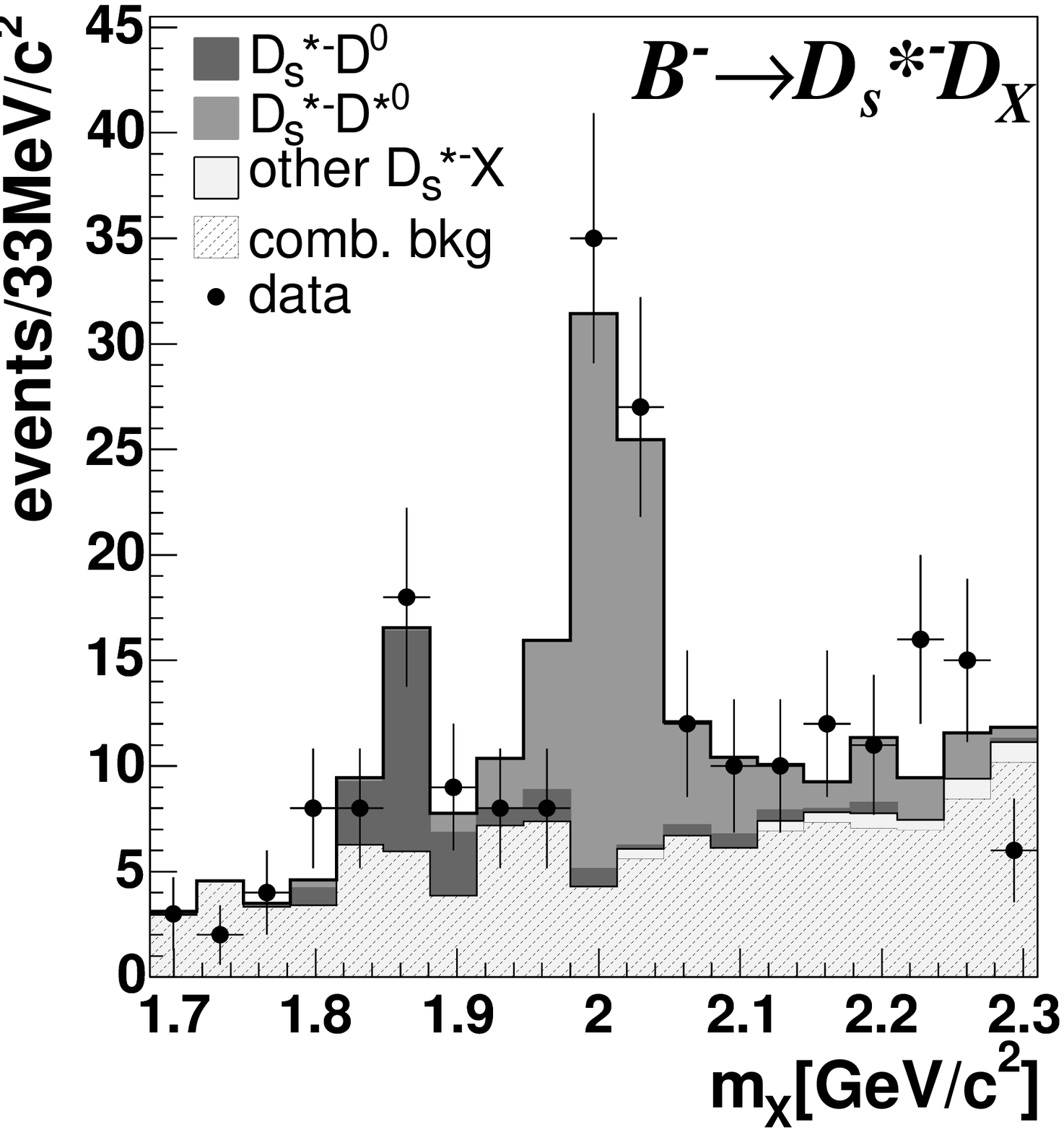}
\end{minipage}
\caption{
The recoil mass against a $D_s$ or  $D_s^*$
From~\textcite{Aubert:2006nm}.
}
\label{fig:babar_ds_recoil}
\end{center}
\end{figure}

More recently, BABAR~\cite{Aubert:2006nm} has
presented results based on 210 fb$^{-1}$ of
data where they use a tag technique in which one $B$ meson is 
fully reconstructed. In events with one fully reconstructed
$B$ meson candidate BABAR reconstructs one additional $D^{(*)}$ or
$D_{s(J)}^{(*)}$ meson. Then they look at the recoil mass against
this reconstructed candidate. The recoil masses are shown
in Figs.~\ref{fig:babar_d_recoil} and~\ref{fig:babar_ds_recoil}.

From these data BABAR extracts ${\cal B}(D_{sJ}(2460)^-\to D_s^{*-}\pi^0)
=(56\pm13\pm9)\%$ and ${\cal B}(D_{sJ}(2460)^-\to D_s^{*-}\gamma)
=(16\pm4\pm3)\%$ in addition to ${\cal B}(D_s^-\to \phi\pi^+)=
(4.52\pm0.48\pm0.68)\%$. BABAR combines this measurement with 
their previous measurement discussed above to obtain
${\cal B}(D_s^-\to \phi\pi^+)=
(4.62\pm0.36\pm0.50)\%$.

\subsection{Study of  $D_s^+\to K^+K^-\pi^+$ in continuum production}

Belle~\cite{Abe:2007jz} has used 552.3 pb$^{-1}$ of $e^+e^-$ data 
to study the process $e^+e^-\to D_s^{*+}D^-_{s1}$
followed by $D^-_{s1}\to D^{*0}K^-$ and $D_s^{*+}\to D_s^+\gamma$.
The very large data sample allow them to study this
exclusive final state in continuum production of 
$D_s$ mesons.
The final state is reconstructed in two ways; either
by partially reconstructing the $D_{s1}$ or the $D_s^*$.
Belle obtains the preliminary branching 
fraction ${\cal B}(D^+_s\to K^+K^-\pi^+)=
(4.0\pm0.4\pm0.4)\%$ which is of comparable statistical
precision to the other methods discussed above.

\subsection{Absolute branching fractions for hadronic $D_s$ decays using double tags}
\label{sect:ds_double_tag}

CLEO-c~\cite{Alexander:2008cqa} has determined 
the absolute hadronic branching fractions
for $D_s$ meson decays using a double tag technique similar to
what was done for the $D$ hadronic branching fractions.
The same technique was used by MARK III~\cite{Adler:1989st}
and BES~\cite{Bai:1994cv}. These initial studies were
limited by statistics; MARK III observed no events and placed
an upper limit while BES observed two events and 
reported a branching fraction of 
$$
{\cal B}(D^+_s\to \phi\pi^+)=(3.9^{+5.1+1.8}_{-1.9-1.1})\%.
$$
The BES analysis used 22.3 \pbinv\ recorded at $E_{\rm cm}=4.03$ GeV.

The CLEO-c analysis used a sample of 298 pb$^{-1}$ of $e^+e^-$
collision data recorded at
a center-of-mass energy of 4170 MeV. At this energy $D_s$ mesons
are produced, predominantly, as $D_s^+D_s^{*-}$ or $D_s^-D_s^{*+}$
pairs. 
The eight hadronic decays considered in this analysis by CLEO-c are
$D^+_s\to K^0_S K^+$,
$D^+_s\to K^0_S K^-\pi^+\pi^+$,
$D^+_s\to K^+K^-\pi^+$,
$D^+_s\to K^+\pi^-\pi^+$,
$D^+_s\to K^+K^-\pi^+\pi^0$,
$D^+_s\to \pi^+\pi^-\pi^+$,
$D^+_s\to \eta\pi^+$, and
$D^+_s\to \eta'\pi^+$. The analysis proceeds similar to the 
$D$ hadronic branching fraction analysis
described in Sect~\ref{sect:cleocdhad}. Yields and
efficiencies for single tags (separately for $D_s^+$ and $D_s^-$)
and double tags are extracted. 
The $\pi^0$ or $\gamma$ from the $D_s^*$ decay is not
reconstructed in this analysis.
The yields, in terms of the
efficiencies, branching fractions, and data sample size are
given by
\begin{eqnarray*}
y_i & = & N_{D_s^*D_s}{\cal B}_i\epsilon_i, \\
y_{\bar j} & = & N_{D_s^*D_s}{\cal B}_j\epsilon_{\bar j}, \\
y_{i\bar j} & =& N_{D_s^*D_s}{\cal B}_i{\cal B}_j\epsilon_{i\bar j},
\end{eqnarray*}
where $i$ indicates a $D_s^+$ and $\bar j$ indicate a 
$D_s^-$. In this analysis a total of 16 single tags
and 64 double tags are used. The event selection is 
detailed in~\textcite{Alexander:2008cqa}. 
A $D_s$ candidate is referred to as 
``indirect'' if it comes from the decay of the $D_s^*$
in the $e^+e^-\to D_s D_s^*$ interaction. Otherwise the
$D_s$ is said to be ``direct''. The $D_s$
candidates are identified based on their momenta and 
invariant mass. The direct $D_s$ has
a fixed momentum in the $e^+e^-$ restframe, whereas the
indirect $D_s$ has momenta in a range due to the 
extra boost from the $D_s^*\to D_s(\gamma,\pi^0)$ decay.
The recoil mass
$M_{\rm rec}$ is defined by
$$
M_{\rm rec}^2=\left(E_0-\sqrt{\mathbf{p}^2_{D_s}+M^2_{D_s}}\right)^2-
                    (\mathbf{p}_0-\mathbf{p}_{D_s})^2,
$$
where $(E_0,\mathbf{p}_0)$ is the $e^+e^-$ center-of-mass four-vector,
$\mathbf{p}_{D_s}$ is the measured $D_s$ momentum and $M_{D_s}$ is
the nominal $D_s$ mass. 
For direct $D_s$ candidates $M_{\rm rec}$ peaks at the 
$D_s^*$ mass of 2.112 GeV, while for indirect $D_s$ candidates $M_{\rm rec}$
is spread about evenly over $\pm 60$ MeV around this peak.
CLEO-c requires that $D_s$ candidates in a double tag, and for most
single tags, satisfies $M_{\rm rec}>2.051$ GeV. For the three single
tag modes, $K^-K^+\pi^+\pi^0$, $\pi^+\pi^0\pi^0$, and $K^+\pi^+\pi^-$, with 
more substantial backgrounds it is required that $M_{\rm rec}$
is greater than (2.099, 2.101, 2.099) GeV, respectively. Note that
this cut eliminates events from $e^+e^-\to D_s^+D_s^-$ as these
events peak at $M_{\rm rec}=M_{D_s}$. A number of vetoes are 
applied to reject fake candidates, primarily from $D^*D^*$ events.

\begin{figure}[tb]
\begin{center}
\includegraphics[width=0.90\linewidth]{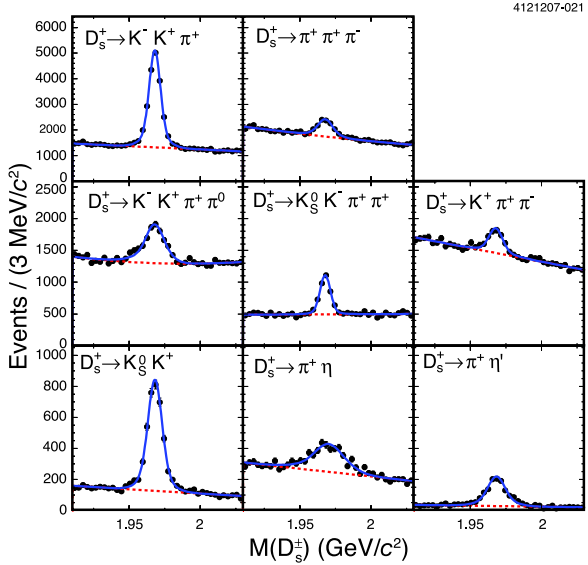}
\caption{Single tag yields for $D_s$ modes used in the CLEO-c 
analysis. Charge conjugate modes are combined.
From~\textcite{Alexander:2008cqa}. 
}
\label{fig:cleoc_ds_st}
\end{center}
\end{figure}

The single tag signal yields  are extracted from the $D_s$ 
invariant mass distributions.
The single tag event yields in data are shown
in Fig.~\ref{fig:cleoc_ds_st}. At most one single tag candidate
per mode and charge are accepted per event. If more than one
candidate pass the selection criteria the candidate with the
value of $M_{\rm rec}$ closest to $M_{D^*_s}$ is selected.
The data is fit to a signal shape and a background shape.
The signal shape is determined from Monte Carlo
simulations, but the $D_s$ mass is allowed to float in the fit.
The background is modeled with a linear function in all modes
except $K^+K^-\pi^+\pi^0$ and $\pi^+\pi^0\pi^0$
where a quadratic form is used instead. The background
shape is constrained to be the same in the two charge conjugate
modes. 

\begin{figure}[tb]
\begin{center}
\includegraphics[width=0.9\linewidth]{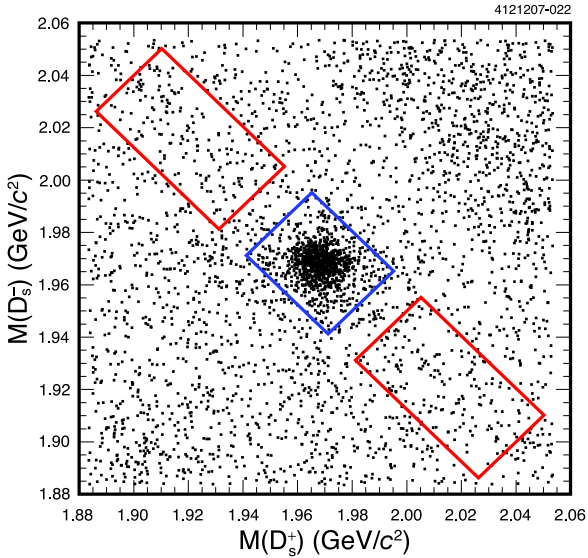}
\caption{Double tag yields for $D_s$ modes used in the CLEO-c 
analysis. The signal region is indicated by the rectangle in
the center and the two sideband regions are the diagonally
offset rectangles. There are 1089 double tag candidates in
the signal region and 339 candidates in the background 
region.
From~\textcite{Alexander:2008cqa}. 
}
\label{fig:cleoc_ds_dt}
\end{center}
\end{figure}

The double tag yields are extracted
by a cut-and-count procedure in the plot of the invariant mass
of the $D^+_s$ {\it vs.} $D^-_s$. All double tag candidates
are shown in Fig.~\ref{fig:cleoc_ds_dt}.
At most one double tag candidate is allowed per event. If there
are more than one candidate the combination with the 
average mass $\hat M\equiv (M(D_s^+)+M(D_s^-))/2$ closest
to the $M_{D_s}$ is kept. The combinatorial background has
structure in $\hat M$, but is more uniform in 
$\Delta M\equiv M(D_s^+)-M(D_s^-)$. The signal region is
defined by $|\hat M-M_{D_s}|<12$ MeV and $|\Delta M|<30$ MeV
and the sideband region is defined by 
$|\hat M-M_{D_s}|<12$ MeV and $50<|\Delta M|<140$ MeV.
In this analysis 
the individual double tag yields and efficiencies are
determined. 
The signal and sideband regions are shown 
in Fig.~\ref{fig:cleoc_ds_dt}

All yields and efficiencies are combined in a
likelihood fit to extract the $D_s$ branching fractions.
The branching fraction results from this fit is presented in 
Table~\ref{tab:cleoc_ds_brfr}. In addition to the branching
fractions, CLEO-c determines the number of $D_sD_s^*$ pairs
produced in their data sample to be 
$N_{D_sD_s^*}=(2.93\pm0.14\pm0.06)\times 10^5$. Combined with
the luminosity, ${\cal L}_{\rm int}=(298\pm3)$ pb$^{-1}$, they obtain 
the  cross-section 
$\sigma_{D_sD_s^*}(E_{\rm cm}=4.17\ {\rm GeV})=(0.983\pm0.046\pm 0.021\pm 0.010)$~nb,
where the last systematic is due to the uncertainty in the 
luminosity.

\begin{table*}[bt]
\caption{Branching fractions for $D_s$ decays determined in the 
CLEO-c analysis.
}
\label{tab:cleoc_ds_brfr}
\begin{center}
\begin{tabular}{lcccc}
\hline\hline
Mode & Branching Fraction ${\cal B}$ (\%) & ${\cal B}/{\cal B}(D^+_s\to K^+K^-\pi^+)$ & ${\cal A}_{CP}$ (\%)\\ \hline
${\cal B}(D^+_s\to K^0_S K^+)$          & $1.49\pm0.07\pm0.05$ & $0.270\pm0.009\pm0.008$ & $+4.9\pm2.1\pm0.9$\\
${\cal B}(D^+_s\to K^+K^-\pi^+)$        & $5.50\pm0.23\pm0.16$ & $1$ & $+0.3\pm1.1\pm0.8$\\
${\cal B}(D^+_s\to K^+K^-\pi^+\pi^0)$   & $5.65\pm0.29\pm0.40$ & $1.03\pm0.05\pm0.08$ & $-5.9\pm4.2\pm1.2$\\
${\cal B}(D^+_s\to K^0_SK^-\pi^+\pi^+)$ & $1.64\pm0.10\pm0.07$ & $0.298\pm0.014\pm0.011$ & $-0.7\pm3.6\pm1.1$\\
${\cal B}(D^+_s\to \pi^+\pi^-\pi^+)$    & $1.11\pm0.07\pm0.04$ & $0.202\pm0.011\pm0.009$ & $+2.0\pm4.6\pm0.7$\\
${\cal B}(D^+_s\to \pi^+\eta)$          & $1.58\pm0.11\pm0.18$ & $0.288\pm0.018\pm0.033$ & $-8.2\pm5.2\pm0.8$\\
${\cal B}(D^+_s\to \pi^+\eta')$         & $3.77\pm0.25\pm0.30$ & $0.69\pm0.04\pm0.06$ & $-5.5\pm3.7\pm1.2$\\
${\cal B}(D^+_s\to K^+\pi^+\pi^-)$      & $0.69\pm0.05\pm0.03$ & $0.125\pm0.009\pm0.005$ & $+11.2\pm7.0\pm0.9$\\
\hline\hline
\end{tabular}
\end{center}
\end{table*}

\begin{figure}[tb]
\begin{center}
\includegraphics[width=0.8\linewidth]{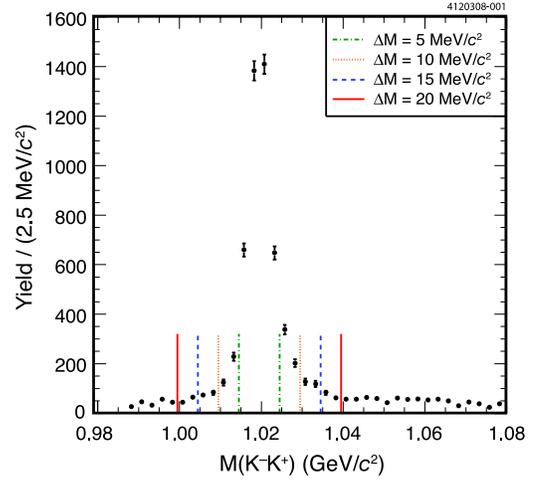}
\caption{
The $K^-K^+$ invariant mass near the $\phi$ resonance
in $D^+_s\to K^-K^+\pi^+$ events from the CLEO-c double 
tag analysis. The single tag fit procedure used in the
CLEO-c analysis is applied to extract the yield in
each $M(K^-K^+)$ bin, hence backgrounds are subtracted
and the yields shown are for the $D^+_s\to K^-K^+\pi^+$
signal. The $\phi$ resonance is clear above an 
additional broad component. Indicated in the plot
are the different mass windows considered by CLEO-c
for their partial branching fractions.
From~\textcite{Alexander:2008cqa}
}
\label{fig:cleoc_ds_kkmass}
\end{center}
\end{figure}

CLEO-c does not quote a $D^+_s\to\phi\pi^+$ branching
fraction. The reason for this is that at the precision of
this measurement the branching fraction for $D^+_s\to\phi\pi^+$
is not a well defined quantity. 
Figure~\ref{fig:cleoc_ds_kkmass} shows the $K^+K^-$ invariant
mass near the $\phi$ resonance.
The combination of the relatively broad $\phi$ resonance and interference
with other resonances, such as the $f_0(980)$, requires a complete
amplitude analysis to determine the different contributions. 
Instead,
CLEO-c provides partial branching fractions in different mass windows
around the $\phi$ resonance. These partial branching fractions,
given in four $K^+K^-$ mass windows centered at the $\phi$ mass are
presented in Table~\ref{tab:cleoc_ds_partialbrfr}. 

\begin{table}[bt]
\small{
\caption{Partial branching fractions in the mode $D_s^+\to K^+K^-\pi^+$
for events with a $K^+K^-$ invariant mass within $\Delta M$ MeV of the 
$\phi$, $|m_{K^-K^+}-m_{\phi}|<\Delta M$.
}
\label{tab:cleoc_ds_partialbrfr}
\begin{center}
\begin{tabular}{lc}
\hline\hline
$\Delta M$   & Partial Branching Fraction (\%)\\
\hline
5    &  $1.69\pm0.08\pm0.06$ \\
10   &  $1.99\pm0.10\pm0.05$ \\
15   &  $2.14\pm0.10\pm0.05$ \\
20   &  $2.24\pm0.11\pm0.06$ \\
\hline\hline
\end{tabular}
\end{center}
}
\end{table}

Systematic uncertainties from tracking efficiencies, $\pi^0$
and $K^0_S$ reconstruction, and particle identification 
are common in this analysis to those from the analysis
of the $D^0$ and $D^+$ absolute branching fractions
discussed in Sect.~\ref{sect:cleocdhad}. In additions,
for modes containing an $\eta$ in the final state an
uncertainty of $\pm 4.0\%$ is applied per $\eta$. 
Other large systematic uncertainties in this analysis
includes the uncertainties from the signal lineshape 
and the background parameterization in the fits for
the yields. These uncertainties are explored by using 
alternative fits. 

The CLEO-c analysis provides the to-date best determination of the
hadronic branching fractions for $D_s$ mesons. 
This analysis is statistics limited; the statistical 
uncertainty in the $D^+_s\to K^+K^-\pi^+$ mode
is 4.2\% and the systematic uncertainty about 3\%.
The largest systematic uncertainties 
come from the yield extraction. 
Both the statistical and systematic uncertainties would improve
with additional data. This analysis was based on 298 \pbinv,
CLEO-c has recorded a total of 589 \pbinv\ of data at this energy.

\subsection{Summary of Cabibbo favored $D^+_s$ decays}

The previous sections discussed the key measurements that
established the absolute branching fraction scale for
$D^+_s$ meson decays. These measurements have evolved
from model dependent determinations, e.g., making use of
equal semileptonic widths as for the $D^+$ decay, to 
model independent measurements using tagging techniques.
Also as the measurements have become more precise we need
to be more precise about what is measured. For example,
the often-used normalization mode $D^+_s\to \phi\pi^+$
suffers from a contamination from the $D^+_s\to f_0(980)\pi^+$
under the $\phi\pi^+$ signal. 

The results for the Cabibbo favored modes are summarized
in Table~\ref{tab:ds_cabibbo_favored}.

\begin{table}[bt]
\small{
\caption{Summary of branching fractions for
Cabibbo favored $D^+_s$ decays.
}
\label{tab:ds_cabibbo_favored}
\begin{center}
\begin{tabular}{lcc}
\hline\hline
Mode  &        & Branching Fraction \\
\hline
$D^+_s\to K^+K^0_S$          &           & $(1.49\pm0.09)\%$ \\
$D^+_s\to K^+K^-\pi^+$      &           & $(5.50\pm0.28)\%$ \\
$D^+_s\to K^+K^-\pi^+\pi^0$ &           & $(5.6\pm0.5)\%$ \\
$D^+_s\to K^0_SK^+\pi^+\pi^-$ &         & $(9.6\pm1.3)\times10^{-3}$ \\
$D^+_s\to K^0_SK^-\pi^+\pi^+$ &         & $(1.64\pm0.12)\%$ \\
$D^+_s\to K^+K^-\pi^+\pi^+\pi^-$ &      & $(8.8\pm1.6)\times10^{-3}$ \\
$D^+_s\to K^0_SK^0_S\pi^+\pi^+\pi^-$ & & $(8.4\pm3.5)\times10^{-4}$ \\
$D^+_s\to \pi^+\pi^+\pi^-$   &          & $(1.11\pm0.08)\%$ \\
$D^+_s\to \pi^+\pi^+\pi^-\pi^0$   &     & $<14\%$ \\
$D^+_s\to \pi^+\pi^+\pi^+\pi^-\pi^-$  & & $(8.0\pm0.9)\times10^{-3}$ \\
$D^+_s\to \pi^+\pi^+\pi^+\pi^-\pi^-\pi^0$ & & $(4.9\pm3.2)\%$ \\
\hline
$D^+_s\to \eta\pi^+$       &             & $(1.58\pm0.21)\%$ \\
$D^+_s\to \omega\pi^+$     &             & $(2.5\pm0.9)\times 10^{-3}$ \\
$D^+_s\to \eta\rho^+$      &             & $(13.0\pm2.2)\%$ \\
$D^+_s\to \eta'\pi^+$      &             & $(3.8\pm0.4)\%$ \\
$D^+_s\to \eta'\rho^+$     &             & $(12.2\pm2.0)\%$ \\
\hline\hline
\end{tabular}
\end{center}
}
\end{table}

\section{CABIBBO SUPPRESSED DECAYS OF $D^0$, $D^+$, AND $D_s^+$ MESONS}
\label{sect:cabibbo}

\subsection{Theoretical issues}

Studies of hadronic singly Cabibbo-suppressed decays of charmed mesons are important for several 
reasons. First, these decays hold the potential for future observation of direct (i.e. not associated with 
$D^0\overline{D}^0$ mixing~\cite{Petrov:2004gs}) $CP$ violation in the $D$-system. In the Standard Model, this is due 
to the fact that the final state particles contain at least one pair of a quark and antiquark of the same flavor, 
making possible a contribution from penguin-type amplitudes. Those amplitudes provide an 
access to the third generation of quarks ($b$-quarks in the loops), needed for observation of $CP$ violation 
in the Standard Model~\cite{Buccella:1992sg, Bianco:2003vb}.
New Physics can also make an entrance through those transitions, affecting both the amplitudes and 
$CP$-violating asymmetries~\cite{Grossman:2006jg}. Second, it offers new ground for studying 
strong dynamics in hadronic decays, in particular, the issue of flavor $SU(3)_F$ breaking in $D$-decays. 
For example, one of the famous failures of the applications of $SU(3)_F$ symmetry involves 
the prediction that the decay rates for $D^0 \to K^+ K^-$ and $D^0 \to \pi^+ \pi^-$ are equal. In reality,
the first rate is about three times larger than the second one. Other puzzles include the fact that 
the rates for decays like $D^+ \to K^{*+} \overline{K}^{*0}$ are so much enhanced by strong dynamics 
that their values appear to be as large as the ones of Cabibbo-favored decays. One popular explanation
for such phenomena include resonant final state interactions~\cite{Chau:1989tk,Kamal:1986tp} that affect 
not only $D$-decays, but  also $D^0\overline{D}^0$ mixing~\cite{Falk:1999ts,Golowich:1998pz}. 
There are also other explanations~\cite{Chau:1991gx,Savage:1991wu}. In order to study those 
phenomena it is convenient to select a base formalism for studies of hadronic transitions.

It is convenient to use the topological diagram approach to predict unknown branching ratios 
for singly-Cabibbo-suppressed decays. The analysis, done in \textcite{Chiang:2002mr} and repeated in 
\textcite{Bhattacharya:2008ss} and \textcite{Bhattacharya:2008ke} with updated experimental data, is 
displayed in Tables \ref{tab:scskpi} through \ref{tab:DtoPiPiandKK}. 

\subsubsection{$D \to PP$ transitions}
\label{sect:theo_Dpp}

Topological diagram approach to singly-Cabibbo-suppressed transitions can 
make use of the information obtained from the fits of CF decays discussed above. In 
particular, the ratio of primed (SCS) to unprimed (CF) amplitudes is fixed, it is just 
$\lambda^\prime = \tan \theta_C = 0.23$.  Table \ref{tab:scskpi} (taken 
from \textcite{Bhattacharya:2008ss}) presents
the most recent compilation of the branching ratios, amplitudes, and
representations in terms of reduced amplitudes for singly-Cabibbo-suppressed
(SCS) charm decays involving pions and kaons. The extracted topological amplitudes,
in units of $10^{-7}$ GeV, are
\bea\label{SCSAmps}
T' & = & 6.44~; \nonumber \\
C' & = & -4.15 - 2.25i~; \\
E' & = & -1.76 + 3.48i~;  \nonumber\\
A' & = &  0.55 - 1.14i~. \nonumber
\eea
The deviations from flavor SU(3) in Table \ref{tab:scskpi} are discussed below.
 
\begin{table*}[btph] 
\begin{center}  
\caption{Branching ratios, amplitudes, decomposition in terms of reduced
amplitudes, and predicted branching ratios for singly-Cabibbo-suppressed
charm decays involving pions and kaons. Predictions for the branching 
ratios are from~\cite{Bhattacharya:2008ss}.
\label{tab:scskpi}}
\begin{center}
\begin{tabular}{lccccccc} \hline \hline
Meson &    Decay    &         ${\cal B}$         & $p^*$ &    $|{\cal A}|$    &
   Rep.    &   Predicted    \\
      &    mode     &     $(10^{-3})$      & (MeV) &($10^{-7}$ GeV)&
           & ${\cal B}~(10^{-3})$ \\ \hline
$D^0$ &$\pi^+ \pi^-$& 1.40$\pm$0.02 & 921.9 & 4.61$\pm$0.03 &
$-(T'+E')$ &    2.23 \\
      &$\pi^0 \pi^0$& 0.80$\pm$0.08 & 922.6 & 3.49$\pm$0.17 &
$-(C'-E')/\sqrt{2}$ & 1.27 \\
      &  $K^+ K^-$  & 3.93$\pm$0.07 & 791.0 & 8.35$\pm$0.08 &
$(T'+E')$  &    1.92 \\
      &  $K^0 \overline{K}^0$  & 0.37$\pm$0.06 & 788.5 & 2.57$\pm$0.35 &
    0      &   0     \\ \hline
$D^+$ &$\pi^+ \pi^0$& 1.24$\pm$0.07 & 924.7 & 2.73$\pm$0.08 &
$-(T'+C')/\sqrt{2}$ & 0.87 \\
      &  $K^+ \overline{K}^0$  & 6.17$\pm$0.20 & 792.6 & 6.58$\pm$0.11 &
$T'-A'$    & 5.12 \\ \hline
$D_s^+$&$\pi^+ K^0$ & 2.44$\pm$0.30 & 915.7 & 5.84$\pm$0.36 &
$-(T'-A')$ & 2.56 \\
      & $\pi^0 K^+$ & 0.75$\pm$0.28 & 917.1 & 3.24$\pm$0.60 &
$-(C'+A')/\sqrt{2}$ & 0.87 \\
\hline\hline
\end{tabular}
\end{center}
\end{center}
\end{table*}

Note that the decay $D^0 \to K^0 \overline{K}^0$ is forbidden by $SU(3)_F$. Estimates of 
$SU(3)_F$-breaking effects lead to predictions for ${\cal B}(D^0 \to K^0 \overline{K}^0)$ 
that are consistent with experimental observations, but are by no means reliable 
\cite{Eeg:2001un,Dai:1999cs,Pham:1987rj,Lipkin:1980es}. We shall discuss those below.

Final states with $\eta$ and $\eta^\prime$ require additional considerations. In particular,
new topological amplitudes, flavor-singlet singlet-exchange $SE^\prime$ and singlet-annihilation
$SA^\prime$. The amplitudes $C$ and $E$ extracted from Cabibbo-favored charm decays imply
values of $C' = \lambda^\prime C$ and $E' = \lambda^\prime E$ which may be used in
constructing amplitudes for singly-Cabibbo-suppressed $D^0$ decays involving
$\eta$ and $\eta'$.  
\begin{table*}[btph]
\caption{Real and imaginary parts of amplitudes for SCS charm decays involving
$\eta$ and $\eta'$, in units of $10^{-7}$ GeV from \cite{Chiang:2002mr}.
\label{tab:scseta}}
\begin{center}
\begin{tabular}{ccrrc} \hline \hline
Amplitude   & Expression & Re & Im & $|{\cal A}_{\rm exp}|$ \\ \hline
$-\sqrt{6} {\cal A}(D^0 \to \pi^0 \eta)$ & $2E'-C' +SE'$
     & 0.63 & 9.21 & $7.79 \pm 0.54$ \\
$\frac{\sqrt{3}}{2} {\cal A}(D^0 \to \pi^0 \eta\,')$ & $\frac12(C' + E') + SE'$
     & $-2.95$ & 0.62 & $3.54 \pm 0.35$ \\
$\frac{3}{2 \sqrt{2}} {\cal A}(D^0 \to \eta \eta)$ & $C'+ SE'$
     & $-4.14$ & $-2.25$ & $5.91 \pm 0.34$ \\
$-\frac{3 \sqrt{2}}{7} {\cal A}(D^0 \to \eta \eta\,')$ & $\frac17(C' + 6E')+SE'$
     & $-2.10$ & 2.66 & $3.48 \pm 0.38$ \\
$\sqrt{3} {\cal A}(D^+ \to \pi^+ \eta)$ & $T'+2C'+2A'+ SA'$
     & $-0.75$ & $-6.77$ & 8.21$\pm$0.26 \\
$-\frac{\sqrt{6}}{4} {\cal A}(D^+ \to \pi^+ \eta\,')$ & $\frac14(T'-C'+2A')+SA'$
     & 2.92 & $-0.01$ & 3.72$\pm$0.15 \\
$-\sqrt{3} {\cal A}(D_s^+ \to \eta K^+)$ & $-(T'+2C')+SA'$
     & 1.85 & 4.50 & 8.05$\pm$0.88 \\
$\frac{\sqrt{6}}{4} {\cal A}(D_s^+ \to \eta\,' K^+)$ & $\frac14(2T'+C'+3A')+SA'$
     & 2.59 & $-1.41$ & 3.43$\pm$0.57 \\ \hline \hline
\end{tabular}
\end{center}
\end{table*}

\subsubsection{$D \to PV$ transitions}

A similar technique can be applied to describe $D \to PV$ transitions.  In this case, similar
topological amplitudes are denoted by a subscript "V". We present the most recent result in
Table \ref{tab:SCSPV}~\cite{Bhattacharya:2008ke}.

\begin{table*}[btph]
\caption{Branching ratios and invariant amplitudes for singly-Cabibbo-suppressed
decays of charmed mesons to one pseudoscalar and one vector meson (from ~\cite{Bhattacharya:2008ke}).
\label{tab:SCSPV}}
\begin{center}
\begin{tabular}{c l c c c c}
\hline \hline
Meson & Decay & Representation
     & ${\cal B}$ \cite{Amsler:2008zzb} & $p^*$ & $|{\cal A}|$ \\
 & mode & & ($\%$) & (MeV) & $(10^{-6})$ \\ \hline \hline
$D^0$ & $\pi^+\, \rho^-$& $-(T_V\,' + E_P\,')$ & 0.497$\pm$0.023 & 763.8
 & 1.25$\pm$0.03 \\
 & $\pi^-\, \rho^+$ & $-(T_P\,' + E_V\,')$ & 0.980$\pm$0.040 & 763.8
 & 1.76$\pm$0.04 \\
 & $\pi^0\, \rho^0$ & $\frac{1}{2}(E_P\,' + E_V\,' - C_P\,' - C_V\,')$
 & 0.373$\pm$0.022 & 764.2 & 1.08$\pm$0.03 \\
 & $K^+\, K^{*-}$ & $T_V\,' + E_P\,'$ & 0.153$\pm$0.015 & 609.8
 & 0.97$\pm$0.05 \\
 & $K^-\, K^{*+}$ & $T_P\,' + E_V\,'$ & 0.441$\pm$0.021 & 609.8
 & 1.65$\pm$0.04 \\
 & $K^0\, \overline{K}^{*0}$ & $E_V\,' - E_P\,'$ & $< 0.18$ & 605.3 & \\
 & $\overline{K}^{0}\, K^{*0}$ & $E_P\,' - E_V\,'$ & $< 0.09$ & 605.3 & \\
 & $\pi^0\, \phi$ & $\frac{1}{\sqrt{2}} C_P\,'$ & 0.124$\pm$0.012& 644.7 & 0.81$\pm$0.04\\
 & $\pi^0\, \omega$ & $\frac{1}{2}(E_P\,' + E_V\,' - C_P\,' + C_V\,')$
 & & 761.2 & \\
 & $\eta\, \rho^0$ & $\frac{1}{\sqrt{6}}(2 C_V\,' - C_P\,' - E_P\,' - E_V\,')$
 & & 652.0 & \\
 & $\eta\, \omega$ & $- \frac{1}{\sqrt{6}}(2 C_V\,' + C_P\,' + E_P\,' +E_V\,')$
 & & 488.8 & \\
 & $\eta\, \phi$ & $\frac{1}{\sqrt{3}}(C_P\,' - E_P\,' - E_V\,')$ & & 648.1
 & \\
 & $\eta\,' \rho^0$ & $\frac{1}{2 \sqrt{3}}(E_P\,' + E_V\,' + C_P\,' +C_V\,')$
 & & 342.5 & \\
 & $\eta\,' \omega$ & $\frac{1}{2 \sqrt{3}}(E_P\,' + E_V\,' + C_P\,' - C_V\,')$
 & & 333.5 & \\ \hline
$D^+$  &$\rho^0\, \pi^+$   &$\frac{1}{\sqrt{2}}(A_P\,'-A_V\,'-C_P\,'-T_V\,')$   &0.082$\pm$0.015&$767$&0.32$\pm$0.03\\
       &$\omega\, \pi^+$   &$-\frac{1}{\sqrt{2}}(A_P\,'+A_V\,'+C_P\,'+T_V\,')$  &$<0.034$&$764$       &\\
       &$\phi\, \pi^+$     &$C_P\,'$                                      &0.620$\pm$0.070&$647$&1.13$\pm$0.06\\
       &$\overline{K}^{*0}\, K^+$&$(T_V\,'-A_V\,')$                             &0.435$\pm$0.048&$611$&1.03$\pm$0.06\\
       &$\pi^0\, \rho^+$   &$\frac{1}{\sqrt{2}}(A_V\,'-A_P\,'-C_V\,'-T_P\,')$   &&$767$               &\\
       &$\eta\, \rho^+$    &$\frac{1}{\sqrt{6}}(A_V\,'+A_P\,'+2C_V\,'+T_P\,')$ &$<0.7$&$656$         &\\
       &$\eta\,' \rho^+$   &$\frac{1}{\sqrt{6}}(C_V\,'-A_V\,'-A_P\,'-T_P\,')$  &$<0.5$&$349$         &\\
       &$\overline{K}^0\, K^{*+}$&$(T_P\,'-A_P\,')$                             &3.18$\pm$1.38&$612$  &2.78$\pm$0.60\\ \hline
$D_s^+$&$\pi^+\, K^{*0}$   &$(A_V\,'-T_V\,')$                             &0.225$\pm$0.039&$773$&0.79$\pm$0.07\\
       &$\pi^0\, K^{*+}$   &$-\frac{1}{\sqrt{2}}(C_V\,'+A_V\,')$                &&$775$&\\
       &$\eta\,  K^{*+}$   &$\frac{1}{\sqrt{3}}(T_P\,'+2C_V\,'+A_P\,'-A_V\,')$ &&$661$&\\
       &$\eta\,' K^{*+}$   &$\frac{1}{\sqrt{6}}(2T_P\,'+C_V\,'+2A_P\,'+A_V\,')$&&$337$&\\
       &$K^0\,   \rho^+$   &$(A_P\,'-T_P\,')$                             &&$743$&\\
       &$K^+\, \rho^0$     &$-\frac{1}{\sqrt{2}}(C_P\,'+A_P\,')$                &0.27$\pm$0.05&$745$&0.92$\pm$0.09\\
       &$K^+\, \omega$     &$-\frac{1}{\sqrt{2}}(C_P\,'-A_P\,')$                &&$741$&\\
       &$K^+\, \phi$       &$T_V\,'+C_P\,'+A_V\,'$                        &$<0.057$&$607$&\\ \hline \hline

\end{tabular}
\end{center}
\end{table*}

\subsection{Cabibbo suppressed $D^0$ and $D^+$ decays}

Experimentally, Cabibbo suppressed or doubly Cabibbo suppressed decays
of $D^0$ or $D^+$ mesons are almost always measured
relative to a Cabibbo favored normalization mode. This includes
most CLEO-c analyses as the branching fractions for
Cabibbo suppressed modes are typically 
suppressed by $|V_{cd}/V_{cs}|^2\approx
0.05$ and the statistics in these modes using a double tag analysis
would be limited. 
In some cases, e.g. the CLEO-c analysis 
of $D^0\to K\bar K$ final states~\cite{Bonvicini:2008nr},
CLEO has normalized against the number of produced $D\bar D$
events and measured directly the branching fraction.

\subsubsection{Two-body decays of $D^0$ and $D^+$}
\label{sect:KK_and_pipi}

There is a substantial amount of data on the two-body decays of $D^0$  and $D+$.
The first measurements of Cabibbo suppressed $D^0$ decays 
were for $D^0\to K^-K^+$ and $D^0\to \pim\pip$ done by the Mark II 
experiment~\cite{Abrams:1979th}. Since the first observation
of these modes they have been measured by many experiments with
increased precision. In these measurements the $D^0\to K^-K^+$
and $D^0\to \pim\pip$ branching fractions are measured relative 
to the $D^0\to K^-\pip$ yield. Experiments operating above the
$c\bar c$ threshold tag the $D^0$ by looking at the $D^0$--$D^{*+}$ mass
difference in the decay $D^{*+}\to D^0\pip$. 

The results for the $D^0\to K^-K^+$ and $D^0\to \pim\pip$ 
decays are summarized in Table~\ref{tab:DtoPiPiandKK}.
The most precise measurement is that of CDF~\cite{Acosta:2004ts}, in
the $D^0\to K^-K^+$ they reconstruct about 16,000 signal
candidates.

As can be seen from Table~\ref{tab:DtoPiPiandKK}, the rate for
$D \to K\overline{K}$ is larger than the rate for $D \to \pi\overline{\pi}$
by a factor of three. In the $SU(3)_F$ (or in the $U$-spin) symmetry limit,
those rates should be the same. $SU(3)_F$ is, in general, expected to
work to 30\%, so this is a rather severe  violation of this symmetry.

While the one popular explanation for this puzzle involves final state interactions 
(i.e. a presence of a resonance that couples stronger to $K^+K^-$ compared to
$\pi^+\pi^-$ state), it might be tempting to try to understand the issue in 
factorization~\cite{Chau:1991gx}. Neglecting for a moment the annihilation diagram 
contribution,
\beq
\frac{{\cal{A}}_{K\overline{K}}}{{\cal{A}}_{\pi\overline{\pi}}} =
\frac{f_K}{f_\pi} \frac{m_D^2-m_K^2}{m_D^2-m_\pi^2}
\frac{F^{DK}(m_K^2)}{F^{D\pi}(m_\pi^2)}.
\eeq
With the recent lattice evaluations $f_K/f_\pi =1.218 \pm 0.002^{+0.011}_{-0.024}$ 
from a recent lattice QCD calculation with domain-wall fermions~\cite{Beane:2006kx}, 
assuming a modified pole dominance for the form-factors $F^{DK}(m_K^2)$ and $F^{D\pi}(m_\pi^2)$, 
and extracting them from semileptonic $D$-decays (see \cite{Artuso:2008vf} for a recent review and
\cite{Besson:2009uv} for recent determination of parameters), we get
\beq
{\cal{A}}_{K\overline{K} }\simeq 1.32 {\cal{A}}_{\pi\overline{\pi}}.
\eeq
In other words, factorization predicts about 30\% breaking of $SU(3)_F$ in spectator
amplitudes (c.f. \cite{Chau:1991gx}). Clearly, this is not sufficient for the resolution of the
puzzle. The presence of final state interaction (FSI)-enhanced exchange amplitude is crucial 
for the explanation of this phenomenon.

\begin{table*}[btph]
\caption{Measurements of $D^0\to K^-K^+$ and $D^0\to \pim\pip$.
The branching fractions have been recalculated using 
${\cal B}(D^0\to K^-\pi^+)=(3.89\pm0.05)\%$.
}
\label{tab:DtoPiPiandKK}
\begin{center}
\begin{tabular}{lcc}
\hline\hline
Experiment  & ${\calB}(D^0\to K^-K^+)\ (10^{-3})$   & ${\calB}(D^0\to \pim\pip)\ (10^{-3})$  \\
\hline
CLEO-c~\cite{Bonvicini:2008nr,Rubin:2005py} & $4.08\pm0.08\pm0.09$  &  $1.41\pm0.04\pm0.03$\\
BES II~\cite{Ablikim:2005ev}   & $4.75\pm0.43\pm0.17$  & \\
CDF~\cite{Acosta:2004ts}       & $3.859\pm0.043\pm0.069$  & $1.40\pm0.02\pm0.03$\\
FOCUS~\cite{Link:2002hi}       & $3.863\pm0.054\pm0.074$  & $1.37\pm0.05\pm0.03$ \\
CLEO II~\cite{Csorna:2001ww}   & $4.05\pm0.13\pm0.13$  & $1.36\pm0.06\pm0.07$ \\
E791~\cite{Aitala:1997ff}      & $4.24\pm0.12\pm0.13$  & $1.56\pm0.08\pm0.12$ \\
CLEO II~\cite{Asner:1996ht}    & $4.51\pm0.27\pm0.28$  & \\
E687~\cite{Frabetti:1993fw}    & $4.24\pm0.27\pm0.35$  & \\
E691~\cite{Anjos:1991dr}       & $4.16\pm0.39\pm0.39$  & \\
CLEO~\cite{Alexander:1990nf}   & $4.55\pm0.39\pm0.28$  & \\ \hline
Average                        & $3.98\pm0.07$ &    $1.40\pm0.03$\\
\hline\hline
\end{tabular}
\end{center}
\end{table*}

A number of other two body final states to pseudo scalars and
have been studied. These decays are summarized in 
Table~\ref{tab:D_to_2body}.

\begin{table}[btph]
\caption{Measurements of $D^0$ and $D^+$ decays to Cabibbo
suppressed, non-strange, two-body final states. The averages are 
from~\textcite{Amsler:2008zzb}, the decay $D^0\to\pi^+\pi^-$
is discussed in Sect.~\ref{sect:KK_and_pipi}.
}
\label{tab:D_to_2body}
\begin{center}
\begin{tabular}{lc}
\hline\hline
Mode                          & ${\calB}\ (10^{-3})$  \\
\hline
$D^0\to\pi^+\pi^-$            & $1.40\pm0.02$  \\
$D^0\to\pi^0\pi^0$            & $0.80\pm0.08$  \\
$D^0\to\eta\pi^0$             & $0.57\pm0.14$  \\
$D^0\to\omega\pi^0$           & $<0.26$  \\
$D^+\to\pi^0\pi^+$            & $1.24\pm0.07$  \\
$D^+\to\pi^+\eta$             & $3.39\pm0.29$  \\
$D^+\to\pi^+\omega$           & $<0.34$  \\
\hline\hline
\end{tabular}
\end{center}
\end{table}

The most complete study of $D$ mesons decays to final states
containing $\eta$ and $\eta'$ mesons is done by CLEO-c~\cite{Artuso:2008ri}.
The excellent electromagnetic calorimeter and the clean environment
near threshold combined with the large data sample collected
at the $\psidprime$ has allowed CLEO-c to measure many modes not
previously seen. 

This analysis uses 281 \pbinv\ of data collected at the $\psidprime$
resonance. In this study CLEO-c makes use of single tags; the modes studied here have
sufficiently small branching fractions that using $D$ tagging is not 
useful. The $\piz$ and $\eta$ mesons are reconstructed in the $\gamma\gamma$
final state. In addition, for modes with two $\eta$ mesons in the 
final state ($\eta\eta$ and $\eta\eta'$) the $\eta\to\pip\pim\piz$
channel is used to reconstruct $\eta$ mesons. The $\eta'$ is
reconstructed in the channel $\eta'\to\eta\pip\pim$. It is required
that $402<M_{\eta\pip\pim}-M_{\eta}<418\ \Mevcsq$.

\begin{figure}[tbph]
\begin{center}
\includegraphics[width=0.99\linewidth]{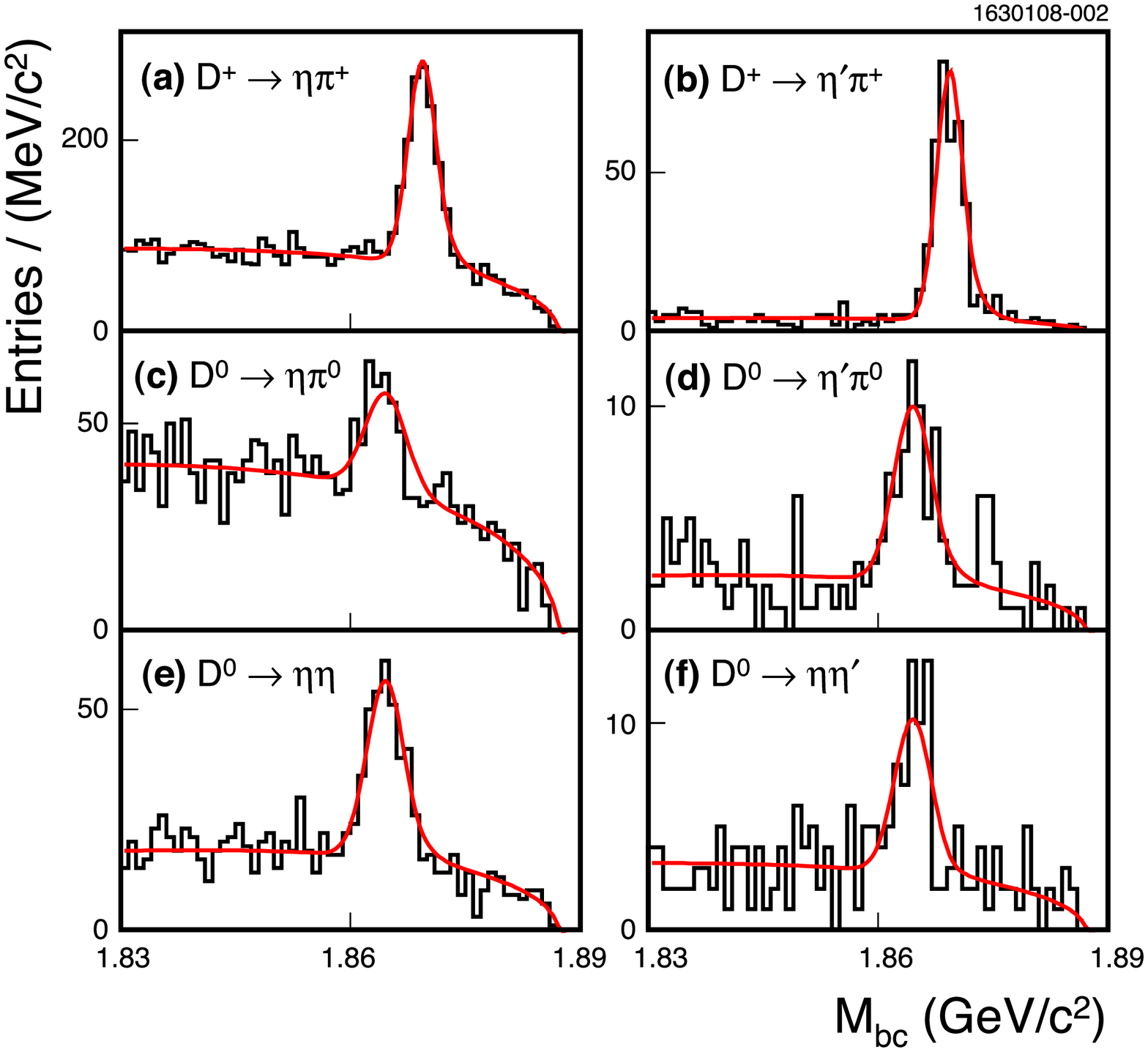}
\caption{Yields for a) $\Dp\to\eta\pip$, b) $\Dp\to\eta'\pip$,
c) $\Dz\to\eta\piz$, d) $\Dz\to\eta'\piz$, e) $\Dz\to\eta\eta$,
and $\Dz\to\eta\eta'$. 
From~\textcite{Artuso:2008ri}.
}
\label{fig:dtoeta1}
\end{center}
\end{figure}

\begin{figure}[tbph]
\begin{center}
\includegraphics[width=0.99\linewidth]{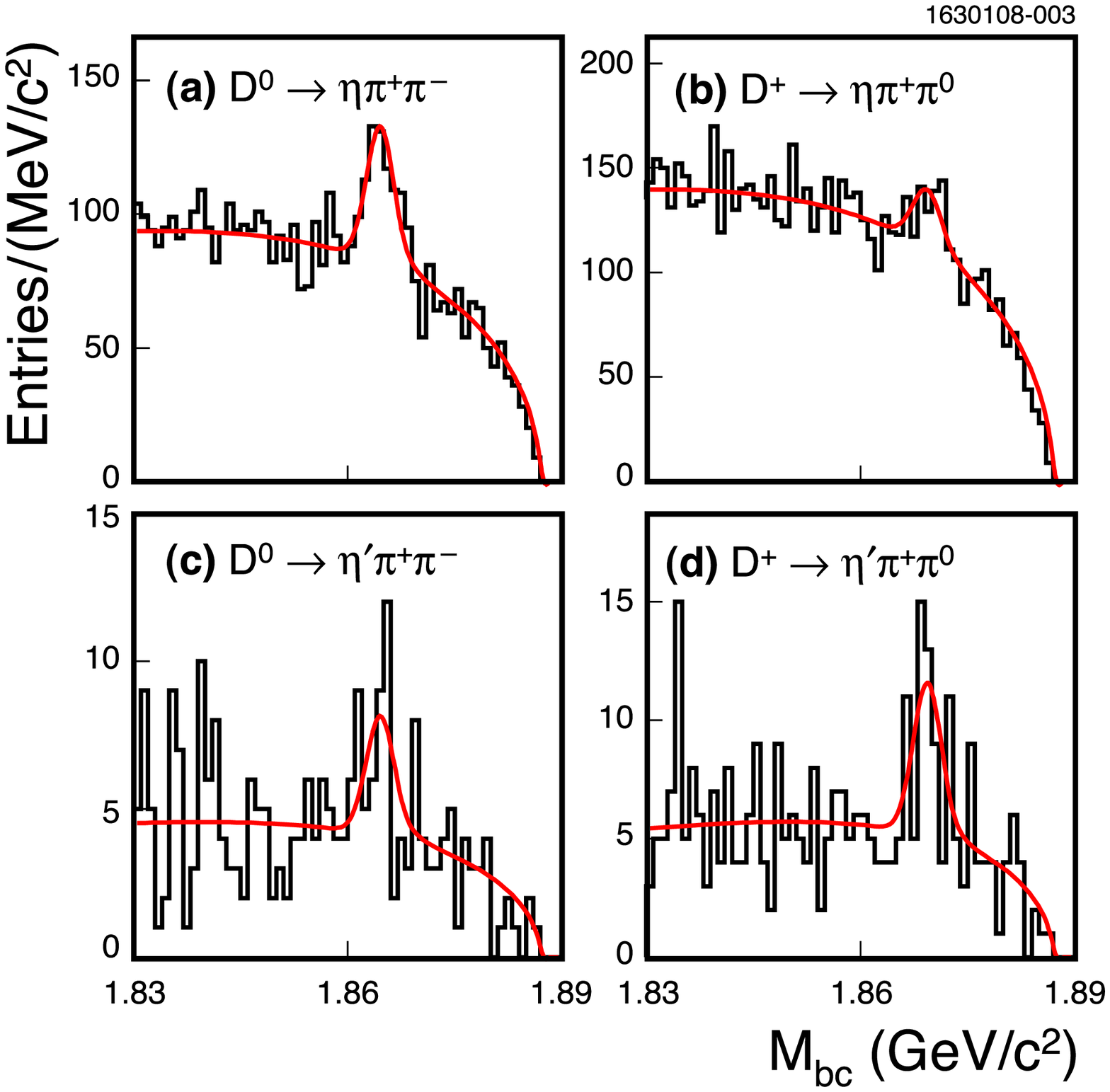}
\caption{Yields for a) $\Dz\to\eta\pip\pim$, b) $\Dp\to\eta\pip\piz$,
c) $\Dz\to\eta'\pip\pim$, and d) $\Dp\to\eta'\pip\piz$.
From~\textcite{Artuso:2008ri}.
}
\label{fig:dtoeta2}
\end{center}
\end{figure}

The yields are extracted by fitting the $\Mbc$ distributions after
selecting events consistent with $\Delta E=0$. In 
Figs.~\ref{fig:dtoeta1},~\ref{fig:dtoeta2} the observed signals are shown.
The significance for all modes are over $4\sigma$ except 
for the $\Dz\to\eta'\pip\pim$ mode where the significance is estimated
to be $3.2\sigma$. The observed yields and branching fractions
are summarized in Tab.~\ref{tab:dtoetaandetaprime}.
This data makes it possible to constrain new singlet exchange $SE^\prime$
amplitudes introduced in Sect.~\ref{sect:theo_Dpp}. In order to do that, one can
rewrite four equations for $D^0$ decay amplitudes to the final states with 
$\eta^{(\prime)}$:
\bea\label{SEdet}
-\sqrt{6} {\cal{A}}(D^0 \to \eta \pi^0) &=& 2 E^\prime - C^\prime + SE^\prime, 
\nonumber \\
\frac{\sqrt{3}}{2} {\cal{A}}(D^0 \to \eta^\prime \pi^0) &=& \frac{1}{2}\left(E^\prime + C^\prime\right) 
+ SE^\prime,
\nonumber \\
\frac{3}{2\sqrt{2}} {\cal{A}}(D^0 \to \eta \eta) &=& C^\prime + SE^\prime, 
\\
-\frac{3\sqrt{2}}{7} {\cal{A}}(D^0 \to \eta^\prime \eta^\prime) &=& \frac{1}{7}\left(C^\prime + 
6 E^\prime\right) + SE^\prime. 
\nonumber
\eea
It is interesting to note that the right-hand side of each of Eqs.(\ref{SEdet}) determines a vector in a complex plane.
Since both amplitudes and phases of $C^\prime$ and $E^\prime$ are known from Eq.~(\ref{SCSAmps}), these 
four equations contain a common complex off-set, $SE^\prime$. Since only the magnitudes of the right-hand sides 
of these equations are known, they each define a circle in the complex plane with the radius given by 
that magnitude. Plotting them on the same graph then determines $SE^\prime$. 

\begin{figure}[tbph]
\begin{center}
\includegraphics[width=0.98\linewidth]{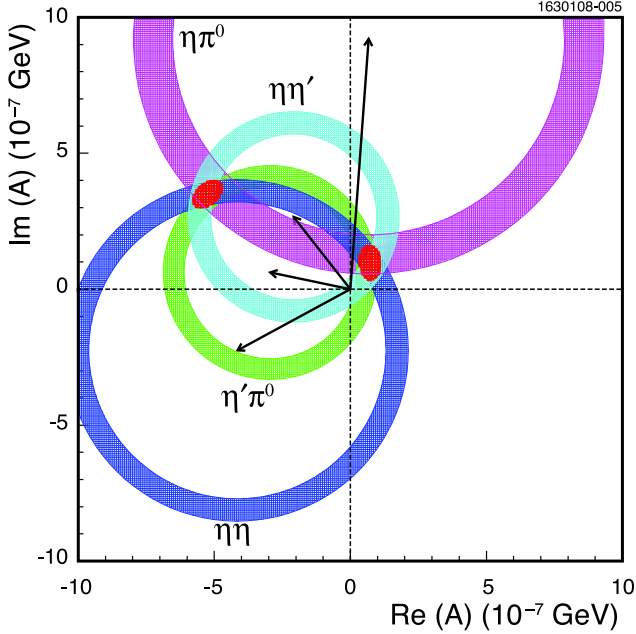}
\caption{Graphical representation of Eqs.~(\ref{SEdet}) used to determine 
$SE^\prime$ from the amplitude analysis for $D\to \eta^{(\prime)} \eta (\pi^0)$ 
\textcite{Artuso:2008ri}. Circles represent absolute values of the decay amplitudes. The
intersection points provide two possible solutions for $SE^\prime$ (see text).
}
\label{fig:dtoetaamps}
\end{center}
\end{figure}

This is done in Fig.~\ref{fig:dtoetaamps}. Notice that all circles intersect in two points, which determine two 
possible solutions for $SE^\prime$. The smaller values for $SE^\prime = (-0.7\pm 0.4) \times 10^{-7}~\mbox{GeV} + 
i (-1.0\pm 0.6) \times 10^{-7}~\mbox{GeV}$ are theoretically preferable, as $SE^\prime$ is an Okubo-Zweig-Iizuka 
(OZI)-suppressed amplitude~\cite{Okubo:1977rk,Zweig:1964jf,Iizuka:1966fk}.
\begin{table}[btph]
\caption{Measurements of $D$ meson decays to final states with $\eta$ and
$\eta'$ mesons. 
}
\label{tab:dtoetaandetaprime}
\begin{center}
\begin{tabular}{lcc}
\hline\hline
Mode                        & Yield           & Branching Fraction ($10^{-4}$)  \\
\hline
${\cal B}(\Dp\to\eta\pip)$  & $1033\pm 42$    & $34.3\pm 1.4\pm 1.7$ \\
${\cal B}(\Dp\to\eta'\pip)$ & $352\pm 20$     & $44.2\pm 2.5\pm 2.9$ \\
${\cal B}(\Dz\to\eta\piz)$  & $156\pm 24$     & $6.4\pm 1.0\pm 0.4$ \\
${\cal B}(\Dz\to\eta'\piz)$ & $50\pm 9$       & $8.1\pm 1.5\pm 0.6$ \\
${\cal B}(\Dz\to\eta\eta)$  & $255\pm 22$     & $16.7\pm 1.4\pm 1.3$ \\
${\cal B}(\Dz\to\eta\eta')$ & $46\pm 9$       & $12.6\pm 2.5\pm 1.1$ \\
${\cal B}(\Dz\to\eta\pip\pim)$ & $257\pm 32$  & $10.9\pm 1.3\pm 0.9$ \\
${\cal B}(\Dp\to\eta\pip\piz)$ & $149\pm 34$  & $13.8\pm 3.1\pm 1.6$ \\
${\cal B}(\Dz\to\eta'\pip\pim)$ & $21\pm 8$   & $4.5\pm 1.6\pm 0.5$ \\
${\cal B}(\Dp\to\eta'\pip\piz)$ & $33\pm 9$   & $15.7\pm 4.4\pm 2.5$ \\
\hline\hline
\end{tabular}
\end{center}
\end{table}
%

\subsubsection{Multi-body decays with kaons and pions}

Multibody decays of $D^0$ and $D^+$ mesons were also extensively studied.
While theoretical studies of those transitions are limited, some of those decays
can be used in the Dalitz-plot analyses of $D^0{\overline D}^0$ mixing~\cite{Artuso:2008vf}. 
The measurements of branching fractions to final states with three of more pions, including 
final states with $\eta$ and $\omega$ mesons could be found in Table~\ref{tab:D_to_multipions}. 
In Section~\ref{sect:multi} Dalitz plot analysis of three-body final states are discussed.

\begin{table}[btph]
\caption{Measurements of $D^0$ and $D^+$ decays to Cabibbo
suppressed final states with three or more pions in the final states.
Final states with $\eta$ and $\omega$ mesons are also included.
Limits are given at 90\% C.L. 
}
\label{tab:D_to_multipions}
\begin{center}
\begin{tabular}{lc}
\hline\hline
Mode                              & ${\calB}\ (10^{-3})$  \\
\hline
$D^0\to\pi^+\pi^-\pi^0$           & $14.1\pm0.6$  \\
$D^0\to\pi^+\pi^-\pi^+\pi^-$      & $7.44\pm0.21$  \\
$D^0\to\pi^+\pi^-\pi^+\pi^-\pi^0$ & $4.2\pm0.5$  \\
$D^0\to\pi^0\pi^0\pi^0$           & $<0.35$ \\
$D^0\to\pi^+\pi^-\pi^0\pi^0$      & $(1.00\pm0.09)$  \\
$D^0\to\pi^+\pi^-\pi^+\pi^-\pi^+\pi^-$ & $4.2\pm1.2$  \\
$D^+\to\pi^+\pi^+\pi^-$           & $3.21\pm0.19$  \\
$D^+\to\pi^+\pi^0\pi^0$           & $4.6\pm0.4$  \\
$D^+\to\pi^+\pi^+\pi^-\pi^0$      & $1.14\pm0.08$  \\
$D^+\to\pi^+\pi^+\pi^-\pi^+\pi^-$ & $1.63\pm0.16$  \\
\hline\hline
\end{tabular}
\end{center}
\end{table}

In addition to the $D^0\to KK$ decays discussed above, many other 
Cabibbo suppressed decays with two kaon in the final states have been studied. 
Dalitz plot analysis has been performed on some three-body final states
as discussed in Sect.~\ref{sect:multi}. The final states with two kaons in the final state are 
summarized in Table~\ref{tab:D_to_multikaons}.

\begin{table}[btph]
\caption{Measurements of $D^0$ and $D^+$ decays to Cabibbo
suppressed final states with two kaons. 
}
\label{tab:D_to_multikaons}
\begin{center}
\begin{tabular}{lc}
\hline\hline
Mode                          & ${\cal B}\ (10^{-3})$  \\
\hline
$D^0\to K^0_SK^-\pi^+$           & $3.5\pm0.5$ \\
$D^0\to K^0_SK^+\pi^-$           & $2.7\pm0.5$ \\
$D^0\to K^+K^-\pi^0$             & $3.29\pm0.14$ \\
$D^0\to K^0_SK^0_S\pi^0$         & $<0.59$ \\
$D^0\to K^+K^-\pi^+\pi^-$        & $2.43\pm0.012$ \\
$D^0\to K^0_SK^0_S\pi^+\pi^-$    & $1.30\pm0.24$ \\
$D^0\to K^0_SK^-\pi^+\pi^+\pi^-$ & $<0.15$ \\
$D^0\to K^+K^-\pi^+\pi^-\pi^0$   & $3.1\pm2.0$ \\
$D^+\to K^+K^-\pi^+$             & $9.63\pm0.31$ \\
$D^+\to K^+K^0_S\pi^+\pi^-$      & $1.69\pm0.18$ \\
$D^+\to K^0_SK^-\pi^+\pi^+$      & $2.32\pm0.18$ \\
$D^+\to K^+K^-\pi^+\pi^+\pi^-$   & $2.3\pm1.2$ \\
\hline\hline
\end{tabular}
\end{center}
\end{table}

\subsection{Cabibbo suppressed $D_s$ decays}

The Cabibbo suppressed $D_s$ decays are final states with one or
three kaons. The measured decays are listed in Table~\ref{tab:Ds_Cabibbo}.
This table also includes the doubly Cabibbo suppressed decay 
$D^+\to K^+K^+\pi^-$. CLEO-c~\cite{Adams:2007mx} has performed a 
systematic study of two-body $D_s$ decays.

\begin{table*}[btph]
\caption{Cabibbo suppressed $D_s^+$ decays.
}
\label{tab:Ds_Cabibbo}
\begin{center}
\begin{tabular}{lcc}
\hline\hline
Mode                                  & Ref.   & ${\cal B}/ 10^{-3}$  \\
\hline
$D_s^+\to K^+ \pi^0$                  & \cite{Adams:2007mx}  & $0.82\pm0.22$ \\
$D_s^+\to K^0_S \pi^+$                & \cite{Link:2007fra,Adams:2007mx}  & $1.25\pm0.15$ \\
$D_s^+\to K^+ \eta$                   & \cite{Adams:2007mx}  & $1.41\pm0.31$ \\
$D_s^+\to K^+ \eta'$                  & \cite{Adams:2007mx}  & $1.6\pm0.5$  \\
$D_s^+\to K^+ \pi^+\pi^-$             & \cite{Alexander:2008cqa} & $6.9\pm0.5$  \\
$D_s^+\to K^0_S \pi^+\pi^+\pi^-$      &  \cite{Link:2007fra} & $3.1\pm1.1$  \\
$D_s^+\to K^+K^+K^-$                  &  \cite{Link:2002iy}  & $0.49\pm0.17$ \\
$D_s^+\to K^+K^+\pi^-$                &  \cite{Link:2005ym}  & $0.29\pm0.11$ \\
\hline\hline
\end{tabular}
\end{center}
\end{table*}

\subsection{Doubly Cabibbo suppressed decays}
\label{sect:doubleCabibbo}

The doubly Cabibbo suppressed decays have two
Cabibbo suppressed weak couplings. Naively,
the rate for the doubly Cabibbo suppressed 
decays are supressed by a factor of $\tan^4\theta_C\approx 2.8\times 10^{-3}$.
The first observation of a doubly Cabibbo suppressed decay 
was in the decay channel $D^0\to K^+\pi^-$~\cite{Cinabro:1993nh}.
Experimentally, the flavor, $D^0$ or $\bar D^0$, of the initial state
is tagged by the charged of the slow pion in the
decay of a $D^{*+}\to D^0\pi^+$. The simplest measurements
observe the time integrated rate of $D^0$ decays
and do not separated direct decay contributions from
mixing, where a $D^0$ oscillated to a $\bar D^0$ and
decayed via a Cabibbo favored decays.

The $D^0$ doubly Cabibbo suppressed decays that have 
been studied are summarized in Table~\ref{tab:D_doublyCabibbo}.
The three most precise measurements of the $D^0\to K^+\pi^-$
decay by CDF~\cite{Aaltonen:2007uc}, 
BABAR~\cite{Aubert:2007wf}, 
and Belle~\cite{Zhang:2006dp} obtained branching rations 
with respect to $D^0\to K^-\pi^+$ of $(4.15\pm0.10)\times 10^{-3}$,
$(3.53\pm0.08\pm0.04)\times 10^{-3}$, and 
$(3.77\pm0.08\pm 0.05)\times 10^{-3}$ respectively. The 
agreement between these  measurements is not very good, the
PDG applies a scale factor of 3.3 for the error on their
average to obtain the average ratio of branching fractions
to be $(3.80\pm0.18)\times 10^{-3}$.

\begin{table}[btph]
\caption{Doubly Cabibbo suppressed $D^0$ decays. The first column
(${\cal B}$) shows the branching fraction for the decay and the
second column ($R$) shows the ratio of the branching fraction
with respect to the corresponding Cabibbo favored decay.
}
\label{tab:D_doublyCabibbo}
\begin{center}
\begin{tabular}{lccc}
\hline\hline
Mode                         & ${\cal B}\ (10^{-4})$   & $R\ (10^{-3})$ \\
\hline
$D^0\to K^+ \pi^-$           & $1.48\pm0.07$        & $3.80\pm0.18$  \\
$D^0\to K^+ \pi^-\pi^0$      & $3.05\pm0.17$        & $2.20\pm0.10$  \\
$D^0\to K^+ \pi^-\pi^+\pi^-$ & $2.62^{+0.21}_{-0.19}$& $3.23^{+0.25}_{-0.22}$\\
\hline\hline
\end{tabular}
\end{center}
\end{table}

The decay $D^0\to K^+\pi^-\pi^0$ was first observed by 
CLEO~\cite{Brandenburg:2001ze}.
The PDG average is dominated by the more recent measurements
from BABAR~\cite{Aubert:2006kt} and Belle~\cite{Tian:2005ik}.

The first significant $D^0\to K^+\pi^-\pi^+\pi^-$ observation
was made by CLEO~\cite{Dytman:2001rx}. The most
recent and precise measurement of this decay was done by
Belle~\cite{Tian:2005ik}.

Both CLEO-c~\cite{Dytman:2006ha} and 
BABAR~\cite{Aubert:2006sh} has studied the doubly Cabibbo suppressed
decay $D^+\to K^+\pi^0$. CLEO-c has reconstructed candidates in a
281 pb$^{-1}$ sample of $e^+e^-$ data recorded at the $\psi(3770)$. BABAR
has used a sample of 124 fb$^{-1}$ recorded at the $\Upsilon(4S)$.
CLEO-c and BABAR finds branching fractions in good agreement with
each other, 
${\cal B}(D^+\to K^+\pi^0)=(2.24\pm0.36\pm0.15\pm0.08)\times 10^{-4}$
and ${\cal B}(D^+\to K^+\pi^0)=(2.52\pm0.46\pm0.24\pm0.08)\times 10^{-4}$,
respectively.
The average branching fraction obtained is $(2.37\pm0.32)\times 10^{-4}$.

The final state $D^+\to K^+\pi^+\pi^-$ has been studied by 
E687~\cite{Frabetti:1995zi},
E791~\cite{Aitala:1997ht}, and
FOCUS~\cite{Link:2004mx}. The average branching fraction from
these measurements is 
${\cal B}(D^+\to K^+\pi^+\pi^-)=(6.2\pm0.7)\times 10^{-4}$.

The decay $D^+\to K^+K^+K^-$ has been observed by
FOCUS~\cite{Link:2002iy}. They measure the ratio of branching fractions
${\cal B}(D^+\to K^+K^+K^-)/{\cal B}(D^+\to K^-\pi^+\pi^+)=
(9.49\pm2.17\pm0.22)\times 10^{-4}$. This gives the 
branching fraction
${\cal B}(D^+\to K^+K^+K^-)=(8.7\pm2.0)\times 10^{-5}$.

\section{FINAL STATE INTERACTIONS AND AMPLITUDE ANALYSIS}
\label{sect:annihilation}

One of the simplest ways to analyze decays of $D$-mesons is to employ 
the flavor flow diagram technique
described earlier. One potential problem with the application of 
this technique\footnote{Similar problem could affect
charm decay analysis using factorization approximation.} to charm decays involves assignment of quark 
amplitudes (${\cal T}$, ${\cal A}$, etc.) to a particular decay. The root of the problem involves inelastic final state 
interactions (FSI).

\subsection{Hadronic decays into meson states}

Historically, the issue came up with decays of the type $D^0 \to \phi K^0$, which have been claimed to 
originate entirely from quark exchange amplitudes. 
Thus, in the topological $SU(3)$ or flavor-flow 
analysis of this transition only an exchange amplitude ${\cal E}$ should be assigned to this decay. 
However, FSI contribution of the type
\beq
D^0 \to \eta^{(\prime)} \bar K^{0*} \to \phi \bar K^0
 \eeq
could proceed through the color-suppressed internal $W$-emission diagram ${\cal C}$ followed by 
strong-interaction rescattering $\eta^{(\prime)} \bar K^{0*} \to \phi \bar K^0$. This contribution is not 
optional, but is, in fact, required by unitarity~\cite{Donoghue:1986nu,Fajfer:2003ag}. While in the 
example above partial cancelation occurs between the intermediate $\eta \bar K^{0*}$ and 
$\eta^\prime \bar K^{0*}$ states~\cite{Lipkin:1986bi}, this cancelation is not generic. If large, the 
contributions of this type could be important in the topological flavor-flow amplitude analysis of charm 
decays~\cite{Cheng:2002wu}.

One way to study the importance of inelastic FSI contributions in charm decays is to seek 
guidance from experimental studies of "annihilation" decays, i.e. decays whose contribution is 
dominated by weak annihilation or exchange amplitudes in the topological flavor-flow analysis.

\begin{figure}[tb]
\begin{center}
\includegraphics[width=0.99\linewidth]{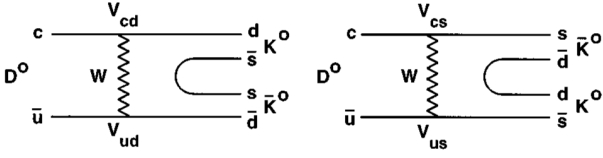}
\caption{
The two quark diagrams that contribute to the decay
$D^0\to K^0_S K^0_S$. Since $V_{cd}=-V_{us}$ the two amplitudes
represented by these diagrams largely cancel. In the limit that the
$d$ and $s$ quark masses where the same the cancellation would have 
been exact.
}
\label{fig:D0_to_K0K0}
\end{center}
\end{figure}

Another related decay mode that is interesting from this perspective is $D^0\to K_S K_S$. 
Naively, there are two $W$ exchange diagrams that contribute to this final state as illustrated in 
Fig.~\ref{fig:D0_to_K0K0}. Since $V_{cd}=-V_{us}$, these amplitudes interfere destructively, so 
in the flavor $SU(3)_F$ limit the branching ratio for this process is zero. Thus, in addition to being the
"pure annihilation" decay,  the rate of $D^0\to K_S K_S$ transition explicitly probes $SU(3)_F$-breaking 
corrections. It should be rather small.

Interestingly enough, a naive calculation of this decay rate in factorization gives exactly zero,
\bea\label{factKK}
{\cal A} (D^0\to K_S K_S) & = & \frac{1}{2} {\cal A} (D^0\to K^0 \overline{K}^0) 
\nonumber \\
& = &
f_D p_D \cdot \left(p_{K^0}-p_{\bar K^0}\right) = 0,
\eea
so ${\cal B}_{{\rm fact}} (D^0\to K_S K_S) =0$. As we discuss later in this section, experimental analyses of 
this transition, however, clearly yield a non-zero result.

The ratio of branching fractions 
${\cal B}(K^0_SK^0_S)/{\cal B}(K^0_S\pi^+\pi^-)$ has been
measured by CLEO~\cite{Alexander:1990nf}, E687~\cite{Frabetti:1994sj},
CLEO II~\cite{Asner:1996ht}, and FOCUS~\cite{Link:2004zc}.
CLEO-c~\cite{Bonvicini:2008nr} has studied this decay using 
a single tag technique and normalized to the number of 
$D^0\bar D^0$ events produced. These measurements are
summarized in Table~\ref{tab:D0_to_K0K0}. Measurements of
the branching ratios ${\cal B}(K^0_SK^0_S)/{\cal B}(K^0_S\pi^+\pi^-)$
has been rescaled using 
${\cal B}(K^0_S\pi^+\pi^-)=(2.99\pm0.17)\%$~\cite{Amsler:2008zzb}.

\begin{table*}[bt]
\caption{
The observed branching fractions for $D^0\to K^0_SK^0_S$.
The errors are statistical, systematic, and from normalization
branching fraction $K^0_S\pi^+\pi^-$ when used. 
}
\label{tab:D0_to_K0K0}
\begin{center}
\begin{tabular}{lcc}
\hline\hline
Experiment          & events   & ${\cal B}(D^0\to K^0_SK^0_S)/10^{-4}$ \\
\hline
CLEO-c~\cite{Bonvicini:2008nr} & $68\pm15$  & $1.46\pm0.32\pm0.09$ \\
FOCUS~\cite{Link:2004zc}       & $79\pm17$  & $4.31\pm0.96\pm0.48\pm0.24$ \\
CLEO II~\cite{Asner:1996ht}    & $26$       & $3.02\pm0.66\pm0.48\pm0.17$ \\
E687~\cite{Frabetti:1994sj}    & $20\pm7$   & $11.7\pm3.9\pm3.9\pm0.7$ \\
CLEO~\cite{Alexander:1990nf}   & $5$        & $6.3^{+3.3}_{-2.4}\pm0.6\pm0.4$\\
\hline
Average                        &            & $1.93\pm0.30$ \\
\hline\hline
\end{tabular}
\end{center}
\end{table*}

The most recent, and most precise, measurement from CLEO-c
gives the smallest central value. Given the large uncertainties
in the earlier measurements there is no strong inconsistency between
the different measurements.  
This clearly points to shortcomings of factorization calculation outlined above.

One way to understand this branching ratio would be to assume that non-factorizable 
pieces, dropped in Eq.~(\ref{factKK}), dominate the branching ratio for 
$D^0\to K_S K_S$. There is, however, no reliable way to estimate those 
(see, however, \cite{Eeg:2001un}). Another way would be to accept that this, and similar
branching ratios are dominated by final state interactions~\cite{Pham:1987rj,Lipkin:1980es}. 
A simple two-channel model estimates give
\beq
\Gamma(D^0\to K^0 \overline{K}^0) = \Gamma(D^0\to K^+ K^-) \tan^2\left(\frac{1}{2}(\delta_0-\delta_1)\right),
\eeq
where $\delta_0$ and $\delta_1$ are the phase shifts for $I=0$ and $I=1$ amplitudes.
Estimates with other models of FSI give comparable results~\cite{Dai:1999cs}. While these
estimates are by no means reliable, they serve as an indication of importance of 
FSI in charm hadronic decays. 

\subsection{Baryonic decay $D_s^+\to p^+\bar n$}

Final states with baryons are not possible for the $D^0$ and $D^+$. The lightest neutral final state
$p\bar p$ has a mass of 1876.54 $\Mevcsq$ and is just above the $D^0$ and $D^+$ mass. However, the
$D_s^+$ is kinematically allowed to decay to $p^+\bar n$.
This decay is also quite interesting because the flavors of all valence quarks that constitute the initial state 
($c\bar s$) differ from the flavors of the final-state quarks composing the $p^+ \bar n$ pair. Thus, it is quite
tempting to declare that the transition $D_s^+ \to p^+ \bar n$ proceeds only via the weak annihilation 
graph~\cite{Pham:1980xe,Pham:1980dc,Chen:2008pf}. 

A factorization ansatz can be employed in order to estimate the branching ratio for this process~\cite{Chen:2008pf}.
It must be emphasized again that contrary to hadronic $B$-decays, simple factorization has not been proven in charm
transitions, especially as applied to annihilation amplitudes. Nevertheless, a factorized decay amplitude is
\beq
A (D_s^+ \to p \bar n) =  \frac{G_F}{\sqrt{2}} V_{cs} V_{ud}^* a_1 f_{D_s} p_{D_s}^\mu
\langle p \bar n | \overline u \gamma_\mu \left(1-\gamma_5\right) d | 0 \rangle,
\eeq
where $p_{D_s}=p_p+p_{\bar n}$ is the four-momentum of a $D_s$-meson. 
The matrix element between the vacuum and the final state can be parametrized. First, let us note that 
vector current conservation implies that
\beq
p_{D_s}^\mu
\langle p \bar n | \overline u \gamma_\mu \left(1-\gamma_5\right) d | 0 \rangle =
(m_p + m_{\bar n}) \langle p \bar n | \overline u \gamma_5 d | 0 \rangle, 
\eeq
so the decay amplitude can be parameterized as
\bea
A (D_s^+ \to p \bar n) &=&  \frac{G_F}{\sqrt{2}} V_{cs} V_{ud}^* a_1 f_{D_s}
\nonumber \\
&\times&  ~\left(2 m_N g_1^{p \bar n} + \frac{m_{D_s}^2}{2 m_N} g_3^{p \bar n}\right)
\overline u_p \gamma_5 v_{\bar n},
\eea
where $g_i^{p^+ \bar n}$ are the formfactors parameterizing the baryon current, and
$m_N$ is the nucleon's mass. The two formfactors $g_1^{p \bar n}$ and $g_3^{p \bar n}$ 
can be related to each other~\cite{Pham:1980xe,Pham:1980dc,Chen:2008pf},
\beq
g_3^{p \bar n} (p_{D_s}^2) = - \frac{4 m_N^2}{p_{D_s}^2-m_\pi^2} g_1^{p \bar n}(p_{D_s}^2),
\eeq
so that the decay amplitude takes the form,
\bea
A (D_s^+ \to p \bar n) &=&  \frac{2 G_F}{\sqrt{2}} V_{cs} V_{ud}^* a_1 f_{D_s}
\nonumber \\
&\times& ~m_N \left(\frac{m_\pi}{m_{D_s}}\right)^2 g_1^{p \bar n}  \ \overline u_p \gamma_5 v_{\bar n}.
\eea
This amplitude leads to the estimate of the decay branching ratio ${\cal B}(D_s^+ \to p \bar n)$  in the factorization 
approximation~\cite{Chen:2008pf},
\beq\label{BarBrRat}
{\cal B}(D_s^+ \to p \bar n)_{\rm th} = (0.4^{+1.1}_{-0.3})\times 10^{-6}.
\eeq
The theoretical error quoted in Eq.~(\ref{BarBrRat}) is entirely due to the uncertainty in the form-factor
value of $g_1^{p \bar n}(m_{D_s}^2)$~\cite{Chen:2008pf}, which was obtained by extrapolation of the nucleon 
data with a particularly assumed shape of $q^2$-dependence.  This estimate gives a rather small branching 
ratio, which nevertheless can be tested experimentally. CLEO-c has studied this final state~\cite{Athar:2008ug}.

As (anti-)neutrons are hard to reconstruct, CLEO-c uses
a missing mass technique to identify this signal. All
particles in the event, except for the (anti-)neutron,
is reconstructed and the signal is extracted by looking
in the missing mass distribution of the events, which
for signal will peak at the neutron mass. 

CLEO-c uses 325 \pbinv of $e^+e^-$ annihilation data 
collected at a center-of-mass energy of 4170 MeV. At
this energy pairs of $D_s^+D_s^{*-}$ and $D_s^-D_s^{*+}$
are produced. CLEO-c uses 8 tag modes ($\Kp\Km\pim$, $\KS\Km$,
$\eta\pim$, $\eta'\pim$, $\phi\rho^-$, $\pim\pip\pim$,
$K^{*-}K^{*0}$, and $\eta\rho^-$) to first reconstruct
a $D_s^-$ candidate. It is required that this $D_s$
candidate has a reconstructed invariant mass which is 
within $2.5\sigma$ of the known $D_s$ mass. Next, this 
candidate is combined with a photon. The recoil mass 
squared against the $D_s^-+\gamma$ is calculated
and required to be consistent with the mass of the
$D_s$. Note here that it does not matter if the photon
came from the $D_s^{*}$ that is the parent of the 
$D_s^-$ or from the parent of the other $D_s$ in the
event. This missing mass squared distribution is fit
to determine the number of tags, CLEO-c reports finding
16,995 $D_s$ tags. This yield will be used as the 
denominator in the branching fraction calculation. 

CLEO-c then search for the proton candidate
in the momentum range from 150 to 550 $\Mevc$. In this
momentum range CLEO-c uses $dE/dx$ to identify
the proton, 550 $\Mevc$ is below Cherenkov 
threshold. Kinematic fits are performed to the $D_s^-$,
photon, and proton candidate. As it is not known
if the photon came from the $D_s^{*-}\to D_s^-\gamma$
decay or the $D_s^{*+}\to D_s^+\gamma$ decay, two
different fits are performed. First the photon is 
added to the $D_s^-$ to form a $D_s^{*-}$. The 
$D_s^{*-}$ momentum is constrained to the known momentum
from the $e^+e^-\to D_s^{*-}D_s^+$ reaction and
the $D_s^{*-}$--$D_s^-$ mass difference. For the
other hypothesis the $D_s^-$ momentum is constrained 
to the known momentum
from the $e^+e^-\to D_s^{*-}D_s^+$ reaction and
the proton is combined with the missing momentum
of the event to make a $D_s^+$ candidate, add
the photon and constrain the $D_s^{*+}$--$D_s^+$
mass difference. The combination with the 
lowest $\chi^2$ is selected. Based on Monte Carlo
it is estimated that the right combination is 
selected 95\% of the time. Applying these kinematic
constraints improve the resolution on the missing 
mass by a factor of two. In addition cuts are applied
on the $\chi^2$ to reject combinatorial background 
not consistent with the signal. If there are 
multiple photon candidates in an event the combination
with the lowest overall $\chi^2$ is selected.

\begin{figure}[tb]
\begin{center}
\includegraphics[width=0.8\linewidth]{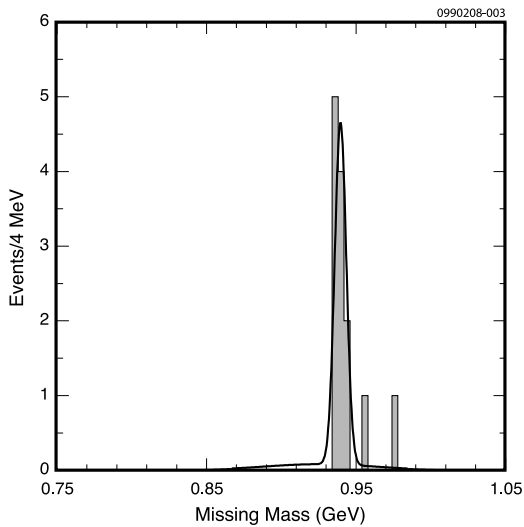}
\caption{The missing mass distribution for all $D_s^+\to p\bar n$ 
candidates. CLEO-c sees 13 signal candidates.
From~\textcite{Athar:2008ug}.
}
\label{fig:CLEOc_dstopn}
\end{center}
\end{figure}

In Fig.~\ref{fig:CLEOc_dstopn} the distribution
of the recoil mass against the proton shown. There
are 13 candidate events consistent with the
$D_s^+\to p\bar n$ signal. From this yield, the
number of tags, and the efficiency for reconstructing
the proton CLEO-c determines the branching fraction
\beq
{\cal B}(D_s^+\to p\bar n)_{\rm exp}=(1.30\pm 0.36^{+0.12}_{-0.16})\times 10^{-3}.
\eeq
This result shows quite unambiguously that the factorization-ansatz 
estimate of Eq.~(\ref{BarBrRat}) fails by 
more than three orders of magnitude! This could be because of the 
following two reasons. First, the use of a
factorization ansatz could be completely misleading for the description 
of $D_s^+ \to p \bar n$. This could be due to the fact that the charm quark is 
too light for the factorization approach to be reliable. In fact, since the
mass of the $D_s$ lies right in the middle of the region populated by 
highly excited light quark resonances, it is possible that the presence of 
nearby states could significantly affect the decay. In addition, 
the decay happens almost at the threshold for $p\bar n$ production, with
no large energy release -- something that factorization-based approaches
usually require. Second, there could be other decay mechanisms that
contribute to this transition besides annihilation. For example, inelastic 
rescattering discussed above could be responsible for the bulk of the result. 
An example of this mechanism would be a 
tree-level transition $D_s^+ \to \eta^{(\prime)} \pi^+$ with subsequent 
rescattering 
$\eta^{(\prime)} \pi^+ \to p\bar n$. It has been argued~\cite{Chen:2008pf} that 
this mechanism can provide a contribution  that is consistent with the experimentally-measured 
branching ratio. More work is definitely needed for complete theoretical understanding 
of this and related processes.

\section{DALITZ DECAYS OF $D$ MESONS}
\label{sect:multi}

In this Section multibody decays of $D$ mesons
are discussed. The most extensive studies of
multibody decays are the Dalitz plot studies 
performed in three-body decays. We give an
overview of the analysis techniques used, 
and discuss some of the final states that have 
been investigated. A few four-body final 
states have also been investigated and they
are discussed next. The last topic is the
study of inclusive distributions in $D$ 
decays.

\subsection{Three-body Dalitz plot analyses}

Many hadronic three-body final states of $D^0$, $D^+$, 
and $D^+_s$ meson decays have been studied using a 
Dalitz plot analysis in which the resonant substructure has been probed. From these analyses we learn
about the amplitudes and phases of the different components that contribute to these final states.
It is seen that most three-body final states are dominated
by pseudo two-body decays.

There is an enormous number of applications of three-body decays of $D$-mesons. Indeed one of the 
most important ones involves proper determination of branching fractions of quasi-two-body decays, such 
as $D \to \rho \pi$. Also, the possibility of determination of all relative decay amplitudes and phases in
the Dalitz analysis of $D^0$ decays allows for novel studies of $D^0\overline{D}^0$ mixing and 
searches of $CP$ violation in the charm system. Finally, Dalitz analyses of $D$-decays offer unique ways to 
study formation of light-quark resonances (such as $\sigma$ and $\kappa$) that are not reachable in 
direct $e^+e^-$-annihilation experiments.

In a Dalitz plot analysis the dynamics of a decay is investigated by analyzing the 
kinematic distributions by plotting the data such that the event density
is proportional to the matrix element squared~\cite{Dalitz:1953cp}.
For the three-body decay $D\to abc$ where $a$, $b$, and $c$ are 
pseudo-scalars the decay rate can be written~\cite{Amsler:2008zzb}
\beq\label{DDalitz}
d \Gamma={1\over 32(2\pi)^3M^3_{D}} |{\cal M}|^2 \ dm^2_{ab} \ dm^2_{bc},
\eeq
where ${\cal M}$ is the decay matrix element and $m_{ij}^2=\left(p_i+p_j\right)^2$ is 
the invariant mass squared of particles $i$ and $j$. Note that for ${\cal M} = \hbox{constant}$,
the Dalitz plot in variables $(m^2_{ab}, m^2_{bc})$ of Eq.~(\ref{DDalitz}) represents a homogeneously-filled shape. 
Any apparent structures would then represent resonant interactions of the final state particles. 

\subsubsection{Formalism for Dalitz plot fits}

In general, the amplitude for the process $D\to Rc$, $R\to ab$ where $R$ is an intermediate resonance 
and $a$, $b$, and $c$ are particles of arbitrary spin,  can be written
\beq\label{DPExpan}
{\cal M}_R(L,m_{ab},m_{bc})=\sum_{\lambda}\langle ab|R_{\lambda}\rangle
                                    T_R(m_{ab})\langle cR_{\lambda}|D\rangle
\eeq
where $L$  is the spin of resonance $R$, and the sum is over the helicity states $\lambda$ of $R$. It is 
customary to break the amplitude of Eq.~(\ref{DPExpan}) into three parts,
\beq
{\cal M}_R(L,m_{ab},m_{bc}) =  Z(L,{\bf p},{\bf q}) \ B_L^D (\left|{\bf p}\right|) \
B_L^R (\left|{\bf q}\right|) \ T_R(m_{ab}), 
\eeq
where $Z$ depends on the spin of resonance $R$ and describes the angular 
distribution of the decay products. If all final state particles are spin-0, which is the case
for all of the decays described here (see Eq.~(\ref{DDalitz})), it reduces to Legendre's polynomials. 
The $B_L$'s are the spin-dependent Blatt-Weisskopf penetration functions that  
incorporate effects due to finite-size of the final-state hadrons, and $T_R$ is a function that 
describes dynamics of the final-state mesons that incorporate a prescription on how to treat the
final-state resonances $R$. The momenta ${\bf p}$ and ${\bf q}$ of $c$ and $a$, respectively, are 
defined in the $R$ rest frame (e.g. $|{\bf q}|=
\sqrt{(m_R^2-(m_a+m_b)^2)(m_R^2-(m_a-m_b)^2)}/2m_R$. 
The main difference between various analyses of Dalitz plots is related to the chosen model for $T_R$.

The most common description of Dalitz plots in three-body decays is the so called {\it isobar model}.
In this model amplitudes are added coherently for each resonance. A nonresonant contribution,
which describes a direct decay of the $D$ into a 3-body final state, is usually added as a
coherent contribution uniformly distributed across the Dalitz plot, making the total amplitude
\beq
{\cal M} = {\cal M}_{NR} + \sum_R  {\cal M}_R(L,m_{ab},m_{bc}).
\eeq
In the isobar model each resonance is described by a Breit-Wigner lineshape,
\beq
T_R(m_{ab}) = \left[m_R^2 - m_{ab}^2 - i m_R \Gamma_{ab} (q) \right]^{-1}.
\eeq
Here $\Gamma_{ab} (q)$ describes a momentum-dependent width of the 
resonance $R$, which generalizes narrow-width approximation,
\beq
\Gamma_{ab} (q) = \Gamma_R \left(\frac{q}{q_0}\right)^{2L+1}
\left(\frac{m_0}{m_{ab}}\right) B_L^\prime (q,q_0)^2.
\eeq
Resonant fractions, or fit fractions, are defined, for each resonance $R$, as
\beq
f_R = \frac{\int |{\cal M}_R|^2}
{\int |{\cal M}_{NR}+\sum_R {\cal M}_R|^2},
\eeq
where the integration above is over the whole Dalitz plot. The sum of fractions, so defined, is 
not required to be unity. One must remember that isobar model is breaking the unitarity and is 
partly the result of interference terms, missing from the denominator, and partly due to kinematic 
limits imposed on the integrals. 

The {\it $K$-matrix model} is used when a proper description of a Dalitz
plot dominated by broad scalar resonances is needed. The $K$-matrix formalism is, by
construction, unitary.  It follows from a specific parameterization of the
scattering matrix,
\beq
S_{if} = \delta_{if} + 2 i T_{if}  = \delta_{if}  + 2i \{\rho_i\}^{1/2} \ \hat T_{if}  \ \{\rho_f\}^{1/2}, 
\eeq
where $\hat T_{if}$ is a Lorentz-invariant scattering amplitude and $\rho_i=2 q_i/m_i$ are the diagonal
elements of the (diagonal) phase space matrix.  Here $q_i = m_i \sqrt{1-4m_i^2/s}$ is the breakup 
momentum for decay channel $i$.

The $K$-matrix represents a particular parameterization of $\hat T$,
\beq
\hat T = \left(\hat I - i \hat K \hat \rho\right)^{-1} \hat K.
\eeq
The final-state resonances appear in the $K$-matrix as a sum of poles. A particular parameterization of 
the $K$-matrix can be chosen, which incorporates data from scattering experiments. 
One useful parameterization of the $K$-matrix can be found in~\cite{Anisovich:2002ij}.
A good description of $K$-matrix formalism can be found in~\cite{Chung:1995dx}. See also
D.~Asner's review in~\cite{Amsler:2008zzb}.

In addition to the isobar model and the $K$-matrix models
presented above, several experiments has use the
Model-Independent Partial Wave Analysis (MIPWA).
This approach was first used by the E791 collaboration~\cite{Aitala:2005yh}. 
Instead of trying to describe the $S$-wave as a sum of 
broad Breit-Wigners resonances, which often leads to
unitarity violation when they overlap, or using the $K$-matrix
parameterization this method
parameterizes the amplitude and phase by dividing the 
$\pi^+\pi^-$ mass spectrum into descrete slices. The amplitude
and phase are interpolated using a Relaxed Cubic 
Spline~\cite{CubicSplines}. 

\subsubsection{Experimental considerations}

When analyzing data using a Dalitz plot analysis there
are several experimental effects to consider. The
reconstruction efficiency for the $D$ candidates is
not uniform across the 
Dalitz plot. The momentum spectrum of the observed 
particles will depend on the position in the Dalitz 
plot and affect the efficiency for finding and reconstructing
the particles. The effect of efficiency variations across the
Dalitz plot is typically incorporated using a
Monte Carlo simulation and parameterization of the 
efficiency as a function of the Dalitz plot variables.

The finite detector resolution is usually neglected
as the resonances studied are mostly broad compared to
the detector resolution. There are a few exceptions
such as $\phi\to K^+K^-$ and $\omega\to \pi^+\pi^-$.
In these cases the resolution function has to be
convolved with the truth level probability distribution. 
A related effect is
resolution effects near the phase-space boundary
in the Dalitz plot. To avoid smearing near the
phase-space boundary the final state particles momenta
can be recalculated using a constraint to the $D$ mass.
This forces the phase-space boundary to be strictly 
respected. 

Experimentally we also have to consider backgrounds that 
pass the event selection criteria. The backgrounds
can be classified into different categories. Combinatorial
backgrounds where the selected particles do not all come
from the decay of a $D$. This background may contain 
resonances, such as a $K^{*}$ or $\rho$. We also have
backgrounds where all candidates come from a $D$ decay
but are not signal. These backgrounds include final 
states with identical particles, e.g. $D^0\to K^0_S\pi^0$
contributing to $D^0\to \pi^+\pi^-\pi^0$ or a $\bar D^0$
decay incorrectly identified as a $D^0$, or misidentified
particles such as $D^+\to\pi^-\pi^+\pi^+$ reconstructed
as $D^+\to K^-\pi^+\pi^+$. 

In the following Sections different Dalitz plot analyses
will be discussed. As in general it is impossible to 
average the results of different analysis the most
recent, or precise, results are discussed in more detail
for each mode.

\subsubsection{$D^0\to K^-\pi^+\pi^0$}

The decay $D^0\to K^-\pi^+\pi^0$ has been studied 
by the tagged photon spectrometer at Fermilab~\cite{Summers:1983nu},
MARK III~\cite{Adler:1987sd}, E691~\cite{Anjos:1992kb}
E687~\cite{Frabetti:1994di}, and CLEO II~\cite{Kopp:2000gv}.
The first of these analyses was a simplified Dalitz
analysis that did not include the interference. The
data was fit to an incoherent sum of $K^-\rho^+$,
$\bar K^{*0}\pi^0$, $K^{*-}\pi^+$, and nonresonant
decays. The latest analysis by CLEO II has about a
factor of 10 higher statistics than any of the earlier 
measurements. 

\begin{figure}[tb]
\begin{center}
\includegraphics[width=0.52\linewidth]{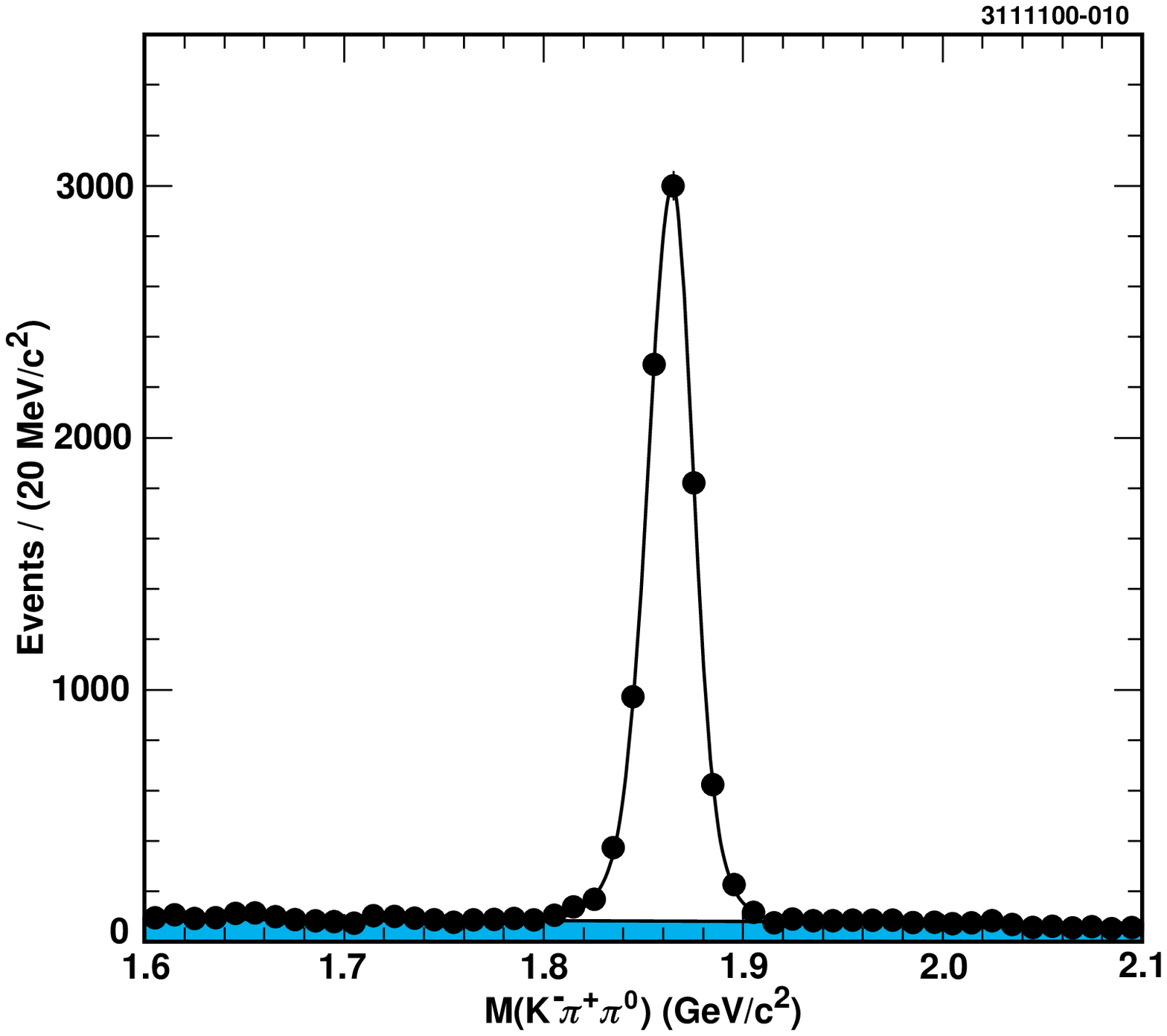}
\includegraphics[width=0.45\linewidth]{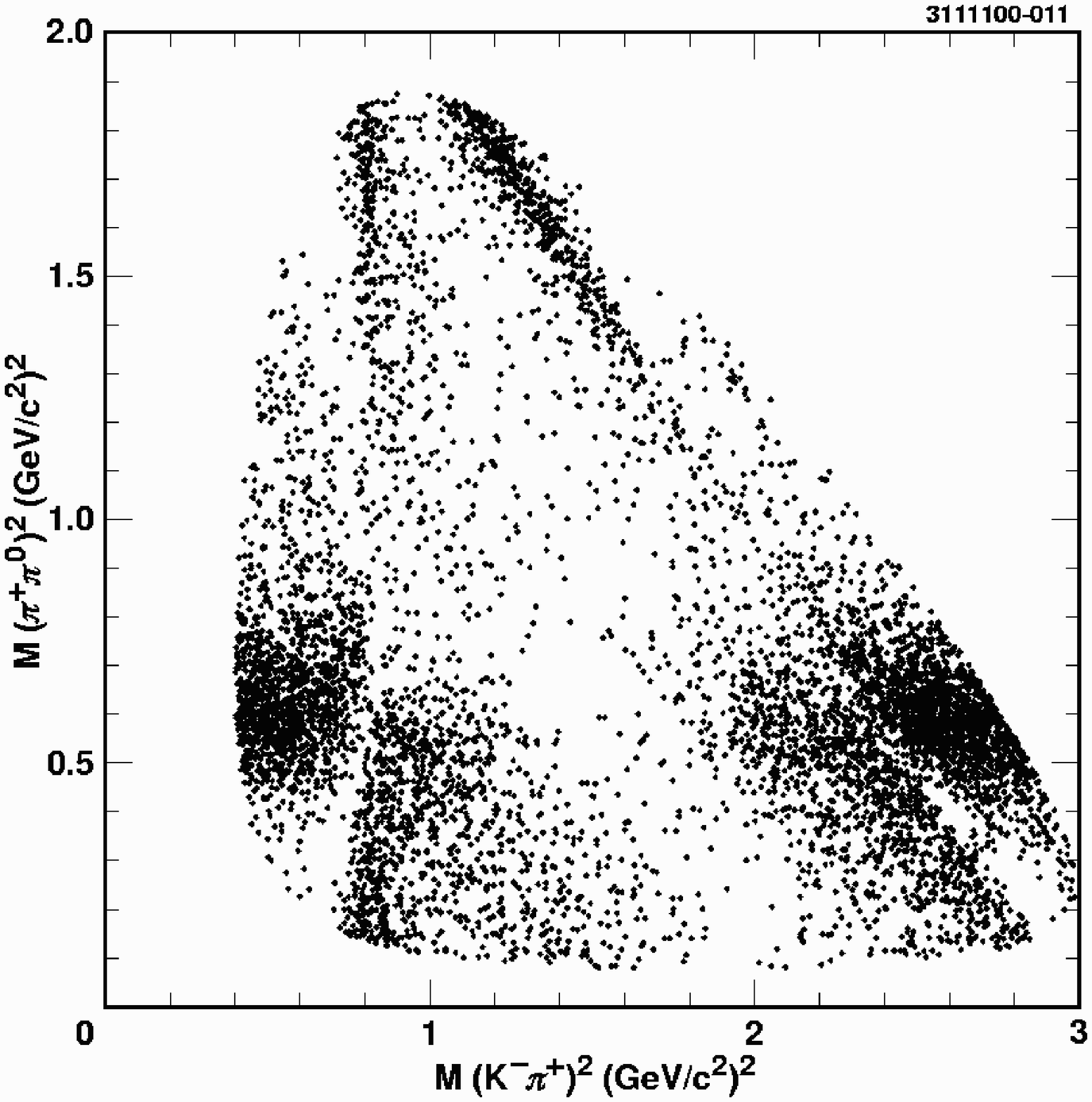}
\caption{The left plot shows the $D^0\to K^-\pi^+\pi^0$ reconstructed
mass distribution. The plot to the right shows
the $M^2(\pi^+\pi^0)$
versus $M^2(K^-\pi^+)$ Dalitz plot
for the 7,070 $D^0\to K^-\pi^+\pi^0$ candidates.
From~\textcite{Kopp:2000gv}.
}
\label{fig:cleoc_Kpipiz_mass_Dalitz}
\end{center}
\end{figure}

The analysis by CLEO II used 4.7 \fbinv\ of $e^+e^-$ collision data
collected at $\sqrt{s}=10.6$ GeV. The $D^0$ candidate
is required to come from a $D^{*+}\to D^0\pi^+$ decay.
The $D^0$ candidate is required to form a $D^{*+}$
candidate which satisfies $144.9<M(D^{*+})-M(D^0)<145.9$ MeV.
The invariant mass distribution of the $K^-\pi^+\pi^0$
candidates and the 
7,070 event selected for the Dalitz plot analysis 
are shown in Fig.~\ref{fig:cleoc_Kpipiz_mass_Dalitz}.
This sample has 
a purity of $96.7\pm1.1\%$.  
The large $K^{*0}$,
$\rho^+$, and $K^{*-}$ resonances and their interference
is easily seen in this plot. 
The results of the Dalitz plot fit are summarized in 
Table~\ref{tab:cleoc_Kpipiz_Dalitz_params}. The $\rho(770)^+$
resonance dominates the Dalitz plot with a 
fit fraction of about 78.8\%.

\begin{figure}[tb]
\begin{center}
\includegraphics[width=0.90\linewidth]{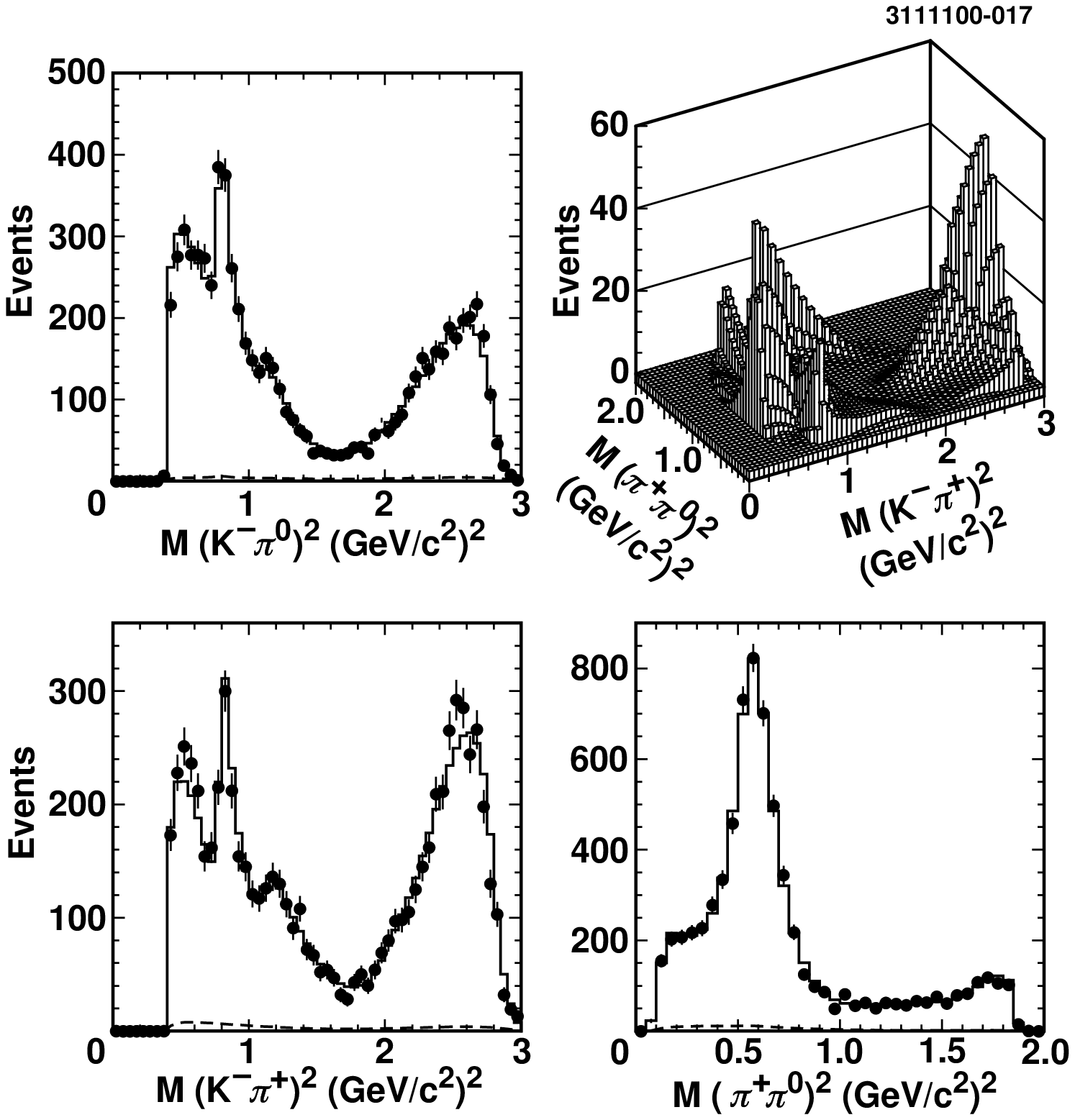}
\caption{The $D^0\to K^-\pi^+\pi^0$ Dalitz fit.
From~\textcite{Kopp:2000gv}.
}
\label{fig:cleoc_Kpipiz_Dalitz_fit}
\end{center}
\end{figure}

\begin{table*}[bt]
\caption{
Dalitz plot parameters from CLEO II analysis of $D^0\to 
K^-\pi^+\pi^0$~\cite{Kopp:2000gv}.
}
\label{tab:cleoc_Kpipiz_Dalitz_params}
\begin{center}
\begin{tabular}{lcc}
\hline\hline
Mode               & Fit fraction   & Phase (deg) \\
\hline
$\rho(770)^+K^-$   & $0.788\pm0.019\pm 0.013\pm 0.046$ & $0.0$ (fixed) \\
$K^*(892)^-\pi^+$  & $0.161\pm0.007\pm 0.007^{+0.026}_{-0.008}$ & $163\pm2.3\pm3.1\pm4.3$ \\
$\bar K^*(892)^0\pi^0$  & $0.127\pm0.009\pm 0.005\pm0.015$ & $-0.2\pm3.3\pm2.2\pm7.0$ \\
$\rho(1700)^+K^-$   & $0.057\pm0.008\pm 0.007\pm0.006$ & $171\pm6\pm5^{+6.1}_{-55}$ \\
$\bar K^*_0(1430)^0\pi^0$  & $0.041\pm0.006\pm 0.007^{+0.031}_{-0.005}$ & $166\pm5\pm4.6\pm12$ \\
$K^*_0(1430)^-\pi^+$  & $0.033\pm0.006\pm 0.007\pm0.012$ & $55.5\pm5.8\pm3.3^{+4.2}_{-13}$ \\
$K^*(1680)^-\pi^+$  & $0.013\pm0.003\pm 0.003\pm0.003$ & $103\pm8\pm7\pm14$ \\
Nonresonant  & $0.075\pm0.009\pm 0.006^{+0.056}_{-0.009}$ & $31\pm4\pm5.5^{+14}_{-3.7}$ \\
\hline\hline
\end{tabular}
\end{center}
\end{table*}

\subsubsection{$D^0\to K^0_S\pi^+\pi^-$}

This decay is of interest for the extraction of the CKM
angle $\gamma$ in the decays $B^{\mp}\to D^{(*)}K^{\mp}$ and
$B^{\mp}\to \bar D^{(*)}K^{\mp}$~\cite{Atwood:2000ck}.
When the decay of the $D^0$
or $\bar D^0$ in these decays is to a common final states,
such as $K^0_S\pi^+\pi^-$, the two decays above interfere and
this allow us to measure the CKM angle $\gamma$
To extract $\gamma$ from this analysis a good understanding
of the $D^0\to K^0_S\pi^+\pi^-$ Dalitz plot is required. 

This final state has been investigated by many experiments.
The first studies were performed 
by~\cite{Adler:1987sd,Frabetti:1992we,Anjos:1992kb,
Albrecht:1993jn,Frabetti:1994di}. 
CLEO was the first experiment to include doubly Cabibbo 
suppressed decays in the
Dalitz plot analysis~\cite{Muramatsu:2002jp} of this decay.
They used 10 resonances
in their fit: $K^0_S\rho^0$, $K^0_S\omega$, $K^0_Sf_0(980)$,
$K^0_Sf_2(1270)$, $K^0_Sf_0(1370)$, 
$K^*(892)^-\pi^+$, $K^*_0(1430)^-\pi^+$, $K^*_2(1430)^-\pi^+$,
$K^*(1680)^-\pi^+$, and the Cabibbo suppressed mode $K^*(892)^+\pi^-$.
CLEO found a very small fit fraction for the nonresonant 
contribution of $(0.9\pm0.4^{+1.0+1.7}_{-0.3-0.2})$\%. They 
also determined
that the phase difference between the Cabibbo allowed 
$K^*(892)^-\pi^+$ and the doubly Cabibbo suppressed decay
$K^*(892)^+\pi^-$ is consistent with 180$^{\circ}$ as
expected from the Cabibbo factors. The
significance of the $K^*(892)^+\pi^-$ resonance is 5.5
standard deviations in the study by CLEO.

\begin{figure}[tb]
\begin{center}
\includegraphics[width=0.48\linewidth]{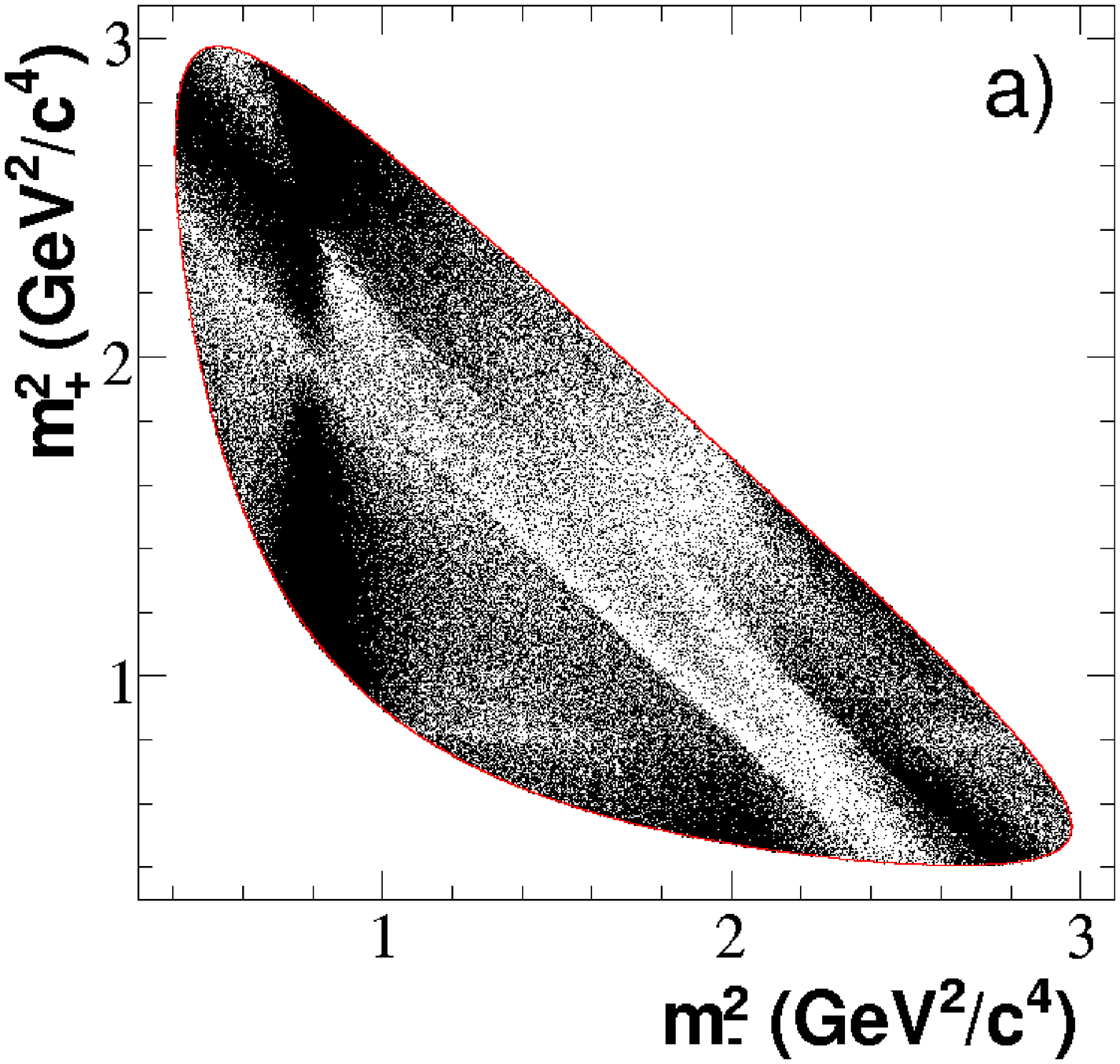}
\includegraphics[width=0.48\linewidth]{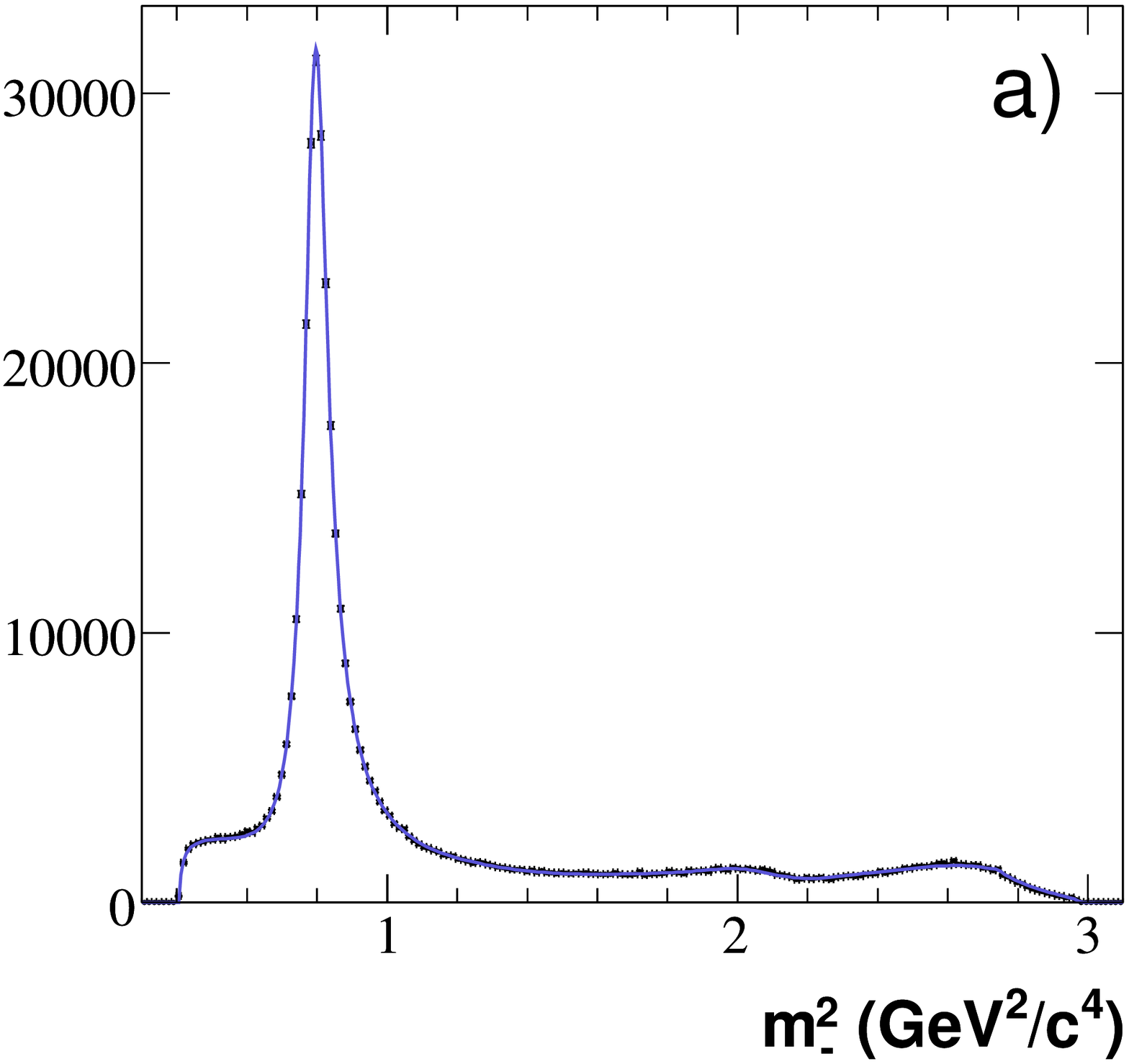}
\includegraphics[width=0.48\linewidth]{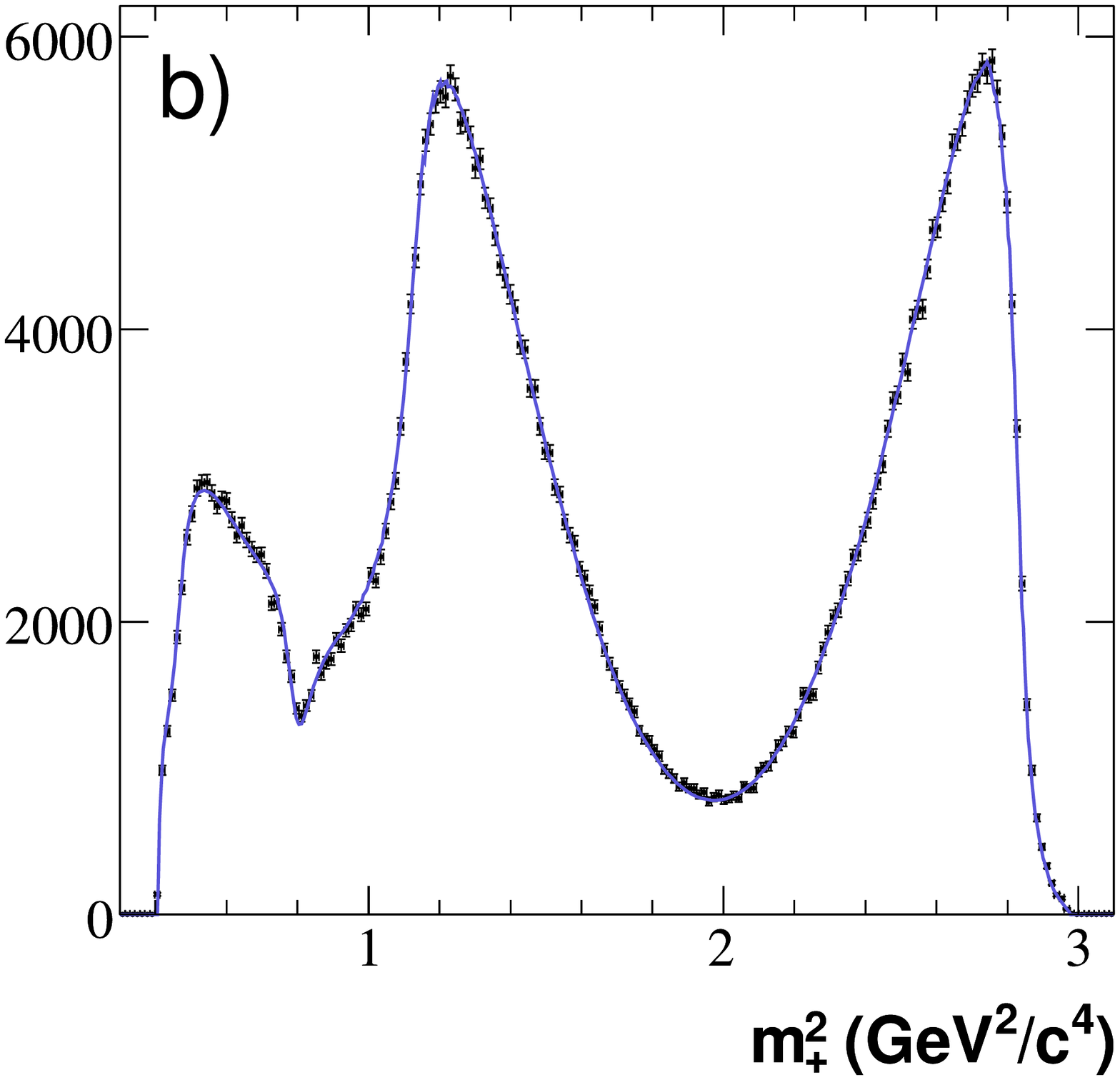}
\includegraphics[width=0.48\linewidth]{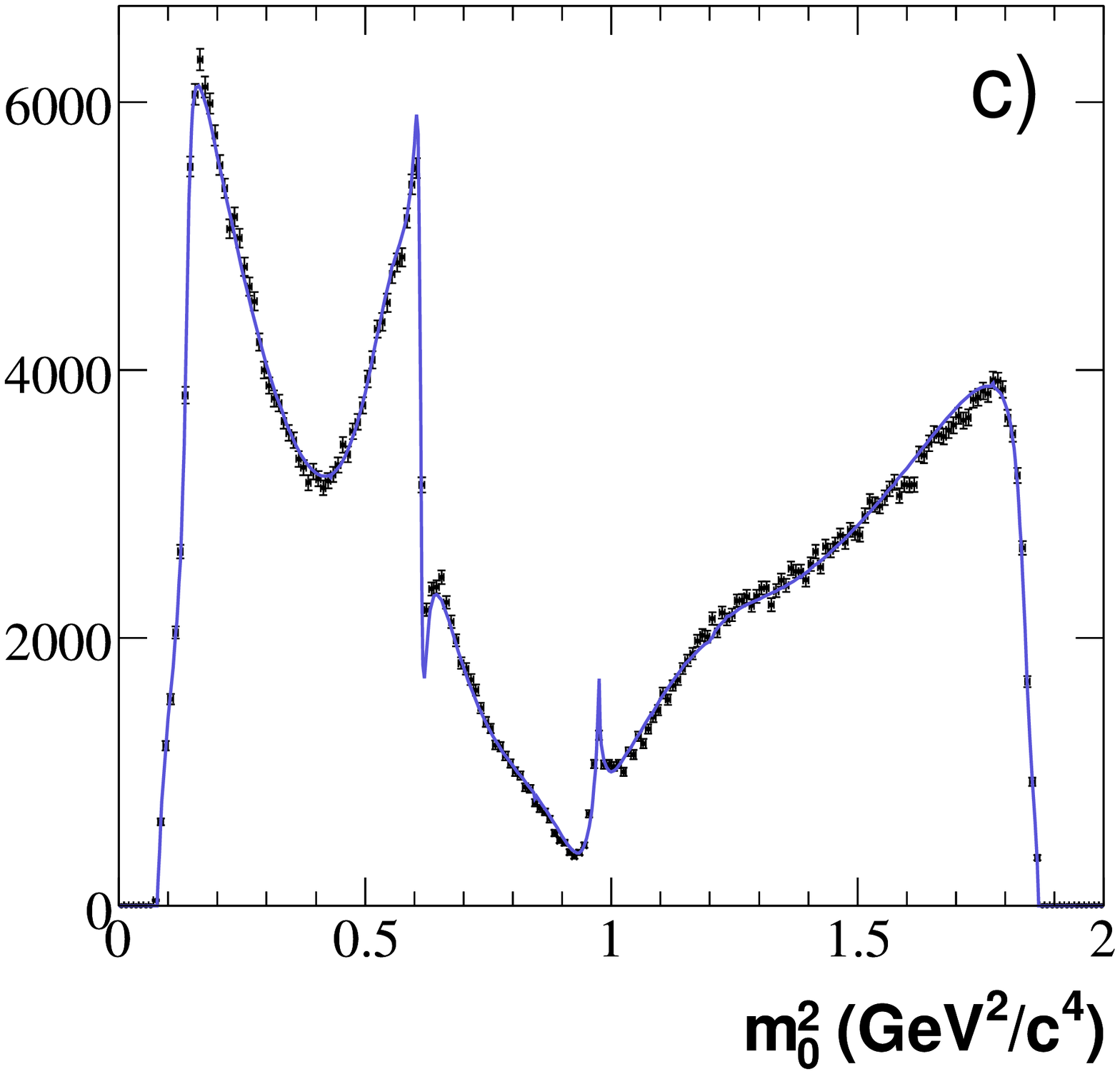}
\caption{BABAR $D^0\to K^0_S\pi^+\pi^-$ Dalitz plot analysis.
From~\textcite{Aubert:2008bd}.
}
\label{fig:BABAR_KSpipi_Dalitz}
\end{center}
\end{figure}

Both
BABAR~\cite{Aubert:2005iz,Aubert:2008bd} and 
Belle~\cite{Poluektov:2006ia,Abe:2008wya} have studied this
decay with samples well over an order of magnitude larger than
CLEO in their program to determine the CKM angle $\gamma$. 
BABAR~\cite{Aubert:2008bd} has used a data sample 
of 351 \fbinv\ collected at
the $\Upsilon(4S)$ to study the $D^0\to K^0_S\pi^+\pi^-$ Dalitz
plot. They reconstruct 487,000 $D^{*+}\to D^0\pi^+$, 
$D^0\to K^0_S\pi^+\pi^-$ decays
with a purity of 97.7\%. 
The Dalitz plot is fit to a sum of eight different $P$ and $D$
wave resonances.
They use three Cabibbo favored resonances $K^*(892)^-$, $K^*(1680)^-$,
and $K^*_2(1430)^-$; two doubly Cabibbo suppressed resonances
$K^*(892)^+$ and $K^*_2(1430)^+$, and three $CP$ eigenstates
$\rho(770)^0$, $\omega(782)$, and $f_2(1270)$. The $K$ matrix
formalism with the $P$-vector approximation is used to describe
the contribution
to the amplitude from the $\pi^+\pi^-$ $S$-wave.
The $K\pi$ $S$-wave includes the $K^*_0(1430)^-$ and $K^*_0(1430)^+$
resonances and a nonresonant component.
The data 
and the fit projections are shown in Fig.~\ref{fig:BABAR_KSpipi_Dalitz}.
The result of the fit is shown in Table~\ref{tab:BABAR_KSpipi_Dalitz}.

\begin{table}[bt]
\caption{
Dalitz plot parameters from BABAR analysis of $D^0\to K^0_S\pi^+\pi^-$.
The errors for the amplitudes and phases include only the statistical
errors. The fit fractions quoted include also the systematic uncertainties.
Upper limits on fit fractions are quoted at 95\% confidence level.
}
\label{tab:BABAR_KSpipi_Dalitz}
\begin{center}
\begin{tabular}{lccc}
\hline\hline
Resonance               & Amplitude   & Phase (deg) & Fit fraction \\
\hline
$K^*(892)^-$            & $1.740\pm0.010$ & $139.0\pm0.3$  & $55.7\pm 2.8$ \\
$K^*_0(1430)^-$         & $8.2\pm0.7$     & $153\pm 8$     & $10.2\pm 1.5$ \\
$K^*_2(1430)^-$         & $1.410\pm0.022$ & $138.4\pm 1.0$ & $2.2\pm 1.6$ \\
$K^*(1680)^-$           & $1.46\pm0.10$   & $-174\pm 4$    & $0.7\pm 1.9$ \\

$K^*(892)^+$            & $0.158\pm0.003$ & $-42.7\pm 1.2$ & $0.46\pm 0.23$ \\
$K^*_0(1430)^+$         & $0.32\pm0.06$   & $143\pm 11$    & $<0.05$ \\
$K^*_2(1430)^+$         & $0.091\pm0.016$ & $85\pm 11$     & $<0.12$ \\

$\rho(770)^0$           & $1$ (fixed)     & $0$ (fixed)    & $21.0\pm1.6$ \\
$\omega(782)$           & $0.0527\pm0.0007$ & $126.5\pm0.9$& $0.9\pm1.0$ \\
$f_2(1270)$             & $0.606\pm0.026$ & $157.4\pm2.2$  & $0.6\pm0.7$ \\
$\pi\pi$ $S$-wave       &                 &                & $11.9\pm2.6$ \\
\hline\hline
\end{tabular}
\end{center}
\end{table}

\begin{figure}[tb]
\begin{center}
\includegraphics[width=0.95\linewidth]{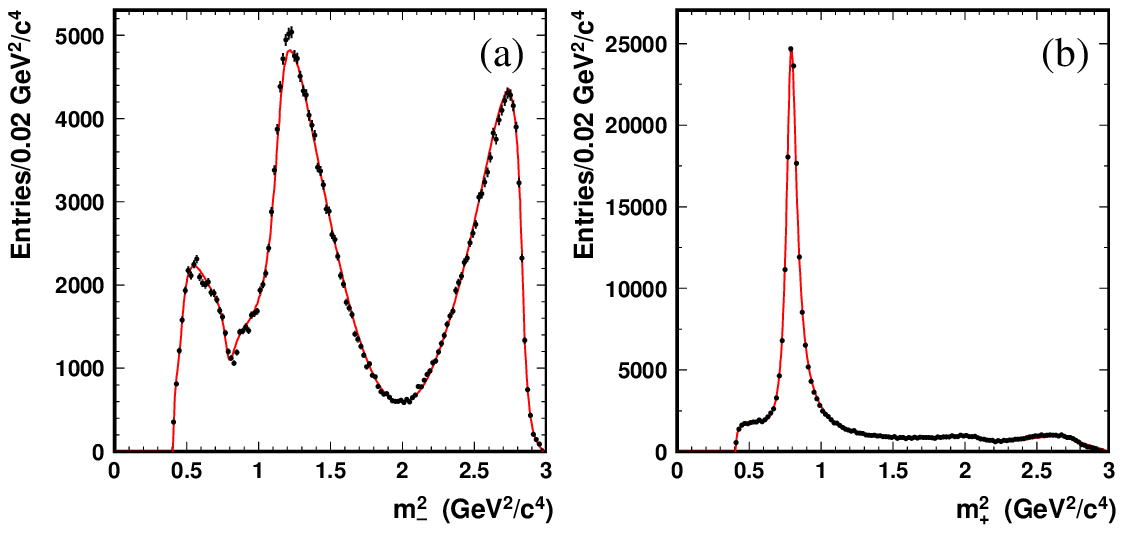}
\includegraphics[width=0.95\linewidth]{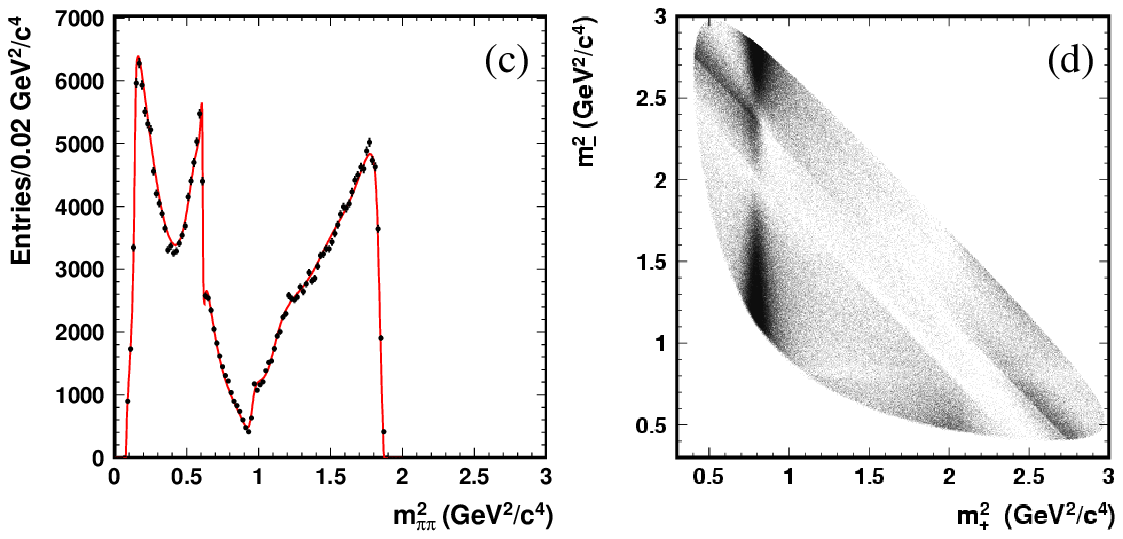}
\caption{Belle $D^0\to K^0_S\pi^+\pi^-$ Dalitz plot analysis.
(a) the $m^2(K^0_S\pi^+)$, 
(b) the $m^2(K^0_S\pi^-)$, and  
(c) the $m^2(\pi^-\pi^+)$ distributions are shown and in (d) 
the Dalitz plot distribution. The points with error bars show the
data and the smooth curve is the result of the fit.
From~\textcite{Poluektov:2006ia}.
}
\label{fig:Belle_KSpipi_Dalitz}
\end{center}
\end{figure}

Belle~\cite{Poluektov:2006ia} has used a 140 \fbinv\ sample collected 
at the $\Upsilon(4S)$
to study the $D^0\to K^0_S\pi^+\pi^-$ Dalitz plot.  They select
a sample of 104,204 events for their analysis with an estimated
purity of 96.9\%. They fit their data to a sum of 15 resonances
plus a nonresonant amplitude. The data and projections of their
fit are shown in Fig.~\ref{fig:Belle_KSpipi_Dalitz}. The
result of their fit is summarized in Table~\ref{tab:Belle_KSpipi_Dalitz}.
For the two $\sigma$ resonances that are included in the fit
Belle obtained $M_{\sigma_1}=539\pm 9$ MeV, $\Gamma_{\sigma_1}=453\pm 16$ MeV,
$M_{\sigma_2}=1048\pm 7$ MeV, and $\Gamma_{\sigma_2}=109\pm 11$ MeV.
The wide $\sigma_1$ resonance is highly correlated with the
nonresonant component. 
Belle has also reported a preliminary
study~\cite{Abe:2008wya} using 605 \fbinv\ of data
to study this Dalitz plot.

\begin{table}[bt]
\caption{
Dalitz plot parameters from Belle analysis of 
$\bar D^0\to K^0_S\pi^+\pi^-$~\cite{Poluektov:2006ia}.
Errors are only statistical. 
}
\label{tab:Belle_KSpipi_Dalitz}
\begin{center}
\begin{tabular}{lcc}
\hline\hline
Resonance               & Amplitude   & Phase (deg)  \\
\hline
$K^*(892)^+$            & $1.656\pm0.012$ & $137.6\pm 0.6$ \\
$K^*(892)^-$            & $0.149\pm0.007$ & $325.2\pm 2.2$  \\
$K^*_0(1430)^+$         & $1.96\pm0.04$   & $357.3\pm 1.5$    \\
$K^*_0(1430)^-$         & $0.30\pm0.05$   & $128\pm 8$     \\
$K^*_2(1430)^+$         & $1.32\pm0.03$   & $313.5\pm 1.8$     \\
$K^*_2(1430)^-$         & $0.21\pm0.03$   & $281\pm 9$ \\
$K^*(1680)^+$           & $2.56\pm0.22$   & $70\pm 6$    \\
$K^*(1680)^-$           & $1.02\pm0.22$   & $103\pm 11$    \\
$\rho(770)^0$           & $1$ (fixed)     & $0$ (fixed)    \\
$\omega(782)$           & $(33.0\pm1.3)\times 10^{-3}$ & $114.3\pm2.3$\\
$f_0(980)$              & $0.405\pm0.008$ & $212.9\pm2.3$  \\
$f_0(1370)$             & $0.82\pm0.10$   & $308\pm 8$  \\
$f_2(1270)$             & $1.35\pm0.06$   & $352\pm 3$  \\
$\sigma_1$              & $1.66\pm0.11$   & $218\pm 4$  \\
$\sigma_2$              & $0.31\pm0.05$   & $236\pm 11$  \\
nonresonant             & $6.1\pm0.3$     & $146\pm 3$  \\
\hline\hline
\end{tabular}
\end{center}
\end{table}

At this point the uncertainties in $\gamma$ are limited by
statistics. Contributions to
the uncertainty on $\gamma$ from these measurements are
not limited by the Dalitz plot uncertainty. But with increased
statistics the $\gamma$ measurement should improve and a 
better understanding of the Dalitz plot will be needed. At
threshold, CLEO-c or in the near future BES III, can
perform a tagged Dalitz plot analysis against $CP$ eigenstates.
These analyses will allow a direct determination of the phase
required for the extraction of $\gamma$. This avoids the
uncertainties from modeling of the Dalitz plot.

\subsubsection{$D^0\to \pi^-\pi^+\pi^0$}

\begin{figure}[tb]
\begin{center}
\includegraphics[width=0.90\linewidth]{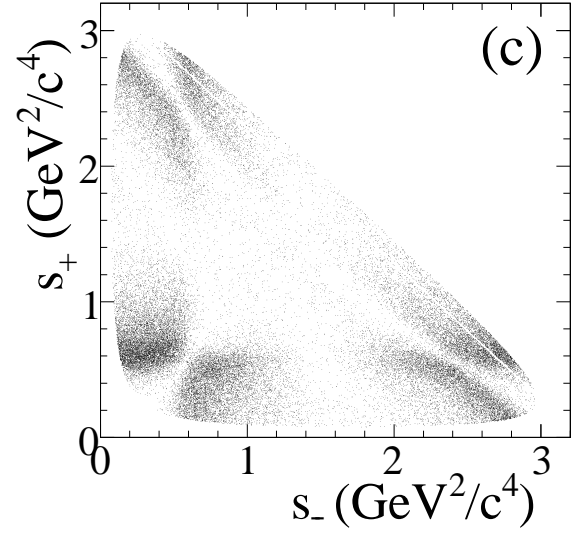}
\caption{BABAR $D^0\to \pi^+\pi^-\pi^0$ Dalitz plot analysis.
}
\label{fig:BABAR_pipipiz_Dalitz}
\end{center}
\end{figure}

The Dalitz plot of $D^0\to \pi^-\pi^+\pi^0$ has been
studied by BABAR as a means to extract information about
the CKM parameter $\gamma$~\cite{Aubert:2007ii} similar to
what was done with $D^0\to K^0_S\pi^+\pi^-$. CLEO has
also studied this decay~\cite{Muramatsu:2002jp}. BABAR 
reconstructs $44,780\pm250$ signal events over a background
of $830\pm70$ events. The Dalitz plot of these events is
shown in Fig.~\ref{fig:BABAR_pipipiz_Dalitz}. The three $\rho$ 
bands are clearly
visible with a strong destructive interference. BABAR 
used 15 resonances plus a nonresonant contribution to fit
the data. The results of the fit are summarized in 
Table~\ref{tab:BABAR_pipipiz_Dalitz}. The $\rho(770)$ 
resonances are clearly the strongest features on the 
Dalitz plot, with fit fractions adding to $(128.6\pm1.6)\%$.
The $\rho(1700)$ resonances contribute with fit fractions
of 3 to 5\% each, much smaller than the dominant contributions.
The remaining amplitudes, including nonresonant, is much
smaller. The large, destructively interfering, $\rho\pi$
amplitudes are suggestive of an $I=0$ dominated final 
state~\cite{Zemach:1965zz}. This is consistent with the 
observation that $D^0\to 3\pi^0$ is strongly suppressed.

\begin{table*}[bt]
\caption{
Dalitz plot parameters from BABAR analysis of $D^0\to \pi^-\pi^+\pi^0$.
}
\label{tab:BABAR_pipipiz_Dalitz}
\begin{center}
\begin{tabular}{lccc}
\hline\hline
Resonance      & Amplitude ratio (\%) & Phase (deg) & Fit fraction  \\
\hline
$\rho^+(770)$  & $100$ (fixed)        & $0$ (fixed)        & $67.8\pm 0.0\pm 0.6$ \\
$\rho^0(770)$  & $58.8\pm0.6\pm0.2$   & $16.2\pm0.6\pm0.4$ & $26.2\pm0.5\pm1.1$\\
$\rho^-(770)$  & $71.4\pm0.8\pm0.3$   & $-2.0\pm0.6\pm0.6$ & $34.6\pm0.8\pm0.3$\\
$\rho^+(1450)$ & $21\pm6\pm13$        & $-146\pm18\pm24$   & $0.11\pm 0.07\pm 0.12$\\
$\rho^0(1450)$ & $33\pm6\pm4$         & $10\pm8\pm12$      & $0.30\pm0.11\pm0.07$\\
$\rho^-(1450)$ & $82\pm5\pm4$         & $16\pm3\pm3$       & $1.79\pm0.22\pm0.12$\\
$\rho^+(1700)$ & $225\pm18\pm14$      & $-17\pm2\pm3$      & $4.1\pm 0.7\pm 0.7$\\
$\rho^0(1700)$ & $251\pm15\pm13$      & $-17\pm2\pm2$      & $5.0\pm0.6\pm1.0$\\
$\rho^-(1700)$ & $100\pm11\pm7$       & $-50\pm3\pm3$      & $3.2\pm0.4\pm0.6$\\
$f_0(980)$     & $1.50\pm0.12\pm0.17$ & $-59\pm5\pm4$      & $0.25\pm0.04\pm0.04$\\
$f_0(1370)$    & $6.3\pm0.9\pm0.9$    & $156\pm9\pm6$      & $0.37\pm0.11\pm0.09$\\
$f_0(1500)$    & $5.8\pm0.6\pm0.6$    & $12\pm9\pm5$       & $0.39\pm0.08\pm0.07$\\
$f_0(1710)$    & $11.2\pm1.4\pm1.7$   & $51\pm8\pm7$       & $0.31\pm0.07\pm0.08$\\
$f_2(1270)$    & $104\pm3\pm21$       & $-171\pm3\pm4$     & $1.32\pm0.08\pm0.10$\\
$\sigma(400)$  & $6.9\pm0.6\pm1.2$    & $8\pm4\pm8$        & $0.82\pm0.10\pm0.10$\\
Nonresonant    & $57\pm7\pm8$         & $-11\pm4\pm2$      & $0.84\pm0.21\pm0.12$ \\
\hline\hline
\end{tabular}
\end{center}
\end{table*}

\subsubsection{$D^0\to K^+K^-\pi^0$}

\begin{figure}[tb]
\begin{center}
\includegraphics[width=0.45\linewidth]{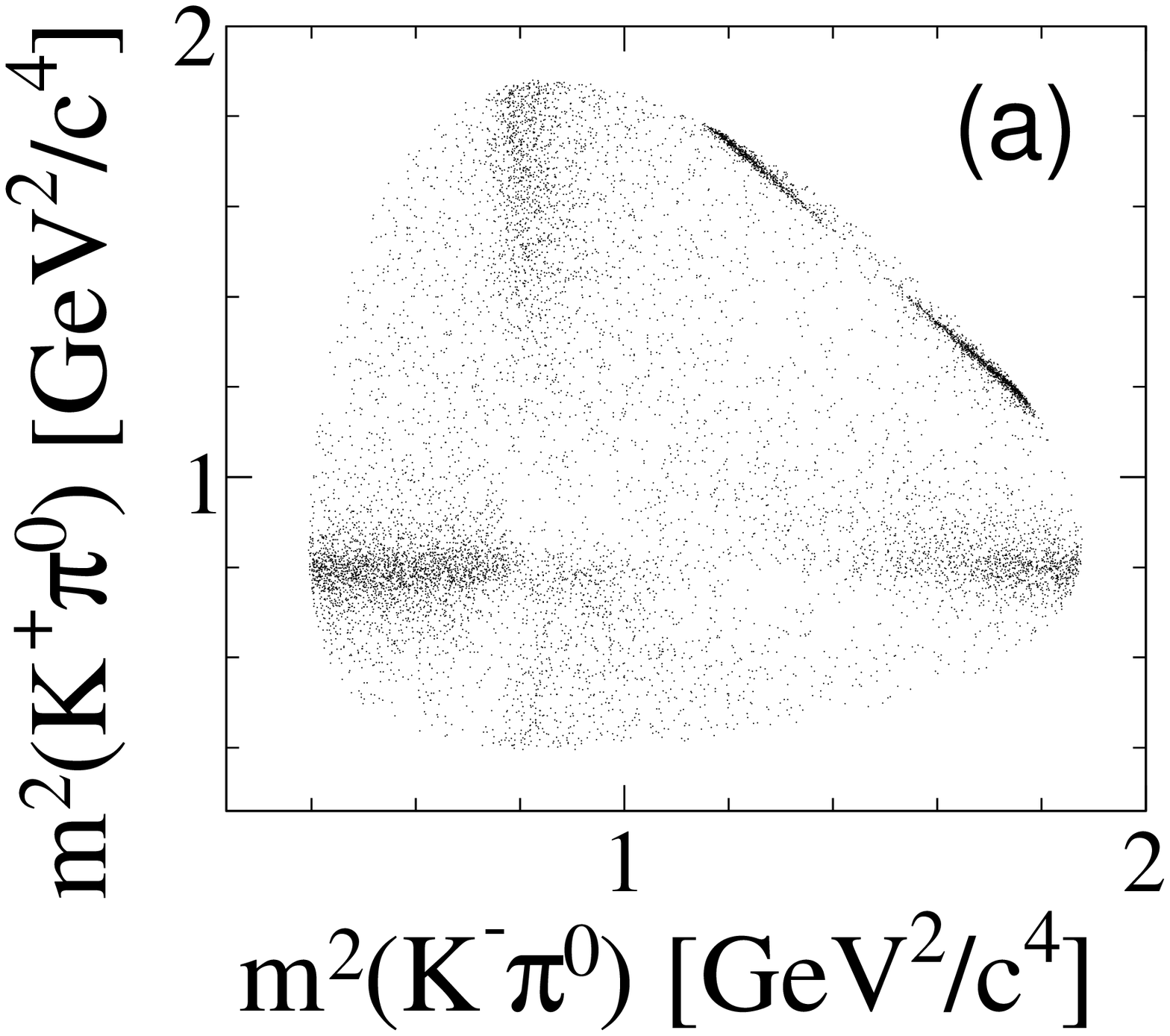}
\includegraphics[width=0.45\linewidth]{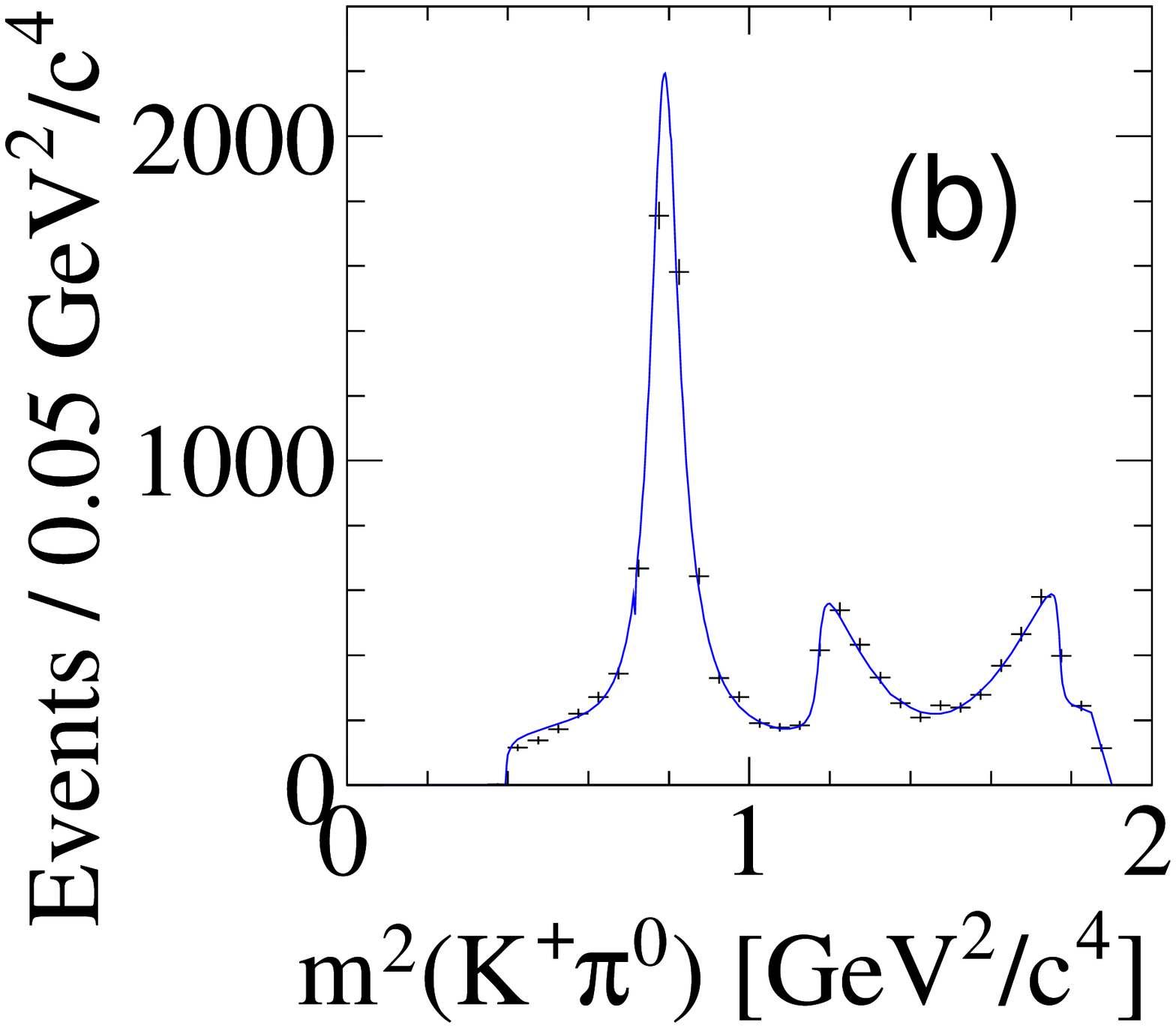}
\includegraphics[width=0.45\linewidth]{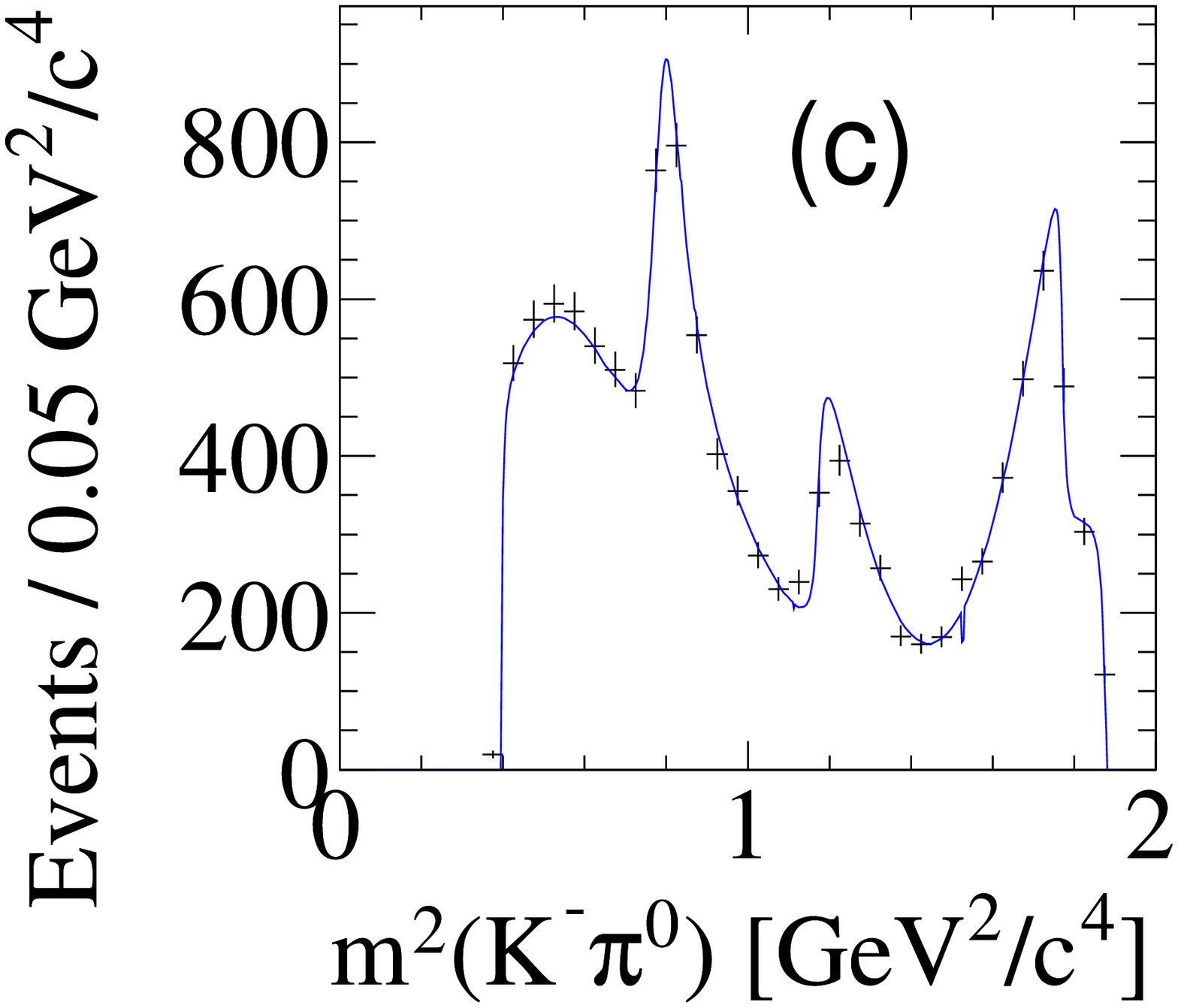}
\includegraphics[width=0.45\linewidth]{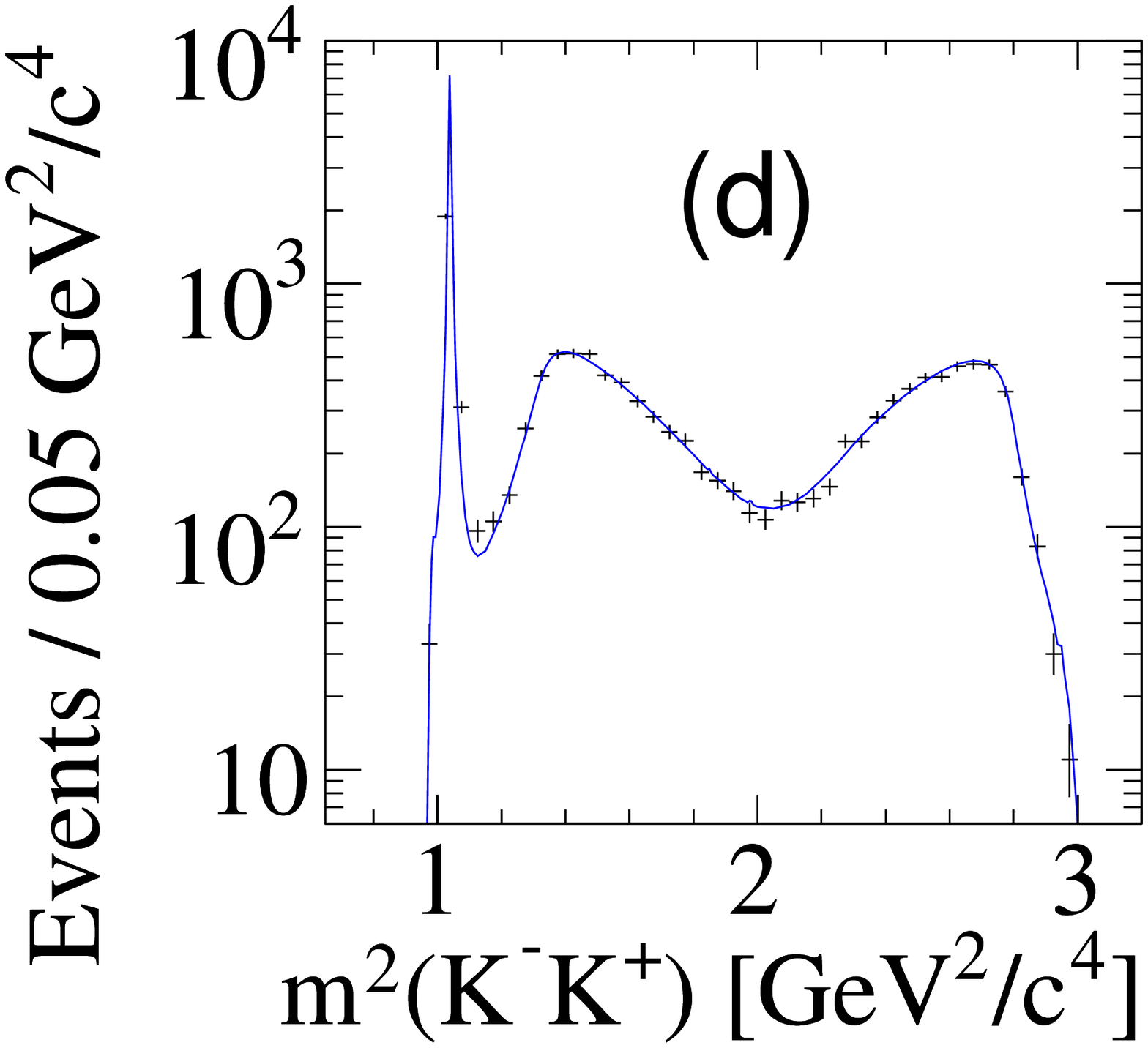}
\caption{BABAR $D^0\to K^+K^-\pi^0$ Dalitz plot analysis. From 
Ref.~\cite{Aubert:2007dc}.
}
\label{fig:BABAR_KKpiz_Dalitz}
\end{center}
\end{figure}

CLEO~\cite{Besson:2005hm} and BABAR~\cite{Aubert:2007dc}
have both studied the Dalitz plot of this decay. 
The BABAR analysis used 358 \fbinv\ of $e^+e^-$ collision 
data collected near the 
$\Upsilon(4S)$ resonance. 
A sample with a high purity of about $98.1\%$
was selected for this study containing $11,278\pm110$
$D^{*+}\to D^0\pi^+$ tagged candidates. The Dalitz plot and the best isobar fit
is shown in Fig.~\ref{fig:BABAR_KKpiz_Dalitz}. The isobar model
allows for several different solutions that each give a similarly
good description of the data. At low $K^+K^-$ invariant mass an
$S$-wave $K^+K^-$ contribution is needed, but the fit can not
distinguish between an $a_0(980)$ and a $f_0(980)$. Similarly,
at intermediate $K^+K^-$ invariant mass either a $f'_2(1525)$ or an $f_0$
with a similar mass works. In the study of this Dalitz 
plot the relative amplitude and phase of the amplitudes
for $D^0\to K^{*-}K^+$ to $D^0\to K^{*+}K^-$ can be
measured. Defining $r_D$ and $\delta_D$ by
$$
r_De^{i\delta_D}\equiv {a_{D^0\to K^{*-}K^+}\over a_{D^0\to K^{*+}K^-}},
$$
BABAR obtains
$$
r_D=0.599\pm0.013({\rm stat})\pm0.011({\rm syst})
$$
and
$$
\delta_D=-35.5^{\circ}\pm1.9^{\circ}({\rm stat})\pm2.2^{\circ}({\rm syst})
$$
consistent with the earlier CLEO results.

\subsubsection{$D^0\to K^+K^-K^0_S$}

\begin{figure}[tb]
\begin{center}
\includegraphics[width=0.90\linewidth]{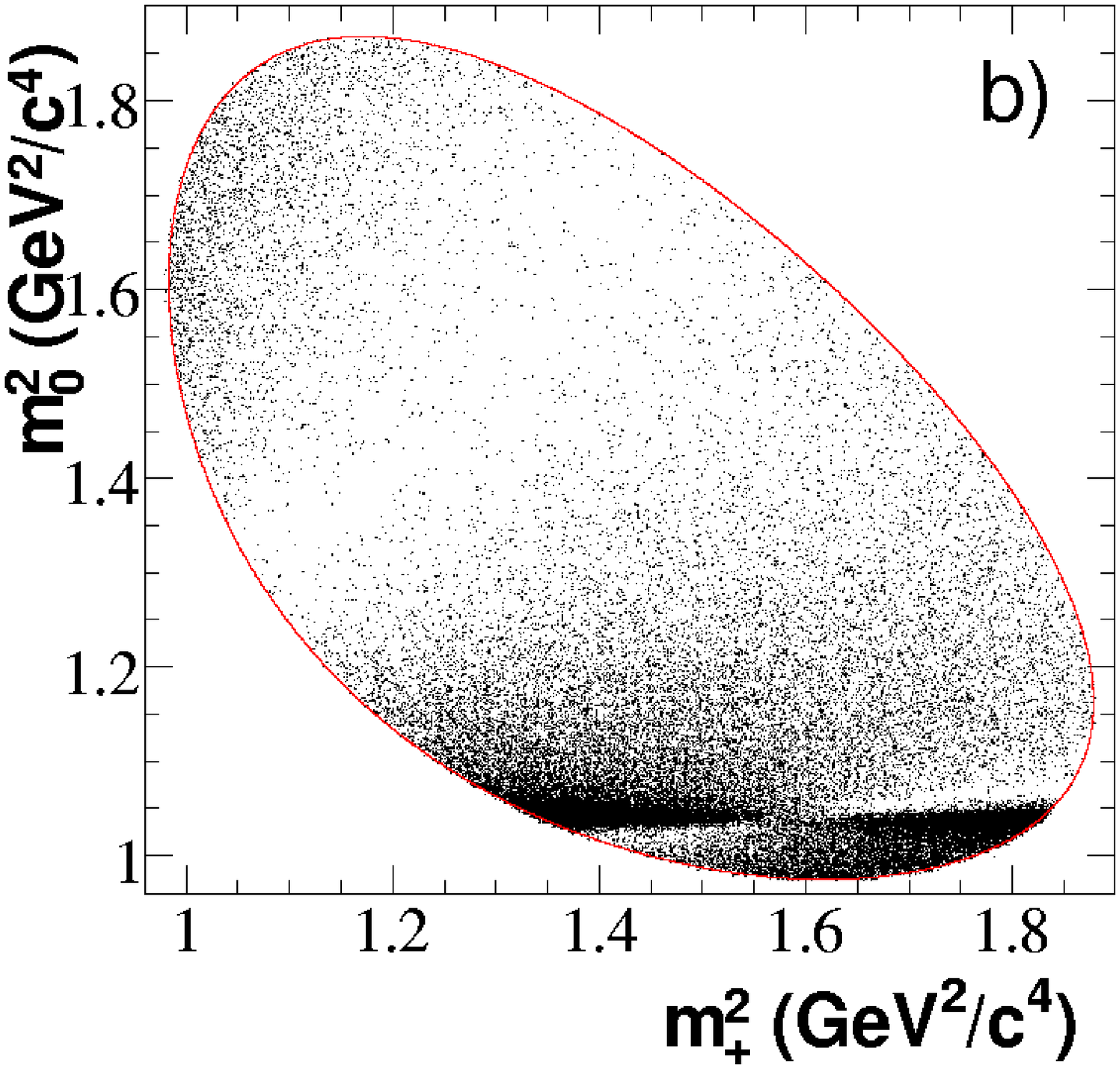}
\caption{BABAR $D^0\to K^+K^-K^0_S$ Dalitz plot analysis.
From~\textcite{Aubert:2008bd}.
}
\label{fig:BABAR_KSKK_Dalitz}
\end{center}
\end{figure}

This mode has been studied by BABAR~\cite{Aubert:2005sm, Aubert:2008bd} as
part of an analysis for $\gamma$ determination. 
BABAR uses a sample of 69,000 reconstructed $D^0\to K^0_SK^+K^-$ decays.
The data, shown in Fig.~\ref{fig:BABAR_KSKK_Dalitz} was fit to an
isobar model which includes eight resonances. The result of this fit
is summarized in Table~\ref{tab:BABAR_KSKK_DalitzFit}.
In the fit BABAR floats the mass and width of the $\phi(1020)$.
The $a_0(980)$ resonance has a mass very close to $KK$ threshold
and decays primarily to $\eta\pi$ and is described by a coupled
channel Breit-Wigner line shape. The data is well described by the
fit, BABAR finds a reduced $\chi^2$ of 1.09 for 6,856 degrees of 
freedom.

\begin{table}[bt]
\caption{
Dalitz plot parameters from BABAR analysis of 
$\bar D^0\to K^0_SK^+K^-$~\cite{Aubert:2008bd}. Errors are only statistical.
}
\label{tab:BABAR_KSKK_DalitzFit}
\begin{center}
\begin{tabular}{lccc}
\hline\hline
Resonance               & Amplitude       & Phase (deg) & Fit fraction (\%) \\
\hline
$K^0_Sa_0(980)^0$       & $1$             & $0$             &  $55.8$ \\
$K^0_S\phi(1020)^0$     & $0.227\pm0.005$ & $56.2\pm 1.0$   &  $44.9$\\
$K^0_Sf_0(1370)^0$      & $0.04\pm0.06$   & $2\pm 80$       &  $0.1$\\
$K^0_Sf_2(1370)^0$      & $0.261\pm0.020$ & $9\pm 6$        &  $0.3$\\
$K^0_Sa_0(1450)^0$      & $0.65\pm0.09$   & $95\pm 10$      &  $12.6$\\
$K^-a_0(980)^+$         & $0.562\pm0.015$ & $179\pm 3$      &  $16.0$\\
$K^-a_0(1450)^+$        & $0.84\pm0.04$   & $97\pm 4$       &  $21.8$\\
$K^+a_0(1450)^-$        & $0.118\pm0.015$ & $138\pm 7$      &  $0.7$\\
\hline\hline
\end{tabular}
\end{center}
\end{table}

\subsubsection{$D^0\to K^0_S\eta\pi^0$}

\begin{figure}[tb]
\begin{center}
\includegraphics[width=0.90\linewidth]{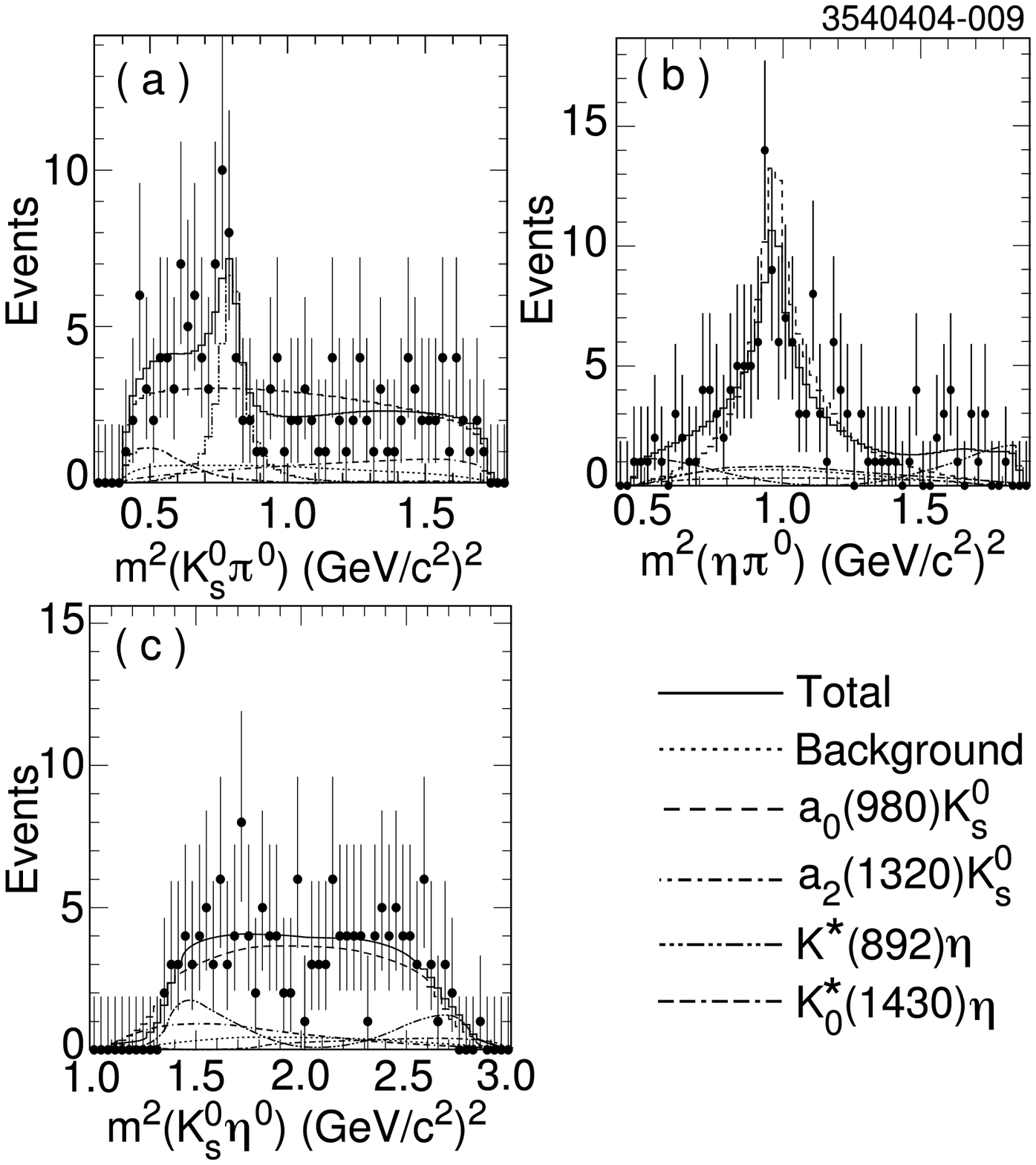}
\caption{CLEO $D^0\to K^0_S\eta\pi^0$ Dalitz plot analysis.
From~\textcite{Rubin:2004cq}. 
}
\label{fig:CLEO_KSetapi0_Dalitz}
\end{center}
\end{figure}

This decay has been studied using a 9.0 \fbinv\ data sample
collected using the CLEO II.V detector in $e^+e^-$ collisions
at the $\Upsilon(4S)$ resonance~\cite{Rubin:2004cq}. The sample contained
155 $D^0\to K^0_S\eta\pi^0$ candidate events. The two 
large contributions to this decay come from $K^*(892)^0\eta$ and
$a_0(980)^0K^0_S$. The projections
of the Dalitz plot fit is shown in Fig.~\ref{fig:CLEO_KSetapi0_Dalitz}.
Fixing the amplitude for $a_0(980)^0K^0_S$
to be 1 with a zero phase CLEO measured
\begin{eqnarray*}
a_{K^*(892)^0\eta} &=& 0.249\pm0.032\pm0.013\pm0.018, \\
\phi_{K^*(892)^0\eta} &=& (259\pm 12\pm 9 \pm 6)^{\circ}, \\
{\rm FF}(K^*(892)^0\eta) & = & 0.293 \pm 0.062 \pm 0.029 \pm 0.019, \\
{\rm FF}(a_0(980)^0K^0_S) & = & 1.19 \pm 0.09 \pm 0.20 \pm 0.16, \\
\end{eqnarray*}
where the errors are statistical, systematic, and model dependence
respectively. For the model dependence CLEO considered alternative
models where they added additional resonances. They considered four
different alternative fits including: a nonresonant component,
$K^*_0(1430)\eta$, 
$K^*_0(1430)\eta$ + $a_2(1320)K^0_S$, and $\kappa\eta$. The fit
probability for these different fits were 6.4\%, 19.4\%, 64.7\%,
and 49.9\% respectively. The fit with only two resonances had
a probability of 0.8\%. From these alternative fits CLEO-c 
derives a fit fraction of $0.246\pm 0.092\pm 0.024\pm 0.087$ for
any additional components beyond the $K^*(892)^0\eta$ and 
$a_0(980)^0K^0_S$.

\subsubsection{$D^+\to K^-\pi^+\pi^+$}

The decay $D^+\to K^-\pi^+\pi^+$ is one of the largest
decays of the $D^+$. CLEO-c has measured the 
branching fraction to be ${\cal B}(D^+\to K^-\pi^+\pi^+)=(\BDpkpipivalue)\%$.
The Dalitz plot for this decay has been studied by
several experiments MARK III~\cite{Adler:1987sd}, 
NA14~\cite{Alvarez:1990sy}, E691~\cite{Anjos:1992kb}, 
E687~\cite{Frabetti:1994di}, E791~\cite{Aitala:2002kr,Aitala:2005yh},
and most recently by CLEO-c~\cite{Bonvicini:2008jw}. 
This Dalitz plot is interesting as the only clear
resonant contribution from $K^{*0}(892)$ only has a 
12\% fit fraction and a contribution of over 60\% from
$K\pi$ $S$-wave. E791~\cite{Aitala:2002kr} obtained 
a good fit including a large low-mass $K^-\pi^+$
scalar resonance $\kappa$. This fit obtained fit fractions
that was significantly different from earlier studies. 
E791~\cite{Aitala:2005yh} re-analyzed the data
using a model independent partial wave analysis.
The CLEO-c analysis also uses the same model
independent partial wave analysis.

\begin{figure}[tb]
\begin{center}
\includegraphics[width=0.90\linewidth]{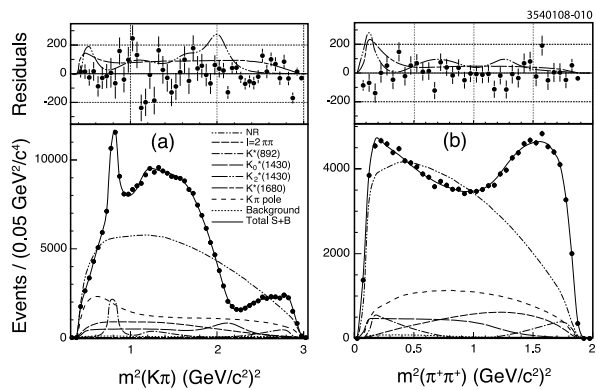}
\caption{Projections of the Dalitz plot fit
in the CLEO-c Dalitz plot 
analysis of $D^+\to K^-\pi^+\pi^+$ in a) for $m^2(K\pi)$ (two entries
per candidate, b) for $m^2(\pi^+\pi^+)$. The data are shown as
points with error bars. The insets on top shows the residuals
between the data and the as points with error bars. The small
contributions in the fit from the $K^*(1680)$ and $K^*_2(1430)$ 
resonances are also shown in the insets enhanced by a factor of 
10.
From~\textcite{Bonvicini:2008jw}.
}
\label{fig:CLEOc_D_Kpipi_Dalitz_fit}
\end{center}
\end{figure}

The CLEO-c study is based on 572 \pbinv\ of $e^+e^-$
collision data collected at the $\psi(3770)$ resonance.
The data sample selected for the Dalitz plot analysis
consists of 140,793 events with a background of about 1.1\%.
The projections of the Dalitz plot is shown in
Fig.~\ref{fig:CLEOc_D_Kpipi_Dalitz_fit}. The CLEO-c analysis
finds that in order to get a good description of the 
data, either in the isobar model or using the model independent
partial wave analysis for the $K\pi$ $S$-wave, they need
to include a $I=2$ $\pi^+\pi^+$ $S$-wave. CLEO-c implements
this $I=2$ $\pi^+\pi^+$ $S$-wave either using an analytic 
form or using a model independent partial wave analysis. The
model independent partial wave analysis results agree with
the analytic form and both give a good fit. CLEO-c finds
a fit fraction of about 10 to 15\% for the $I=2$ $\pi^+\pi^+$ $S$-wave.



\subsubsection{$D^+\to \pi^+\pi^+\pi^-$}

\begin{figure}[tb]
\begin{center}
\includegraphics[width=0.90\linewidth]{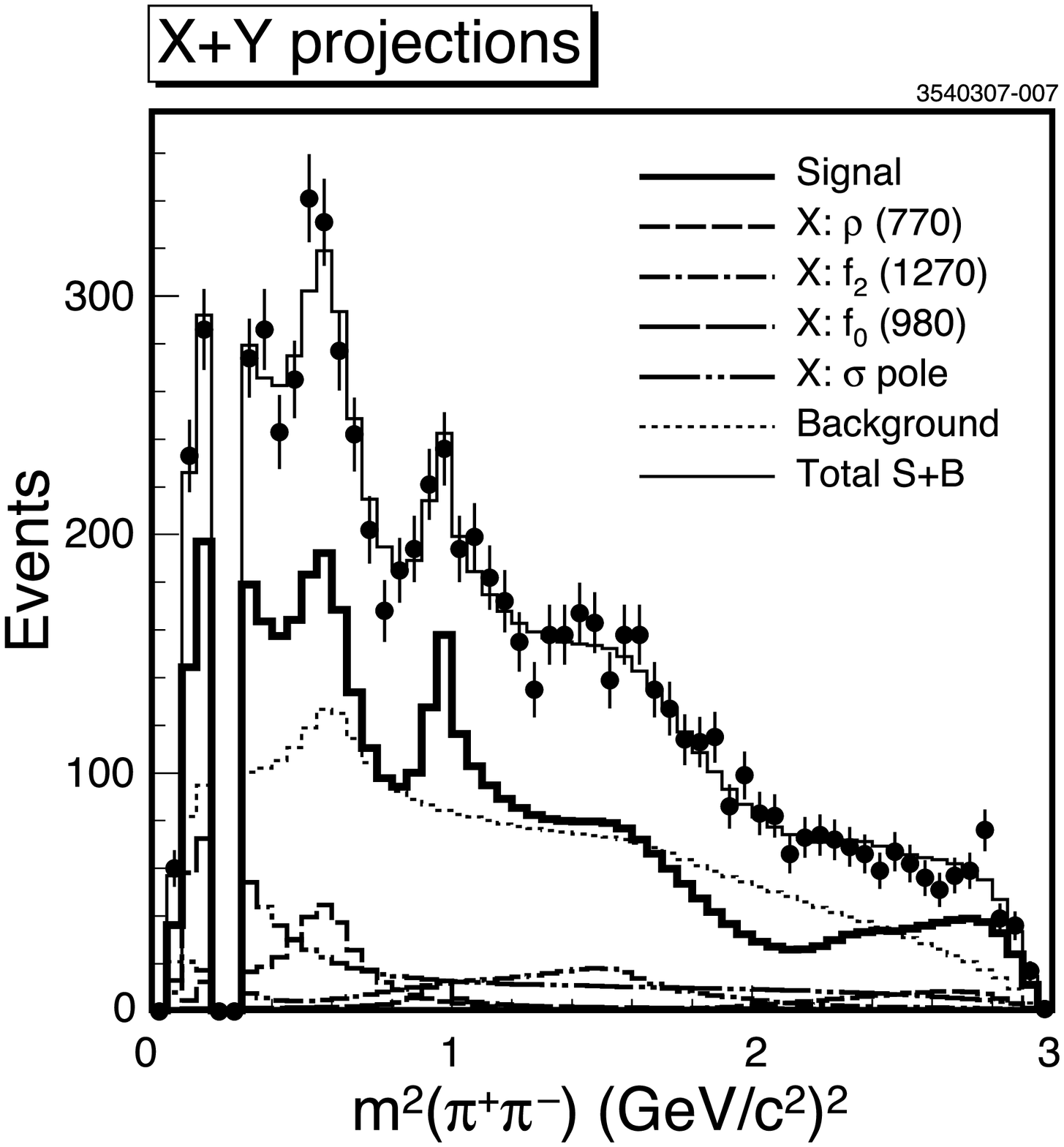}
\caption{CLEO-c $D^+\to \pi^+\pi^-\pi^+$ Dalitz plot analysis.
From~\textcite{Bonvicini:2007tc}. 
}
\label{fig:CLEOc_pipipi_Dalitz}
\end{center}
\end{figure}

The $D^+\to \pi^+\pi^+\pi^-$ decay has been studied 
by E687~\cite{Frabetti:1997sx},
E691~\cite{Anjos:1988pu},
E791\cite{Aitala:2000xu}, FOCUS~\cite{Link:2003gb}, 
and CLEO-c~\cite{Bonvicini:2007tc}. 
The most
recent analysis, with the largest data sample, is the CLEO-c analysis.
The earlier analysis by E791 had reported the need to add a $\sigma(500)$
Breit-Wigner to the $\pi^+\pi^-$ $S$-wave in order to get an acceptable fit.
FOCUS analyzed this mode using a $K$-matrix description of the $\pi^+\pi^-$
$S$-wave. They obtained an acceptable fit, but did not rule out the need for
a $\sigma(500)$. CLEO-c has studied these decays with a sample of about 2,600
signal events, excluding the $K^0_S$ events. The nominal fit using the
isobar model supports the need for a $\sigma\pi^+$ component. The fit
to the isobar model is shown in Figure~\ref{fig:CLEOc_pipipi_Dalitz}
and the result from the fit is summarized in 
Table~\ref{tab:CLEOc_pipipi_Dalitz}.

\begin{table}[bt]
\caption{
Dalitz plot parameters from CLEO-c analysis~\cite{Bonvicini:2007tc} of $D^+\to \pi^+\pi^-\pi^+$.
}
\label{tab:CLEOc_pipipi_Dalitz}
\begin{center}
\begin{tabular}{lccc}
\hline\hline
Resonance               & Amplitude   & Phase (deg) & Fit fraction (\%)  \\
\hline
$\rho(770)\pi^+$        & $1$ (fixed)       & $0$ fixed    & $20.0\pm2.3\pm0.9$ \\
$f_0(980)\pi^+$         & $1.4\pm0.2\pm0.2$ & $12\pm10\pm5$& $4.1\pm0.9\pm0.3$ \\
$f_2(1270)\pi^+$        & $2.1\pm0.2\pm0.1$ & $-123\pm6\pm3$&$18.2\pm2.6\pm0.7$ \\
$f_0(1370)\pi^+$        & $1.3\pm0.4\pm0.2$ & $-21\pm15\pm14$&$2.6\pm1.8\pm0.6$ \\
$f_0(1500)\pi^+$        & $1.1\pm0.3\pm0.2$ & $-44\pm13\pm16$&$3.4\pm1.0\pm0.8$ \\
$\sigma$ pole           & $3.7\pm0.3\pm0.2$ & $-3\pm4\pm2$ & $41.8\pm1.4\pm2.5$ \\
\hline\hline
\end{tabular}
\end{center}
\end{table}

\subsubsection{$D^+_s\to K^+K^-\pi^+$}
\label{sect:DstoKKpiDalitz}

The Dalitz plot for $D^+_s\to K^+K^-\pi^+$ is of interest
as it contains the large $D^+_s\to \phi\pi^+$ contribution
that traditionally has been the reference branching fraction
for $D_s^+$ decays. The decay $D^+_s\to K^+K^-\pi^+$ has
been studied by E687~\cite{Frabetti:1995sg} using a sample
of 701 events. This analysis showed evidence for a large
$D_s^+\to f_0(980)\pi^+$ contribution. FOCUS has also
reported a preliminary study of this Dalitz plot~\cite{Malvezzi:2002xt}.  
Most recently CLEO-c~\cite{Mitchell:2009tr} has reported
preliminary results from their study of the Dalitz plot
in this decay. 

\begin{figure}[tb]
\begin{center}
\includegraphics[width=0.90\linewidth]{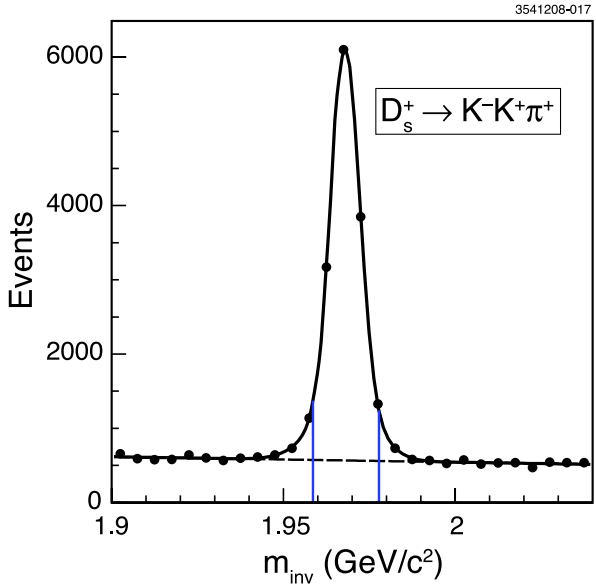}
\caption{The $K^+K^-\pi^+$ invariant mass for the signal candidates
in the CLEO-c Dalitz plot analysis of $D^+_s\to K^+K^-\pi^+$.
From~\textcite{Mitchell:2009tr}.
}
\label{fig:CLEOc_Ds_KKpi_Dalitz_mass}
\end{center}
\end{figure}

\begin{figure}[tb]
\begin{center}
\includegraphics[width=0.90\linewidth]{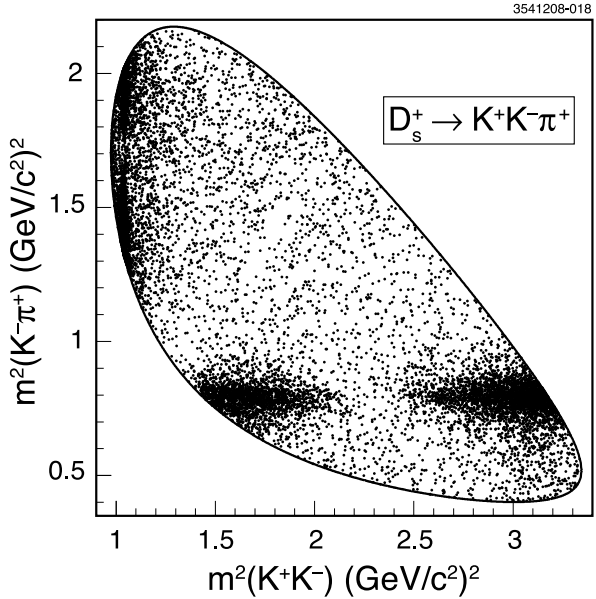}
\caption{The Dalitz plot for $D^+_s\to K^+K^-\pi^+$ candidates 
in the CLEO-c analysis of $D^+_s\to K^+K^-\pi^+$.
From~\textcite{Mitchell:2009tr}.
}
\label{fig:CLEOc_Ds_KKpi_Dalitz}
\end{center}
\end{figure}

The CLEO-c analysis uses 586 \pbinv\ of $e^+e^-$ collision data
collected at $\sqrt{s}=4.17$ GeV. This sample corresponds to
about $0.57\times 10^6$ $D^{\pm}_sD^{*\pm}_s$ pairs. In this analysis
about 14,400 $D^+_s\to K^+K^-\pi^+$ candidates are reconstructed
with a background of about 15\%. The invariant mass distribution
for the $K^+K^-\pi^+$ candidates are shown in 
Fig.~\ref{fig:CLEOc_Ds_KKpi_Dalitz_mass}. The Dalitz plot is 
shown in Fig.~\ref{fig:CLEOc_Ds_KKpi_Dalitz}. Clearly visible
in this plot are the $\phi$ and $K^{*0}$ resonances. 

\begin{figure}[tb]
\begin{center}
\includegraphics[width=0.49\linewidth]{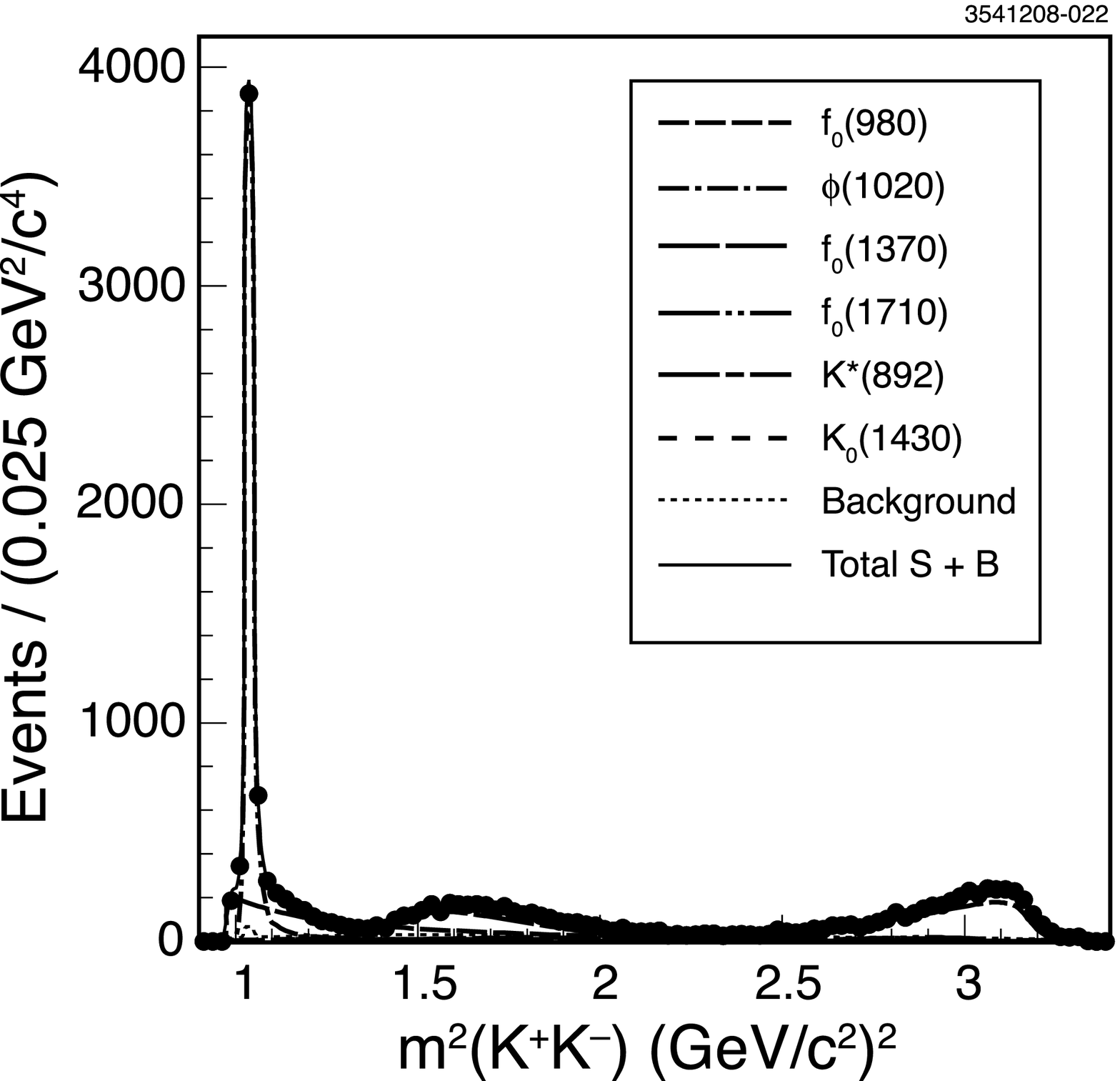}
\includegraphics[width=0.49\linewidth]{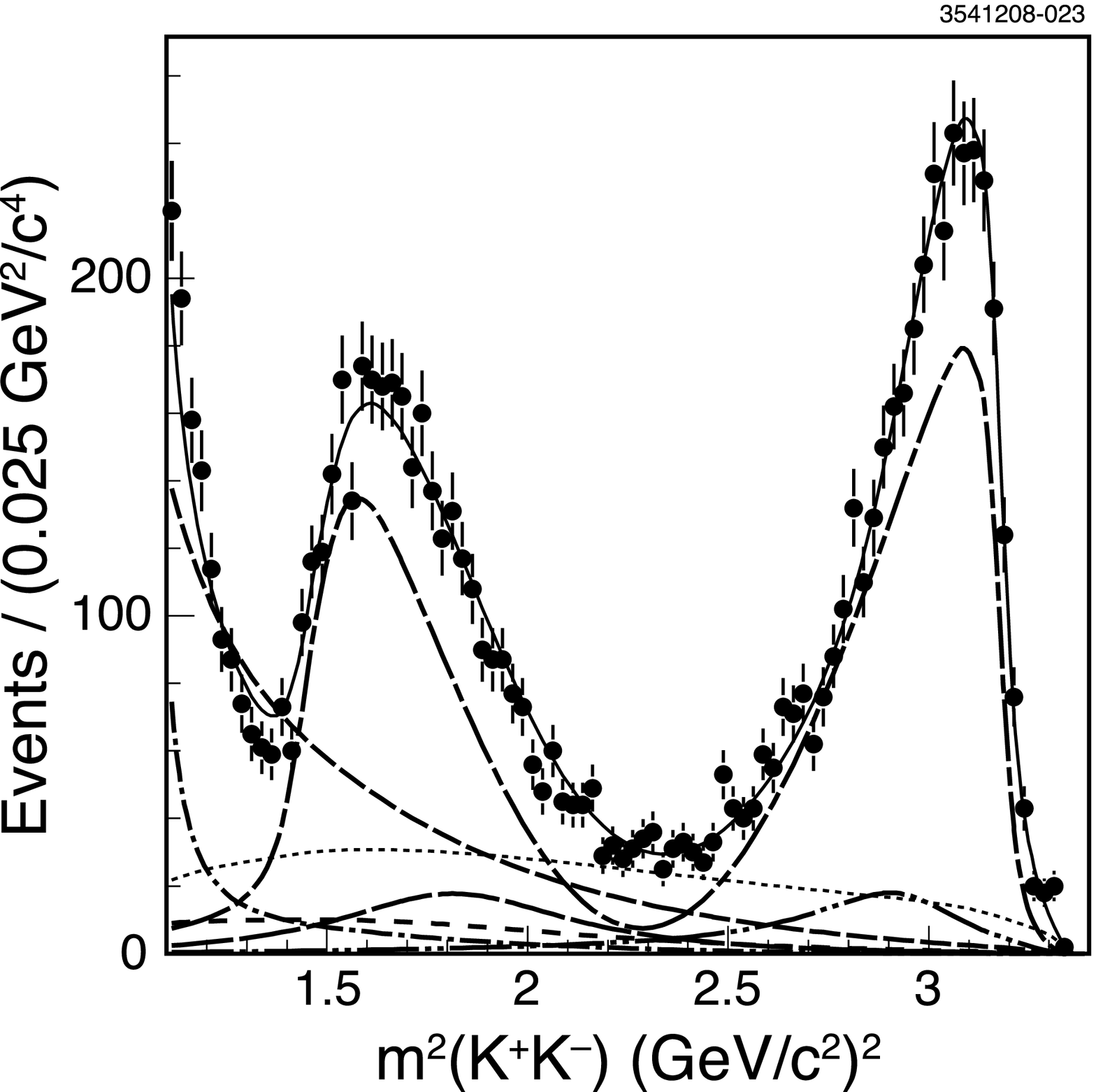}
\includegraphics[width=0.49\linewidth]{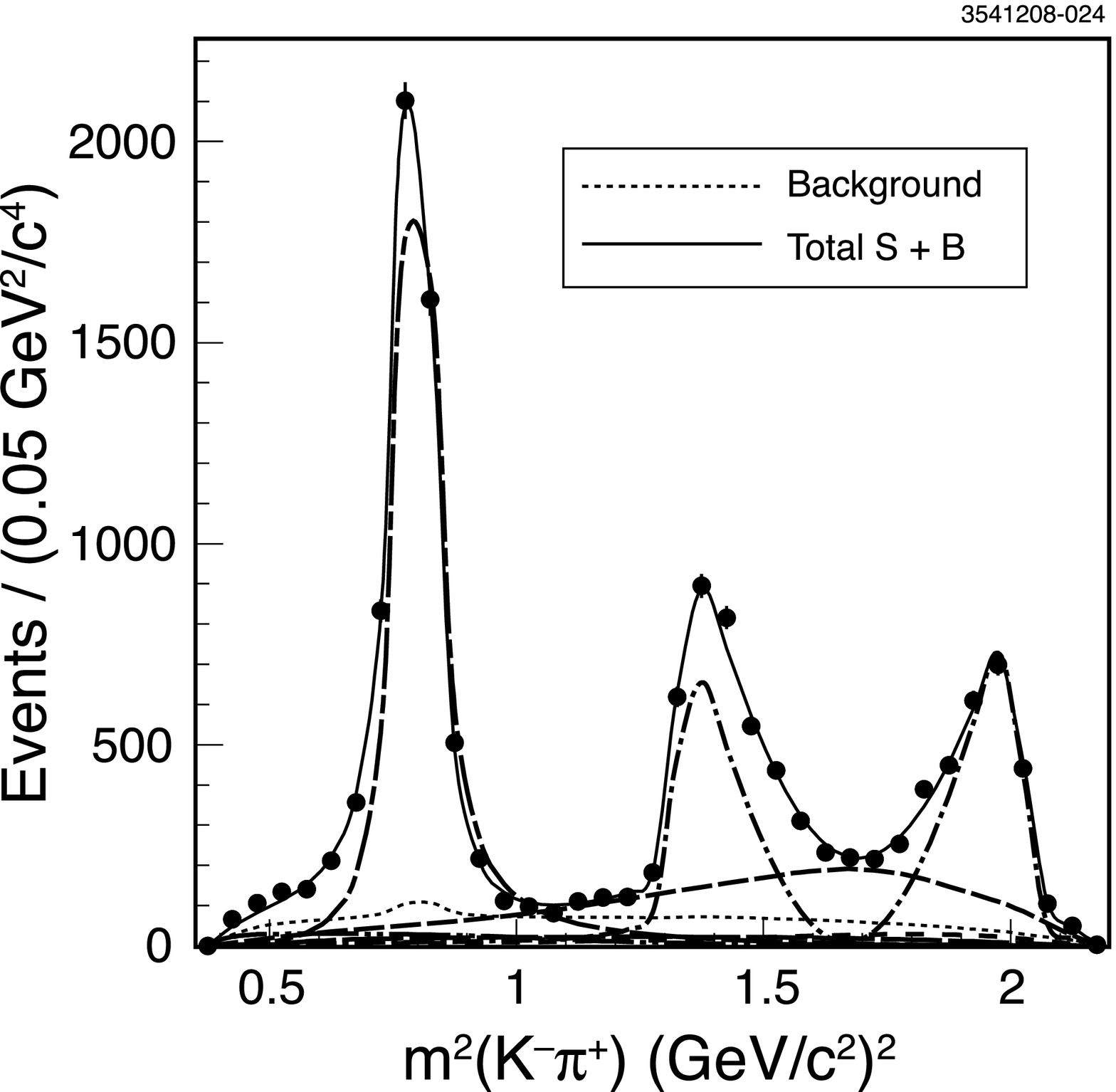}
\includegraphics[width=0.49\linewidth]{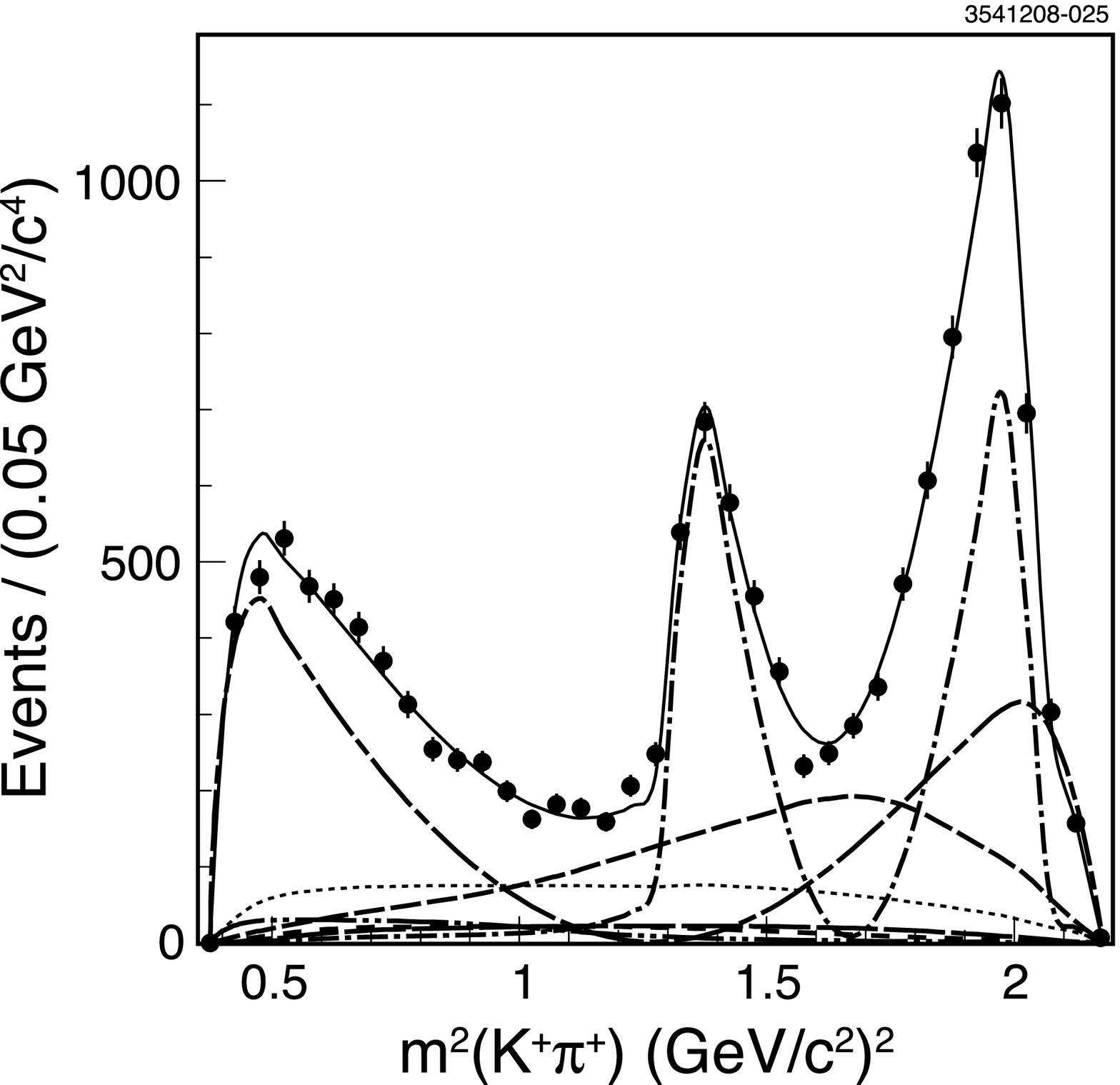}
\caption{The CLEO-c Dalitz plot fit for $D^+_s\to K^+K^-\pi^+$ candidates.
From~\textcite{Mitchell:2009tr}.
}
\label{fig:CLEOc_Ds_KKpi_Dalitz_fit}
\end{center}
\end{figure}

The data are fit to an isobar model including the $f_0(980)$,
$\phi$, $f_0(1370)$, $f_0(1710)$, $K^*(892)$, and $K^*_0(1430)$
resonances. 
CLEO-c finds that 
all resonances studied by E687 are significant, but that in
order to obtain a good fit they need to add an additional
$K^+K^-$ resonance. Several resonant, or nonresonant, contributions
gives a similar improvement of the fit quality, though the 
$f_0(1370)$ gives the best fit and is used in the main result. 
The result of this fit is shown in 
Fig.~\ref{fig:CLEOc_Ds_KKpi_Dalitz_fit}. A summary of the
amplitudes and phases extracted from this fit is shown
in Table~\ref{tab:CLEOc_Ds_KKpi_Dalitz_Tab}. CLEO-c 
obtains a reasonably good fit, $\chi^2/{\rm d.o.f}=178/117$,
using these resonances.

\begin{table}[bt]
\caption{
Dalitz plot parameters from CLEO-c analysis of $D^+_s\to K^-K^+\pi^+$.
}
\label{tab:CLEOc_Ds_KKpi_Dalitz_Tab}
\begin{center}
\begin{tabular}{lccc}
\hline\hline
Resonance               & Amplitude   & Phase (deg) & Fit fraction (\%)  \\
\hline
$\bar K^*(892)^0K^+$    & $1$ (fixed) & $0$ (fixed) & $47.4\pm1.5$ \\
$\bar K^*_0(1430)K^+$   & $1.51\pm 0.11$ & $146\pm 8$ & $3.9\pm 0.5$ \\
$\bar f_0(980)\pi^+$   & $4.72\pm 0.18$ & $157\pm 3$ & $28.2\pm 1.9$ \\
$\bar \phi(1020)\pi^+$   & $1.13\pm 0.02$ & $-8\pm 4$ & $42.2\pm 1.6$ \\
$\bar f_0(1370)\pi^+$   & $1.15\pm 0.09$ & $53\pm 5$ & $4.3\pm 0.6$ \\
$\bar f_0(1710)\pi^+$   & $1.11\pm 0.07$ & $89\pm 5$ & $3.4\pm 0.5$ \\
\hline\hline
\end{tabular}
\end{center}
\end{table}

\subsubsection{$D^+_s\to \pi^+\pi^-\pi^+$}

\begin{figure}[tb]
\begin{center}
\includegraphics[width=0.90\linewidth]{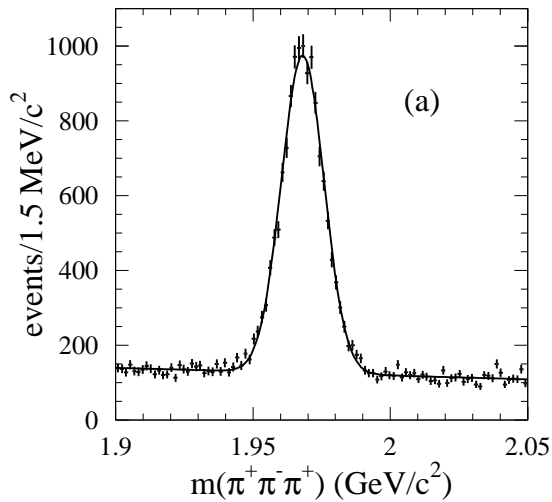}
\caption{The $\pi^+\pi^-\pi^+$ invariant mass for the signal candidates
in the BABAR Dalitz plot analysis of $D^+_s\to \pi^+\pi^-\pi^+$.
From~\textcite{Aubert:2008tm}.
}
\label{fig:BABAR_Ds_pipipi_Dalitz_mass}
\end{center}
\end{figure}

\begin{figure}[tb]
\begin{center}
\includegraphics[width=0.90\linewidth]{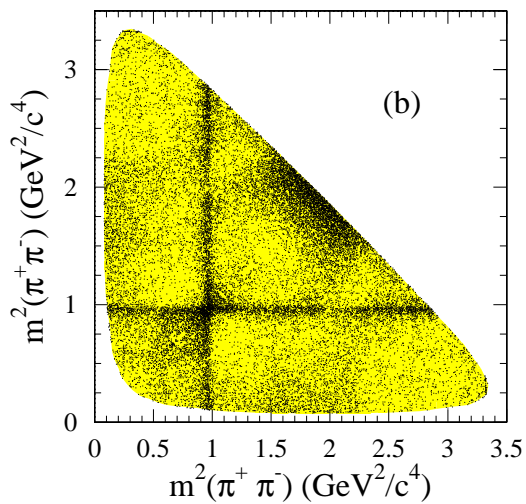}
\caption{The Dalitz plot for signal candidates
in the BABAR Dalitz plot analysis of $D^+_s\to \pi^+\pi^-\pi^+$.
From~\textcite{Aubert:2008tm}.
}
\label{fig:BABAR_Ds_pipipi_Dalitz}
\end{center}
\end{figure}

The decay $D^+_s\to \pi^+\pi^-\pi^+$ has been
studied by E791~\cite{Aitala:2000xt}, FOCUS~\cite{Link:2003gb}, and
BABAR~\cite{Aubert:2008tm}. The BABAR analysis
selects 13,179 events with a purity of 80\%.
The invariant mass distribution of the $D^+_s\to \pi^+\pi^-\pi^+$
candidates is shown in Fig.~\ref{fig:BABAR_Ds_pipipi_Dalitz_mass}
and the symmetrized Dalitz plot distribution is
shown in Fig.~\ref{fig:BABAR_Ds_pipipi_Dalitz}. The symmetrized
plot shows two entries in the Dalitz plot for each candidate.
The analysis by BABAR includes three resonances, $f_2(1270)\pi^+$,
$\rho(770)\pi^+$, and $\rho(1450)\pi^+$. In addition to these
$P$- and $D$-wave resonances the MIPWA is used for the $\pi^+\pi^-$ 
$S$-wave. This method
parameterizes the amplitude and phase by dividing the 
$\pi^+\pi^-$ mass spectrum into 29 slices. The results for the
amplitudes and phases from the fit for the 
parameterization of the $S$-wave clearly show the 
$f_0(980)$ resonance. There is also some evidence
for the $f_0(1370)$ and $f_0(1500)$. 
In Table~\ref{tab:BABAR_Ds_pipipi_Dalitz_Tab} the summary of 
the fit is given. The $S$-wave parameterization accounts
for a fit fraction of $(83.0\pm 0.9\pm1.9)\%$. This decay
also has an important contribution from a spin-2 resonance,
$D_s^+\to f_2(1270)\pi^+$.

\begin{table}[bt]
\caption{
Dalitz plot parameters from BABAR analysis of $D^+_s\to \pi^+\pi^-\pi^+$.
}
\label{tab:BABAR_Ds_pipipi_Dalitz_Tab}
\begin{center}
\begin{tabular}{lccc}
\hline\hline
Resonance               & Amplitude   & Phase (rad) & Fit fraction (\%)  \\
\hline
$f_2(1270)\pi^+$   & $1$ (fixed) & $0$ (fixed) & $10.1\pm1.5\pm1.1$ \\
$\rho(770)\pi^+$   & $0.19\pm0.02\pm0.12$  & $1.1\pm0.1\pm0.2$ & $1.8\pm0.5\pm1.0$ \\
$\rho(1450)\pi^+$ & $1.2\pm0.3\pm1.0$  & $4.1\pm0.2\pm0.5$ & $2.3\pm0.8\pm1.7$ \\
$S$-wave         & \multicolumn{2}{c}{See Ref.~\cite{Aubert:2008tm}} & $83.0\pm0.9\pm1.9$ \\
\hline\hline
\end{tabular}
\end{center}
\end{table}

\subsection{Four-body decays}

Similar to the three-body decays discussed in the previous
section the resonant substructure can be studied in higher
multiplicity final states. A four-body final state 
has a five-dimensional phase space which is hard to 
visualize.

MARK III~\cite{Adler:1989gy}
studied the decay $D^0\to K^-\pi^+\pi^-\pi^+$. They
performed an unbinned maximum likelihood fit in
the five-dimensional phase space to extract 
amplitudes for two-body decays. MARK III 
selected a sample of $1,281\pm45$ $D^0\to K^-\pi^+\pi^-\pi^+$
candidates.
The result of the fit to this samples is
summarized in Table~\ref{tab:MARKIII_D_Kpipipi_Dalitz_Tab}.
The largest two-body decay contributing to this final
state is $D^0\to K^-a_1(1260)^+$ with a fit fraction
of $0.492\pm0.024\pm0.08$. The fit gives a fit fraction
of $0.242\pm0.025\pm0.06$ for nonresonant four-body
final states, but it is likely that this includes contributions
from other wide resonances. 

\begin{table}[bt]
\caption{
Fit fractions and phases from the MARK III analysis~\cite{Adler:1989gy}
of the decay  $D^0\to K^-\pi^+\pi^-\pi^+$.
}
\label{tab:MARKIII_D_Kpipipi_Dalitz_Tab}
\begin{center}
\begin{tabular}{lcc}
\hline\hline
Resonance              & Fit fraction (\%) & Phase (rad) \\
\hline
$\bar K^{*0}\rho^0$ Transverse ($S$-wave) & $0.142\pm0.016\pm0.05$ & $-1.39\pm0.09$\\
$K^-a_1(1260)^+$                          & $0.492\pm0.024\pm0.08$ & $0$\\
$K_1(1270)^-\pi^+$                        & $0.066\pm0.019\pm0.03$ & $0.71\pm0.25$\\
$\bar K^{*0}\pi^+\pi^-$                   & $0.140\pm0.018\pm0.04$ & $3.07\pm0.09$\\
$K^{-}\rho^0\pi^+$                        & $0.084\pm0.022\pm0.04$ & $-0.30\pm0.13$\\
Four-body nonresonant                     & $0.242\pm0.025\pm0.06$ & $-1.07\pm0.08$\\
\hline\hline
\end{tabular}
\end{center}
\end{table}

The decay $D^0\to K^+K^-\pi^+\pi^-$ has been studied by 
E687~\cite{Frabetti:1995za},
E791~\cite{Aitala:1997gy}, and FOCUS~\cite{Link:2004wx}. The FOCUS study used 
$1,279\pm48$ events. They performed an unbinned maximum likelihood 
fit including 10 resonances. The amplitudes are summarized in
Table~\ref{tab:FOCUS_KKpipi}. The dominant contribution to the
decay rate, about 55\%, comes from decays to intermediate states
with an axial vector and a pseudo scalar. The second larges contribution,
about 30\%, comes from intermediate states with two vectors mesons.
The remaining contributions are from three body decays $D\to VPP$
and $D\to SPP$.

\begin{table*}[bt]
\caption{
Fit fractions and phases from the FOCUS~\cite{Link:2004wx} analysis
of the decay  $D^0\to K^-\pi^+\pi^-\pi^+$.
}
\label{tab:FOCUS_KKpipi}
\begin{center}
\begin{tabular}{lccc}
\hline\hline
Mode                        & Magnitude & Phase ($^\circ$) & Fraction (\%) \\
\hline
$K_1(1270)^+K^-,\ K_1\to\rho(770)^0K^+$  & $1$ (fixed) & $0$ (fixed) & $18\pm6\pm3$  \\
$K_1(1270)^+K^-,\ K_1\to K^*_0(1430)\pi^+$  & $0.27\pm0.08\pm0.06$ & $354\pm19\pm19$ & $2\pm1\pm0$  \\
$K_1(1270)^+K^-,\ K_1\to K*(892)^0\pi^+$  & $0.94\pm0.16\pm0.13$ & $12\pm12\pm15$ & $16\pm4\pm5$  \\
$K_1(1270)^+K^-$  & -- & -- & $33\pm6\pm4$  \\
$K_1(1400)^+K^-$  & $1.18\pm0.19\pm0.09$ & $259\pm11\pm13$ & $22\pm3\pm4$  \\
$K^*(892)^0{\bar K}^*(892)^0$  & $0.39\pm0.09\pm0.11$ & $28\pm13\pm10$ & $3\pm2\pm1$  \\
$\phi(1020)\rho(770)^0$  & $1.30\pm0.11\pm0.07$ & $49\pm11\pm12$ & $29\pm2\pm1$  \\
$\rho(770)^0K^+K^-$  & $0.33\pm0.12\pm0.16$ & $278\pm26\pm20$ & $2\pm2\pm2$  \\
$\phi(1020)\pi^+\pi^-$  & $0.30\pm0.06\pm0.06$ & $163\pm16\pm15$ & $1\pm1\pm0$  \\
$K^*(892)^0K^+\pi^-$  & $0.83\pm0.09\pm0.10$ & $234\pm10\pm11$ & $11\pm2\pm1$  \\
$f_0(980)\pi^+\pi^-$  & $0.91\pm0.13\pm0.05$ & $240\pm11\pm17$ & $15\pm3\pm2$  \\
\hline\hline
\end{tabular}
\end{center}
\end{table*}

\section{CONCLUSIONS}
\label{sect:conclusions}

Charm decays remain an exciting field for both theoretical and experimental investigations.
In fact, most discoveries in heavy flavor physics in the last five years involved charm 
quarks one way or another. These include $D^0{\overline D}^0$ mixing, 
new open-charm $D_{sJ}$ states, charmonium states $X, Y, Z$ states with ordinary and exotic 
quantum numbers, etc.

In this review, we touched only a part of a vast field of charm physics, the hadronic transitions of
charmed mesons. We did not review many other exciting developments in charm physics. For 
example, a set of hadronic resonant states with new and exciting properties has been discovered in
both open- and hidden-charm quark systems, many exciting results were obtained in theoretical 
(lattice) computations and experimental measurements of leptonic and semi-leptonic decays of 
charmed mesons, $D^0\overline{D}^0$-mixing was discovered and used to constrain New Physics
at the scales of several TeV~\cite{Golowich:2007ka}, etc. Also, experimental search for CP-violation 
in charm transitions remains one of the primary ways of probing New Physics in low-energy 
interactions~\cite{Grossman:2006jg}. Finally, we did not discuss inclusive charm decays, lifetimes 
of charmed states~\cite{Bianco:2003vb,Gabbiani:2004tp}, as well as charmed 
spectroscopy and decays of charmed baryons. 

Our knowledge of hadronic charm decays has improved significantly over the
last few years. The $B$-factory experiments, BABAR and Belle, has very 
large charm data samples that has allowed them to do very precise studies,
including the absolute hadronic branching fractions for but $D^0$ and 
$D_s^+$ mesons. In addition, the unique CLEO-c data samples allow
detailed studies of $D^0$, $D^+$, and $D_s^+$ decays. In this review
we have covered the status of the determination of the absolute
branching fractions first for $D^0$ and $D^+$ mesons. These measurements
are dominated by results from CLEO-c and BABAR and have statistical
uncertainties now below $\pm 1\%$ and systematic uncertainties of about 
$\pm 1.8\%$.
The determination
of the $D_s^+$ branching fractions is dominated by CLEO-c. The previously
commonly used normalization mode $D_s^+\to \phi\pi^+$ is not used by
CLEO-c any more as it is ambiguous at the level of precision now 
obtained by CLEO-c. CLEO-c instead quotes partial branching fractions 
for a range of different $K^+K^-$ mass ranges around the $\phi$ resonance.
These partial branching fractions do not try to disentangle the contributions
from the $\phi$ or other resonance contributing to the rate. The CLEO-c
measurement obtains a statistical precision of about 4.2\% and systematic
uncertainties of about 3\% in the $D_s^+\to K^+K^-\pi^+$ mode. This
result should improve when CLEO-c includes their full data sample. In
the future BES III should also be able to contribute to the determination
of the absolute hadronic $D$ branching fractions.
The larger samples have allowed more detailed studies of Cabibbo 
suppressed $D$ and $D_s$ decays. Decays with smaller branching fractions
have been explored as well as final states with $\pi^0$ and $\eta$
mesons that traditionally has been harder to reconstruct, but thanks
the excellent electro-magnetic calorimeters of the BABAR, Belle,
and CLEO-c are now accessible. Finally, a summary of Dalitz decays
of $D$ mesons is given. Many of the three-body final states have now
been analyzed for their resonant substructure, and also a few final states
with more than three particles in the final state
have been studied. These studies show that
most of the $D$ decays proceed via pseudo two-body decays. A few
inclusive measurements of $D$ and $D_s$ decays are also presented.

We are confident that charm quarks will bring us 
new and exciting discoveries, particularly
in the field of $CP$ violation. While the current experimental data sets are not large enough 
to probe Standard Model-level $CP$ violation in the charm sector, the precision of asymmetry 
measurements is not limited by systematics uncertainties. This means that new samples
of charm data from Belle, LHCb and Super-B factories will allow for even more precise 
measurements of $CP$-violating observables. In addition, multibody channels, which could be 
less prone  to systematic uncertainties, show great potential for searches for $CP$ violation.

\section*{Acknowledgments}
We would like to thank David Cinabro and Rob Harr for careful reading the manuscript and 
insightful comments. A.R.~was supported in part by the U.S.~National Science Foundation 
under Grant PHY-0757894 and CAREER Award PHY-0846388. A.R.~also thanks
the Alfred P.~Sloan foundation for their support.  
A.A.P.~was supported in part by the U.S.~National Science 
Foundation under CAREER Award PHY-0547794, and by the U.S.~Department 
of Energy under Contract DE-FG02-96ER41005. 

\bibliographystyle{apsrmp}
\bibliography{DHadRMP}


\end{document}